\documentstyle[11pt,epsfig]{report}

\def\htype#1{\vspace*{-0.6cm}\centerline{\huge\bf #1}\vspace{1cm}}

\def\slash#1{\mbox{/\llap #1}}

\def\be{\begin{equation}}
\def\ee{\end{equation}}
\def\ba{\begin{eqnarray}}
\def\ea{\end{eqnarray}}
\def\bann{\begin{eqnarray*}}
\def\eann{\end{eqnarray*}}
\def\mref#1{Eq.~(\ref{eq:#1})}
\def\nref#1{(\ref{eq:#1})}
\def\oref#1{\ref{eq:#1}}
\def\mrefb#1#2{Eqs.~(\ref{eq:#1}, \ref{eq:#2})}
\def\mlab#1{\label{eq:#1}}
\def\nn{\nonumber}
\def\Comment#1{}

\def\G{{\cal G}}
\def\S{\Sigma}
\def\F{{\cal F}}

\def\deltay{{\delta y}}
\def\dS{\delta\S}
\def\im{i}
\def\e{\mbox{e}}

\def\psib{\bar{\psi}}
\def\etab{\bar{\eta}}
\def\psie{\hat{\psi}}
\def\psibe{\hat{\bar{\psi}}}
\def\Ae{\hat{A}}
\def\Order{{\cal O}}
\def\Trace{\mbox{Tr}}
\def\Beta{B}

\def\l{\left}
\def\r{\right}
\def\logten{\log_{10}}
\def\mvec#1{{\bf #1}}
\def\tol{{\cal T}}
\def\norm#1{\l\|#1\r\|}
\def\parder#1#2{\frac{\partial #1}{\partial #2}}
\def\boxeq#1{\fbox{$\displaystyle#1$}}
\def\NS{N_\S}
\def\NF{N_\F}
\def\NG{N_\G}
\def\Ds#1{{#1+\S^2(#1)}}
\def\Proj{{\cal P}}
\def\compop{\overline{\psi}\psi}
\def\Tp{\tilde{T}}
\def\Self{\mbox{E}_f}
\def\ten#1{\mbox{e}{#1}}
\def\oten#1{1\ten{#1}}
\def\GeV{\mbox{GeV}}
\def\mvecF{{\mbox{\boldmath$\F$}}}
\def\bG{{\mbox{\boldmath$\G$}}}

\newcommand{\D}{\displaystyle}

\def\lijndikte{\thinlines}

\newsavebox{\ifbox}
\savebox{\ifbox}(0,0)[bl]{
\setlength{\unitlength}{1mm}
\lijndikte
\put(0,12) {\line(-2,-1){24.0}}
\put(0,12) {\line(2,-1) {24.0}}
\put(0,-12){\line(-2,1) {24.0}}
\put(0,-12){\line(2,1)  {24.0}} 
}

\newsavebox{\statement}
\savebox{\statement}(0,0){ \thicklines
\put(-12.5,-6){\line(1,0) {25.0}}
\put(12.5,-6) {\line(0,1) {12.0}}
\put(12.5,6)  {\line(-1,0){25.0}}
\put(-12.5,6) {\line(0,-1){12.0}}
}

\begin{document}

\pagenumbering{roman}

\baselineskip=20pt

\begin{titlepage}
\begin{center}

\vspace*{0mm}
\Huge{\bf Numerical Investigation\\[1mm]of\\[1mm]Fermion Mass Generation in QED}

\vspace{5cm}

\large
A thesis submitted for the degree of\\[1mm]
Doctor of Philosophy\\[1mm]
by

\LARGE{\bf Jacques Christophe Rodolphe Bloch}

\vspace{8cm}

\large 
University of Durham\\[1mm]
Department of Physics\\[1mm]
November 1995

\normalsize

\end{center}
\end{titlepage}

\begin{titlepage}

\htype{Abstract}

We investigate the dynamical generation of fermion mass in quantum
electrodynamics (QED). This non-perturbative study is performed using a
truncated set of Schwinger-Dyson equations for the fermion and the photon
propagator.

First, we study dynamical fermion mass generation in quenched QED with the
Curtis-Pennington vertex, which satisfies the Ward-Takahashi identity and
moreover ensures the multiplicative renormalizability of the fermion
propagator. We apply bifurcation analysis to determine the critical point
for a general covariant gauge.

In the second part of this work we investigate the dynamical generation of
fermion mass in full, unquenched QED. We develop a numerical method to
solve the system of three coupled non-linear equations for the dynamical
fermion mass, the fermion wavefunction renormalization and the photon
renormalization function. Much care is taken to ensure the high accuracy of
the solutions. Moreover, we discuss in detail the proper numerical
cancellation of the quadratic divergence in the vacuum polarization
integral and the requirement of using smooth approximations to the
solutions. To achieve this, we improve the numerical method by introducing
the Chebyshev expansion method. We apply this method to the bare vertex
approximation to unquenched QED to determine the critical coupling for a
variety of approximations. This culminates in the detailed, highly
accurate, solution of the Schwinger-Dyson equations for dynamical fermion
mass generation in QED including both, the photon renormalization function
and the fermion wavefunction renormalization in a consistent way, in the
bare vertex approximation and, for the first time, using improved
vertices. We introduce new improvements to the numerical method, to achieve
the accuracy necessary to avoid unphysical quadratic divergences in the
vacuum polarization with the Ball-Chiu vertex.

\end{titlepage}

\begin{titlepage}

\htype{Declaration}

I declare that no material in this thesis has previously been submitted for
a degree at this or any other university.

\vspace{3mm}
All the research in this work has been carried out in collaboration with
Dr.~M.R.~Pennington. 

\vspace{3mm}
Part of Chapter~\ref{QuenQED} has been published in\\ {\it Critical
coupling in strong QED with weak gauge dependence}\\ D. Atkinson,
J.C.R.~Bloch, V.P. Gusynin, M.R.~Pennington and M.~Reenders,
Phys.Lett. {\bf B329} (1994) 117.

\vspace{3mm}
The discussion of Chapter~\ref{Paper2} has been published in\\
{\it Can the photon quadratic divergence be cancelled numerically?}\\
J.C.R.~Bloch and M.R.~Pennington, Mod. Phys. Lett. {\bf
A10} (1995) 1225.

\vspace{20mm}
\copyright {\it The copyright of this thesis rests with the author. No
quotation from it should be published without his prior written consent and
information derived from it should be acknowledged.}

\end{titlepage}

\begin{titlepage}

\htype{Acknowledgements}

I would like to acknowledge Mike Pennington for his supervision during my
research work.

This thesis would not have been possible without the endless patience,
encouragements and love of Caroline, Manon and Rapha{\"e}l. Many thanks to
my parents for their unconditional support throughout my education and
especially during our stay in Durham. I am also very grateful to my
parents-in-law for being so helpful and hospitable, even after I
abducted their daughter to the high north of England.

I would not have started this research was it not for my
friend and artist Royden Rabinowitch whose enthusiasm gave me the
confidence I needed. I remember very well the very first of our many
absorbing discussions, in his studio in front of numerous cups of tea
(premonition?) with his wife Liz.

I wish to thank my non-perturbative friends Adnan, Ayse, and Kirsten for
keeping me efficiently from working, for baby-sitting for our little devils
and most importantly for their support in difficult moments.  Thanks also
to my first-year office mate Richard for the entertaining discussions in
our first steps in elementary particle theory, to Charlie for putting up
with me for three years in Room 303 and finally, to all my colleague
students and staff members of the physics department for the agreeable
working atmosphere they contributed to.

\vspace{4.2mm}
\begin{center}
\parbox{75mm}{
\epsfig{file=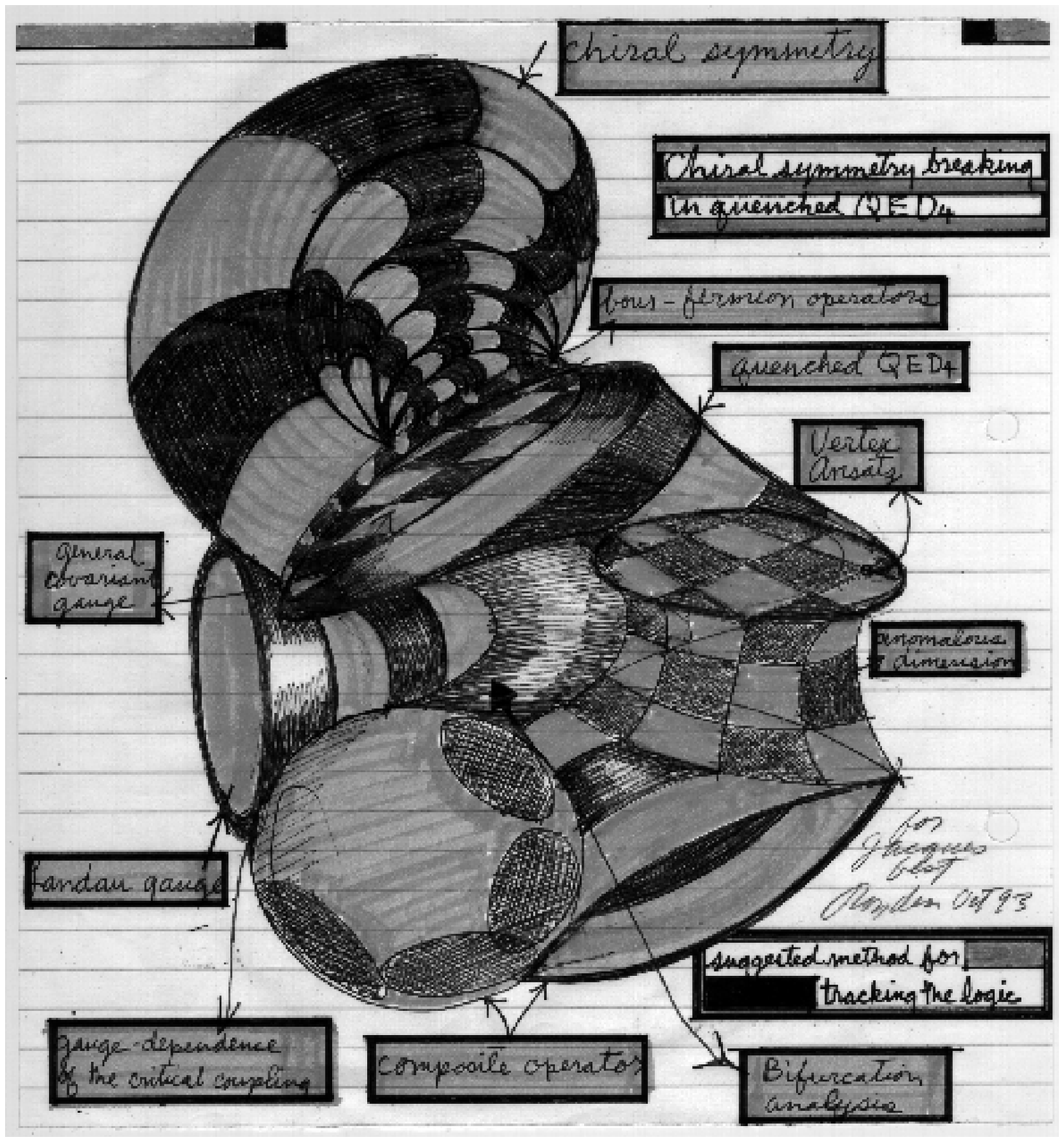,height=8cm}
\vspace{-1mm}
\parbox{73mm}{\tiny 'Chiral symmetry breaking in quenched (QED)$_4$ --
suggested method for tracking the logic', R.~Rabinowitch, Oct 93.}
}
\end{center}

\end{titlepage}

\addtocounter{page}{4}

\tableofcontents

\clearpage
\pagenumbering{arabic}

\chapter{Introduction}



The fundamental laws of nature are believed to be described by quantum
field theory and with the exception of gravity are embodied in the Standard
Model (SM) of strong, weak and electromagnetic interactions. Despite the
enormous success of the model when comparing its predictions with
experimental data, one can argue about its large number of parameters.  One
of the outstanding problems in particle physics is the {\it origin of
mass}. Is it just a parameter which has to be measured experimentally and then
inserted in the theoretical model describing the particles and their
interactions or is there an underlying mechanism through which these
particles acquire their mass ?

For the first possibility to be consistent with the quantum field theory of
the Standard Model things are not as easy as they may seem at first
sight. Merely inserting the experimentally determined fermion masses in the
Lagrangian of the SM is not allowed because such mass terms would break the
gauge invariance and hence ruin the renormalizability of the theory. For
fermions to be massive the concept of spontaneous symmetry breaking has to
be introduced. The original Lagrangian is $SU(2)_L\times U(1)_Y$ symmetric
but a new, fundamental scalar, {\it Higgs} field is introduced which
explicitly breaks this symmetry to $U(1)_{EM}$. The non-zero vacuum
expectation value of this field is directly responsible for the mass of the
W and Z intermediate bosons. To generate a mass for the fermions we have to
introduce additional Yukawa interaction terms between the fermions and the
Higgs boson in the Lagrangian. Then, the vacuum expectation value of the
Higgs field yields fermion masses which are proportional to the Yukawa
coupling.

Although the Standard Model remains renormalizable after the introduction
of the Higgs field, the idea is somehow unattractive because of new
quadratic mass divergences. Renormalizing these requires a very sharp fine
tuning to keep the fermion masses to the scales at which they are
experimentally measured. Without such fine tuning the quantum corrections
would raise the fermion masses to the scale of new physics which is
expected to be between $10^{15}\,\GeV$ and $10^{19}\,\GeV$.

An attractive alternative to the spontaneous symmetry breaking mechanism
engendered by the Higgs field is that of fermion mass generation through
dynamical symmetry breaking. There, the initially massless fermions acquire
their mass through a non-perturbative dynamical mechanism without the need
for any fundamental scalar field with a non-zero vacuum expectation value
to be introduced. We know from quantum field theory that the mass of a
particle receives loop corrections because of its interactions with the
gauge field. In perturbation theory each term in the perturbative expansion
of the corrected fermion mass is proportional to its bare mass. Hence, if
the theory is originally massless, it remains so at each order in
perturbation theory.  Of course this argument is only valid as long as the
perturbative series makes sense. One can imagine that when the coupling is
of order unity, an expansion in powers of the coupling constant does not
necessarily give us relevant information about the theory. Indeed, it has
been shown~\cite{Nam61} that provided the coupling is larger than some
critical value one can generate a non-zero fermion mass dynamically, even in
a theory without any bare mass in the Lagrangian.

Furthermore, not only the fermions but also the intermediate gauge bosons
can acquire mass by the dynamical breaking of chiral symmetry.
Unfortunately, in quantum chromodynamics (QCD), the only strong interaction
in the SM, the running coupling becomes strong at a scale which is far too
low to account for the measured W and Z masses. Therefore, the mechanism of
dynamical mass generation can only explain the experimentally measured
masses if a new interaction, with a higher scale, is introduced, as has
been proposed in the Technicolor~(TC)~\cite{Sus79} and Extended Technicolor
theories~(ETC)~\cite{Dim79}. One of the major problems encountered by these
theories is the excess of flavour changing neutral currents(FCNC). A
possible solution for this has been suggested by Holdom in the Walking
Technicolor theory~\cite{Hol85}. To make realistic predictions in these
theories, the non-perturbative phenomenon of dynamical fermion mass
generation in gauge field theories has to be well understood. The research
undertaken for this purpose can be divided into two main categories:
phenomenological studies where the basic concepts of dynamical fermion mass
generation are applied to realistic models of gauge groups, constructed to
reproduce the experimental results; and theoretical studies, concerned with
the fundamental aspects of the dynamical generation of fermion mass, which
are needed to provide the correct ideas and numbers for phenomenologists to
refine their calculations.

The research presented in this thesis belongs to these theoretical studies.
We will investigate the dynamical generation of fermion mass in strong
coupling quantum electrodynamics~(QED), the simplest gauge field theory in
nature.  There are several motivations for this study. In the realm of TC
and ETC theories it can serve as a toy model for more complicated gauge
theories. However, it is also important in its own right, to study the
consistency of QED as a quantum field theory, as discussed by
Landau~\cite{Lan55} as early as in 1955. Furthermore, it could be of
interest considering the possibility of a new phase transition of nature's
QED in strong electromagnetic fields, as might be suggested by some
unexplained narrow peaks in $\mbox{e}^+\mbox{e}^-$ coincidence spectra in
heavy ion collision experiments~\cite{Sch89,Cal89}.


Because of the intrinsically non-perturbative nature of dynamical fermion
mass generation, we need an appropriate framework to conduct our
investigation.  The two methods most frequently used to study
non-perturbative aspects of quantum field theory are the continuum method
using the Schwinger-Dyson equations and lattice gauge theory, where the
field theory is solved on a discretized lattice.  In these lattice studies
one cannot take the bare mass identically to zero. Therefore, one has to
compute results for various, finite values of this mass, in order to
extrapolate to the zero-mass situation. This extrapolation procedure can be
a source of difficulties in the correct interpretation of the lattice
results.

The Schwinger-Dyson (SD) equations are an infinite set of coupled integral
equations derived from the functional integral formalism, relating all the
Green's functions of the quantum field theory. If these equations could be
solved, all the Green's functions would be known and the S-matrix for all
physical processes could be calculated exactly. However, because there are
an infinite number of coupled equations, such solution is not possible and
one must truncate the system in some way. The most common procedure is to
expand the equations in powers of the coupling and to truncate the series
at a certain order. This method is just perturbation theory. However, if we
are to investigate the non-perturbative aspects of the theory, this clearly
will not suffice and we have to devise other ways of truncating the
infinite tower of equations. In the study of the dynamical generation of
fermion mass, one is primarily interested in the behaviour of the fermion
propagator. The fermion SD equation determines how the fermion propagator
is altered by the self-energy generated by the interactions. In this
equation the fermion propagator is related to the photon propagator and the
QED-vertex. The photon SD equation describes how the vacuum polarization
corrects the photon propagator and again relates the fermion and photon
propagator and the vertex. An infinity of other SD equations relate higher
order Green's functions. To study the fermion mass generation we will
decouple the first two equations from all the others by choosing some
suitable vertex Ansatz and investigate the possibility of these equations
having a non-trivial mass solution in a theory without bare mass.

Several studies of fermion mass generation in QED have been performed in
the rainbow approximation, where the full photon propagator and the full
vertex are replaced by their bare quantities. In this approximation, it has
been shown that QED does undergo a phase transition and the originally
massless fermions acquire a mass, when the value of the fixed coupling is
larger than a critical value, which is of order unity in the Landau
gauge~\cite{Fuk76}\nocite{Fomin83,Mir85}-\cite{Mir85b}. However the bare
vertex does not satisfy the Ward-Takahashi identity, which is a consequence
of gauge invariance. Therefore, in the first part of this work, we will
study the dynamical generation of fermion mass in quenched QED with the
Curtis-Pennington vertex Ansatz~\cite{CP90}, which satisfies the
Ward-Takahashi identity and moreover ensures the multiplicative
renormalizability of the fermion propagator. We will apply bifurcation
analysis to determine the critical point for a general covariant gauge.

In the second part of the study we will investigate the dynamical
generation of fermion mass in full, unquenched QED. All studies performed
so far in the Landau gauge have used the bare vertex approximation.
Furthermore, various additional approximations were introduced to simplify
the analytical and numerical
calculations~\cite{Kondo91}\nocite{Gusynin,Kondo89,Kondo92b,Oli90,Rak91,Atk92}%
-\cite{Kondo92}.  The most frequently encountered approximations are:
replacing the full photon propagator by its 1-loop perturbative result,
removing the angular dependence of the vacuum polarization and setting the
fermion wavefunction renormalization to one.  To avoid these approximations
we will develop a numerical method to solve the system of coupled,
non-linear integral equations for the fermion and the photon propagator,
paying special attention to achieve high degree of accuracy.  We will also
give a detailed discussion about the proper numerical cancellation of the
quadratic divergence in the vacuum polarization integral. We will apply
this method to the bare vertex approximation to unquenched QED to determine
the critical coupling for a variety of approximations to the system of
coupled integral equations, and will compare our results with those found
in the literature. We will give detailed, highly accurate results of
dynamical fermion mass generation in QED, including both the photon
renormalization function and the fermion wavefunction renormalization in a
consistent way. Finally, we will produce the first results of fermion mass
generation in unquenched QED using improved vertices and will discuss in
detail how to avoid unphysical quadratic divergences in the vacuum
polarization integral, with the Ball-Chiu vertex.

In Chapter~\ref{SDeq} we formulate the Schwinger-Dyson equations for the
fermion and photon propagator. We discuss how fermion mass can be generated
dynamically from these equations when the QED coupling is sufficiently
large. We derive three coupled non-linear integral equations for the
dynamical fermion mass $\S$, the fermion wavefunction renormalization $\F$
and the photon renormalization function $\G$, in the bare vertex
approximation to the full vertex and with the Curtis-Pennington vertex
Ansatz.

In Chapter~\ref{QuenQED} we study the dynamical fermion mass generation in
quenched QED, where the full photon propagator is replaced by the bare
one. We determine the critical point in the Curtis-Pennington approximation
using bifurcation analysis. We also derive the Miransky scaling law,
specific to quenched QED, for the bare vertex approximation and with the
Curtis-Pennington vertex.

In Chapter~\ref{Sec:UnqQED} we give a literature survey of the various
approximations introduced in the investigation of dynamical fermion mass
generation in unquenched QED. 

In Chapter~\ref{Numprog} we develop a numerical method to solve the system
of coupled non-linear integral equations, describing fermion mass
generation in QED, taking special care to achieve high accuracy and
convergence rate. This is done by discretizing the unknown functions and
solving a system of non-linear algebraic equations for the function values
at a finite number of points, using the natural iterative procedure. After
a first attempt to apply the method to the problem of fermion mass
generation in QED, simplified to the solution of a sole non-linear integral
equation for $\S$, the poor convergence of the procedure will be improved
by introducing Newton's iterative method.  We discuss how the numerical
method can only be satisfactory if a suitable choice of integration rule is
made. The scene being set, we apply the numerical method to the
$\S$-equation and show the main results.

In Chapter~\ref{Paper2} we apply this method to the system of coupled
integral equations for $\S$ and $\G$ (neglecting the corrections to the
wavefunction renormalization $\F$). We compare our results with those of
Kondo et al.~\cite{Kondo92} and discuss the improper cancellation of the
quadratic divergence in the vacuum polarization, which generates an
unphysical behaviour in the photon renormalization function $\G$. We
suggest that this could be remedied by introducing smooth approximations to
the functions $\S$, $\F$ and $\G$.

In Chapter~\ref{Cheby} we realize this by introducing Chebyshev expansions
for the unknown functions and modifying the numerical method of
Chapter~\ref{Numprog} accordingly.

In Chapter~\ref{Chap:Numres} we apply the Chebyshev expansion method to
various approximations to the system of coupled non-linear integral
equations for $\S$, $\F$ and $\G$ in the bare vertex approximation. In the
1-loop approximation to the vacuum polarization we first solve the
$\S$-equation with and without the LAK-approximation, then we solve the
coupled ($\S$, $\F$)-system. Consequently, we redo the calculation of
Chapter~\ref{Paper2} for the coupled ($\S$, $\G$)-system, finding that
indeed the unphysical behaviour of $\G$ disappears. Finally, we solve the
complete ($\S$, $\F$, $\G$)-system of integral equations.

In Chapter~\ref{Sec:Impvx} we perform the first calculations of fermion
mass generation in unquenched QED using improved vertices. We solve the
coupled ($\S$, $\F$, $\G$)-system with various vertex approximations.  The
specific structure of the Ball-Chiu vertex leads to accuracy problems to
cancel the quadratic photon divergence properly. These problems are dealt
with in detail, since this is crucial for further numerical studies.

Finally, in Chapter~\ref{Concl} we summarize our results and give some
suggestions for future studies.

\def\funcint{\int{\cal D}\psib\, {\cal D}\psi \,{\cal D}A \, }
\def\dd#1#2{\frac{\delta#1}{\delta#2}}

\chapter{Schwinger-Dyson equations}
\label{SDeq}

In this chapter we formulate the Schwinger-Dyson equations for the fermion
and photon propagator. These equations are just two of the infinite tower
of integral equations relating the Green's functions of the quantum field
theory~\cite{Itzyk}. We discuss how fermion mass generation can be studied
using these equations and derive three coupled, non-linear algebraic
integral equations for the dynamical mass $\S$, the fermion wavefunction
renormalization $\F$ and the photon renormalization function $\G$, first
using the bare vertex approximation, then with the Curtis-Pennington vertex
Ansatz.

\section{QED Lagrangian}

The Lagrangian for a free Dirac field with bare mass $m_0$ is:
\be
{\cal L}_0 = \im \psib\gamma^\mu\partial_\mu\psi -
m_0\psib\psi.
\mlab{freeL}
\ee

The fermion field $\psi$ will transform under a local U(1) gauge
transformation as:
\be
\psi \rightarrow \psi'(x) = \e^{- \im e \lambda(x)} \psi(x).
\mlab{gtf}
\ee

The Lagrangian of \mref{freeL} is not invariant under the transformation
\mref{gtf}. 

Local U(1) gauge invariance of the Lagrangian can be achieved by
introducing a vector field $A_\mu$, called the gauge field, which
transforms as:
\be
A_\mu \rightarrow A'_\mu = A_\mu + \partial_\mu \lambda
\mlab{Atf}
\ee

and replacing the ordinary derivative $\partial_\mu$ in the free
Lagrangian of \mref{freeL} by a covariant derivative $D_\mu$:
\be
D_\mu = \partial_\mu + \im e A_\mu.
\ee

The new Lagrangian is now given by:
\be
{\cal L} = \im \psib\gamma^\mu\partial_\mu\psi - e
\psib\gamma^\mu A_\mu \psi  - m_0\psib\psi.
\ee

To this Lagrangian one has to add a kinetic term for the gauge field
$A_\mu$, which has to be invariant under the transformation
\mref{Atf}. The full QED Lagrangian for a fermion field $\psi$ with 
charge $e$ in an electromagnetic field $A_\mu$ is given by:
\be
{\cal L} = \im \psib\gamma^\mu\partial_\mu\psi - e
\psib\gamma^\mu A_\mu \psi  - m_0\psib\psi -
\frac{1}{4}F_{\mu\nu} F^{\mu\nu}
\ee
where
\be
F_{\mu\nu} = \partial_\mu A_\nu - \partial_\nu A_\mu.
\ee

The quantum field theory defined by this Lagrangian can be derived by
applying the functional integral method \cite{Itzyk,Roberts,Ryder}
using the following generating functional:
\be
Z[\etab,\eta,J] = \frac{1}{N} \funcint \exp \left[\, \im \int
d^4x \, ({\cal L} + \psib\eta +
\etab\psi + J^\mu A_\mu)\right]
\mlab{Zgen}
\ee
where $\etab$, $\eta$ and $J$ are the source fields for the
fermion, antifermion and gauge boson, and the normalization factor $N$
is given by:
\[
N = \funcint \exp \left[\, \im \int d^4x \, {\cal L}\right].
\]

A peculiarity of gauge theories is that there are orbits of gauge
fields $A_\mu$ which are just gauge transforms of each other. Since
the Lagrangian is gauge invariant the functional integral over a
complete orbit of gauge fields will automatically be infinite.  To
avoid this we must pick out one representative on each orbit and
integrate over these representative gauge fields. To do this we impose
a gauge condition which is only satisfied by one field per orbit. In
QED this is done by introducing a gauge fixing term in the
Lagrangian. A common choice for this is the covariant gauge fixing
term $-1/2\xi(\partial_\mu A^\mu)^2$. The full QED Lagrangian then becomes:
\be
{\cal L}_{QED} = \im \psib\gamma^\mu\partial_\mu\psi - e
\psib\gamma^\mu A_\mu \psi  - m_0\psib\psi -
\frac{1}{4}F_{\mu\nu} F^{\mu\nu} - \frac{1}{2\xi} (\partial^\mu A_\mu)^2.
\mlab{LQED}
\ee

It can be shown \cite{Itzyk} that the generating functional of
connected Green's functions, $G[\etab,\eta,J^\mu]$ can be
defined from the generating functional, \mref{Zgen}, by:
\be
Z[\etab,\eta,J] = \exp(G[\etab,\eta,J]).
\mlab{eq4}
\ee

Let us now define effective fields $\psie$, $\psibe$, $\Ae$ by:
\be
\psie \equiv \frac{\delta G}{i\delta\etab}  \, ; \hspace{1cm}
\psibe \equiv -\frac{\delta G}{i\delta\eta}  \, ; \hspace{1cm}
\Ae_\nu \equiv \frac{\delta G}{i\delta J^\nu} \, .
\ee

Next we define an effective action $\Gamma[\psie,\psibe,\Ae]$, as the
Legendre transform of the generating functional of connected Green's
functions $G[\etab,\eta,J]$:
\be
i\Gamma[\psie,\psibe,\Ae] \equiv G[\etab,\eta,J] - i \int d^4y \, 
(\etab \psie + \psibe \eta + J^\mu \Ae_\mu).
\mlab{effac}
\ee
One can prove that the effective action $\Gamma[\psie,\psibe,\Ae]$ is the
generating functional of the one-particle-irreducible (1PI) Green's
functions (see pp.~289-294 of Ref.~\cite{Itzyk}) .

We now define the Green's functions which will be used in the investigation
of fermion mass generation. The connected 2-point fermion Green's function
or {\it fermion propagator} $iS(x,y)$ is:
\be
iS_{ab}(x,y) \equiv \left. -\frac{\delta^2 G}{\delta\eta_b(y)\delta\etab_a(x)} 
\right|_{\eta,\etab,J=0} \;.
\mlab{eq14}
\ee

We define the connected 2-point photon Green's function or {\it
photon propagator} $iD^{\mu\nu}(x,y)$ as:
\be
iD_{\mu\nu}(x,y) \equiv \left. -\frac{\delta^2 G}
{\delta J^\nu(y)\delta J^\mu(x)}
\right|_{\eta,\etab,J=0} \; .
\mlab{eq19}
\ee

The 1PI 3-points Green's function or {\it vertex} $e\Gamma(x,y;z)$ is
defined by:
\be
e\Gamma^\mu_{ab}(x,y;z) \equiv 
\left. -\frac{\delta^3\Gamma}{\delta\Ae_\mu(z)\delta\psie_b(y)\delta\psibe_a(x)}
\right|_{\eta,\etab,J=0} \;.
\mlab{eq15}
\ee 

The Schwinger-Dyson equations can be derived by applying the functional
integral formalism to the QED Lagrangian (see pp.~475-481 of
Ref.~\cite{Itzyk}).

\section{Fermion SD equation}

The Schwinger-Dyson equation for the fermion
propagator in coordinate space is given by:
\be
\left[S^{-1}\right](x,y) = 
\left(i\gamma^\mu\partial_\mu - m_0 \right)\delta^4(x-y)
- i e^2 \int d^4x_1 \, d^4x_2 \, 
\gamma^\mu S(x,x_1) \Gamma^\nu(x_1,y;x_2) D_{\nu\mu}(x_2,x) 
\, .
\mlab{eq40}
\ee
 
After Fourier transforming the various Green's functions, the
Schwinger-Dyson equation for the fermion propagator in momentum
space is given by:
\be
\boxeq{
\left[S^{-1}\right](p) = p_\mu \gamma^\mu - m_0 - \Self(p) }
\mlab{eq44}
\mlab{a1}
\ee

where the fermion self-energy $\Self(p)$ is defined by:
\be
\boxeq{ \Self(p) \equiv \frac{i e^2}{(2\pi)^4} \int d^4k \,
\gamma^\mu \, S(k) \, \Gamma^\nu(k,p) \, D_{\nu\mu}(k-p) \:. }
\mlab{eq44.1}
\ee

Here, we simplified the notation with:
\ba
S(p) &\equiv& S(p,-p) \mlab{simplif1} \nn\\
D^{\mu\nu}(q) &\equiv& D^{\mu\nu}(q,-q)  \\
\Gamma^\mu(k,p) &\equiv& \Gamma^\mu(k,p;k-p) \nn \, , 
\ea
where in the vertex, $\Gamma^\mu(k,p)$, $k$ is the incoming fermion momentum,
$p$ is the outgoing fermion momentum and the photon momentum is taken
outgoing.

\begin{figure}[htbp]
\begin{center}
\mbox{\epsfig{file=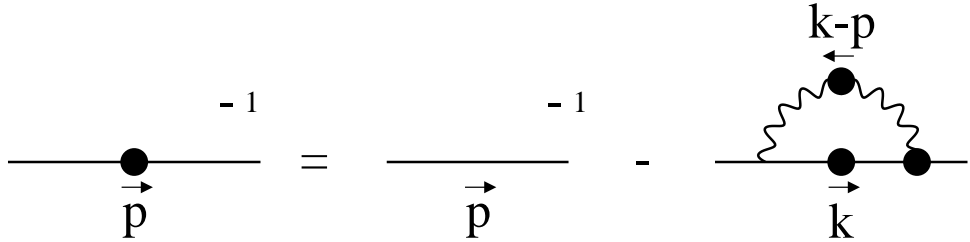,width=14cm}}
\end{center}
\vspace{-1cm}
\caption{Schwinger-Dyson equation for the fermion propagator.}
\label{fig:SDferm}
\end{figure}

\mref{eq44} is represented diagrammatically in Fig.~\ref{fig:SDferm}.

Because of the spinor structure of the fermion propagator $S(p)$, its
most general form is:
\be
S(p) = A(p^2) \, p_\mu \gamma^\mu + B(p^2) \;.
\ee

We rewrite this as:
\be
S(p) = \frac{\F(p^2)}{\slash{p} - \S(p^2)} 
= \frac{\F(p^2)}{p^2 - \S^2(p^2)}\l(\slash{p} + \S(p^2)\r) \, ,
\mlab{eq44.2}
\mlab{3.5}
\ee
where $\F(p^2)$ is called the {\it fermion wavefunction renormalization},
$\S(p^2)$ is the {\it dynamical fermion mass} and we introduced the notation
$\slash{p} \equiv p_\mu \gamma^\mu$.

From \mref{eq44} we see that the fermion propagator for a free fermion
field or {\it bare fermion propagator} is given by:
\be
S^0(p) = \frac{1}{\slash{p} -m_0} \, .
\ee

\section{Photon SD equation}

The Schwinger-Dyson equation for the photon propagator in coordinate space
is:
\ba
\left[D^{-1}\right]^{\rho\lambda}\!(x,y)  &=&
\left[ g^{\rho\lambda}\partial^2 + \left(\frac{1}{\xi}-1\right) 
\partial^\rho\partial^\lambda \right]\delta^4(x-y) \nn \\
&& \hspace{-2cm} + iN_fe^2 \int d^4x_1 \, d^4x_2 \, 
\Trace \left[ \gamma^\rho \, S(x,x_1) \,
\Gamma^\lambda(x_1,x_2;y) \, S(x_2,x) \right] \;.
\mlab{eq21}
\ea

To derive this equation in momentum space we Fourier transform the various
Green's functions. The Schwinger-Dyson equation for the photon propagator
in momentum space is:
\be
\boxeq{
\left[D^{-1}\right]^{\mu\nu}\!(q) =
- q^2 \left [g^{\mu\nu}  + \left(\frac{1}{\xi}-1\right)\frac{q^\mu q^\nu}{q^2}
\right] + \Pi^{\mu\nu}(q) }
\mlab{eq28}
\mlab{a2}
\ee

where the vacuum polarization tensor $\Pi^{\mu\nu}(q)$ is defined by:
\be
\boxeq{ \Pi^{\mu\nu}(q) \equiv
\frac{iN_fe^2}{(2\pi)^4} \int
d^4k\, \Trace \left[ \gamma^\mu \, S(k) \,
\Gamma^\nu(k,k-q) \, S(k-q) \right] \:. } 
\mlab{eq28.1}
\mlab{a2.1}
\mlab{191}
\ee

\begin{figure}[htbp]
\begin{center}
\mbox{\epsfig{file=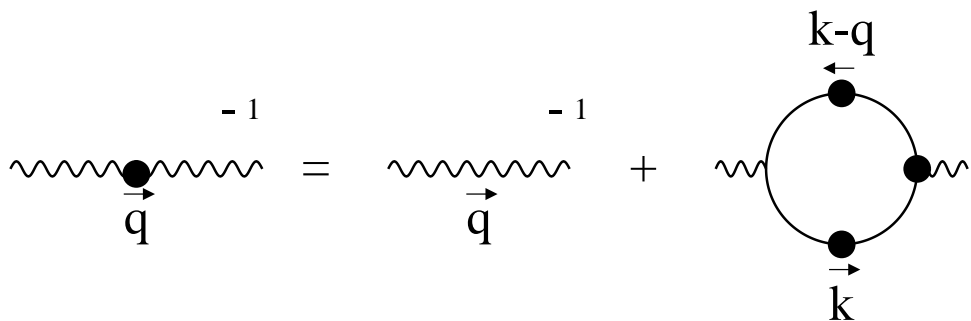,width=14cm}}
\end{center}
\vspace{-1cm}
\caption{Schwinger-Dyson equation for the photon propagator.}
\label{fig:SDphot}
\end{figure}

This equation is represented diagrammatically in Fig.~\ref{fig:SDphot}.

The number of fermion flavours $N_f$ in the vacuum polarization integral,
\mref{eq28.1}, accounts for the number of distinct flavour loops which can
occur in the photon propagator. We assume here that all fermion flavours
couple with the same strength $e$ to the electromagnetic field.  Because of
fermion flavour conservation, there is no factor of $N_f$ multiplying the
fermion self-energy integral, \mref{eq44.1}.

To study further the structure of the photon propagator, we
will use the following Ward-Takahashi identity, which tells us that
the vacuum polarization is transverse to the photon momentum:
\be
q_\mu \Pi^{\mu\nu}(q) = 0 \, .
\ee

Therefore the vacuum polarization tensor can be written as:
\be
\Pi^{\mu\nu}(q) = -q^2\left[g^{\mu\nu} - \frac{q^\mu
q^\nu}{q^2}\right] \Pi(q^2) \, .
\mlab{eq29}
\ee

Substituting \mref{eq29} in \mref{eq28} gives:
\be
\left[D^{-1}\right]^{\mu\nu}\!(q) =
- q^2 \left [ \left( g^{\mu\nu} - \frac{q^\mu q^\nu}{q^2} \right) 
\left(1+\Pi(q^2)\right)
  + \frac{1}{\xi}\frac{q^\mu q^\nu}{q^2} \right] \, .
\mlab{eq30}
\ee

To find the photon propagator we must invert the previous expression.
The definition of inverse in momentum space (which can be deduced by
Fourier transforming the definition for inverse in coordinate space)
is:
\be
D_{\mu\lambda}(q) \left[D^{-1}\right]^{\lambda\nu}\!(q) = \delta_\mu^\nu \, .
\mlab{eq31}
\ee

Substituting the most general tensor form $D_{\mu\nu}(q) = A g_{\mu\nu}
+ B q_\mu q_\nu/q^2$ in \mref{eq31} and using \mref{eq30} we find: 
\be
D_{\mu\nu}(q) = - \frac{1}{q^2} 
\left[ \G(q^2)\left(g_{\mu\nu} - \frac{q_\mu q_\nu}{q^2}\right)
+ \xi \frac{q_\mu q_\nu}{q^2} \right] \, ,
\mlab{photprop}
\mlab{3.6}
\ee
where we defined the {\it photon renormalization function} $\G(q^2)$ as:
\be
\G(q^2) \equiv \frac{1}{1+\Pi(q^2)} \, .
\mlab{photfunc}
\ee

From \mref{photprop} one finds that the photon propagator 
in a pure gauge theory or {\it bare photon propagator} is given by:
\be
D_{\mu\nu}^0(q) = - \frac{1}{q^2} 
\left[ \left(g_{\mu\nu} - \frac{q_\mu q_\nu}{q^2}\right)
+ \xi \frac{q_\mu q_\nu}{q^2} \right] \, .
\ee

\section{Fermion mass generation}

From the Schwinger-Dyson equations we now derive the algebraic integral
equations necessary for the investigation of dynamical fermion mass
generation in QED.

After inserting the fermion propagator, \mref{3.5}, in the fermion SD equation,
\mref{a1} we can write:
\be
\frac{\slash{p} - \S(p^2)}{\F(p^2)} = \slash{p} - m_0 - \Self(p) \, .
\mlab{3.7}
\ee   

This integral equation contains the two unknown propagator functions,
$\F(p^2)$ and $\S(p^2)$, and the unknown vertex $\Gamma^\nu(k,p)$. From the
spinor equation \mref{3.7} one can derive two algebraic integral equations.
Taking the trace of \mref{3.7} and dividing the equation by $(-4)$ gives:
\be
\frac{\S(p^2)}{\F(p^2)}
= m_0 + \frac{1}{4}\Trace{\Big[\Self(p)\Big]} .
\mlab{156}
\ee

A second, independent equation is derived by multiplying \mref{3.7}
with $\slash{p}$, taking the trace and dividing by $4p^2$:
\be
\frac{1}{\F(p^2)} = 1 - \frac{1}{4p^2}
\Trace{\Big[\slash{p}\:\Self(p)\Big]} \;,
\mlab{157}
\ee

where we recall the fermion self-energy, \mref{eq44.1}:
\be
\Self(p) \equiv \frac{i e^2}{(2\pi)^4} \int d^4k \,
\gamma^\mu \, S(k) \, \Gamma^\nu(k,p) \, D_{\nu\mu}(k-p) \;.
\mlab{a1.1}
\ee

We will now derive the equation for the photon renormalization function
$\G(q^2)$. We have seen in \mref{eq29} that the photon Ward-Takahashi
identity requires the vacuum polarization tensor to have the following
form:
\be
\Pi^{\mu\nu}(q) = -q^2\left[g^{\mu\nu} - \frac{q^\mu
q^\nu}{q^2}\right] \Pi(q^2) \, .
\mlab{192}
\ee

It is important to note that unless the vertex $\Gamma^{\mu}(k,p)$
satisfies the fermion Ward-Takahashi identity and the regularization of the
loop integrals is translation invariant, the vacuum polarization integral,
\mref{191}, will {\bf not} have the correct Lorentz structure of \mref{192}
with the coefficients of $g^{\mu\nu}$ and $q^{\mu}q^{\nu}$ being related to
a single function $\Pi(q^2)$. When these conditions are satisfied, we can
extract the vacuum polarization function $\Pi(q^2)$ by contracting \mref{192}
with the operator $\Proj_{\mu\nu} = g_{\mu\nu}- n {q_\mu q_\nu}/{q^2}$ (with
any value of $n$)~:
\be
\Proj_{\mu\nu}\Pi^{\mu\nu}(q) = - 3 q^2 \Pi(q^2) .
\mlab{193}
\ee

The integral equation for $\Pi(q^2)$ can then be derived by applying the
operator $\Proj_{\mu\nu}$ to the vacuum polarization integral, \mref{191}, and
equating this to \mref{193}. This gives:
\be
\Pi(q^2) = - \frac{iN_f e^2\Proj_{\mu\nu}}{3(2\pi)^4 q^2} \int d^4k\, 
\Trace \Big[ \gamma^\mu \, S(k) \,
\Gamma^\nu(k,p) \, S(p) \Big] \, ,
\mlab{194}
\ee
where we defined the fermion momentum $p\equiv k-q$.

From the photon SD equation and the WT-identity we know that the photon
renormalization function $\G$ can be written as (\mref{photfunc}):
\be
\G(q^2) \equiv \frac{1}{1+\Pi(q^2)} \, .
\mlab{195}
\ee

Combining \mrefb{194}{195} yields the integral equation for $\G$:
\be
\frac{1}{\G(q^2)} = 1 
- \frac{iN_f e^2\Proj_{\mu\nu}}{3(2\pi)^4 q^2} \int d^4k\, 
\Trace \Big[ \gamma^\mu \, S(k) \,
\Gamma^\nu(k,p) \, S(p) \Big] \;.
\mlab{196}
\ee

We want to investigate the dynamical generation of fermion mass with
the use of the coupled integral equations, Eqs.~(\oref{156}, \oref{157},
\oref{196}), for $\S$, $\F$ and $\G$ when $m_0=0$.
However, these three integral equations still contain the full QED vertex,
which is itself coupled to higher order Green's functions through other SD
equations. To make the problem tractable, we want to decouple the three
equations for $\S$, $\F$ and $\G$ from the rest of the infinite tower of SD
equations.  This can be achieved by introducing an Ansatz for the QED
vertex. The choice of vertex Ansatz can be dictated by reasons of
simplicity or better by physical motivations. In the past, many additional
approximations have been introduced in the ($\S$, $\F$, $\G$)-system of
equations to simplify the search for its solution. The key equation for the
study of dynamical fermion mass generation is the $\S$-equation also called
gap-equation, \mref{156}, as this is the one which generates the purely
non-perturbative solution for the fermion mass when the coupling constant
is sufficiently large. It is easy to verify that the trivial solution,
$\S\equiv 0$, is always a solution of \mref{156}, when the bare mass is
zero. This is the solution which corresponds with perturbation
theory. However, it has been demonstrated that a non-trivial solution
exists in the quenched approximation to QED ($N_f=0$), when the coupling
constant is sufficiently large~\cite{Fuk76}. When the coupling exceeds a
certain critical value, the non-zero mass solution bifurcates away from the
trivial one. Above this critical point, the generated fermion mass will
increase further with increasing values of the coupling. In this work we
will investigate the dynamical generation of fermion mass and determine the
value of the critical coupling for quenched QED with the Curtis-Pennington
vertex and in unquenched QED in a variety of approximations. For this
purpose we will now derive the three coupled integral equations for the
bare vertex approximation and with the Curtis-Pennington vertex.

\section{The bare vertex approximation}
\label{Sec:Coupeq}

\subsection{The fermion equations}

In the bare vertex approximation the fermion self-energy, \mref{a1.1},
becomes:
\be
\Self(p) = 
\frac{i e^2}{(2\pi)^4} \int d^4k \,
\gamma^\mu \, S(k) \, \gamma^\nu \, D_{\nu\mu}(k-p) \;.
\mlab{185}
\ee

In this approximation Eqs.~(\oref{156}, \oref{157}) now are:
\ba
\frac{\S(p^2)}{\F(p^2)}
&=& m_0 + \frac{i e^2}{4(2\pi)^4} \int d^4k \,
\Trace{\left[\gamma^\mu \, S(k) \, \gamma^\nu \right]} \,
D_{\nu\mu}(k-p) \mlab{3.10} \\
\frac{1}{\F(p^2)}
&=& 1 - \frac{i e^2}{4p^2(2\pi)^4} \int d^4k \,
\Trace{\left[\slash{p} \: \gamma^\mu \, S(k) \, \gamma^\nu \right]} \,
D_{\nu\mu}(k-p) \, .
\mlab{3.11}
\ea

Substituting the fermion propagator, \mref{3.5}, in \mrefb{3.10}{3.11}
yields:
\ba
\frac{\S(p^2)}{\F(p^2)}
&=& m_0 + \frac{i e^2}{4(2\pi)^4} \int d^4k \, 
\frac{\F(k^2)}{k^2 - \S^2(k^2)} \,
\Trace{\left[\gamma^\mu \left(\slash{k} + \S(k^2)\right) 
\gamma^\nu \right]} \, D_{\nu\mu}(k-p) \mlab{3.12} \\
\frac{1}{\F(p^2)}
&=& 1 - \frac{i e^2}{4p^2(2\pi)^4} \int d^4k \, 
\frac{\F(k^2)}{k^2 - \S^2(k^2)} \,
\Trace{\left[\slash{p} \: \gamma^\mu \left(\slash{k} + \S(k^2)\right)
\gamma^\nu \right]}  \, D_{\nu\mu}(k-p) .
\mlab{3.13}
\ea
where the photon momentum is given by $q=k-p$.

To evaluate the traces in \mrefb{3.12}{3.13} we will need to compute
traces of products of gamma matrices. The Dirac gamma matrices obey the
anticommutation relation:
\be
\{\gamma^\mu, \gamma^\nu\} = 2 g^{\mu\nu} \, .
\mlab{anticom}
\ee

From \mref{anticom} it is easy to prove that in 4 dimensions:
\ba
\setlength{\jot}{3mm}
\parbox{14cm}{\vspace{-3ex}\bann
&&\Trace{[I]} = 4 \nn\\
&&\Trace{[\gamma^\mu\gamma^\nu]} = 4 \, g^{\mu\nu} \nn\\
&&\Trace{[\slash{k}\,\slash{p}]} = 4 \, k.p \\
&&\Trace{[\slash{k}_1\slash{k}_2\slash{k}_3\slash{k}_4]} =
4 \, [(k_1.k_2)(k_3.k_4) - (k_1.k_3)(k_2.k_4) + (k_1.k_4)(k_2.k_3)] \nn\\ 
&& \Trace{[\slash{k}_1,\ldots,\slash{k}_n]} = 0 \quad , 
\qquad\mbox{if $n$ is odd} .\nn
\eann}
\mlab{traces}
\ea

Applying those rules to \mrefb{3.12}{3.13} and substituting the photon
propagator, \mref{3.6}, gives us:
\ba
\frac{\S(p^2)}{\F(p^2)}
&=& m_0 + \frac{i e^2}{(2\pi)^4} \int d^4k \, 
\frac{\F(k^2)\S(k^2)}{k^2 - \S^2(k^2)} \, g^{\mu\nu} \left\{ - \frac{1}{q^2}  
\left[ \G(q^2) \left(g_{\mu\nu} - \frac{q_\mu q_\nu}{q^2}\right)
+ \xi \frac{q_\mu q_\nu}{q^2} \right] \right\} \hspace{5mm} \\
\frac{1}{\F(p^2)}
&=& 1 - \frac{i e^2}{p^2(2\pi)^4} \int d^4k \, 
\frac{\F(k^2)}{k^2 - \S^2(k^2)} \,
\l(p^\mu k^\nu + k^\mu p^\nu - k.p \, g^{\mu\nu}\r) \\ 
&& \hspace{5cm} \times \left\{ - \frac{1}{q^2} 
\left[ \G(q^2) \left(g_{\mu\nu} - \frac{q_\mu q_\nu}{q^2}\right)
+ \xi \frac{q_\mu q_\nu}{q^2} \right] \right\} \, . \nn
\ea

Executing the Lorentz-contractions with the photon propagator and
substituting $q=k-p$ yields:
\ba
\frac{\S(p^2)}{\F(p^2)}
&=& m_0 - \frac{i e^2}{(2\pi)^4} \int d^4k \, 
\frac{\F(k^2)\S(k^2)}{q^2(k^2 - \S^2(k^2))} \, \left\{ 3\G(q^2) + \xi \right\}
\mlab{3.19} \\
\frac{1}{\F(p^2)}
&=& 1 + \frac{i e^2}{p^2(2\pi)^4} \int d^4k \, 
\frac{\F(k^2)}{q^2(k^2 - \S^2(k^2))} \mlab{3.20.1} \\
&& \times 
\l\{\G(q^2) \left[ 2\l(\frac{k^2p^2 - (k.p)^2}{q^2}\r) - 3 k.p\right] 
+ \xi \left[\frac{(k^2+p^2)\,k.p - 2 k^2p^2}{q^2}\right]\r\}. \nn
\ea

In order to enable us to compute the 4-dimensional integral we will
now perform a change of coordinates, called {\it Wick rotation}. The
transformation consists of $k_0 \rightarrow i k_0$ and $k_j
\rightarrow k_j$, $j=1,\ldots,3$. By doing so, the phase space is transformed
from a Minkowski space to a Euclidean space as the original metric, which
was $k^2 = k_0^2 -k_1^2 -k_2^2 - k_3^2$, has been transformed to $-k^2 = -
(k_0^2 +k_1^2 + k_2^2 + k_3^2)$. The Wick rotation in fact consists of
changing from real time to imaginary time and then rotating back the
integration interval over $90^\circ$ to integrate over the real time
axis. One can prove that in most cases the value of the integral remains
unchanged after a Wick rotation. In Minkowski space the mass of a particle
is defined as the pole of its propagator. From
\mref{3.5} we see that this pole occurs at the timelike momentum $m^2$
which solves the equation $m^2=\S^2(m^2)$. After the Wick rotation the mass
of the fermion, still defined as pole of the propagator, will be given by
$m^2=-p_E^2=\S^2(-p_E^2)$, which will be satisfied by some $p_E^2<0$. When
solving the integral equation in Euclidean space, one only finds solutions
for $p_E^2\ge 0$. To determine the mass of the particle one has to
analytically continue the mass function $\S(p_E^2)$ to negative values of
$p_E^2$. In our study we will refrain from doing so and will only consider
$\S(p_E^2)$ for positive values of Euclidean momentum. In the further
discussion we will omit the subscript {\small\it E} for Euclidean space to
simplify the notation.

After the Wick rotation to Euclidean space , \mrefb{3.19}{3.20.1} are given
by:
\ba
\frac{\S(p^2)}{\F(p^2)}
&=& m_0 + \frac{e^2}{(2\pi)^4} \int d^4k \, 
\frac{\F(k^2)\S(k^2)}{q^2(k^2 + \S^2(k^2))} \, \l\{ 3\G(q^2) + \xi \r\} 
\mlab{3.21} \\
\frac{1}{\F(p^2)}
&=& 1 - \frac{e^2}{p^2(2\pi)^4} \int d^4k \, 
\frac{\F(k^2)}{q^2(k^2 + \S^2(k^2))} \mlab{3.22} \\
&& \times 
\l\{\G(q^2) \left[ 2\l(\frac{k^2p^2 - (k.p)^2}{q^2}\r) - 3 k.p\right] 
+ \xi \left[\frac{(k^2+p^2)k.p - 2 k^2p^2}{q^2}\right]\r\}. \nn
\ea

Once in Euclidean space we can now change to spherical coordinates:
\be
\left\{
\begin{array}{r@{\quad}c@{\quad}l}
k_0 &=& k\cos\theta \\
k_1 &=& k\sin\theta\cos\phi \\
k_2 &=& k\sin\theta\sin\phi\cos\psi \\
k_3 &=& k\sin\theta\sin\phi\sin\psi \, ,
\end{array}
\right.
\ee
where $k = (k_0^2+k_1^2+k_2^2+k_3^2)^{1/2}$ and $\theta$ is taken to be
the angle between the incoming fermion momentum $p$ and the fermion
loop momentum $k$. The volume element $d^4k$ now becomes
$k^3\sin^2\theta\sin\phi \, dk \, d\theta \, d\phi \, d\psi$. The
integration ranges of the new variables are: $k \in [0,\infty]$, 
$\theta, \phi \in [0,\pi]$ and $\psi \in [0,2\pi]$.

The angular integrals over the angles $\phi$ and $\psi$ can always be
separated yielding:
\be
\int_0^\pi d\phi \sin\phi \int_0^{2\pi} d\psi = 4\pi \, . 
\ee

If we now define the coupling constant $\alpha\equiv e^2/4\pi$ and
introduce the notation $x=p^2$, $y=k^2$ and $z=q^2$, then
\mrefb{3.21}{3.22} in spherical coordinates are given by:
\ba
\frac{\S(x)}{\F(x)}
&=& m_0 + \frac{\alpha}{2\pi^2} \int dy \,
\frac{y\F(y)\S(y)}{y + \S^2(y)} \,  
\int d\theta \, \frac{\sin^2\theta}{z} \Big\{ 3\G(z) + \xi \Big\} 
\mlab{3.25} \\
\frac{1}{\F(x)}
&=& 1 - \frac{\alpha}{2\pi^2 x} \int dy \, 
\frac{y\F(y)}{y + \S^2(y)} \mlab{3.26} \\
&& \hspace{-0.5cm} \times  \int d\theta \, \frac{\sin^2\theta}{z}
\l\{\G(z) \left[ \frac{2xy\sin^2\theta}{z} - 3 \sqrt{yx}\cos\theta\right] 
+ \xi \left[\frac{(y+x)\sqrt{yx}\cos\theta - 2 yx}{z}\right]\r\}. \nn
\ea

Here, the angular integrals of the $\xi$-part can be computed analytically,
as shown in Appendix~\ref{App:angint}. Substituting Eqs.~(\oref{A1},
\oref{A10},
\oref {A11}) in \mrefb{3.25}{3.26} yields:
\ba
\fbox{\parbox{14cm}{\bann
\frac{\S(x)}{\F(x)}
&=& m_0 + \frac{3\alpha}{2\pi^2} \int dy \,
\frac{y\F(y)\S(y)}{y + \S^2(y)} \,  
\int d\theta \, \sin^2\theta \, \frac{\G(z)}{z} \hspace{4cm} \\ 
&& + \frac{\alpha\xi }{4\pi} \int dy \,
\frac{\F(y)\S(y)}{y + \S^2(y)} \,  
\l[\frac{y}{x}\theta(x-y)+\theta(y-x)\r] 
\eann}}
\mlab{189} \\
\fbox{\parbox{14cm}{\bann
\frac{1}{\F(x)}
&=& 1 - \frac{\alpha}{2\pi^2 x} \int dy \, 
\frac{y\F(y)}{y + \S^2(y)}
\int d\theta \, \sin^2\theta \, \G(z)
\left[ \frac{2xy\sin^2\theta}{z^2} - \frac{3 \sqrt{yx}\cos\theta}{z}\right]
\\
&&  + \frac{\alpha\xi}{4\pi} \int dy \, 
\frac{\F(y)}{y + \S^2(y)}
\left[\frac{y^2}{x^2}\theta(x-y) + \theta(y-x) \right] \;.
\eann}}
\mlab{190} 
\ea

\subsection{The photon equation}
\label{Sec:photeq} 

We introduce the bare vertex approximation in \mref{196}:
\be
\frac{1}{\G(q^2)} = 1 - 
\frac{iN_f e^2\Proj_{\mu\nu}}{3(2\pi)^4 q^2} \int d^4k\, 
\Trace \Big[ \gamma^\mu \, S(k) \,
\gamma^\nu \, S(p) \Big] \, .
\mlab{197}
\ee

Substituting the fermion propagator, \mref{3.5}, in \mref{197}, gives:
\be
\frac{1}{\G(q^2)} = 1 - 
\frac{iN_f e^2}{3(2\pi)^4 q^2} \int d^4k\, 
\frac{\F(k^2)\F(p^2)}{\l(k^2 - \S^2(k^2)\r)\l(p^2 - \S^2(p^2)\r)} 
\, \Proj_{\mu\nu}T^{\mu\nu}
\mlab{198}
\ee
where
\be
T^{\mu\nu} \equiv \Trace \Big[ \gamma^\mu \, (\slash{k} + \S(k^2)) \,
\gamma^\nu \,  (\slash{p} + \S(p^2)) \Big] \:.
\mlab{198.1}
\ee

We now compute the trace, \mref{198.1}, using \mref{traces}. This gives:
\be
T^{\mu\nu} 
= 4 \l[ k^\mu p^\nu + p^\mu k^\nu - \l(k.p-\S(k^2) \S(p^2)\r) g^{\mu\nu} \,
\r] \;.
\mlab{1100}
\ee

To simplify \mref{198} we first work out the Lorentz contraction of
$T^{\mu\nu}$ with $\Proj_{\mu\nu} = g_{\mu\nu}- n {q_\mu q_\nu}/{q^2}$, and
substitute $p=k-q$:
\be
\Proj_{\mu\nu} T^{\mu\nu} = 
4 \l[ (n-2)k^2 - \frac{2n(k.q)^2}{q^2} + (n+2)k.q - (n-4)\S(k^2) \S(p^2)
\r] \;.
\mlab{1102}
\ee

Substituting \mref{1102} in \mref{198} gives:
\ba
\frac{1}{\G(q^2)} &=& 1 - \frac{4iN_f e^2}{3(2\pi)^4 q^2} \int d^4k\, 
\frac{\F(k^2)\F(p^2)}{\l(k^2 - \S^2(k^2)\r)\l(p^2 - \S^2(p^2)\r)}\mlab{199} \\
&& \times 
\l[ (n-2)k^2 - \frac{2n(k.q)^2}{q^2} + (n+2)k.q - (n-4)\S(k^2) \S(p^2) \r] \, .
\nn
\ea

As for the fermion equation we perform a Wick rotation to Euclidean
space. We have:
\ba
\frac{1}{\G(q^2)} &=& 1 + \frac{4N_f e^2}{3(2\pi)^4 q^2} \int d^4k\, 
\frac{\F(k^2)\F(p^2)}{\l(k^2 + \S^2(k^2)\r)\l(p^2 + \S^2(p^2)\r)}\mlab{1103} \\
&& \times 
\l[ (n-2)k^2 - \frac{2n(k.q)^2}{q^2} + (n+2)k.q + (n-4)\S(k^2) \S(p^2) \r] \, .
\nn
\ea

Changing to spherical coordinates, substituting $\alpha=e^2/4\pi$ and
defining $x\equiv q^2$, $y\equiv k^2$ and $z\equiv p^2$ we find:
\ba
\frac{1}{\G(x)} &=& 1 + \frac{2N_f \alpha}{3\pi^2 x} \int dy \, 
\frac{y\F(y)}{y + \S^2(y)}
\int d\theta \, \sin^2\theta \, \frac{\F(z)}{z + \S^2(z)}\mlab{1105} \\
&& \times 
\l[ (n-2)y - 2ny\cos^2\theta + (n+2)\sqrt{yx}\cos\theta
+ (n-4)\S(y) \S(z) \r] \, .
\nn
\ea
 
In general, if we regularize the theory using an ultraviolet cutoff, the
vacuum polarization integral in \mref{1105} contains a quadratic divergence
which has to be removed, since such a photon mass term is not allowed in
more than 2 dimensions. One can show that the $q_\mu q_\nu/q^2$ term of the
vacuum polarization tensor cannot receive any quadratically divergent
contribution. Consequently, if we choose the operator $\Proj_{\mu\nu}$ of
\mref{194} with $n=4$, the resulting integral will be free of quadratic
divergences because the contraction $\Proj_{\mu\nu} g^{\mu\nu}$
vanishes. Setting $n=4$ in the photon equation \mref{1105} yields:
\label{Sec:n=4} 
\be
\fbox{\parbox{14.3cm}{
\[
\frac{1}{\G(x)} = 1 + \frac{4 N_f \alpha}{3\pi^2 x} \int dy 
\frac{y\F(y)}{y+\S^2(y)} \int d\theta \, \sin^2\theta \, 
\frac{\F(z)}{z+\S^2(z)}
\l[y(1-4\cos^2\theta) + 3\sqrt{yx}\cos\theta\r] \:.
\]}}
\mlab{104}
\ee

\section{Improving the vertex Ansatz}

In the previous section we have derived the integral equations for the
study of dynamical fermion mass generation with the bare vertex
approximation. This vertex Ansatz has the advantage of being very simple
and therefore it makes the manipulation of the Schwinger-Dyson equations
easier.  However, this approximation does not satisfy the Ward-Takahashi
identity relating the QED vertex with the fermion propagator, which is a
consequence of the gauge invariance of the theory. Therefore, the bare
vertex approximation does not ensure that the calculated physical
quantities are gauge invariant, as they should be.

In this section we will introduce the Ball-Chiu vertex~\cite{BC} which is
the exact longitudinal part of the full QED vertex, uniquely determined by
the Ward-Takahashi identity relating the vertex with the fermion
propagator. However, the transverse part of the vertex is still
arbitrary. We then consider the Curtis-Pennington vertex~\cite{CP90} in
which the transverse part of the vertex is constructed by requiring the
multiplicative renormalizability of the fermion propagator and the
reproduction of the perturbative results in the weak coupling limit.

\subsection{Ball-Chiu Vertex}

The Ward-Takahashi identity relating the QED vertex and the fermion
propagator is:
\be
(k-p)_\mu \, \Gamma^\mu(k,p) = S^{-1}(k) - S^{-1}(p) \;.
\mlab{1.1000}
\ee

In the limit $p\to k$, \mref{1.1000} becomes the Ward identity:
\be
\Gamma^\mu(k,k) = \frac{\partial S^{-1}(k) }{\partial k_\mu} .
\mlab{1.1000.1}
\ee

In general, the full QED vertex can be written as the sum of a longitudinal
and a transverse part:
\be
\Gamma^\mu(k,p) = \Gamma_{L}^\mu(k,p) + \Gamma_{T}^\mu(k,p) . 
\mlab{1.1003}
\ee

The longitudinal part, $\Gamma_L^\mu(k,p)$, of the vertex is determined by
the Ward-Takahashi identity, \mref{1.1000}, and the Ward identity,
\mref{1.1000.1}, as demonstrated by Ball and Chiu~\cite{BC}, and is given
by:
\ba
\Gamma^\mu_{L}(k,p) &=& 
\frac{1}{2}\l[\frac{1}{\F(k^2)}+\frac{1}{\F(p^2)}\r]\gamma^\mu
+ \frac{1}{2}\l[\frac{1}{\F(k^2)}-\frac{1}{\F(p^2)}\r]
\frac{(k+p)^\mu(\slash{k}+\slash{p})}{k^2-p^2} \mlab{1.1002} \\[1mm]
&& - \l[\frac{\S(k^2)}{\F(k^2)} - \frac{\S(p^2)}{\F(p^2)}\r]
\frac{(k+p)^\mu}{k^2-p^2} \nn .
\ea

The transverse part of the vertex, which has to satisfy the transversality
condition 
\[
(k-p)_\mu \, \Gamma_T^\mu(k,p) = 0 \;,  \qquad \mbox{and} \qquad
\Gamma_T^\mu(p,p) = 0 \;,
\]
is not constrained by the Ward-Takahashi identity. However, other
properties of gauge theories can be used to restrict its form. These
constraints are mainly multiplicative renormalizability, reproduction of
perturbative results in weak coupling, absence of kinematical singularities
and gauge invariance of physical observables~\cite{CP90,Bashir94,Ayse}.

The most general form for the transverse part of the vertex can be given
by~\cite{BC,Ayse95}:
\be
\Gamma_T^\mu(k,p) = \sum^8_{i=1}\,\tau_i(k^2,p^2,q^2) \, T^{\mu}_i(k,p) ,
\ee
where the $T^{\mu}_i$ form a tensor basis in spinor space and are defined
as:
\ba
T^{\mu}_{1}(k,p) &=& p^{\mu}(k\cdot q)-k^{\mu}(p\cdot q)\nonumber\\
T^{\mu}_{2}(k,p) &=& \left[p^{\mu}(k\cdot q)-k^{\mu}(p\cdot q)\right](\slash{k}
+\slash{p})\nonumber\\
T^{\mu}_{3}(k,p) &=& q^2\gamma^{\mu}-q^{\mu}\slash{q}\nonumber\\
T^{\mu}_4(k,p) &=& q^2\left[\gamma^{\mu}(\slash{p}+\slash{k})-p^{\mu}-k^{\mu}\right]
+2(p-k)^{\mu}k^{\lambda}p^{\nu}\sigma_{\lambda\nu}\nonumber\\
T^{\mu}_{5}(k,p) &=& q_{\nu}{\sigma^{\nu\mu}} \mlab{1.betensor} \\
%
T^{\mu}_{6}(k,p) &=& \gamma^{\mu}(k^2-p^2) 
- (k+p)^{\mu}(\slash{k}-\slash{p})\nonumber\\
T^{\mu}_{7}(k,p) &=& \frac{1}{2}(p^2-k^2)\left[\gamma^{\mu}(\slash{p}+\slash{k})
-p^{\mu}-k^{\mu}\right]
+\left(k+p\right)^{\mu}k^{\lambda}p^{\nu}\sigma_{\lambda\nu}\nonumber\\
T^{\mu}_{8}(k,p) &=& -\gamma^{\mu}k^{\nu}p^{\lambda}{\sigma_{\nu\lambda}}
+k^{\mu}\slash{p}-p^{\mu}\slash{k} \nonumber
\ea
with $q=k-p$ and $\sigma_{\mu\nu} = \frac{1}{2}[\gamma_{\mu},\gamma_{\nu}]$.

\subsection{Curtis-Pennington vertex}
\label{Sec:CP-vertex}

In Ref.~\cite{CP90} Curtis and Pennington have proposed a vertex Ansatz
which ensures the multiplicative renormalizability of the fermion
propagator, reproduces the perturbative results in the weak coupling limit
and is free of kinematical singularities in the massive case. As these
requirements do not constrain the transverse part uniquely, they have
chosen a simple form satisfying them and which is only composed of
$T_6^\mu$:
\be
\Gamma^\mu_T(k,p) = \frac{1}{2}\l[\frac{1}{\F(k^2)}-\frac{1}{\F(p^2)}\r]
\frac{(k^2+p^2)\l[\gamma^{\mu}(k^2-p^2)-(k+p)^{\mu}(\slash{k}-\slash{p})\r]}
{(k^2-p^2)^2+\l(\S^2(k^2)+\S^2(p^2)\r)^2} .
\mlab{1.1004}
\ee

Substituting \mrefb{1.1002}{1.1004} in \mref{1.1003} yields the full {\it
Curtis-Pennington vertex Ansatz}:
\def\CP{{\mbox{\tiny CP}}}
\ba
\Gamma^\mu_{\CP}(k,p) &=&
\frac{1}{2}\l[\frac{1}{\F(k^2)}+\frac{1}{\F(p^2)}\r]\gamma^\mu
+ \frac{1}{2}\l[\frac{1}{\F(k^2)}-\frac{1}{\F(p^2)}\r]
\frac{(k+p)^\mu(\slash{k}+\slash{p})}{k^2-p^2} \mlab{1.1005} \\[1mm]
&& - \l[\frac{\S(k^2)}{\F(k^2)} - \frac{\S(p^2)}{\F(p^2)}\r]
\frac{(k+p)^\mu}{k^2-p^2} \nn\\
&& + \frac{1}{2}\l[\frac{1}{\F(k^2)}-\frac{1}{\F(p^2)}\r]
\frac{(k^2+p^2)\l[\gamma^{\mu}(k^2-p^2)-(k+p)^{\mu}(\slash{k}-\slash{p})\r]}
{(k^2-p^2)^2+\l(\S^2(k^2)+\S^2(p^2)\r)^2} \nn .
\ea

\section{The Curtis-Pennington equations}

\subsection{The fermion equations}

We now derive the equations necessary for the study of
dynamical mass generation in QED with the Curtis-Pennington vertex
Ansatz. The fermion self-energy integral, \mref{a1.1}, with the
Curtis-Pennington vertex is:
\be
\Self(p) = 
\frac{i e^2}{(2\pi)^4} \int d^4k \,
\gamma^\mu \, S(k) \, \Gamma^\nu_{\CP}(k,p) \, D_{\nu\mu}(k-p) \;.
\mlab{1106}
\ee

If we substitute the photon propagator, \mref{3.6}, in the self-energy
integral, \mref{1106}, we find:
\be
\Self(p) = - \frac{i e^2}{(2\pi)^4} \int d^4k \,
\gamma^\mu \, S(k) \, \Gamma^\nu_{\CP}(k,p) \, 
\left[ \frac{\G(q^2)}{q^2}\left(g_{\nu\mu} - \frac{q_\nu q_\mu}{q^2}\right)
+ \xi \frac{q_\nu q_\mu}{q^4} \right] \;,
\mlab{150}
\ee
where we defined $q \equiv k-p$.

Because the CP-vertex satisfies the Ward-Takahashi identity,
\mref{1.1000}, in the same way as the full vertex does, it is useful to 
substitute this identity in the $\xi$-part of \mref{150}, to ensure
translational invariance. This yields:
\be
\Self(p) = - \frac{i e^2}{(2\pi)^4} \int d^4k 
\bigg\{ \frac{\G(q^2)}{q^2}\left(g_{\nu\mu} - \frac{q_\nu q_\mu}{q^2}\right)
\gamma^\mu \, S(k) \, \Gamma^\nu_{\CP}(k,p)
+ \frac{\xi}{q^4}  \slash{q} \, \Big[1 - S(k) S^{-1}(p)\Big] \bigg\} \;.
\mlab{153} 
\ee

From translational invariance we know:
\be
\int_{-\infty}^{+\infty} d^4k \; \frac{\slash{q}}{q^4}
 = \int_{-\infty}^{+\infty} d^4q \; \frac{\slash{q}}{q^4} = 0 .
\mlab{154}
\ee

So that \mref{153} becomes:
\be
\Self(p) = - \frac{i e^2}{(2\pi)^4} \int d^4k 
\bigg\{ \frac{\G(q^2)}{q^2}\left(g_{\nu\mu} - \frac{q_\nu q_\mu}{q^2}\right)
\gamma^\mu \, S(k) \, \Gamma^\nu_{\CP}(k,p)
- \frac{\xi}{q^4}  \slash{q} \, S(k) S^{-1}(p) \bigg\} \;. \nn
\mlab{155}
\ee

Substituting \mref{155} in the $\S$-equation, \mref{156}, and introducing
$\alpha \equiv e^2/4\pi$ we find
\be
\frac{\S(p^2)}{\F(p^2)}
= m_0 - \frac{i \alpha}{16\pi^3} \int d^4k 
\bigg\{ \frac{\G(q^2)}{q^2}\left(g_{\nu\mu} - \frac{q_\nu q_\mu}{q^2}\right)
\Trace{\Big[\gamma^\mu \, S(k)\, \Gamma^\nu_{\CP}(k,p)\Big]} 
- \frac{\xi}{q^4}  \Trace{\Big[\slash{q} \, S(k) S^{-1}(p)\Big]} \bigg\} .
\mlab{158}
\ee

We now substitute the fermion propagator, \mref{3.5}, in the integral of
\mref{158}, yielding
\be
\frac{\S(p^2)}{\F(p^2)}
= m_0 - \frac{i \alpha}{16\pi^3} \int d^4k \; \frac{\F(k^2)}{k^2 -
\S^2(k^2)} \bigg\{ 
\frac{\G(q^2)}{q^2}\left(g_{\nu\mu} - \frac{q_\nu q_\mu}{q^2}\right)
T^{\mu\nu}_\G - \frac{\xi}{\F(p^2)q^4} T_\xi \bigg\}
\mlab{159} 
\ee

where we defined\vspace{-3mm}
\ba
T^{\mu\nu}_\G &\equiv& 
\Trace{\Big[\gamma^\mu \, (\slash{k} + \S(k^2)) \, \Gamma^\nu_{\CP}(k,p)\Big]} 
\mlab{160}\\
T_\xi &\equiv& \Trace{\Big[\slash{q} \, (\slash{k} + \S(k^2)) \,
(\slash{p} - \S(p^2))\Big]} \;. \mlab{161}
\ea

We first consider the $\xi$-part of the integral in
\mref{159}, which we call $I_\xi$:
\be
I_\xi \equiv \frac{i \alpha\xi}{16\pi^3\F(p^2)} \int d^4k \;
\frac{\F(k^2)}{(k^2 - \S^2(k^2))\,q^4} \; T_\xi \;.
\mlab{1107}
\ee
 
We compute the trace, \mref{161}, using \mref{traces}, and substitute
$q=k-p$. This gives:
\be
T_\xi = 4 \bigg[ \Big(\S(k^2)+\S(p^2)\Big) \, k \cdot p - \S(k^2) \, p^2 
- \S(p^2) \, k^2 \bigg] \;.
\mlab{163}
\ee

Substituting the trace, \mref{163}, in the $\xi$-integral, \mref{1107},
we find:
\be
I_\xi = \frac{i \alpha\xi}{4\pi^3\F(p^2)} \int d^4k \; \frac{\F(k^2)}{k^2 -
\S^2(k^2)} \frac{1}{q^4} 
\bigg[ \Big(\S(k^2)+\S(p^2)\Big) \, k \cdot p - \S(k^2) \, p^2 
- \S(p^2) \, k^2 \bigg] \;.
\mlab{164}
\ee

Performing a Wick rotation to go from Minkowski to Euclidean space, as
explained in Section~\ref{Sec:Coupeq}, gives:
\be
I_\xi = -\frac{\alpha\xi}{4\pi^3\F(p^2)} \int d^4k \; \frac{\F(k^2)}{k^2 +
\S^2(k^2)} \frac{1}{q^4} 
\bigg[ \Big(\S(k^2)+\S(p^2)\Big) \, k \cdot p - \S(k^2) \, p^2 
- \S(p^2) \, k^2 \bigg] \;.
\mlab{165}
\ee

We can now change to spherical coordinates. Two angles can be integrated
out straight away leaving us with one angle, $\theta$, and a Jacobian which is
$2\pi dy\,y \;d\theta \,\sin^2\theta$ where we denoted $y\equiv k^2$. If we
also define $x\equiv p^2$ and $z\equiv q^2$, \mref{165} becomes:
\be
I_\xi = -\frac{\alpha\xi}{2\pi^2\F(x)} \int dy \; \frac{y\F(y)}{y +
\S^2(y)} \int d\theta \, \frac{\sin^2\theta}{z^2} 
\bigg[ \Big(\S(y)+\S(x)\Big) \, \sqrt{yx}\cos\theta - \S(y) \, x 
- \S(x) \, y \bigg] \;.
\mlab{166}
\ee

The angular integrals of \mref{166} can be computed analytically as shown
in Appendix~\ref{App:angint}. Substituting \mrefb{A10}{A11} in \mref{166}
finally gives:
\be
I_\xi = \frac{\alpha\xi}{4\pi\F(x)} \int dy \; \frac{\F(y)}{y + \S^2(y)} 
\bigg[ \frac{y\S(y)}{x}\theta(x-y) + \S(x)\theta(y-x) \bigg] \;.
\mlab{170}
\ee

We now consider the $\G$-part of the integral in the $\S$-equation, \mref{159}:
\be
I_\G \equiv 
- \frac{i \alpha}{16\pi^3} \int d^4k \; \frac{\F(k^2)}{k^2 - \S^2(k^2)} 
\frac{\G(q^2)}{q^2}\left(g_{\nu\mu} - \frac{q_\nu q_\mu}{q^2}\right)
T^{\mu\nu}_\G 
\mlab{171} 
\ee
where we defined $T^{\mu\nu}_\G$ in \mref{160} as:
\be
T^{\mu\nu}_\G \equiv 
\Trace{\Big[\gamma^\mu \, (\slash{k} + \S(k^2)) \,
\Gamma^\nu_{\CP}(k,p)\Big]} \;.
\mlab{172}
\ee

In contrast to the $\xi$-part of the $\S$-equation, the $\G$-part depends
on the specific vertex Ansatz used. We recall the CP-vertex,
\mref{1.1005}:
\ba
\Gamma^\mu_{\CP}(k,p) &=& A(k^2,p^2) \gamma^\mu
+ B(k^2,p^2) (k+p)^\mu(\slash{k}+\slash{p}) + C(k^2,p^2) (k+p)^\mu
\mlab{1108} \\
&& + \tau_6(k^2,p^2)
\l[\gamma^{\mu}(k^2-p^2)-(k+p)^{\mu}(\slash{k}-\slash{p})\r] \nn
\ea
where we define:
\ba
\parbox{15cm}{
\ba
A(k^2,p^2) &\equiv& \frac{1}{2}\l[\frac{1}{\F(k^2)}+\frac{1}{\F(p^2)}\r] \nn\\
B(k^2,p^2) &\equiv& \frac{1}{2(k^2-p^2)}\l[\frac{1}{\F(k^2)}-\frac{1}{\F(p^2)}\r]\nn\\
C(k^2,p^2) &\equiv& - \frac{1}{k^2-p^2}
\l[\frac{\S(k^2)}{\F(k^2)} - \frac{\S(p^2)}{\F(p^2)}\r]\nn\\
\tau_6(k^2,p^2) &\equiv& 
\frac{(k^2+p^2)}{2\l[(k^2-p^2)^2+\l(\S^2(k^2)+\S^2(p^2)\r)^2\r]}
\l[\frac{1}{\F(k^2)}-\frac{1}{\F(p^2)}\r] \;. \nn
\ea}
\mlab{1111}
\ea
 
Substituting the CP-vertex, \mref{1108}, in \mref{172} and computing the
traces using \mref{traces} yields: 
\ba
T_\G^{\mu\nu} &=& 4\S(k^2) \, \bigg\{A(k^2,p^2)\,g^{\mu\nu}
+ B(k^2,p^2) (k+p)^\mu (k+p)^\nu \mlab{1112}\\
&& + \tau_6(k^2,p^2)
\l[(k^2-p^2)\,g^{\mu\nu} - (k-p)^\mu (k+p)^{\nu}\r]\bigg\}
 + 4 C(k^2,p^2)\,k^\mu (k+p)^\nu \;. \nn
\ea

We now contract $T_\G^{\mu\nu}$ with the transverse tensor,
$g_{\mu\nu}-q_\mu q_\nu/q^2$, of the photon propagator. This
gives:\vspace{-3mm}
\ba
\l(g_{\mu\nu}-\frac{q_\mu q_\nu}{q^2}\r) T_\G^{\mu\nu} 
&=& 4\, \Bigg\{3\S(k^2)\l[A(k^2,p^2) + \tau_6(k^2,p^2)(k^2-p^2)\r] 
\mlab{1113}\\
&& + 2\l[2 B(k^2,p^2)\S(k^2) + C(k^2,p^2)\r]  \l[\frac{k^2p^2 - (k.p)^2}{q^2}\r]
\Bigg\} \;. \nn
\ea

Substituting \mref{1113} in the integral, \mref{171}, we find:
\ba
I_\G &=&
 - \frac{i \alpha}{4\pi^3} \int d^4k \; \frac{\F(k^2)}{k^2 - \S^2(k^2)} 
\frac{\G(q^2)}{q^2} 
\,\Bigg\{3\S(k^2)\l[A(k^2,p^2) + \tau_6(k^2,p^2)(k^2-p^2)\r] \mlab{1114}\\
&& \hspace{2cm} 
+ 2\l[2 B(k^2,p^2)\S(k^2) + C(k^2,p^2)\r]  \l[\frac{k^2p^2 - (k.p)^2}{q^2}\r]
\Bigg\} \;. \nn
\ea

We now perform a Wick rotation to Euclidean space. Note, from \mref{1111},
that $A\to A$, $B\to -B$, $C\to -C$ and $\tau_6\to -\tau_6$. This gives:
\ba
I_\G &=&
\frac{\alpha}{4\pi^3} \int d^4k \; \frac{\F(k^2)}{k^2 + \S^2(k^2)} 
\frac{\G(q^2)}{q^2} 
\,\Bigg\{3\S(k^2)\l[A(k^2,p^2) + \tau_6(k^2,p^2)(k^2-p^2)\r] \mlab{1115}\\
&& \hspace{2cm} 
+ 2\l[2 B(k^2,p^2)\S(k^2) + C(k^2,p^2)\r]  \l[\frac{k^2p^2 - (k.p)^2}{q^2}\r]
\Bigg\} \;. \nn
\ea

Introducing spherical coordinates and defining $x=p^2$, $y=k^2$, $z=q^2$
yields:
\ba
I_\G &=&
\frac{\alpha}{2\pi^2} \int dy \; \frac{y\F(y)}{y + \S^2(y)} \,
\int d\theta \; \sin^2\theta \, 
\frac{\G(z)}{z} 
\,\Bigg\{3\S(y)\l[A(y,x) + \tau_6(y,x)(y-x)\r] \mlab{1116}\\
&& \hspace{2cm} 
+ \l[2 B(y,x)\S(y) + C(y,x)\r] \frac{2yx\sin^2\theta}{z}
\Bigg\} \;. \nn
\ea

From \mref{1111} we check that:
\be
2 B(y,x)\S(y) + C(y,x) = 
- \frac{1}{\F(x)}\l[\frac{\S(y)-\S(x)}{y-x}\r] \;. 
\mlab{1117}
\ee

Substituting Eqs.~(\oref{170}, \oref{1116}, \oref{1117}) in the
$\S$-equation, \mref{159}, yields:
\ba
\fbox{\parbox{14cm}{\ba
\frac{\S(x)}{\F(x)}
&=& m_0 
+ \frac{\alpha}{2\pi^2} \int dy \; \frac{y\F(y)}{y + \S^2(y)}  \, 
\int d\theta \; \sin^2\theta \, \frac{\G(z)}{z} \nn \\
&& \times \Bigg\{3\S(y)\l[A(y,x) + \tau_6(y,x)(y-x)\r]
- \frac{1}{\F(x)}\l[\frac{\S(y)-\S(x)}{y-x}\r] \frac{2yx\sin^2\theta}{z}
\Bigg\}  \nn\\
&& + \frac{\alpha\xi}{4\pi\F(x)} \int dy \; \frac{\F(y)}{y + \S^2(y)} 
\bigg[ \frac{y\S(y)}{x}\theta(x-y) + \S(x)\theta(y-x) \bigg] \;. \nn
\ea}}\hspace{5mm}
\mlab{1118} 
\ea

Next, we derive the $\F$-equation in a similar way. Substituting the
self-energy, \mref{155}, in the $\F$-equation, \mref{157}, and introducing
$\alpha \equiv e^2/4\pi$ we have
\be\hspace{-2pt}
\frac{1}{\F(p^2)} = 1 +  
\frac{i\alpha}{16 p^2 \pi^3} \int \hspace{-2pt} d^4k 
\bigg\{ \frac{\G(q^2)}{q^2}\left(g_{\nu\mu} - \frac{q_\nu q_\mu}{q^2}\right)
\Trace{\Big[\slash{p} \, \gamma^\mu \, S(k) \, \Gamma^\nu_{\CP}(k,p)\Big]}
- \frac{\xi}{q^4}  \Trace{\Big[\slash{p} \, \slash{q} \, S(k) S^{-1}(p)\Big]}
\bigg\} .
\mlab{1119}
\ee

After introducing the fermion propagator, \mref{3.5}, in \mref{1119} we find:
\be
\frac{1}{\F(p^2)} = 1 +  
\frac{i\alpha}{16 p^2 \pi^3} \int d^4k  \; \frac{\F(k^2)}{k^2 - \S^2(k^2)} 
\bigg\{ \frac{\G(q^2)}{q^2}\left(g_{\nu\mu} - \frac{q_\nu q_\mu}{q^2}\right)
T_\G^{\mu\nu} - \frac{\xi}{\F(p^2)q^4} T_\xi
\bigg\}
\mlab{1120}
\ee
where we defined the traces:\vspace{-3mm}
\ba
T_\G^{\mu\nu} &\equiv& \Trace{\Big[\slash{p} \, \gamma^\mu \, (\slash{k} +
\S(k^2)) \, \Gamma^\nu_{\CP}(k,p)\Big]} \mlab{1121}\\
T_\xi &\equiv& \Trace{\Big[\slash{p} \, \slash{q} \,  (\slash{k} + \S(k^2)) \, 
(\slash{p} - \S(p^2))\Big]} \;. \mlab{1122}
\ea

We first compute the $\xi$-part of the integral in \mref{1120}, which
we call $I_\xi$:
\be
I_\xi \equiv  - \frac{i\alpha\xi}{16 p^2 \F(p^2)\pi^3} \int d^4k  \; 
\frac{\F(k^2)}{(k^2 - \S^2(k^2))\,q^4} \; T_\xi \;.
\mlab{1123}
\ee

We compute the trace, \mref{1122}, using \mref{traces} and substitute
$q=k-p$:
\be
T_\xi = 4 \Big[ p^2 (k^2-k.p) - \S(k^2) \S(p^2) (k.p-p^2) \Big] \;.
\mlab{1124}
\ee

Substituting this trace in \mref{1123} gives:
\be
I_\xi =  - \frac{i\alpha\xi}{4 \pi^3\F(p^2)} \int d^4k  \; 
\frac{\F(k^2)}{(k^2 - \S^2(k^2))\,q^4} \; 
\l[ k^2-k.p - \S(k^2) \S(p^2) \l(\frac{k.p}{p^2}-1\r) \r] \;.
\mlab{1125}
\ee

After performing a Wick rotation on \mref{1125}, we find in Euclidean space:
\be
I_\xi =  \frac{\alpha\xi}{4 \pi^3\F(p^2)} \int d^4k  \; 
\frac{\F(k^2)}{(k^2 + \S^2(k^2))\,q^4} \; 
\l[ k^2-k.p + \S(k^2) \S(p^2) \l(\frac{k.p}{p^2}-1\r) \r] \;.
\mlab{1126}
\ee

We can now change the integration variables to spherical coordinates, again
introducing $x=p^2$, $y=k^2$ and $z=q^2$. This gives:
\be
I_\xi =  \frac{\alpha\xi}{2 \pi^2\F(x)} \int dy  
\frac{y\F(y)}{y + \S^2(y)} \; \int d\theta \;
\frac{\sin^2\theta}{z^2} \,
\l[ y - \sqrt{yx}\cos\theta 
+ \S(y) \S(x) \l(\sqrt{\frac{y}{x}}\cos\theta-1\r) \r] \;.
\mlab{1127}
\ee

The angular integrals of \mref{1127} can be computed analytically and are
given in Appendix~\ref{App:angint}. Substituting \mrefb{A10}{A11} in
\mref{1127} yields:
\be
I_\xi =  - \frac{\alpha\xi}{4 \pi\F(x)} \int dy  
\frac{\F(y)}{y + \S^2(y)} \; 
\l[\frac{y\S(y)\S(x)}{x^2} \theta(x-y) - \theta(y-x)\r] \;.
\mlab{1130}
\ee

Next, we consider the $\G$-part of the integral in \mref{1120}:
\be
I_\G \equiv \frac{i\alpha}{16 p^2 \pi^3} \int d^4k  \; 
\frac{\F(k^2)}{k^2 - \S^2(k^2)} 
\frac{\G(q^2)}{q^2}\left(g_{\nu\mu} - \frac{q_\nu q_\mu}{q^2}\right) 
T_\G^{\mu\nu}
\mlab{1131}
\ee
where the trace $T_\G^{\mu\nu}$ has been defined in \mref{1121} as:
\be
T_\G^{\mu\nu} = \Trace{\Big[\slash{p} \, \gamma^\mu \, (\slash{k} +
\S(k^2)) \, \Gamma^\nu_{\CP}(k,p)\Big]} \;. \mlab{1132}
\ee

Substituting the vertex expression, \mref{1108}, in \mref{1132}, and
computing the traces using \mref{traces}, gives:
\ba
T_\G^{\mu\nu} &=& 
4 A(k^2,p^2) (p^\mu k^\nu + p^\nu k^\mu - k.p \, g^{\mu\nu} )
+ 4 B(k^2,p^2) (k+p)^\nu (p^2 k^\mu + k^2 p^\mu) \mlab{1133}\\
&+&  4\tau_6(k^2,p^2) \l[(k-p)^\mu (p^2 k^\nu + k^2 p^\nu)
- (k^2-p^2) k.p \,g^{\mu\nu}\r]
+  4 C(k^2,p^2)\S(k^2) p^\mu (k+p)^\nu \;. \nn
\ea

We now contract $T_\G^{\mu\nu}$ of \mref{1133} with the transverse tensor
$g_{\mu\nu}-q_\mu q_\nu/q^2$ and substitute $q=k-p$. This gives:
\ba
\l(g_{\mu\nu}-\frac{q_\mu q_\nu}{q^2}\r) T_\G^{\mu\nu} 
&=& 4 \Bigg\{ A(k^2,p^2) \bigg[ 2 \l(\frac{k^2p^2 -(k.p)^2}{q^2}\r) - 3
k.p\bigg]  \mlab{1134}\\
&& \hspace{-2.8cm} + 2 \l[ B(k^2,p^2)(k^2+p^2) + C(k^2,p^2)\S(k^2)\r]
\l(\frac{k^2 p^2 - (k.p)^2}{q^2}\r)
- 3 \tau_6(k^2,p^2) (k^2-p^2)\, k.p
\Bigg\} \;. \nn
\ea

Substituting \mref{1134} in the integral \mref{1131} yields:
\ba
I_\G &=& \frac{i\alpha}{4 p^2 \pi^3} \int d^4k  \; 
\frac{\F(k^2)}{k^2 - \S^2(k^2)} \frac{\G(q^2)}{q^2}
\Bigg\{ A(k^2,p^2) \bigg[ 2 \l(\frac{k^2p^2 -(k.p)^2}{q^2}\r) - 3
k.p\bigg] \mlab{1135}\\
&+& 2 \l[ B(k^2,p^2)(k^2+p^2) + C(k^2,p^2)\S(k^2)\r]
\l(\frac{k^2 p^2 - (k.p)^2}{q^2}\r) 
- 3 \tau_6(k^2,p^2) (k^2-p^2)\, k.p
\Bigg\} \;. \nn
\ea

After a Wick rotation to Euclidean space, \mref{1135} becomes: 
\ba
I_\G &=& - \frac{\alpha}{4 p^2 \pi^3} \int d^4k  \; 
\frac{\F(k^2)}{k^2 + \S^2(k^2)} \frac{\G(q^2)}{q^2}
\Bigg\{ A(k^2,p^2) \bigg[ 2 \l(\frac{k^2p^2 -(k.p)^2}{q^2}\r) - 3
k.p\bigg] \mlab{1136}\\
&+& 2 \l[ B(k^2,p^2)(k^2+p^2) - C(k^2,p^2)\S(k^2)\r]
\l(\frac{k^2 p^2 - (k.p)^2}{q^2}\r) 
- 3 \tau_6(k^2,p^2) (k^2-p^2)\, k.p
\Bigg\} \;. \nn
\ea

Introducing spherical coordinates and defining $x=p^2$, $y=k^2$, $z=q^2$
yields:
\ba
I_\G &=& - \frac{\alpha}{2 x \pi^2} \int dy  \; 
\frac{y\F(y)}{y + \S^2(y)} \int d\theta \; \sin^2\theta \,
\frac{\G(z)}{z}
\Bigg\{ A(y,x) \bigg[ \frac{2 yx\sin^2\theta}{z}
- 3 \sqrt{yx}\cos\theta \bigg] \hspace{0.5cm} \mlab{1137}\\
&+& \l[ B(y,x)(y+x) - C(y,x)\S(y)\r]
\l(\frac{2yx\sin^2\theta}{z}\r) - 3 \tau_6(y,x) (y-x)\, \sqrt{yx}\cos\theta
\Bigg\} \;. \nn
\ea

Substituting \mrefb{1130}{1137} in the $\F$-equation, \mref{1120}, finally
gives:
\ba
\fbox{\parbox{14.14cm}{\ba
\frac{1}{\F(x)} &=& 1  
- \frac{\alpha}{2 x \pi^2} \int dy  \; 
\frac{y\F(y)}{y + \S^2(y)} \int d\theta \; \sin^2\theta \,
\frac{\G(z)}{z} \nn\\
&& \times \Bigg\{ A(y,x) \bigg[ \frac{2 yx\sin^2\theta}{z}
- 3 \sqrt{yx}\cos\theta \bigg] \hspace{0.5cm} \nn\\
&& + \l[ B(y,x)(y+x) - C(y,x)\S(y)\r]
\frac{2yx\sin^2\theta}{z}
- 3 \tau_6(y,x) (y-x)\, \sqrt{yx}\cos\theta
\Bigg\} \nn\\
&& - \frac{\alpha\xi}{4 \pi\F(x)} \int dy  
\frac{\F(y)}{y + \S^2(y)} \; 
\l[\frac{y\S(y)\S(x)}{x^2} \theta(x-y) - \theta(y-x)\r] \;. \nn
\ea}}\hspace{0.5cm}
\mlab{1138} 
\ea

\subsection{The photon equation}

We will now derive the integral equation for the photon renormalization
function using the Curtis-Pennington vertex Ansatz.  We introduce the
CP-vertex in the integral equation for $\G$, \mref{196}:
\be
\frac{1}{\G(q^2)} = 1
- \frac{iN_f e^2\Proj_{\mu\nu}}{3(2\pi)^4 q^2} \int d^4k\, 
\Trace \Big[ \gamma^\mu \, S(k) \, 
\Gamma_{\CP}^\nu(k,p) \, S(p) \Big] \, .
\mlab{1139}
\ee

When substituting the fermion propagator, \mref{3.5}, in \mref{1139}, we
find:
\be
\frac{1}{\G(q^2)} = 1
- \frac{iN_f e^2}{3(2\pi)^4 q^2} \int d^4k\, 
\frac{\F(k^2)\F(p^2)}{(k^2-\S^2(k^2))(p^2-\S^2(p^2))} \, 
\Proj_{\mu\nu} T^{\mu\nu}
\mlab{1140}
\ee
where
\be
T^{\mu\nu} \equiv \Trace \Big[ \gamma^\mu \, (\slash{k}+\S(k^2)) \, 
\Gamma_{\CP}^\nu(k,p) \, (\slash{p}+\S(p^2)) \Big] \, .
\mlab{1141}
\ee

Now, insert the CP-vertex, \mref{1108} in \mref{1141}:
\ba
T^{\mu\nu} &=& 4 \Bigg\{ 
A(k^2,p^2) \Big[k^\mu p^\nu + p^\mu k^\nu - k.p\,g^{\mu\nu}
+ \S(k^2)\S(p^2)\,g^{\mu\nu}\Big] \mlab{1142}\\
&&\hspace{2mm} + B(k^2,p^2) \Big[ p^2 k^\mu + k^2 p^\mu 
+ \S(k^2) \S(p^2) (k+p)^\mu \Big] (k+p)^\nu \nn \\ 
&&\hspace{2mm} + C(k^2,p^2) \Big[ \S(p^2) k^\mu + \S(k^2) p^\mu \Big] (k+p)^\nu \nn \\
&&\hspace{2mm} + \tau_6(k^2,p^2) \Big[ (k-p)^\mu (k^2p^\nu + p^2 k^\nu) 
- (k^2-p^2) k.p\,g^{\mu\nu}) \nn\\
&& \hspace{18mm} 
+ \S(k^2)\S(p^2) ( (k^2-p^2)\,g^{\mu\nu} - (k-p)^\mu (k+p)^{\nu} \Big] \Bigg\}
. \nn
\ea

When contracting $T^{\mu\nu}$ with the operator
$\Proj_{\mu\nu}=g_{\mu\nu}-nq_\mu q_\nu/q^2$ and substituting $p=k-q$, we
find:
\ba
\parbox{15cm}{\vspace{-2ex}\bann
\Proj_{\mu\nu} T^{\mu\nu} &=& 4 \Bigg\{
A(k^2,p^2) \l[(n-2) k^2 + (n+2) k.q - \frac{2n(k.q)^2}{q^2} 
+ (4-n) \S(k^2)\S(p^2) \r] \\
&& + B(k^2,p^2) \bigg[ \Big(k^2 + p^2 + 2\S(k^2)\S(p^2)\Big) 
\l(2k^2 - \frac{2n(k.q)^2}{q^2}  + (n-1) k.q\r) \\
&& \hspace{2cm} + (n-1)(k^2-p^2)\Big(k^2+\S(k^2)\S(p^2)\Big)\bigg] \\
&& + C(k^2,p^2) \bigg[
\Big(\S(k^2)+\S(p^2)\Big) \l(2k^2 -\frac{2n(k.q)^2}{q^2} + (n-1)k.q\r) \\
&& \hspace{2cm} + (n-1)(k^2-p^2)\S(k^2)\bigg] \\
&& - 3\tau_6(k^2,p^2) (k^2-p^2) \Big(k^2 - k.q - \S(k^2)\S(p^2)\Big)
\Bigg\} \;.
\eann}
\mlab{1143}
\ea

Now, substitute \mref{1143} in the $\G$-equation, \mref{1140}:
\ba
\parbox{15cm}{\vspace{-2ex}\bann
\frac{1}{\G(q^2)} &=& 1 - \frac{4iN_f e^2}{3(2\pi)^4 q^2} \int d^4k\, 
\frac{\F(k^2)\F(p^2)}{(k^2-\S^2(k^2))(p^2-\S^2(p^2))}\\
&&\times \Bigg\{
A(k^2,p^2) \l[(n-2) k^2 + (n+2) k.q - \frac{2n(k.q)^2}{q^2} 
+ (4-n) \S(k^2)\S(p^2) \r] \\
&& + B(k^2,p^2) \bigg[ \Big(k^2 + p^2 + 2\S(k^2)\S(p^2)\Big) 
\l(2k^2 - \frac{2n(k.q)^2}{q^2}  + (n-1) k.q\r) \\
&& \hspace{2cm} + (n-1)(k^2-p^2)\Big(k^2+\S(k^2)\S(p^2)\Big)\bigg] \\
&& + C(k^2,p^2) \bigg[
\Big(\S(k^2)+\S(p^2)\Big) \l(2k^2 -\frac{2n(k.q)^2}{q^2} + (n-1)k.q\r) \\
&& \hspace{2cm} + (n-1)(k^2-p^2)\S(k^2)\bigg] \\
&& - 3\tau_6(k^2,p^2) (k^2-p^2) \Big(k^2 - k.q - \S(k^2)\S(p^2)\Big) \Bigg\} \;.
\eann}
\mlab{1144}
\ea

After performing a Wick rotation to Euclidean space, \mref{1144} becomes:
\ba
\parbox{15cm}{\vspace{-2ex}\bann
\frac{1}{\G(q^2)} &=& 1 + \frac{4N_f e^2}{3(2\pi)^4 q^2} \int d^4k\, 
\frac{\F(k^2)\F(p^2)}{(k^2+\S^2(k^2))(p^2+\S^2(p^2))}\\
&&\times \Bigg\{
A(k^2,p^2) \l[(n-2) k^2 + (n+2) k.q - \frac{2n(k.q)^2}{q^2} 
- (4-n) \S(k^2)\S(p^2) \r] \\
&& + B(k^2,p^2) \bigg[ \Big(k^2 + p^2 - 2\S(k^2)\S(p^2)\Big) 
\l(2k^2 - \frac{2n(k.q)^2}{q^2}  + (n-1) k.q\r) \\
&& \hspace{2cm} + (n-1)(k^2-p^2)\Big(k^2-\S(k^2)\S(p^2)\Big)\bigg] \\
&& - C(k^2,p^2) \bigg[
\Big(\S(k^2)+\S(p^2)\Big) \l(2k^2 -\frac{2n(k.q)^2}{q^2} + (n-1)k.q\r) \\
&& \hspace{2cm} + (n-1)(k^2-p^2)\S(k^2)\bigg] \\
&& - 3\tau_6(k^2,p^2) (k^2-p^2) \Big(k^2 - k.q + \S(k^2)\S(p^2)\Big)
\Bigg\} \;.
\eann}
\mlab{1145}
\ea

Finally, we introduce spherical coordinates and substitute
$\alpha=e^2/4\pi$ in \mref{1145}. The equation for the photon
renormalization function $\G$ becomes:
\ba
\parbox{15cm}{\vspace{-2ex}\bann
\frac{1}{\G(x)} &=& 1 + 
\frac{2N_f\alpha}{3\pi^2 x} \int dy \, \frac{y\F(y)}{y+\S^2(y)}
\int d\theta \, \sin^2\theta \, 
\frac{\F(z)}{z+\S^2(z)}\\
&&\times \Bigg\{
A(y,z) \l[(n-2) y + (n+2) \sqrt{yx}\cos\theta - 2ny\cos^2\theta 
- (4-n) \S(y)\S(z) \r] \\
&& + B(y,z) \bigg[ \Big(y + z - 2\S(y)\S(z)\Big) 
\l(2y - 2ny\cos^2\theta  + (n-1) \sqrt{yx}\cos\theta\r) \\
&& \hspace{2cm} + (n-1)(y-z)\Big(y-\S(y)\S(z)\Big)\bigg] \\
&& - C(y,z) \bigg[
\Big(\S(y)+\S(z)\Big) \l(2y - 2ny\cos^2\theta
+ (n-1)\sqrt{yx}\cos\theta\r) \\
&& \hspace{2cm} + (n-1)(y-z)\S(y)\bigg] \\
&& - 3\tau_6(y,z) (y-z) \Big(y - \sqrt{yx}\cos\theta + \S(y)\S(z)\Big)
\Bigg\} \;.
\eann}
\mlab{1147}
\ea

As discussed in Section~\ref{Sec:n=4}, if we take $n=4$ in the operator
$\Proj_{\mu\nu}$, the vacuum polarization integral should be free of quadratic
divergences. Then, \mref{1147} becomes:
\ba
\fbox{\parbox{14.05cm}{\bann
\frac{1}{\G(x)} &=& 1 + 
\frac{2N_f\alpha}{3\pi^2 x} \int dy \, \frac{y\F(y)}{y+\S^2(y)}
\int d\theta \, \sin^2\theta \, 
\frac{\F(z)}{z+\S^2(z)}\\
&\times& \Bigg\{
2A(y,z) \l[y(1 - y\cos^2\theta) + 3\sqrt{yx}\cos\theta \r] \\
&& + B(y,z) \bigg[ \Big(y + z - 2\S(y)\S(z)\Big) 
\l(2y(1 - 4\cos^2\theta)  + 3\sqrt{yx}\cos\theta\r) \\
&& \hspace{2cm} + 3(y-z)\Big(y-\S(y)\S(z)\Big)\bigg] \\
&& - C(y,z) \bigg[
\Big(\S(y)+\S(z)\Big) \l(2y(1 - 4\cos^2\theta)
+ 3\sqrt{yx}\cos\theta\r) + 3(y-z)\S(y)\bigg]\\
&& - 3\tau_6(y,z) (y-z) \Big(y - \sqrt{yx}\cos\theta + \S(y)\S(z)\Big)
\Bigg\} \;.
\eann}}\hspace{5mm}
\mlab{1148}
\ea

\clearpage

\subsection{Summary}

\vspace{-2mm}
The set of coupled integral equations (in Euclidean space) using the
Curtis-Pennington vertex Ansatz, Eqs.~(\oref{1118}, \oref{1138},
\oref{1148}) are now summarized:
\ba
\fbox{{\small\parbox{14.14cm}{\vspace{-3mm}
\bann
\frac{\S(x)}{\F(x)}
&=& m_0 
+ \frac{\alpha}{2\pi^2} \int dy \; \frac{y\F(y)}{y + \S^2(y)}  \, 
\int d\theta \; \sin^2\theta \, \frac{\G(z)}{z} \nn \\
&& \times \Bigg\{3\S(y)\l[A(y,x) + \tau_6(y,x)(y-x)\r]
- \frac{1}{\F(x)}\l[\frac{\S(y)-\S(x)}{y-x}\r] \frac{2yx\sin^2\theta}{z}
\Bigg\}  \nn\\
&& + \frac{\alpha\xi}{4\pi\F(x)} \int dy \; \frac{\F(y)}{y + \S^2(y)} 
\bigg[ \frac{y\S(y)}{x}\theta(x-y) + \S(x)\theta(y-x) \bigg] \\[2mm]
\hline\\[2mm]
\frac{1}{\F(x)} &=& 1  
- \frac{\alpha}{2 x \pi^2} \int dy  \; 
\frac{y\F(y)}{y + \S^2(y)} \int d\theta \; \sin^2\theta \,
\frac{\G(z)}{z} \nn\\
&& \times \Bigg\{ A(y,x) \bigg[ \frac{2 yx\sin^2\theta}{z}
- 3 \sqrt{yx}\cos\theta \bigg] \hspace{0.5cm} \nn\\
&& + \l[ B(y,x)(y+x) - C(y,x)\S(y)\r]
\frac{2yx\sin^2\theta}{z}
- 3 \tau_6(y,x) (y-x)\, \sqrt{yx}\cos\theta
\Bigg\} \nn\\
&& - \frac{\alpha\xi}{4 \pi\F(x)} \int dy  
\frac{\F(y)}{y + \S^2(y)} \; 
\l[\frac{y\S(y)\S(x)}{x^2} \theta(x-y) - \theta(y-x)\r] \\[2mm]
\hline\\[2mm]
\frac{1}{\G(x)} &=& 1 + 
\frac{2N_f\alpha}{3\pi^2 x} \int dy \, \frac{y\F(y)}{y+\S^2(y)}
\int d\theta \, \sin^2\theta \, 
\frac{\F(z)}{z+\S^2(z)}\\
&\times& \Bigg\{
2A(y,z) \l[y(1 - y\cos^2\theta) + 3\sqrt{yx}\cos\theta \r] \\
&& + B(y,z) \bigg[ \Big(y + z - 2\S(y)\S(z)\Big) 
\l(2y(1 - 4\cos^2\theta)  + 3\sqrt{yx}\cos\theta\r) \\
&& \hspace{2cm} + 3(y-z)\Big(y-\S(y)\S(z)\Big)\bigg] \\
&& - C(y,z) \bigg[
\Big(\S(y)+\S(z)\Big) \l(2y(1 - 4\cos^2\theta)
+ 3\sqrt{yx}\cos\theta\r) + 3(y-z)\S(y)\bigg]\\
&& - 3\tau_6(y,z) (y-z) \Big(y - \sqrt{yx}\cos\theta + \S(y)\S(z)\Big)
\Bigg\}
\eann\vspace{-3mm}}}}\hspace{0.5cm}\nn\\[-14cm]
\mlab{1.1006}\\[3.5cm]
\mlab{1.1007}\\[4.8cm]
\mlab{1.1008}\\[1.5cm]\nn
\ea

\ba
\parbox{15cm}{{\footnotesize
\bann \mbox{\small{where}}\qquad
A(y,x) &=& \frac{1}{2}\l[\frac{1}{\F(y)}+\frac{1}{\F(x)}\r] \\[1mm]
B(y,x) &=& \frac{1}{2(y-x)}\l[\frac{1}{\F(y)}-\frac{1}{\F(x)}\r] \\[1mm]
C(y,x) &=& -\frac{1}{y-x}\l[\frac{\S(y)}{\F(y)}-\frac{\S(x)}{\F(x)}\r] \\[1mm]
\tau_6(y,x) &=& \frac{y+x}{2\l[(y-x)^2+(\S^2(y)+\S^2(x))^2\r]}
\l[\frac{1}{\F(y)}-\frac{1}{\F(x)}\r] \:. \hspace{3cm}
\eann}}
\mlab{105} \ea

\chapter{Fermion mass generation in quenched QED}
\label{QuenQED}
\def\arg{\mbox{arg}}

\section{Introduction}

In the previous chapter we derived the system of equations describing the
dynamical generation of fermion mass in QED. To truncate the infinite set
of integral equations we introduced a suitable vertex Ansatz which reduces
the system of equations to three integral equations relating $\S$, $\F$ and
$\G$. A way to reduce the number of simultaneous equations even more is to
consider what is called the {\it quenched} approximation to QED.  In this
approximation the full photon propagator is replaced by the bare one,
neglecting any fermion loops, and the two fermion equations now form an
independent system of two coupled integral equations for $\S$ and
$\F$.

Formally, this approximation is obtained by setting the number of flavours
$N_f$ equal to zero. Then, the vacuum polarization contribution to the
photon propagator, \mref{a2}, will vanish and the full photon propagator
will be identical to the bare one. In this way the photon equation is now
decoupled from the fermion equation and the photon propagator occurring in
the fermion self-energy integral, \mref{a1.1}, is known. Although it can
seem bizarre to put $N_f \equiv 0$ and still consider the behaviour of the
fermion propagator and its self-energy, this limit is mathematically
perfectly sound (in the same way as $N_f$ could be given any non-integer
value) as $N_f$ is a free parameter, which occurs only in the photon
equation, while it is absent from the fermion equation. However, one can
wonder to what extent the results obtained in the quenched approximation
will reflect the physical reality of the theory with one or more fermion
flavours. In this context it is useful to note that in gauge theories, the
coupling will run with momentum as a consequence of quantum
corrections. This, in turn, brings about the need to renormalize the
theory. It is exactly this renormalization procedure which introduces a
scale in the theory, to which the generated fermion mass will be related.
An important consequence of quenching the theory is that the running of the
coupling disappears, the coupling in quenched QED is constant and no
renormalization is needed. It is therefore not clear what can set the scale
of the generated fermion mass, other than the ultraviolet cutoff, in
quenched QED without bare mass.

\section{Bare vertex}

\begin{figure}[htbp]
\begin{center}
\parbox{2.6cm}{\huge $E_f(p) =$}\hspace{-3mm}
\parbox{12cm}{\epsfig{file=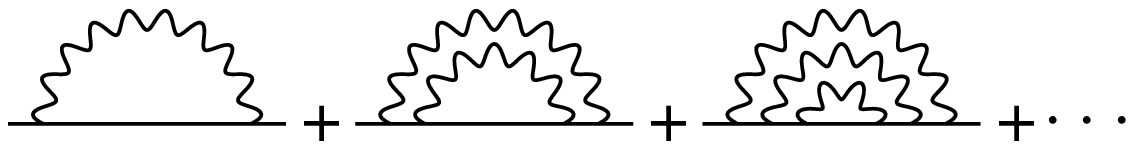,width=330pt}}
\end{center}
\vspace{-5mm}
\caption{Fermion self-energy in the rainbow approximation.}
\label{fig:Rainbow}
\end{figure}

In the simplest calculation in quenched QED, we replace the full vertex
$\Gamma^\mu(k,p)$ by the bare vertex $\gamma^\mu$. This is
called the {\it rainbow} approximation, which is obvious if we look at the
Feynman diagram decomposition of the fermion self-energy $\Self(p)$ shown in
Fig.~\ref{fig:Rainbow}. It is well known from the literature that fermion
mass is generated dynamically in the Rainbow approximation to QED provided
that the coupling is larger than a critical value, which is
$\alpha_c=\pi/3$ in the Landau
gauge~\cite{Fuk76,Fomin83,Mir85,Mir85b}. Because the bare vertex violates
the Ward-Takahashi identity, the critical coupling varies wildly if we go
to other gauges as shown in Ref.~\cite{CP93} by Curtis and Pennington.
They find $\alpha_c=1.69$ in the Feynman gauge~($\xi=1$) and $\alpha_c=
2.04$ in the Yennie gauge~($\xi=3$). Because of this strong dependence of
$\alpha_c$ on the covariant gauge parameter, we will investigate the
behaviour of the critical coupling using the Curtis-Pennington vertex
Ansatz in the next section.

Comparing these results with those of numerical lattice studies is not
straightforward. In fact the study of the SD equations of quenched QED
shows that the large anomalous dimension of the $\compop$ operator makes
this operator renormalizable~\cite{Bar86}. Therefore a four-fermion
interaction should in principle be included in the Lagrangian of quenched
QED. In Ref.~\cite{Kon89} Kondo et al.\ found the critical line, describing
the phase transition in quenched QED, in the $(\alpha,G)$-plane where
$\alpha$ is the usual QED coupling constant and $G$ is the strength of a
four-fermion interaction, and in Ref.~\cite{Bar90} Bardeen et al.\ studied
the corresponding critical scaling laws. This is important for the
comparison with lattice studies as the numerical simulation of quenched
non-compact QED appears to automatically incorporate the four-fermion
interaction in the calculation.  The lattice calculations also find a phase
transition but the critical point is situated somewhere on the critical
line~\cite{Koc90} rather than in the pure QED point~($G=0$).  Therefore the
critical coupling found in the lattice calculations is not directly
comparable with the value of $\pi/3$ found in the SD treatment of ``pure''
quenched QED. This is also true for the scaling law which is of Miransky
type for the SD treatment of quenched QED while the power-law scaling in
the lattice calculation coincides with a mixture of QED and four fermion
interaction.  Moreover, there is an additional problem as the chiral limit,
$m_0\to 0$, can only be retrieved through extrapolation in lattice studies.
Recently Kogut et al.~\cite{Kog94} have introduced the momentum space
lattice method in contrast to the conventional position space formulation.
They indicate that the method could avoid the contamination of QED by
four-fermion interactions and that it would then be possible to locate the
critical point of pure QED. \label{LatQuen}

\section{Curtis-Pennington vertex}
\label{Sec:abgpr}

\subsection{Introduction}

In this section we will discuss the dynamical generation of fermion mass in
quenched QED with the Curtis-Pennington vertex.  This study has been
performed independently in Durham and in Groningen and the common results
obtained, have been merged and published in {\it Critical Coupling in
Strong QED with Weak Gauge Dependence} by D. Atkinson, J.C.R.~Bloch,
V.P. Gusynin, M.R.~Pennington and M.~Reenders in Ref.~\cite{ABGPR}.  As
observed by Dong et al.~\cite{Dong94}, the regularization scheme used in
that paper was not translationally invariant and a spurious additional term
appeared in the equation for the fermion wavefunction renormalization. In
this section we will use the corrected equations. 

The Curtis-Pennington vertex not only ensures satisfaction of the
Ward-Takahashi identity and avoids singularities that would imply the
existence of a scalar, massless particle, but it also respects the
requirement of multiplicative renormalizability, a property of exact QED
that is destroyed by the rainbow approximation.  It agrees moreover with
perturbative results in the weak coupling limit.

Our study has been motivated by the previous numerical work performed by
Curtis and Pennington in Ref.~\cite{CP93} where the system of non-linear
equations for $\S$ and $\F$ was solved numerically in the Landau, Feynman
and Yennie gauge. They find a critical coupling $\alpha_c \approx 0.92$
which is almost exactly gauge independent, in complete contrast to the
rainbow approximation.  In Ref.~\cite{Atk93} Atkinson et al.\ use
bifurcation analysis to determine the critical coupling analytically using
the Curtis-Pennington vertex Ansatz in the Landau gauge. They introduce
various approximations to simplify the bifurcation equations and find
similar results, although with a much larger inaccuracy, with a critical
coupling ranging from 0.910 to 1.047, depending on the approximation.

We will consider the Schwinger-Dyson equations in a general covariant
gauge, with the Curtis-Pennington Ansatz, and apply bifurcation analysis to
them.  This involves calculating the functional derivative of the nonlinear
mapping of the mass function into itself.  Thanks to the scale-invariance
of the problem, the bifurcation equation can be solved by inspection, in
the limit that the ultraviolet cutoff is taken to infinity.  A solution
for the mass function is a power of the momentum that has to satisfy a
certain transcendental equation. The onset of criticality is heralded by
the coming together of two solutions of this transcendental equation, for
that is the indication that oscillatory takes over from non-oscillatory
behaviour.  

We find the gauge dependence of the critical coupling to be slight,
varying by only a few percent over a relatively large range of the
gauge parameter. This confirms the previous wholly numerical findings
of Curtis and Pennington~\cite{CP93}, which covered only small changes
of gauge. This weak gauge dependence is in marked contrast to the
rainbow approximation, for which the critical coupling changes by 60\%
between just the Landau and Feynman gauges~\cite{CP93}.

\subsection{Gauge independence of fermion mass and critical coupling}

The physical mass of the fermion is defined to be the lowest position at
which the denominator function in the fermion propagator,
\[
S(p)=\F(p^2)\frac{\gamma^\mu p_\mu + \S(p^2)}{p^2-\S^2(p^2)}\, ,
\]
has a zero, which is therefore a solution, $m$, of 
\[
m=\S(m^2)\, .
\]
On physical grounds, this singularity should be on the real timelike axis
of $p^2$ and should be gauge-independent. When we work in Euclidean space
we can either choose to determine the `Euclidean mass', which is the lowest
solution of $M=\S(M^2)$, and is not the same as the physical mass $m$, or
one might perhaps take $\S(0)$ as an ersatz {\it effective mass}. Both
approximations are not expected to be exactly gauge-invariant, but one
might hope them to be approximately so, on the grounds that they
should be close to the physical mass $m$, which {\em is} gauge-invariant,
at least in exact QED, or in a quenched approximation in which the first
two Ward-Takahashi identities are respected~\cite{Fry}.

The value of the wavefunction at an arbitrarily selected renormalization
point, $\mu$, is defined to be the wavefunction renormalization constant,
which is conventionally dubbed $Z_2$:
\[
Z_2=\F(\mu^2)\, .
\] 
It is convenient to choose the renormalization point to be Euclidean; 
the renormalized wave function is specified by 
\be
\tilde\F(x)=Z_2^{-1}\F(x) \, .
\mlab{Zren}
\ee
The Curtis-Pennington Ansatz defines a renormalizable scheme, so that in
it  $\tilde\F(x)$ has a finite limit as the ultraviolet
regularization is removed. The renormalized wavefunction contains no
explicit cutoff, but it is dependent on the renormalization point, and
on the gauge parameter. 

Chiral symmetry breaking occurs if the coupling, $\alpha$, is greater than
a certain critical value, $\alpha_c$. This critical coupling is potentially
a physically measurable quantity, since it signals a change of phase, and
so it should be gauge invariant. Although this is not exactly true in the
Curtis-Pennington system, it is approximately so.  Indeed, the requirement
that $\alpha_c$ be gauge-invariant could perhaps be used to specify further
the form of the Ansatz for the vertex function. The transverse part of the
vertex is not uniquely determined, and the above requirement might with
profit be used to refine this transverse part of the vertex as discussed in
Ref.~\cite{Bashir94}.

\subsection{Bifurcation analysis and critical point} 

The basic coupled integral equations for quenched QED with the
Curtis-Pennington vertex Ansatz are now derived by putting $N_f=0$ in
Eqs.~(\oref{1.1006}, \oref{1.1007}, \oref{1.1008}). The photon equation,
\mref{1.1008}, yields $\G(x) = 1$, so \mrefb{1.1006}{1.1007}) now become:
\ba
\frac{\S(x)}{\F(x)} &=& m_0 + \frac{\alpha}{2\pi^2} \int dy \, 
\frac{y\F(y)}{\Ds{y}} \int d\theta \, \frac{\sin^2\theta}{z} \, \mlab{1.1009} \\
&& \hspace{10mm} \times \l\{  3\S(y) \l[A(y,x) + \tau_6(y,x)(y-x)\r]
- \frac{1}{\F(x)}\l[\frac{\S(y)-\S(x)}{y-x}\r]
\frac{2yx\sin^2\theta}{z} \r\} \nn \\
&& + \frac{\alpha\xi}{4\pi\F(x)} \int dy \, \frac{\F(y)}{\Ds{y}}
\l\{\frac{y\S(y)}{x}\theta(x-y) + \S(x)\theta(y-x) \r\} \nn\\[10mm]
\frac{1}{\F(x)} &=& 1 - \frac{\alpha}{2\pi^2 x} \int dy \,
\frac{y\F(y)}{y+\S^2(y)} \int d\theta \, \sin^2 \theta \, \mlab{1.1010} \\
&& \times \l\{
A(y,x)\l[\frac{2yx\sin^2\theta}{z^2} - \frac{3\sqrt{yx}\cos\theta}{z}\r] 
\right. \nn \\
&& \hspace{6mm}+\l[B(y,x)(y+x)-C(y,x)\S(y)\r]\frac{2yx\sin^2\theta}{z^2} \nn \\
&& \hspace{-12mm} \left. \phantom{\frac{2yx\sin^2\theta}{z^2}} 
- \tau_6(y,x)(y-x)\frac{3\sqrt{yx}\cos\theta}{z} 
\r\} \nn \\
&& - \frac{\alpha\xi}{4\pi\F(x)} \int dy \, 
\frac{\F(y)}{y+\S^2(y)} 
\l\{ \frac{y\S(x)\S(y)}{x^2}\theta(x-y) - \theta(y-x) \r\} \;. \nn
\ea

The complicated kernels are explicit functions of $\S$ and $\F$. In the
quenched case the angular integrals of the fermion equations,
\mrefb{1.1009}{1.1010}, can be computed analytically and are given in 
Appendix~\ref{App:angint}. Substituting \mrefb{A1}{A2} in \mref{1.1009} and
putting the bare mass $m_0=0$, yields:
\ba
\frac{\S(x)}{\F(x)} &=& \frac{3\alpha}{8\pi} \int dy \, 
\frac{y\F(y)}{\Ds{y}} \mlab{1.1017} \\
&& \hspace{5mm} \times \l\{ 2\l[A(y,x) + \tau_6(y,x)(y-x)\r] \S(y)
\l[\frac{\theta(x-y)}{x}+\frac{\theta(y-x)}{y}\r] \r. \nn\\
&& \hspace{10mm} \l. - \frac{1}{\F(x)}\frac{\S(y)-\S(x)}{y-x}
\l[\frac{y}{x}\theta(x-y)+\frac{x}{y}\theta(y-x)\r] \r\} \nn \\
&& + \frac{\alpha\xi}{4\pi\F(x)} \int dy \, \frac{\F(y)}{\Ds{y}}
\l\{\frac{y\S(y)}{x}\theta(x-y)+\S(x)\theta(y-x)\r\} \;. \nn
\ea

When substituting \mrefb{A2}{A3} in \mref{1.1010}, the $A(y,x)$ term
vanishes (which is similar to the bare vertex case) and \mref{1.1010} now
becomes:
\ba
\frac{1}{\F(x)} &=& 1 - \frac{3\alpha}{8\pi x} \int dy \,
\frac{y\F(y)}{y+\S^2(y)} \, \mlab{1.1016} \\
&& \times \l\{
\Big[B(y,x)(y+x)-C(y,x)\S(y) - \tau_6(y,x)(y-x)\Big]
\l[\frac{y}{x}\theta(x-y)+\frac{x}{y}\theta(y-x)\r] \r\} \nn \\
&& - \frac{\alpha\xi}{4\pi\F(x)} \int dy \, 
\frac{\F(y)}{y+\S^2(y)} 
\l[\frac{y\S(x)\S(y)}{x^2}\theta(x-y)-\theta(y-x)\r] \;. \nn
\ea

As can be seen from \mrefb{1.1017}{1.1016}, the complete Curtis-Pennington
equations are nonlinear and complicated. Clearly $\S(x)\equiv 0$ is always
a possible solution; but it is not the one in which we are interested.
However, the equations simplify at the critical point, where a nontrivial
solution {\em bifurcates} away from the trivial one. To investigate this
critical point, we have to take the 
functional derivative of the nonlinear operators with respect to $\S(x)$
and evaluate it at the trivial `point', $\S(x)\equiv 0$. This amounts in
fact simply to throwing away all terms that are quadratic or higher in the
mass function. It must be emphasized that this is {\em not} an
approximation: it is a precise manner to locate the critical point by
applying bifurcation theory.

We now apply bifurcation analysis to \mrefb{1.1017}{1.1016}.  After
substitution of the expressions for $A$, $B$, $C$ and $\tau_6$,
\mref{105}, and neglecting the terms of $\Order(\S^2)$, the $\F$-equation,
\mref{1.1016}, is reduced to
\be
\frac{1}{\F(x)} = 1 
+ \frac{\alpha\xi}{4\pi} \int_0^{\Lambda^2} \frac{dy}{y} \,
\frac{\F(y)}{\F(x)} \theta(y-x) 
\mlab{109}
\ee
where the UV-cutoff $\Lambda^2$ has been introduced to regularize the
integral.  It is important to note that now the $\F$-equation is
independent of $\S$. After multiplying both sides with $\F(x)$ and applying
the step function, this gives:
\be
\F(x) = 1 - \frac{\alpha\xi}{4\pi} \int_x^{\Lambda^2} dy \, \frac{\F(y)}{y} \;.
\mlab{110}
\ee

It is easy to check that the unique solution to this equation is:
\be
\F(x) = \l(\frac{x}{\Lambda^2}\r)^\nu
\mlab{111}
\ee

where
\be
\nu = \frac{\alpha\xi}{4\pi} .
\mlab{112}
\ee

Next we apply the bifurcation analysis to the $\S$-equation,
\mref{1.1017}, neglecting terms of $\Order(\S^2)$, and find: 
\ba
\frac{\S(x)}{\F(x)} &=& \frac{3\alpha}{8\pi} \int_0^{\Lambda^2} dy \, 
\l\{\l(1+\frac{\F(y)}{\F(x)} 
+ \frac{y+x}{y-x}\l[1-\frac{\F(y)}{\F(x)}\r] \r) \S(y)
\l[\frac{\theta(x-y)}{x}+\frac{\theta(y-x)}{y}\r] \r.  \nn \\
&& \hspace{20mm} \l. - \frac{\F(y)}{\F(x)}\frac{\S(y)-\S(x)}{y-x}
\l[\frac{y}{x}\theta(x-y)+\frac{x}{y}\theta(y-x)\r] \r\} \mlab{114} \\
&& + \frac{\alpha\xi}{4\pi} \int_0^{\Lambda^2} \frac{dy}{y} \, 
\frac{\F(y)}{\F(x)}
\l\{\frac{y}{x}\S(y)\theta(x-y)+\S(x)\theta(y-x)\r\} \;. \nn
\ea

The second term of the $\xi$-part of \mref{114} is identical to the
integral in the $\F$-equation, \mref{109}, and can be replaced by
$(1/\F(x)-1)$. Then, the left hand side of \mref{114} cancels, the
integrals are now finite and need not be regularized anymore, as a
consequence of the multiplicative renormalizability of the fermion
propagator in the Curtis-Pennington approximation. However, the limit
$\Lambda \rightarrow \infty$ has to be taken in a proper way to respect the
axial current conservation~\cite{Fomin83,CP93,Atkin87}. If not, one will
wrongly find that chiral symmetry breaking occurs for all values of the
coupling~\cite{CP92}.  Therefore the UV-cutoff can only be taken to
infinity if the boundary conditions imposed by \mref{114}
at $x=\Lambda^2$ are satisfied, as will be ensured later in this
section. After substituting the solution \mref{111} for $\F(x)$ and
eliminating $\alpha$ using $\alpha = 4\pi\nu/\xi$, the $\S$-equation,
\mref{114}, becomes:
\ba
\S(x) &=& \frac{3\nu}{2\xi} \int_0^\infty dy \, 
\l\{\l(1+\l(\frac{y}{x}\r)^\nu + \frac{y+x}{y-x}
\l[1-\l(\frac{y}{x}\r)^\nu \r] \r) \S(y) 
\l[\frac{\theta(x-y)}{x}+\frac{\theta(y-x)}{y}\r] \r. \mlab{116} \\
&& \l. - \l(\frac{y}{x}\r)^\nu\frac{\S(y)-\S(x)}{y-x}
\l[\frac{y}{x}\theta(x-y)+\frac{x}{y}\theta(y-x)\r] \r\} 
 + \nu \int_0^\infty dy \, \l(\frac{y}{x}\r)^\nu
\S(y)\frac{\theta(x-y)}{x} \;.
\nn
\ea

The last equation is scaling invariant, and is solved by 
\be
\S(x) = x^{-s}
\mlab{117}
\ee
as will be shown below.

After substituting \mref{117} in \mref{116} we find:
\ba
x^{-s} &=& \frac{3\nu}{2\xi} \int_0^\infty dy \, 
\l\{\bigg(y^{-s}+\frac{y^{\nu-s}}{x^\nu}  + \frac{y+x}{y-x}
\bigg[y^{-s}-\frac{y^{\nu-s}}{x^\nu}\bigg] \bigg) 
\l[\frac{\theta(x-y)}{x}+\frac{\theta(y-x)}{y}\r] \r. \mlab{118} \\
&& \hspace{10mm} \l. - \l(\frac{y}{x}\r)^\nu\frac{y^{-s}-x^{-s}}{y-x}
\l[\frac{y}{x}\theta(x-y)+\frac{x}{y}\theta(y-x)\r] \r\} 
+ \nu \int_0^\infty dy \, \frac{y^{\nu-s}}{x^{\nu+1}} \theta(x-y) \;. \nn
\ea

We now divide \mref{118} by $x^{-s}$, change variables to $t=y/x$, and
apply the step functions, giving:
\ba
1 &=& 
\frac{3\nu}{2\xi} \int_0^1 dt \, \l\{ 
\l[t^{-s}+t^{\nu-s}\r]
+ \frac{t^{-s+1}+t^{-s}-2t^{\nu-s+1}-t^{\nu-s}+t^{\nu+1}}{t-1}
\r\} \mlab{120}\\
&+& \frac{3\nu}{2\xi} \int_1^\infty dt \, \l\{
\l[t^{-s-1}+t^{\nu-s-1}\r]
+ \frac{t^{-s}+t^{-s-1}-t^{\nu-s}-2t^{\nu-s-1}+t^{\nu-1}}{t-1} 
\r\} \nn\\
&+& \nu \int_0^1 dt \, t^{\nu-s} \;. \nn
\ea

After putting the terms on a common denominator, we get:
\ba
1 &=& 
\frac{3\nu}{2\xi} \int_0^1 dt \, \l\{ 
\frac{2t^{-s+1}-t^{\nu-s+1}-2t^{\nu-s}+t^{\nu+1}}{t-1}
\r\} \mlab{121}\\
&+& \frac{3\nu}{2\xi} \int_1^\infty dt \, \l\{
\frac{2t^{-s}-3t^{\nu-s-1}+t^{\nu-1}}{t-1} 
\r\} 
+ \nu \int_0^1 dt \, t^{\nu-s} \;. \nn
\ea

To solve these integrals we will use Eq.~(3.231.5) of Ref.~\cite{Grads},
which can be written as:
\be
\int_0^1 dt \; \frac{t^{\mu-1} - t^{\nu-1}}{t-1} =
\psi(\mu) - \psi(\nu) \qquad [\mbox{Re}(\mu)>0, \mbox{Re}(\nu)>0] . 
\mlab{122}
\ee

From this integral we also derive the following integral:
\be
\int_1^\infty du \; \frac{u^{-\mu} - u^{-\nu}}{u-1} =
-\psi(\mu) + \psi(\nu)  \qquad [\mbox{Re}(\mu)>0, \mbox{Re}(\nu)>0] . 
\mlab{123}
\ee

To show this, we change variables $t=1/u$ in \mref{123}:
\be
\int_1^\infty du \; \frac{u^{-\mu} - u^{-\nu}}{u-1} 
= \int_1^0 -\frac{dt}{t^2} \; \frac{t^{\mu} - t^{\nu}}{t^{-1}-1}
= - \int_0^1 dt \; \frac{t^{\mu-1} - t^{\nu-1}}{t-1} 
\mlab{124}
\ee
and apply \mref{122} to \mref{124}, yielding \mref{123}.

We now substitute the integral evaluations \mrefb{122}{123} in
\mref{121}. This gives:
\ba
1 &=& \frac{3\nu}{2\xi} \Big[ 
2\psi(-s+2) - \psi(\nu-s+2) - 2\psi(\nu-s+1) + \psi(\nu+2)
 \mlab{125}\\
&& \hspace{5mm} {} - 2\psi(s) + 3\psi(-\nu+s+1) - \psi(-\nu+1)
\Big] + \frac{\nu}{\nu-s+1} \nn
\ea
where the region of the $(s,\nu)$-plane for the convergence of the
integrals in \mref{121} is specified by:
\vspace{-8mm}
\ba
0  <& s &< 2 \nn\\[-8pt]
-2 <& \nu &< 1 \mlab{126}\\[-8pt]
-1 <& \nu-s &< 1 \;. \nn
\ea
We note that each of these inequalities has one limit imposed by requiring
the integrals to be infrared finite, while the other comes from the
ultraviolet side.

We now mention two properties of the $\psi$-function(see Eq.~(8.365.1,
8.365.8) of Ref.~\cite{Grads}):
\ba
\psi(z+1) &=& \psi(z) + \frac{1}{z} \mlab{127} \\
\psi(1-z) &=& \psi(z) + \pi \cot z\pi . \mlab{129}
\ea

Applying \mrefb{127}{129} to \mref{125} and bringing the term in the left hand
side to the right gives:
\ba
f(\xi,\nu, s) &\equiv& 
\frac{3\nu}{2\xi} \Big[ 
2\pi\cot s\pi -\pi\cot\nu\pi + 3\pi\cot(\nu-s)\pi \mlab{132}\\
&& + \frac{2}{1-s}  + \frac{1}{\nu+1} + \frac{1}{\nu}
 - \frac{3}{\nu-s} - \frac{1}{\nu-s+1} \Big] 
- \frac{1-s}{\nu-s+1} = 0. \nn
\ea

This means that \mref{117} is a solution to \mref{116} if $s$ satisfies
\mref{132} together with the convergence conditions \mref{126}.

In a chosen gauge specified by $\xi$, this equation defines roots $s$ for
any value of the coupling $\alpha$. Bifurcation occurs when two of these
roots (with $\nu$ and $s$ satisfying \mref{126}) are equal. Then $\alpha
\equiv \alpha_c$. A necessary condition for equality of two roots is 
${\partial f(\xi,s,\nu)}/{\partial s}=0$, i.e.
\ba
g(\xi,s,\nu) \equiv \parder{f(\xi,s,\nu)}{s} &=&
\frac{3\nu}{2\xi} \Big[ 
-2\pi^2\csc^2 s\pi + 3\pi^2\csc^2(\nu-s)\pi \mlab{134}\\
&& + \frac{2}{(1-s)^2}  - \frac{3}{(\nu-s)^2} - \frac{1}{(\nu-s+1)^2} \Big] 
+ \frac{\nu}{(\nu-s+1)^2} = 0. \nn
\ea

To find the critical point we have to solve the coupled system of
transcendental equations \mrefb{132}{134} for $\nu_c$ and $s_c$, together with
the convergence conditions \mref{126}. The numerical
program~\cite{Numrecep} used to solve this system of equations requires a
realistic starting guess in order to find the solution. For this purpose, as
well as to understand how and when bifurcation happens, it is useful
to consider first the situation in the Landau gauge, $\xi=0$ i.e. $\nu =0$
and $\nu/\xi = \alpha/4\pi$.  Then
\mref{132} simplifies to:
\be
f(\xi=0, \nu, s) \equiv f_0(\alpha,s) =
\frac{3\alpha}{8\pi} \Big[ - \pi\cot s\pi \mlab{133} + \frac{1}{1-s}
 + \frac{3}{s} + 1 \Big] - 1 = 0, \nn
\mlab{135}
\ee
and the conditions, \mref{126}, with $\nu=0$, are reduced to $ 0 < s < 1$.

\begin{figure}[htbp]
\begin{center}
\mbox{\epsfig{file=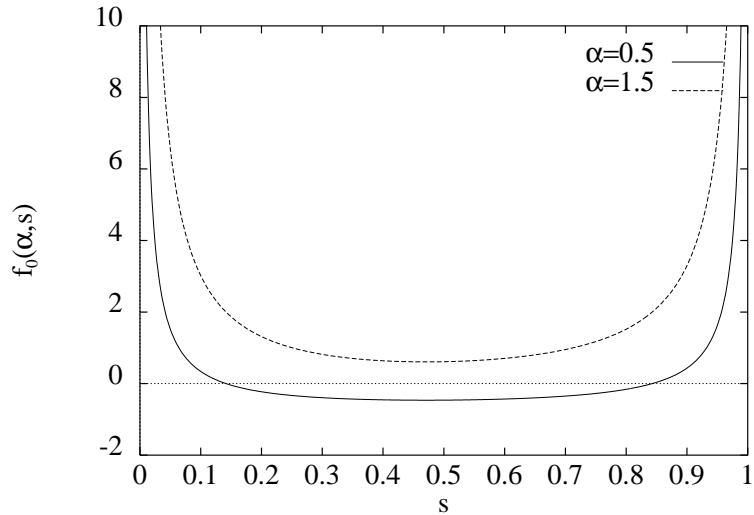,height=8cm,angle=-90}}
\end{center}
\vspace{-0.5cm}
\caption{Function $f_0(\alpha,s)$ versus exponent $s$. To satisfy the
integral equation \protect\mref{116} in the Landau gauge, the exponent $s$
has to satisfy $f_0(\alpha,s)=0$.}
\label{Fig:f0}
\end{figure}

From the plot of this function for $0<s<1$, Fig.~\ref{Fig:f0}, we see that
it has just two real roots in this interval when $\alpha$ is small. As
$\alpha$ is increased, they approach one another, becoming equal at
criticality. To find this critical point we take the derivative of
\mref{135} with respect to $s$:
\be
g(\xi=0,s,\nu) \equiv g_0(s) =
\pi^2\csc^2 s\pi + \frac{1}{(1-s)^2} - \frac{3}{s^2}  = 0.
 \mlab{136}
\ee

\begin{figure}[htbp]
\begin{center}
\mbox{\epsfig{file=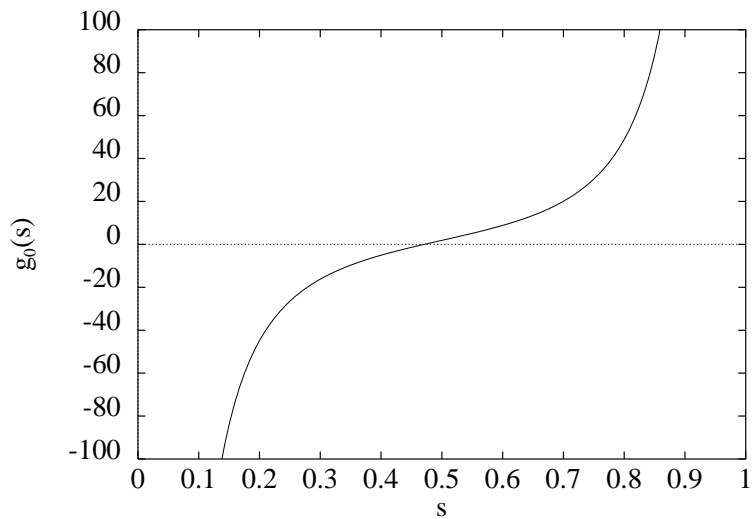,height=8cm,angle=-90}}
\end{center}
\vspace{-0.5cm}
\caption{Function $g_0(s)$ versus exponent $s$. To determine the critical
point of the integral equation \protect\mref{116} in the Landau gauge, the
exponent $s_c$ has to satisfy $g_0(s_c)=0$.}
\label{Fig:g0}
\end{figure}

We plot the function $g_0(s)$ from \mref{136} in Fig.~\ref{Fig:g0}.
\mref{136} is a single transcendental equation which determines the
exponent $s_c$ in the critical point. The numerical program used to solve
it, finds $s_c=0.470966$. Substituting this value for $s_c$ in \mref{135}
allows us to compute the value of the critical coupling, which is
$\alpha_c(\xi=0)=0.933667$. The boundary conditions imposed by \mref{114}
at $x=\Lambda^2$ demand that the behaviour of the mass function be
oscillatory, and that implies that the roots in \mref{135} are
complex. Thus only for $\alpha$ greater than $\alpha_c$ does \mref{114}
have a non-zero solution for $\S(x)$: only then can chiral symmetry
breaking occur.

To find the solutions in other than the Landau gauge we will look for
solutions of the system of equations, \mrefb{132}{134}, which are
continuously connected to the one found in the Landau gauge. We will start
from values of the gauge close to zero and work our way up and down to
positive and negative values of $\xi$, using the solution at the previous
$\xi$-value as starting guess for the new calculation.

The solutions for $\nu_c$ and $s_c$ are shown in
Fig.~\ref{Fig:nu_c,s_c}. We only find solutions satisfying the convergence
conditions, \mref{126}, as long as $\xi>-3$. For $\xi=-3$ one can show from
\mref{121} that the $\xi$-term causes an additional cancellation and the
integrals of \mref{116} are still convergent. Below this, for $\xi<-3$, the
condition $\nu-s>-1$ is not satisfied anymore: the transcendental
equation, \mref{132}, is not equivalent anymore to the integral equation,
\mref{116}, which becomes infrared divergent. For positive values of $\xi$,
however large, we always find a solution which satisfies the conditions,
\mref{126}, needed for the convergence of the integrals of \mref{116}.

\Comment{However, for
$\alpha >\alpha_c$ this potential divergence is suppressed by terms
quadratic in the mass-function: $y$ is replaced by $y+\S^2(y)$ at crucial
places in the denominators~\cite{Atk93}.  The solution then no longer has
exactly the power form of \mref{117}, but asymptotically $[\, x >>
\S^2(x)\, ]$ this behaviour is still valid, and this is all we need to
extend the bifurcation analysis. BUT THERE IS NO CRITICAL COUPLING FOR
$\xi<-3$, SO HOW CAN $\alpha>alpha_c$ ? or
The infrared singularity for $\xi<-3$ is in fact a consequence of
oversimplifying the use of bifurcation analysis. In this case one is not
allowed to drop the $\M^2$ term in the $k^2+\M^2$ denominator of
the Curtis-Pennington equations. This means that the non-linear
equations will not hold anymore and have to be replaced by other
conditions.
}

\begin{figure}[htbp]
\begin{center}
\mbox{\epsfig{file=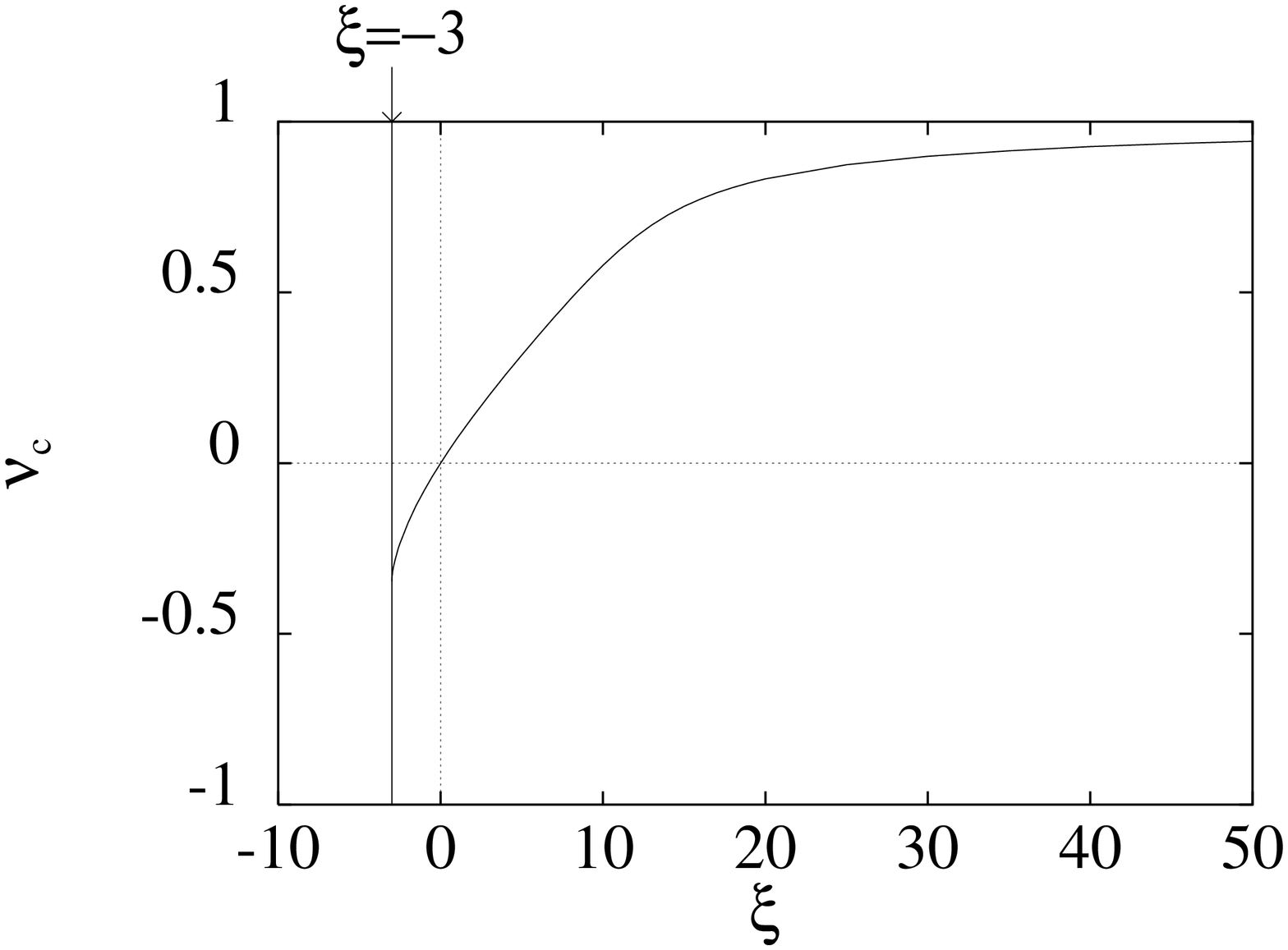,height=5.5cm,angle=-90}
\epsfig{file=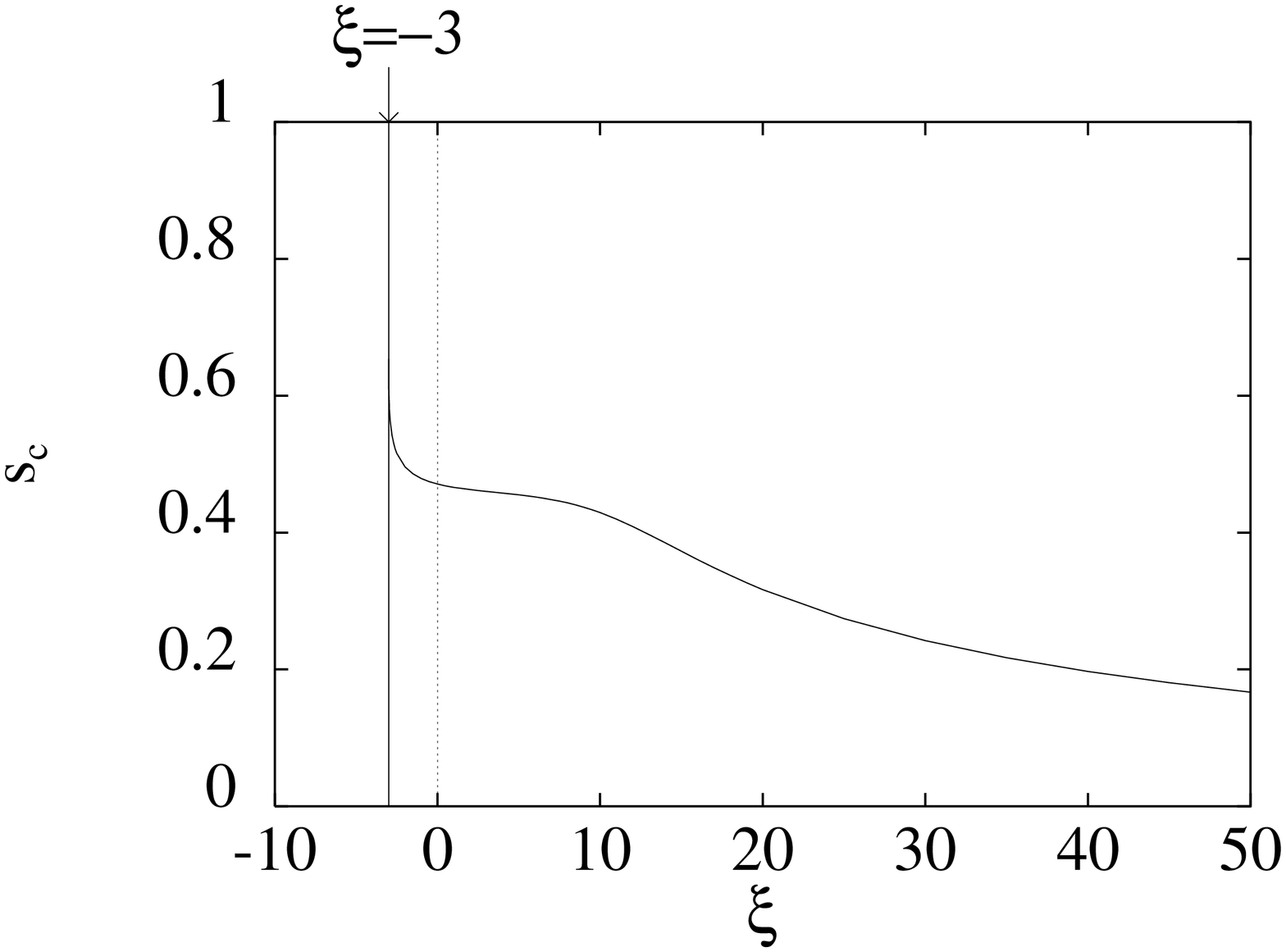,height=5.5cm,angle=-90}}
\end{center}
\vspace{-0.5cm}
\caption{Solutions $\nu_c$ and $s_c$ versus covariant gauge parameter $\xi$.}
\label{Fig:nu_c,s_c}
\end{figure}

\begin{figure}[htbp]
\begin{center}
\mbox{\epsfig{file=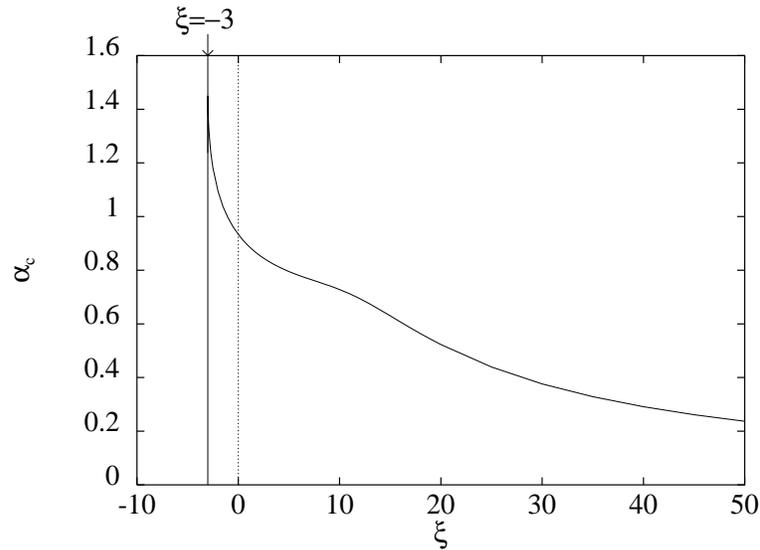,height=8cm,angle=-90}}
\end{center}
\vspace{-0.5cm}
\caption{Critical coupling, $\alpha_c$, as a function of the covariant gauge
parameter, $\xi$.}
\label{Fig:alpha_c}
\end{figure}

From the solution $\nu_c$ and \mref{112} we compute $\alpha_c(\xi)$ as a
function of $\xi$. The variation of the critical coupling with the
covariant gauge parameter is shown in Fig.~\ref{Fig:alpha_c}, where we have
plotted the critical coupling $\alpha_c$ against $\xi$ over the rather
large domain $-3\le \xi \le 50$. The results are in agreement with those of
Ref.~\cite{CP93} where the system of non-linear coupled equations,
\mrefb{1.1017}{1.1016}, was solved numerically at $\xi =0,\, 1$ and $3$.  
We can compare these results with those obtained in the rainbow
approximation~\cite{CP93}, where the bare vertex is used instead of the
Curtis-Pennington vertex. The values of the critical coupling for
$\xi=0,1,3$ are compared in Table~\ref{Tab:comp_alpha}. We note the
reassuringly weak gauge dependence of the critical coupling in the CP-case.

\begin{table}[htbp]
\begin{center}
\begin{tabular}{|c|c|c|}
\hline
$\xi$ & CP-vertex & Rainbow \\
\hline
0 & 0.933667 & 1.047 \\
1 & 0.890712 & 1.690 \\
3 & 0.832927 & 2.040 \\
\hline
\end{tabular}
\caption{Critical coupling, $\alpha_c$, for $\xi = 0,1,3$ with the 
Curtis-Pennington vertex and in the rainbow approximation.}
\label{Tab:comp_alpha}
\end{center}
\end{table}

In solving the bifurcation equation, we have at the same time found the
exponent $s$ of \mref{117}. This too is only weakly gauge dependent in a
sizeable region of $\xi$, as can be seen in Fig.~\ref{Fig:nu_c,s_c}. For
instance, in the Landau gauge ($\xi=0$), $s=0.4710$, while with $\xi=5$,
$s=0.4551$. This exponent determines the ultraviolet behaviour of the
mass-function $\S(x)$ and is consequently related to $\gamma_m$, the
anomalous dimension of the $\overline{\psi}\psi$ operator by
$\gamma_m=2(1-s)$.  Thus in the Landau gauge $\gamma_m = 1.058$, close to
the value 1 that holds in the rainbow approximation and Holdom claims is
exactly true in all gauges~\cite{Bob}.

The fact that the variation of the critical coupling is small over a
sizeable region of the gauge parameter indicates the superiority of the
Curtis-Pennington vertex over the bare vertex as well as over previous
Ans\"atze for the vertex function~\cite{Rem90,Kondo,Hae91,me} made in
the past in an attempt to improve on the ladder approximation.

In this section we have determined the critical point, where the generation
of fermion mass sets in, in quenched QED with the Curtis-Pennington
vertex. In the next section we will investigate how this generated mass
scales when we increase the coupling above its critical value.

\section{Scaling law: mass generation versus coupling}
\def\half{\frac{1}{2}}

One of the most interesting features in the study of mass generation as a
consequence of chiral symmetry breaking is the scaling law relating the
scale of the generated fermion mass M to the coupling $\alpha$, when this
coupling is larger than, but still very close to its critical value
$\alpha_c$. It is generally thought that this scaling in quenched QED can
be described by what is often referred to as the {\it Miransky scaling
law}~\cite{Mir85b}:
\be
\frac{\Lambda}{M} =
\exp{\left(\frac{A}{\sqrt{\frac{\alpha}{\alpha_c}-1}} - B\right)} \;.
\mlab{1200}
\ee

The scaling law is important as it can be related to the triviality of the
theory. It is thought that a mean field scaling law indicates that the
theory is trivial, while any departure from it opens the possibility for
the continuum theory to remain interactive. From the Miransky scaling law,
one can show that the critical coupling $\alpha_c$ can be interpreted as an
ultraviolet fixed point of the continuum theory.

We will first prove this formula in the rainbow approximation. Then, we
will apply the same method to the Curtis-Pennington vertex to show that
\mref{1200} remains valid and to determine the coefficients $A$ and $B$.

\subsection{Bare vertex}
\label{Sec:scalbare}

In the rainbow approximation to QED we know that $\F(x)\equiv 1$ and the 
mass equation is given by:
\be
\S(x) = \frac{3\alpha}{4\pi} \left[ \frac{1}{x} \int_0^{x}
dy \frac{y \S(y)}{y + \S^2(y)} + \int_{x}^{\Lambda^2}
dy \frac{\S(y)}{y + \S^2(y)}\right] \;.
\mlab{SD}
\ee

We now want to linearize this equation to make it tractable. By doing this
\mref{SD} becomes scale invariant and all information about the scale
of the generated mass is lost. This can be remedied by introducing an
IR-cutoff in the linearized equation. To retain the correct scale of the
generated mass in the linearized equation, the IR-cutoff $\kappa$ has to
satisfy $\S(\kappa^2) = \kappa$. This can be understood by noting that
below this cutoff the original integral in \mref{SD} is negligible while
the linearized equation would give a big contribution; above the cutoff
both integrals will be similar. The linearized $\S$-equation is:
\be
\S(x) = \frac{3\alpha}{4\pi} \left[ \frac{1}{x} \int_{\kappa^2}^{x}
dy \S(y) + \int_{x}^{\Lambda^2}
dy \frac{\S(y)}{y}\right] \;.
\mlab{SD-IR}
\ee

To solve this integral equation it is usual to transform it to a
differential equation, through successive differentiation, with boundary
conditions derived from the integral equation.  Differentiating
\mref{SD-IR} once with respect to $x$ gives:
\be
\S'(x) = - \frac{3\alpha}{4\pi} \frac{1}{x^2} \int_{\kappa^2}^{x}
dy \S(y) \;.
\mlab{SD'}
\ee

Multiplying \mref{SD'} by $x^2$ and differentiating once more with respect
to $x$ gives the following differential equation:
\be
x^2 \S''(x) + 2x \S'(x) + \frac{3\alpha}{4\pi} \S(x) = 0 \;.
\mlab{Diffeq}
\ee

This is a standard differential equation which has solutions of the
form:
\be
\S(x) = x^{-s} \;.
\mlab{power}
\ee

Substituting the solution \nref{power} in \mref{Diffeq} gives the following
condition for $s$:
\be
s^2 - s + \frac{3\alpha}{4\pi} = 0 \;.
\ee

Therefore, the exponent $s$ of the solution \mref{power} has the following
values:
\be
s_{1,2} = \frac{1}{2} \pm \frac{1}{2} \sqrt{1-\frac{\alpha}{\alpha_c}}
\ee
with $\alpha_c = \pi/3$.

The general solution for $\S(x)$ can thus be written as:
\be
\S(x) = C_1 \,x^{-s_1} + C_2\, x^{-s_2} \;.
\ee

If the value of the coupling $\alpha$ is larger than the critical
value $\alpha_c$ this solution can be written as:
\be
\S(x) = C_1 \,x^{-\frac{1}{2} - \frac{i}{2}\tau} 
+ C_2 \,x^{-\frac{1}{2} + \frac{i}{2}\tau}
\mlab{sol}
\ee
where $\tau = \sqrt{\frac{\alpha}{\alpha_c} - 1}$ is real.

The coefficients $C_1$ and $C_2$ of \mref{sol} have to be determined from the
boundary conditions which are derived from \mrefb{SD-IR}{SD'}. These
boundary conditions are:
\be
[x\S(x)]'(\Lambda^2) = 0
\mlab{BC1}
\ee
and
\be
\S'(\kappa^2) = 0 \, .
\mlab{BC2}
\ee

Substituting the solution, \mref{sol}, in the UV and IR boundary
conditions, \mrefb{BC1}{BC2}, yields:
\be
C_1 \l(\frac{1}{2}-\frac{i \tau}{2}\r) \Lambda^{-i \tau} + C_2 \l(\frac{1}{2}+
\frac{i \tau}{2}\r) \Lambda^{i \tau} = 0
\mlab{Lambda}
\ee
and
\be
C_1 \l(-\frac{1}{2}- \frac{i \tau}{2}\r) \kappa^{-i \tau} + C_2 \l(-\frac{1}{2}+
\frac{i \tau}{2}\r) \kappa^{i \tau} = 0 \;.
\mlab{kappa}
\ee

Eliminating $C_1$ and $C_2$ from \mrefb{Lambda}{kappa} gives:
\be
\left( \frac{\Lambda}{\kappa} \right) ^ {2i \tau} =
\frac{(\frac{1}{2}-\frac{i\tau}{2})^2}{(\frac{1}{2}+\frac{i\tau}{2})^2}
= \frac{r^2 \exp(-2i\theta)}{r^2 \exp(2i\theta)}
= \exp(-4i\theta)
\mlab{1201}
\ee
where $r=\frac{1}{2}\sqrt{1 + \tau^2}$ and $\theta = \arctan(\tau)$.

From \mref{1201} we find:
\be
2\tau \ln\left(\frac{\Lambda}{\kappa}\right) = -4\arctan(\tau) + 2k\pi
\, .
\ee

If $\alpha$ is close to $\alpha_c$ we can expand the $\arctan$ in a
Taylor series, this yields:
\be
\ln\left(\frac{\Lambda}{\kappa}\right) = \frac{k\pi}{\tau} - 2 \;.
\ee

The ground state of the system is found for k=1, and so the generated mass
is related to the coupling as:
\be
\frac{\Lambda}{\kappa} =
\exp\left(\frac{\pi}{\sqrt{\frac{\alpha}{\alpha_c}-1}} - 2  \right)
\, .
\mlab{scaling-1.2}
\ee

The use of a hard IR-cutoff $\kappa$ in \mref{SD-IR} is in fact quite crude,
as the low momentum behaviour of the mass function, \mref{SD}, is not
approximated very well.  A better approximation consists in replacing the
quadratic term $\S^2(y)$ in the denominator of
\mref{SD} by a constant term $m^2 = \S^2(0)$. This seems more realistic, as the
mass term is approximately constant at low momentum and is negligible at
large momentum:
\be
\S(x) = \frac{3\alpha}{4\pi} \left[ \frac{1}{x} \int_0^{x}
dy \frac{y \S(y)}{y + m^2} + \int_{x}^{\Lambda^2}
dy \frac{\S(y)}{y + m^2}\right] \;.
\ee

From this integral equation one derives the following differential
equation:
\be
x^2 \S''(x) + 2 x \S'(x) + \frac{3\alpha}{4\pi}
\frac{x}{x+m^2} \S(x) = 0 \;.
\mlab{diffeq2}
\ee

To find the solutions of this differential equation we introduce the new
variable $z=-x/m^2$ and find:
\be
z(1-z)\S''(z) + 2(1-z)\S'(z) - \frac{3\alpha}{4\pi}\S(z) = 0 \;.
\ee

From Section~9.15 of Ref.~\cite{Grads} we know that the hypergeometric
function $F(a,b;c;z)$ satisfies the following differential equation:
\be
\left[z(1-z)\frac{d^2}{dz^2} + [c-(a+b+1)z]\frac{d}{dz} - ab\right] 
F(a,b;c;z) = 0 \;.
\ee

This means that, using the normalization condition $\S(0) = m$, the
mass function is given by:
\be
\S(x) = m F\l(\frac{1}{2}+\sigma,\frac{1}{2}-\sigma; 2;
-\frac{x}{m^2}\r)
\mlab{sol2}
\ee
where $\sigma = \frac{1}{2}\sqrt{1-\frac{\alpha}{\alpha_c}}$.

From \mref{diffeq2} we derive the UV boundary condition, which is
\be
[x\S(x)]'(\Lambda^2) = 0 \;.
\mlab{1202}
\ee

A property of hypergeometric functions is~\cite{Abram}:
\be
\frac{d}{dz}[z^{c-1}F(a,b;c;z)] = (c-1)z^{c-2} F(a,b;c-1;z) \;.
\mlab{1202.1}
\ee

Substituting the solution, \mref{sol2}, in the boundary condition,
\mref{1202}, and applying the property, \mref{1202.1}, gives:
\be
 m F(\frac{1}{2}+\sigma,\frac{1}{2}-\sigma; 1;-\frac{\Lambda^2}{m^2})
= 0 \;.
\mlab{UVBC}
\ee

To simplify this last expression we will make use the following
equality~\cite{Abram}:
\ba
F(a,b,c;z) &=&
\frac{\Gamma(c)\Gamma(b-a)}{\Gamma(b)\Gamma(c-a)}(-z)^{-a}
F(a,1+a-c;1+a-b;z^{-1}) \mlab{1203}\\
&+& \frac{\Gamma(c)\Gamma(a-b)}{\Gamma(a)\Gamma(c-b)}(-z)^{-b}
F(1+b-c,b;1+b-a;z^{-1}) \;. \nn
\ea

Applying \mref{1203} to the UV boundary condition, \mref{UVBC}, when
$\Lambda^2 \gg m^2$ and keeping the leading order terms yields:
\be
m\frac{\Gamma(-2\sigma)}{\Gamma^2(\frac{1}{2}-\sigma)}
\left(\frac{\Lambda^2}{m^2}\right)^{-\frac{1}{2}-\sigma} + m\frac{\Gamma(2\sigma)}{\Gamma^2(\frac{1}{2}+\sigma)}
\left(\frac{\Lambda^2}{m^2}\right)^{-\frac{1}{2}+\sigma} = 0 \, .
\mlab{eq1}
\ee

When $\alpha > \alpha_c$ we write $\sigma=\frac{i}{2}\tau$ where
$\tau=\sqrt{\frac{\alpha}{\alpha_c}-1}$. \mref{eq1} then becomes:
\be
\left(\frac{\Lambda^2}{m^2}\right)^{i\tau} 
= -\frac{\Gamma(-i\tau)\Gamma^2(\frac{1}{2}+\frac{i}{2}\tau)}
{\Gamma(i\tau)\Gamma^2(\frac{1}{2}-\frac{i}{2}\tau)} \;.
\ee

Using the equality $\Gamma(\overline{z})=\overline{\Gamma(z)}$ and
defining $\Gamma(i\tau)\equiv r_1\exp(i\theta_1)$ and 
$\Gamma(\frac{1}{2}+\frac{i}{2}\tau)\equiv r_2\exp(i\theta_2)$ we get:
\vspace{-3mm}
\be
\left(\frac{\Lambda^2}{m^2}\right)^{i\tau}
=\exp(i\theta)
\mlab{scaling-2.1}
\ee
where $\theta = \pi - 2\theta_1 + 4\theta_2$ and 
$\theta_1=\mbox{arg}\l(\Gamma(i\tau)\r)$,  
$\theta_2=\mbox{arg}\l(\Gamma(\frac{1}{2}+\frac{i}{2}\tau)\r)$.

We want to approximate $\theta$ for small values of $\tau$. Therefore, we
Taylor expand the Gamma functions:
\be
\Gamma(i\tau) \approx \frac{1}{i\tau}(1+i\tau\psi(1)) = \psi(1)-\frac{i}{\tau}
\ee
and
\be
\Gamma\l(\frac{1}{2}+\frac{i}{2}\tau\r) 
\approx \Gamma\l(\frac{1}{2}\r)\l(1 + \frac{i}{2}\tau\psi\l(\frac{1}{2}\r) \r) 
\;.
\ee

Thus, the argument $\theta$ from \mref{scaling-2.1} becomes:
\ba
\theta 
&=&\pi - 2\mbox{arg}(\Gamma(i\tau)) +
4\mbox{arg}\l(\Gamma\l(\frac{1}{2}+\frac{i}{2}\tau\r)\r) \nn\\
&\approx& \pi - 2\arctan\left(-\frac{1}{\tau\psi(1)}\right) +
4\arctan\left(\frac{\tau}{2}\psi\l(\half\r)\right) \nonumber \\
&\approx&\pi - 2\left(\frac{\pi}{2}+\tau\psi(1)\right) +
2\tau\psi\l(\half\r) \nonumber \\
&\approx& - 4\tau\ln2 \mlab{1204}
\ea
knowing that $\psi(1)=-\gamma$ and $\psi(\frac{1}{2})=-\gamma-2\ln2$.

After substituting \mref{1204} in \mref{scaling-2.1} and inverting the
exponential, we find:
\be
\ln\left(\frac{\Lambda}{m}\right) = \frac{1}{2\tau}(-4\tau\ln2 + 2k\pi) \;.
\ee

The ground state of the system is found for k=1, and so the generated mass
is related to the coupling as:
\be
\frac{\Lambda}{m} = \exp\left( \frac{\pi}{\sqrt{\frac{\alpha}{\alpha_c}-1}}
 - 2\ln2 \right) 
\mlab{scaling-2.2}
\ee
where we note that $2\ln2 \approx 1.386$.

We see from \mrefb{scaling-1.2}{scaling-2.2} that both cutoff methods yield
a similar scaling law, in agreement with \mref{1200}. The coefficients of
the scaling law are $A=\pi$ in both approximations, but $B=2$ in the
first case and $B=2\ln2$ in the second, more realistic, case.

\subsection{Curtis-Pennington vertex}
\label{Sec:scaleCP}

We now want to make an analogous calculation using the Curtis-Pennington
vertex Ansatz. In this case, the linearized equation for the fermion mass
in an arbitrary covariant gauge is given by the bifurcation equation,
\mref{116}. In the Landau gauge~($\xi=0$), the $\S$-equation becomes:
\be
\S(x) = \frac{3\alpha}{8\pi} \int_{\kappa^2}^{\Lambda^2} \left\{
2\S(y)\l[\frac{\theta(x-y)}{x}+\frac{\theta(y-x)}{y}\r]
- \frac{\S(y)-\S(x)}{y-x}
\l[\frac{y}{x}\theta(x-y)+\frac{x}{y}\theta(y-x)\r]
\right\}
\mlab{gap3}
\ee
where we introduced again an infrared cutoff $\kappa^2$, satisfying
$\kappa=\S(\kappa^2)$, to retain the scale of the generated fermion mass
when $\alpha$ is larger, but very close, to the critical coupling
$\alpha_c$. We know from the discussion of Section~\ref{Sec:abgpr} that in
the critical point, where the infrared cutoff $\kappa$ is zero, the
solution of this equation has a power behaviour. Therefore we assume that
close to the critical point, the solution to \mref{gap3} is still power
behaved:
\be
\S(x) \sim x^{-s} \;.
\mlab{1205}
\ee

After substituting this solution in \mref{gap3} we get:
\ba
x^{-s} &=& \frac{3\alpha}{8\pi} \Bigg\{\left[ -\pi\cot{\pi s} +\frac{3}{s}
+\frac{1}{1-s} + 1\right] x^{-s} \mlab{gap3.1} \\
&& - \frac{2}{1-s} \frac{(\kappa^2)^{1-s}}{x} + \int_0^{\kappa^2} {dy
\frac{y}{x} \frac{y^{-s}-x^{-s}}{y-x}}
- \frac{2}{s} (\Lambda^2)^{-s} + \int_{\Lambda^2}^{\infty} {dy \frac{x}{y}
\frac{y^{-s}-x^{-s}}{y-x}} \Bigg\} \;. \nn
\ea

The exponent $s$ has to satisfy the transcendental equation derived by
equating the terms in $x^{-s}$ in \mref{gap3.1} ($\kappa\to 0$ and
$\Lambda\to\infty$):
\be
1 = \frac{3\alpha}{8\pi} \left[ -\pi\cot{\pi s} +\frac{3}{s}
+\frac{1}{1-s} + 1 \right] \, .
\mlab{1206}
\ee

We want to investigate the behaviour of $s$ in the neighbourhood of the
critical point. Therefore we rewrite \mref{1206} as:
\be
\frac{3\alpha}{8\pi} = h(s) = \frac{1}{f(s)}
\mlab{1207}
\ee
where
\be
f(s) = -\pi\cot{\pi s} +\frac{3}{s} + \frac{1}{1-s} + 1 \;.
\mlab{f}
\ee

We now make a Taylor expansion of the function $h(s)$ around the critical
point:
\be
h(s) = h(s_c) + (s-s_c) h'(s_c) + \half (s-s_c)^2 h''(s_c) +
\Order(s-s_c)^3 \;.
\mlab{g}
\ee

From \mref{1207} we know that $h(s_c) = 3\alpha_c/8\pi$ and in
Section~\ref{Sec:abgpr} we have shown that $h'(s_c)=0$ in the critical point.
Therefore, \mref{g} becomes:
\be
\frac{3\alpha}{8\pi} = \frac{3\alpha_c}{8\pi} - 
\half (s-s_c)^2 \frac{f''(s_c)}{f^2(s_c)} \;.
\ee

The behaviour of $s$ in the neighbourhood of the critical point is thus
given by,
\be
s_{1,2} = s_c \pm \beta_c\sqrt{1-\frac{\alpha}{\alpha_c}}
\mlab{s}
\ee
where
\be
\beta_c \equiv \sqrt{\frac{2f(s_c)}{f''(s_c)}} \;.
\mlab{betac}
\ee

Differentiating \mref{f} twice yields:
\be
f''(s) = -2\pi^3 \csc^2{\pi s} \cot{\pi s} + \frac{6}{s^3} +
\frac{2}{(1-s)^3} \;.
\mlab{f2}
\ee

Substituting the value $s_c = 0.4710$, found in Section~\ref{Sec:abgpr}, in
the previous equations, Eqs.~(\oref{f}, \oref{f2}, \oref{betac}), gives
\fbox{$\beta_c = 0.5246$}.

The general solution of \mref{gap3.1}, in the neighbourhood of the critical
point is:
\be
\S(x) = C_1 \, x^{-s_c-i\beta_c\tau} + C_2 \, x^{-s_c+i\beta_c\tau} \;.
\mlab{solCP}
\ee

The scale of the generated mass will be determined by the boundary
conditions which are derived from \mref{gap3.1}. The IR boundary
condition is found by taking $x=\kappa^2$ in \mref{gap3.1}
($\Lambda\to\infty$):
\ba
C_1 \left[ - \frac{2}{1-s_1} (\kappa^2)^{-s_1}
+ \int_0^{\kappa^2} dy
\frac{y}{\kappa^2} \frac{y^{-s_1}-(\kappa^2)^{-s_1}}{y-\kappa^2} \right] &&
\mlab{IR} \\
+ C_2 \left[ - \frac{2}{1-s_2} (\kappa^2)^{-s_2}
 + \int_0^{\kappa^2} dy
\frac{y}{\kappa^2} \frac{y^{-s_2}-(\kappa^2)^{-s_2}}{y-\kappa^2} \right] &=& 0 \nn \;.
\ea

Substituting $t=y/\kappa^2$ in the integral of \mref{IR} yields:
\be
C_1 \, (\kappa^2)^{-s_1} \left[ - \frac{2}{1-s_1}
+ \int_0^{1} dt \, \frac{t^{-s_1+1}-t}{t-1} \right]
+ C_2 \, (\kappa^2)^{-s_2} \left[ - \frac{2}{1-s_2}
 + \int_0^{1} dt \, \frac{t^{-s_2+1}-t}{t-1} \right] = 0 \;.
\mlab{IR.1}
\ee

The integrals in \mref{IR.1} are given in \mref{122}. Substituting these, we
find:
\be
C_1 \, (\kappa^2)^{-s_1} \left[ - \frac{2}{1-s_1}
+ \psi(-s_1+2) - \psi(2) \right]
+ C_2 \, (\kappa^2)^{-s_2} \left[ - \frac{2}{1-s_2}
 + \psi(-s_2+2) - \psi(2) \right] = 0 \;.
\mlab{IR.2}
\ee

Taking $x=\Lambda^2$ in \mref{gap3.1} gives the UV boundary
condition ($\kappa\to 0$):
\ba
C_1 \left[ - \frac{2}{s_1} (\Lambda^2)^{-s_1} + \int_{\Lambda^2}^{\infty}
{dy \frac{\Lambda^2}{y} \frac{y^{-s_1}-(\Lambda^2)^{-s_1}}{y-\Lambda^2}} \right] && \mlab{UV} \\
+ C_2  \left[ - \frac{2}{s_2} (\Lambda^2)^{-s_2} + \int_{\Lambda^2}^{\infty}
{dy \frac{\Lambda^2}{y} \frac{y^{-s_2}-(\Lambda^2)^{-s_2}}{y-\Lambda^2}}
\right] &=& 0 \;. \nn
\ea

Substituting $t=y/\kappa^2$ in the integral of \mref{UV} yields:
\be
C_1 \, (\Lambda^2)^{-s_1}\left[ - \frac{2}{s_1}  + \int_{1}^{\infty}
dt \,  \frac{t^{-s_1-1}-t^{-1}}{t-1} \right]
+ C_2 \, (\Lambda^2)^{-s_2}\left[ - \frac{2}{s_2}  + \int_{1}^{\infty}
dt \,  \frac{t^{-s_2-1}-t^{-1}}{t-1} \right] = 0 \;.
\mlab{UV.1}
\ee

Substituting the integral \mref{123} in \mref{UV.1} yields:
\be
C_1 \, (\Lambda^2)^{-s_1}\left[ - \frac{2}{s_1}
-\psi(s_1+1) + \psi(1) \right]
+ C_2 \, (\Lambda^2)^{-s_2}\left[ - \frac{2}{s_2} 
-\psi(s_2+1) + \psi(1) \right] = 0 \;.
\mlab{UV.2}
\ee

Now, eliminate $C_1$ and $C_2$ from the boundary conditions,
\mrefb{IR.2}{UV.2}, and substitute $s_{1,2}$ from \mref{s}:
\be
\l(\frac{\Lambda^2}{\kappa^2}\r)^{2i\beta_c\tau} =
\frac{\left[ - \frac{2}{s_c+i\beta_c\tau}-\psi(s_c+i\beta_c\tau+1) 
+ \psi(1)
\right]\left[ - \frac{2}{1-s_c+i\beta_c\tau}+ \psi(-s_c+i\beta_c\tau+2) 
- \psi(2) \right]}
{\left[ - \frac{2}{s_c-i\beta_c\tau}-\psi(s_c-i\beta_c\tau+1)
 + \psi(1) \right] 
\left[ - \frac{2}{1-s_c-i\beta_c\tau} + \psi(-s_c-i\beta_c\tau+2) 
- \psi(2) \right]}.
\mlab{elim}
\ee

After noting that $\psi(\overline{z})=\overline{\psi(z)}$, we rewrite
\mref{elim} as:
\be
\l(\frac{\Lambda^2}{\kappa^2}\r)^{2i\beta_c\tau} = \exp(i\theta) =
\exp\Big(2i(\theta_1+\theta_2)\Big) ,
\mlab{theta}
\ee
where
\ba
\theta_1 &\equiv& \arg{\left[ - \frac{2}{s_c+i\beta_c\tau}-\psi(s_c+i\beta_c\tau+1) 
+ \psi(1)\right]} \\
\theta_2 &\equiv& 
\arg{\left[ - \frac{2}{1-s_c+i\beta_c\tau} + \psi(-s_c+i\beta_c\tau+2) 
- \psi(2) \right]}\;.
\ea

For small values of $\tau$, we compute $\theta_1$ and $\theta_2$ to
$\Order(\tau)$:
\ba
\theta_1 &\approx& \arg{\left[ - \frac{2}{s_c} -\psi(s_c+1) + \psi(1)
+i \beta_c\tau \l( \frac{2}{s_c^2} - \psi'(s_c+1)\r) \right]} \nn\\
&\approx& \l[ \frac{\frac{2}{s_c^2} - \psi'(s_c+1) }
{ - \frac{2}{s_c} -\psi(s_c+1) + \psi(1)} \r] \beta_c\tau  \mlab{theta1}\\[3mm]
\theta_2 &\approx& \arg{\left[ - \frac{2}{1-s_c} + \psi(-s_c+2) - \psi(2) 
+ i\beta_c\tau \l( \frac{2}{(1-s_c)^2} + \psi'(-s_c+2) \r)
\right]} \nn \\
&\approx& \l[\frac{ \frac{2}{(1-s_c)^2} + \psi'(-s_c+2) }
{ - \frac{2}{1-s_c} + \psi(-s_c+2) - \psi(2) } \r]\beta_c\tau \mlab{theta2}
\;.
\ea

We compute $\theta$ by inserting \mrefb{theta1}{theta2} in Maple:
\be
\theta = 2(\theta_1+\theta_2) = -7.2269 \: \beta_c\tau \;.
\mlab{maple}
\ee

From \mref{theta} we find:
\be
\frac{\Lambda}{\kappa} = \exp\l( \frac{k\pi}{2\beta_c\tau}
+ \frac{\theta}{4\beta_c\tau} \r) \;,
\mlab{scaling5}
\ee
and substituting \mref{maple} for $\theta$ in this equation yields
precisely the Miransky scaling law, close to the critical point, where $\tau
\ll 1$~(k=1):
\be
\frac{\Lambda}{\kappa} =
\exp\left(\frac{A}{\sqrt{\frac{\alpha}{\alpha_c}-1}} - B \right)
\ee
where
\be
\fbox{$\D A = \frac{\pi}{2\beta_c} = 0.9531 \, \pi$}  \qquad \mbox{and} \qquad
%
\fbox{\raisebox{0pt}[3.8mm][3mm]{$\D B = 1.8067$}} \;.
\mlab{AB}
\ee

\subsection{Summary}

The previous results can be compared with the numerical results of Curtis and
Pennington~\cite{CP93} who solve the non-linear integral equation for
various values of $\alpha$ and fit their results to the form of
\mref{1200}, in the rainbow approximation and with the Curtis-Pennington
vertex. In the rainbow approximation they find $A=0.9886\,\pi$, $B=1.4883$,
while, in Section~\ref{Sec:scalbare}, we found $A = \pi$ and $B = 2$ or
$B=2\ln2\approx1.386$, depending on the IR treatment of the linearized
integral. There is good agreement between our analytical and their numerical
results. The parameter $A$ agrees extremely well and
seems independent of the way the IR part of the integral is treated.
The deviations for the parameter $B$ can be explained as it sets
the scale of the generated mass and is therefore sensitive to the IR
approximation introduced in the linearized equation.

This is confirmed for the CP-vertex where Curtis and Pennington find
$A=0.9326\,\pi$ and $B = 1.2606$. Again, the parameter $A$ is in very good
agreement with our analytical results, \mref{AB}, while the parameter $B$
deviates from it.  It is clear from our discussion in
Section~\ref{Sec:scaleCP} that the parameter $A$ is independent of the
boundary conditions and the IR treatment of the linearized integral, and
will therefore have the correct value of the true scaling law for the
original non-linear equation.

\chapter {Fermion mass generation in unquenched QED: a survey}
\label{Sec:UnqQED}
\def\ordpar{<\!\!\compop\!\!>}

In the previous chapter we discussed dynamical fermion mass generation in
quenched QED. We found that, provided the coupling is larger than some
critical value, this indeed happens. The value of the critical coupling
depends on the vertex Ansatz used to truncate the infinite set of SD
equations. Furthermore, the scale of the generated mass in quenched QED
seems to follow the Miransky scaling law. 

As the phenomenon of mass generation is thought to be governed by the
dynamics at low momentum it has long been the question if it also occurs in
unquenched QED. Here, the coupling runs as a consequence of screening
effects due to fermion-antifermion pair productions, such that the
interaction becomes weak at long distances. It is this opposition between
short and long distance aspects which makes the study of fermion mass
generation in unquenched QED essentially different from that in quenched
QED.

We will now give a review of the work accomplished prior to this study.  It
is not so long ago since the first studies about fermion mass generation in
unquenched QED, using the Schwinger-Dyson equation approach, were
published.
The integral equations describing the dynamical generation of fermion mass,
Eqs.~(\oref{156}, \oref{157}, \oref{196}), are given by:
\ba
\frac{\S(p^2)}{\F(p^2)}
&=& m_0 + \frac{i e^2}{4(2\pi)^4} \int d^4k \:
\Trace{\Big[\gamma^\mu \, S(k) \, \Gamma^\nu(k,p) \Big]} \, D_{\nu\mu}(k-p) \\
\frac{1}{\F(p^2)} &=& 1 -
\frac{i e^2}{4p^2(2\pi)^4} \int d^4k \:
\Trace{\Big[\slash{p} \, \gamma^\mu \, S(k) \, \Gamma^\nu(k,p) \Big]} 
\, D_{\nu\mu}(k-p) \\
\frac{1}{\G(q^2)} &=& 1 
- \frac{iN_f e^2\Proj_{\mu\nu}}{3(2\pi)^4 q^2} \int d^4k\, 
\Trace \Big[ \gamma^\mu \, S(k) \,
\Gamma^\nu(k,p) \, S(p) \Big] \;.
\mlab{1301}
\ea

In many studies the full vacuum polarization is replaced by its 1-loop
approximation, replacing all the full quantities in the integral of
\mref{1301} by bare quantities. The integral can then be computed
analytically and the 1-loop vacuum polarization function is given by:
\be
\Pi(q^2) = \frac{N_f\alpha}{3\pi}\l[\ln{\frac{\Lambda^2}{q^2}}+C\r]
\mlab{vacpol}
\ee
where C a constant dependent on the regularization scheme.
As we can write the photon renormalization function as
\[
\G(q^2) \equiv \frac{1}{1+\Pi(q^2)} \,,
\]
the full photon propagator itself is then approximated by an infinite sum of
bare loops. 

In Ref.~\cite{Kondo91} Kondo et al.\ perform an analytical calculation using
the 1-loop approximation to the vacuum polarization. Furthermore they
introduce the so-called LAK-approximation (in analogy to Landau, Abrikosov
and Khalatnikov~\cite{LAK56}):
\be
\Pi\l((k-p)^2\r) \approx \Pi\l(\max(k^2,p^2)\r)
\mlab{lakvacpol}
\ee
to allow them to compute the angular integrals of the fermion equations
analytically. The full vertex $\Gamma^\mu(k,p)$ is approximated by the bare
vertex $\gamma^\mu$. To determine the {\it critical point} in the Landau
gauge they then linearize the mass equation, applying bifurcation analysis.
Consequently, they derive a differential equation with boundary conditions,
which cannot be solved by any known special functions.  Therefore they use
an asymptotic expansion method which is only valid for the number of
flavours $N_f=1$,$2$. The main results published in this paper are
$\alpha_c(N_f=1)=1.99972$, $\alpha_c(N_f=2)=2.71482$. The generated dynamical
mass and the order parameter $\ordpar$ scale according to the
mean field scaling law (in contrast with the Miransky scaling in the
quenched case):
\be
\frac{m}{\Lambda} \sim \sqrt{\frac{\alpha}{\alpha_c}-1} \;,
\mlab{MF}
\ee
and the anomalous dimension of the composite operator is $\gamma_m=0$.
 
In another analytical paper, Gusynin~\cite{Gusynin} follows almost the same
path as Kondo et al.~\cite{Kondo91}. He introduces the same approximations
and hence, finds the same differential equation as they do. His treatment
differs in the way he solves the differential equation.  His solution
method is only valid when $3\pi/N_f\alpha \gg 1$ and is thus limited to
$N_f=1$. For the critical coupling he finds $\alpha_c(N_f=1)\approx
1.95$. Also, the scale of the generated mass follows the mean field law,
\mref{MF}.

In Ref.~\cite{Kondo}, Kondo discusses how to recover the gauge invariance
of the critical coupling and the scaling law in quenched and unquenched QED
by a specific choice of vertex. The vertex is constructed such that the
Landau gauge results, in quenched and unquenched QED, remain unchanged in
any arbitrary gauge. However, it is clear, in the light of the most recent
investigations about the construction of the full QED
vertex~\cite{BC,Bashir94,Dong94}, that Kondo's choice of vertex shows
unphysical properties.

As mentioned before, all the analytical studies of unquenched QED required
the introduction of several approximations to make the problem tractable.
Therefore, new, numerical studies were undertaken as in
Ref.~\cite{Kondo92b} and its preliminary report, Ref.~\cite{Kondo89}, by
Kondo et al. They solve the integral equations numerically in the Landau
gauge. Again, they use the bare vertex approximation, while the vacuum
polarization is still taken at 1-loop, \mref{vacpol}. In the first part of
their calculation they introduce the LAK-approximation, \mref{lakvacpol},
as in the analytical calculations. This is useful to verify the analytical
results and to simplify the numerical calculation.  In this approximation
the $\F$-equation decouples, yielding $\F(x)=1$. The angular integrals of
the $\S$-equation can be calculated analytically leaving a one-dimensional
non-linear integral equation for $\S(x)$ to be solved numerically. They
plot the variation of the generated mass and the order parameter,
$\ordpar$, versus coupling for $N_f=1,2,3,4$. According to them, all the
scaling laws are of the mean field type.  The values of the critical
couplings are: $\alpha_c(N_f=1)=1.9989$, $\alpha_c(N_f=2)=2.7517$,
$\alpha_c(N_f=3)=3.5062$, $\alpha_c(N_f=4)=4.3177$. The numerical results
very much agree with the analytical results of Ref.~\cite{Kondo91}.  In the
second part of Ref.~\cite{Kondo92b} they relax the LAK-approximation,
keeping the angular dependence of the vacuum polarization. Then, the
two-dimensional integrals, i.e. radial and angular, in the fermion
equations have to be solved numerically and the two coupled integral
equations for $\F$ and $\S$ have to be solved simultaneously.  As an
intermediate step they again take $\F(x)\equiv 1$ (as this is a good
approximation in the Landau gauge) and solve the integral equation for
$\S(x)$. They find $\alpha_c(N_f=1)=2.0728$ and $\alpha_c(N_f=2)=2.8209$.
Finally, they solve the system of coupled equations for $\S$ and $\F$. They
do not mention any critical coupling but observe that indeed $\F(x)\approx
1$. They state that the scaling laws remain of the mean field type.
However, surprisingly, they only find a phase transition for
$N_f=1,2$. Their iterative procedure does not converge for $N_f \ge 3$. The
reason for this, is that the positivity of the vacuum polarization
$\Pi(q^2)$ is not guaranteed anymore as can be seen from \mref{vacpol}. If
$k^2$, $p^2$ $\in[0,\Lambda^2]$ then $(k-p)^2
\in [0,4\Lambda^2]$. It seems inconsistent to use the vacuum
polarization for momenta up to $4\Lambda^2$ when the integrals used to
compute the vacuum polarization were computed using an UV cutoff $\Lambda^2$.
The trouble only occurs for $N_f \ge 3$ because then the photon momentum
in the integral of the fermion equation becomes larger than the Landau
pole.

In Ref.~\cite{Oli90} Oliensis and Johnson derive a non-linear differential
equation from the integral equations and solve this numerically. They use
the bare vertex approximation, but their vacuum polarization is slightly
different from \mref{vacpol}, being
\be
\Pi(q^2) = \frac{N_f\alpha}{3\pi}\ln{\frac{\Lambda^2}{q^2 + \S^2(0)}} \;,
\ee
where they incorporated an infrared cutoff through the term $\S^2(0)$,
which suppresses the effects of the vacuum polarization below this scale.
They fix $\S(0)$ and solve their differential equation for increasing
values of $\Lambda$, finding a critical coupling
$\alpha_c(N_f=1)=1.999534163$. Furthermore, the generated mass follows a
mean field scaling law.

In Ref.~\cite{Rak91} Rakow investigates the renormalization group flow in
QED using the Schwinger-Dyson equations. He includes the effect of fermion
loops in the photon propagator by considering the fermion and photon SD
equations simultaneously. The full vertex is still approximated by the bare
one. He solves the coupled set of non-linear integral equations
numerically. He determines the critical coupling by investigating the
dependence of the chiral condensate $\ordpar$ on the coupling $\alpha$ and
the bare mass $m_0$. He finds a second order phase transition for
$\alpha_c(N_f=1)=2.25$. However, we think that this value has not been
determined accurately as can be inferred from the data of Ref.~\cite{Rak91}
and we will show in our own calculations later on. He then goes on to show
that the renormalized coupling is zero at the critical point. We note that
Rakow renormalizes the coupling at zero momentum. It is clear that
the unrenormalized, running coupling goes to zero at zero momentum in the
critical point, when the fermion mass generation disappears, and therefore
it seems obvious that his renormalized coupling, defined as
$\alpha_r=\alpha\G(0)\F^2(0)$, is zero. Nevertheless, we do not agree
with Rakow's claim that this proves the triviality of QED, as will be
explained later.

In Ref.~\cite{Atk92} Atkinson et al.\ solve the coupled integral equations
in the Landau gauge. They use the bare vertex approximation and assume
$\F(x)\equiv 1$. Furthermore they use the LAK-approximation,
\mref{lakvacpol}, for the vacuum polarization. They make an analogous
approximation for the mass function in the vacuum polarization integral to
allow the analytic calculation of the angular integrals. The integral
equations for $\S$ and $\Pi$ are transformed into differential equations
(using some more simplifications) which are then solved numerically. They
find $\alpha_c(N_f=1)=2.100286$ and a mean field type scaling law.

A more detailed numerical investigation of fermion mass generation can be
found in Ref.~\cite{Kondo92}. There, Kondo et al.\ start from the coupled
system of integral equations for $\S$, $\F$ and $\G$. They simplify the
numerical problem by assuming $\F(x)\equiv 1$, as this is thought to be a
good approximation. Then the system of two coupled integral equations for
$\S$ and $\G$ is solved by an iterative procedure. Their discussion of the
numerical aspect of the calculation is very interesting as this is often
disregarded in publications. They compare the results obtained for $\S(x)$
and $\G(x)$ in the self-consistent treatment and in the 1-loop
approximation to the vacuum polarization. The scaling law is consistent
with the mean field type scaling, although they very pertinently point out
that the exact scaling law is very difficult to pin down numerically
because the scaling window is very narrow, i.e. the scaling law is only
valid very close to the critical point. Analytical studies seem more
appropriate for this purpose. They find a critical coupling
$\alpha_c(N_f=1)=2.084$. We note the peculiar behaviour of the full photon
renormalization function compared to its 1-loop approximation,
as we will discuss later.

We will end our review with alternative methods to the Schwinger-Dyson
approach. In Ref.~\cite{Uki90} Ukita et al.\ suggest a gauge invariant way
to study the strong coupling phase of QED by applying the inversion method
to the chiral condensate $\ordpar$. The lowest order inversion method leads
to the gauge-independent critical point $\alpha_c=2\pi/3=2.094395$.  To the
lowest order inversion this value is independent of the number of flavours
$N_f$. Although $\alpha_c$ seems reasonably well approximated for $N_f=1$,
it is wrong by a factor two for the quenched case. In Ref.~\cite{Kondo94}
Kondo et al.\ compare the inversion method with their SD approach from
Ref.~\cite{Kondo91b}. They show that their asymptotic solution to the
lowest order is the same as that of the inversion method. They then go on
showing that including higher orders in the asymptotic solution recovers
the known result in quenched QED, i.e. $\alpha_c=\pi/3$, for which the
series converges. For $N_f=1$ the series is asymptotic and they find
$\alpha_c(N_f=1)=1.9995$. They conclude that therefore the inversion
method, which is very useful because gauge invariance is guaranteed and no
approximations to vertex nor vacuum polarization are necessary, has to be
computed to higher orders to find a critical coupling which depends on the
number of flavours.

Finally we will want to know what lattice studies can tell us about the
dynamical generation of fermion mass. In Section~\ref{LatQuen} we discussed
briefly the lattice calculations in quenched QED. There have been quite few
lattice investigations of unquenched QED. In Ref.~\cite{Dag88} Dagotto et
al.\ showed that there is a second order phase transition from a massless to
a massive phase. Since then, the discussion about the scaling law, which is
related to the triviality of the theory, is still active~\cite{Koc93,Goc94}.

The values of the critical coupling $\alpha_c$ for $N_f$=1, 2, determined
in the above discussed papers are tabulated in 
Table~\ref{Tab:survey}.

\begin{table}[htbp]
\begin{center}
\begin{tabular}{|l|l|l|}
\hline
Ref. & $\alpha_c(N_f=1)$ & $\alpha_c(N_f=2)$ \\
\hline
\cite{Kondo91} & 1.99972 & 2.71482 \\
\cite{Gusynin} & 1.95 & \\
\cite{Kondo92b} a) & 1.9989 & 2.7517 \\
\cite{Kondo92b} b) & 2.0728 & 2.8209 \\ 
\cite{Oli90} & 1.999534163 &\\
\cite{Rak91} & 2.25 & \\
\cite{Atk92} & 2.100286 & \\
\cite{Kondo92} & 2.084 & \\
\cite{Uki90} & 2.094395 & \\
\cite{Kondo94} & 1.9995 & \\
\hline
\end{tabular}
\caption{Literature survey of critical couplings for $N_f=1$ and $N_f=2$
in unquenched QED.}
\label{Tab:survey}
\end{center}
\end{table}

We have seen that various approximations have been introduced, in the
analytical as well as in the numerical calculations, producing results with
varying accuracy. In the next chapters, we will develop a numerical method
to make a unified, {\bf highly accurate} numerical study of the various
approximations to the Schwinger-Dyson equations for dynamical fermion mass
generation in unquenched QED. The final aim is the solution of the system
of three coupled, non-linear integral equations for $\S$, $\F$ and
$\G$. This will first be achieved with the bare vertex, then, we will for
the first time take the study of fermion mass generation in unquenched QED
beyond the bare vertex approximation by introducing improved vertices.

\def\maal{$\times$}
\chapter{Numerical Solution of Schwinger-Dyson Equations}
\label{Numprog}

The aim of this chapter is to set up the formalism needed to solve the
integral equations numerically. We will start by considering a single
integral equation determining one unknown function. The type of
integral equations which are of interest to us are called {\it
non-linear Fredholm equations of the second kind}~\cite{Delves}:
\be
x(s) = y(s) + \lambda\int_a^b K(s,t,x(s),x(t)) \: dt ,
\ee 
where $y(s)$ is a known function and $x(s)$ is the unknown function we
want to determine.

Unfortunately, the major part of the literature about numerical
methods to solve integral equations is only concerned with linear
integral equations. The linear Fredholm equation of the second kind
is:
\be
x(s) = y(s) + \lambda\int_a^b K(s,t) \, x(t) \: dt .
\ee

For these linear equations there exists convergence criteria related
to the behaviour of the kernel $K(s,t)$. Moreover several different
solution approaches exist: Nystrom method, expansion methods as
Ritz-Galerkin method, ... Very little can be found in the literature
about non-linear equations so that the researcher confronted with such
an equation has to gather together bits and pieces from a number of
different areas of numerical analysis to tackle this problem.

The basic approximation introduced to solve an integral equation
numerically resides in the numerical method used to evaluate the
integrals involved in the problem.

If one needs to calculate numerically the integral $I[f]$ of some
function $f(s)$ given by:
\be
I[f] = \int_a^b f(s) \:ds ,
\ee
one will generally use a {\it quadrature formula} to approximate the
exact value of the integral.  Most quadrature formulae approximating
the integral value by the value $R[f]$ can be expressed as:
\be
R[f] = \sum_{i=1}^N w_i f(t_i) .
\ee

The error $E[f]$ introduced by this approximation is:
\be
E[f] = I[f] - R[f] .
\ee

Well-known quadrature formulae are for instance the Newton-Cotes
formulae as the midpoint rule, the trapezoidal rule, Simpson's rule,
... , or the various Gauss rules of which the Gauss-Legendre rule is
the best-known. Several other quadrature formulae exist which can be
used depending on the behaviour of the integration kernel.

\section{Linear Fredholm equation of the second kind}

\subsection{Linear Fredholm equation and the Neumann series}

Let us consider a linear Fredholm equation of the second kind:
\be
x(s) = y(s) + \lambda\int_a^b K(s,t) \, x(t) \: dt .
\mlab{Fred0}
\ee

Formally this can also be written in operator form as:
\be
x = y + \lambda K x .
\mlab{Fred1}
\ee

Starting from an initial guess $x_0=y$ for the function $x$ we will
define an iterative procedure:
\be
x_{n+1} = y + \lambda K x_n .
\mlab{it1}
\ee

If this procedure converges when $n\rightarrow\infty$, then from
\mref{it1} this limit can be written as a series called the {\it Neumann}
series. One can show that the Neumann series converges to the solution of
\mref{Fred1}:
\be
x = \lim_{n\rightarrow\infty} x_n = \sum_{i=0}^\infty \lambda^i K^i y .
\ee

If the integral $K x_n = \int_a^b K(s,t) x_n(t) dt$ can be computed
analytically, the iterative procedure \mref{it1} can be used to find a good
approximation to the function $x$ by truncating it after $n$
iterations and taking $x\approx x_n$.

Can we compute an error bound on the approximate solution~?  Let us define
the error function $e_n$ on the $n^{th}$ approximation by:
\be
e_n = x - x_n  .
\mlab{err0}
\ee

If we subtract \mref{it1} from \mref{Fred1} we find:
\be
x - x_{n+1} = \lambda K (x - x_n)
\ee
or
\be
e_{n+1} = \lambda K e_n.
\mlab{err1}
\ee

However we also find:
\be
x_{n+1} - x_n = (x_{n+1} - x) + (x - x_n) = e_n - e_{n+1} ,
\ee
giving
\be
e_n = e_{n+1} + (x_{n+1} - x_n) . 
\ee

Taking the norm (using some appropriate function norm) of the previous
equation gives:
\be
\|e_n\| \le \|e_{n+1}\| + \|x_{n+1} - x_n\| .
\mlab{err2}
\ee

If we now substitute \mref{err1} in \mref{err2},
\be
\|e_n\| \le \|\lambda K\|.\|e_n\| + \|x_{n+1} - x_n\| .
\ee

After rearranging terms,
\be
(1-\|\lambda K\|)\|e_n\| \le \|x_{n+1} - x_n\| ,
\ee
and provided $\|\lambda K\| < 1$,
\be
\|e_n\| \le \frac{\|x_{n+1} - x_n\|}{1-\|\lambda K\|} . 
\mlab{err3}
\ee

One can show that the condition $\|\lambda K\| < 1$ is sufficient for
the Neumann series to converge to the solution of the integral
equation \mref{Fred1}. Furthermore \mref{err3} then gives an {\it a
posteriori} error bound which can be computed given two successive
iterations $x_n$ and $x_{n+1}$.

\subsection{Numerical solution using the Neumann series}
\label{NumNeumann}

In the previous section we defined an iterative procedure to solve the
integral equation \mref{Fred0} assuming that the integrals $K x_n =
\int_a^b K(s,t) \, x_n(t) \: dt$ could be computed analytically,
\be
x_{n+1}(s) = y(s) + \lambda\int_a^b K(s,t) \, x_n(t) \: dt .
\mlab{it1b}
\ee

In most problems this will not be possible and the integrals will have
to be evaluated using some numerical quadrature formula.  If we use
the quadrature formula R on the interval [a,b] with weights $w_i$ and
nodes $t_i$ the integral will be written as:
\be
I[f] \equiv \int_a^b f(t) dt   
= \sum_{i=1}^N w_i f(t_i) + E[f].
\mlab{errEf}
\ee

To compute the Neumann series numerically we approximate the integral of
\mref{it1b} by introducing some suitable quadrature formulae and truncating
the error term $E[f]$:
\be
\int_a^b K(s,t) \, x(t) \: dt \approx \sum_{j=1}^N w_j K(s,t_j) x(t_j) .
\ee

The iterative procedure \mref{it1b} can now be replaced
by:
\be
x_{n+1}(s) = y(s) + \lambda \sum_{j=1}^N w_j K(s,t_j) x_n(t_j) .
\mlab{approx1}
\ee

It is interesting to note from this last equation that if we confine
ourselves to values $s=t_j$, the nodes of the integration formula,
\mref{approx1} only involves the function values $x_n(t_j)$ in the
successive iteration steps. The notation can then be simplified by
introducing vector notation,
\be 
{\bf x_{n+1}} = {\bf y} + \lambda K {\bf x_n} \;,
\mlab{it2}
\ee
where we define,
\ba
({\bf x_n})_i &=& x_n(t_i) \nn \\
{\bf y}_i &=& y(t_i) \nn \\
K_{ij} &=& w_j K(t_i,t_j) ,
\ea
with $i,j = 1, \ldots, N$ .

One can show that if the iterative procedure \mref{it2} converges, it
converges to the solution of the set of linear algebraic equations
\be
(I- \lambda K) {\bf x_R} = {\bf y} ,
\mlab{Fred2}
\ee
where the subscript $R$ shows the explicit dependence of the solutions
of \mref{Fred2} on the quadrature rule~$R$.
 
If we now define an error vector ${\bf e_n}$
\be
{\bf e_n} = {\bf x_R} - {\bf x_n} ,
\mlab{defen}
\ee
then, repeating the error analysis from Eqs.~(\oref{err0}-\oref{err3}), on
\mrefb{Fred2}{it2} and replacing the function norm by some suitable
matrix norm gives us an error bound on the truncated solution:
\be
\|{\bf e_n}\| \le \frac{\|{\bf x_{n+1}} - {\bf x_n}\|}{1-\|\lambda K\|} ,
\mlab{erren}
\ee
provided $\|\lambda K\| < 1$. This error is the error on the
solution ${\bf x_R}$ to the set of linear algebraic equations introduced by
truncating the iterative procedure and approximating the solution by
${\bf x_n}$, it is {\bf not} the error with respect to the exact
solution $x(s)$ from the integral equation.

\subsection{The Nystrom method}

In the previous section we have shown how to find an approximate
solution to the integral equation \mref{Fred0} by applying the Neumann
series and the corresponding iterative procedure, truncating the
procedure after $n$ steps and approximating the integrals in each step
by a quadrature rule $R$. 

Instead one could approximate the integral equation \mref{Fred0}
straight away by replacing the integral in \mref{Fred0} by a
quadrature rule $R$,
\be
x(s) = y(s) + \lambda \sum_{j=1}^N w_j K(s,t_j) x(t_j) .
\mlab{1.9}
\ee

If we only consider \mref{1.9} in the integration nodes $s=t_i$,
\mref{1.9} becomes a system of $N$ linear equations with $N$ 
unknowns $x_i=x(t_i)$ with solution vector $\mvec{x_R}$,
\be
(x_R)_i = y_i + \lambda \sum_{j=1}^N w_j K(t_i,t_j) (x_R)_j ,
\mlab{2}
\ee
which is equivalent to \mref{Fred2}. This is called the {\it Nystrom}
method.

The iterative procedure from Section~\ref{NumNeumann} is, if it converges,
just one possible numerical method which can be used to solve the set of
linear algebraic equations \mref{2}. As a matter of fact there are several
other methods to solve sets of linear algebraic equations and one applies
the method which suits the problem best. If we denote the approximate
numerical solution of the set of linear equations \mref{2} by ${\bf x_S}$
(generalizing the notation $\mvec{x_n}$ of Section~\ref{NumNeumann} to an
arbitrary convergent method), then the error estimate ${\bf e_S}$ on the
solution ${\bf x_R}$ of \mref{2} is defined as,
\be
{\bf e_S} = {\bf x_R} - {\bf x_S} .
\mlab{2.1}
\ee

The error ${\bf e_S}$ will have to be determined by analysing the
specific numerical method used to solve the set of linear algebraic
equations.

Of course this error estimate is only part of the error if we want to
use the solution ${\bf x_S}$ as an approximation to the solution
$x(s)$ of the integral equation at the nodes $s=t_i$. For this purpose
we will also have to investigate the error caused when replacing the
integral by a finite sum using the quadrature formula $R$. If we could
solve the set of linear equations \mref{Fred2} exactly, then $({\bf
x_R})_i$ are approximations to the function $x(s)$ at the grid points
$s=t_i$. \mref{1.9} then defines an approximating function $x_R(s)$ to
$x(s)$ for all $s$ occurring in the left hand side of the
equation. Can we find an error estimate on this approximation~?

We define the error function $e_R(s)$ as,
\be
e_R(s) = x(s) - x_R(s) .
\mlab{defeR}
\ee

Using \mref{errEf} in \mref{1.9} we see that the function $x_R(s)$
satisfies the integral equation
\be
x_R(s) = y(s) + \lambda\int_a^b K(s,t) x_R(t) dt - E[\lambda K x_R](s) .
\mlab{approx2}
\ee

Subtracting \mref{approx2} from the original integral equation
\mref{Fred0} yields:
\be
x(s) - x_R(s) = \lambda\int_a^b K(s,t) \, \l(x(t)-x_R(t)\r) \:dt
 + E[\lambda K x_R](s),
\ee
or
\be
e_R(s) = \lambda\int_a^b K(s,t) \, e_R(t) \: dt  + E[\lambda K x_R](s) .
\mlab{erreR}
\ee

This means that the error function $e_R(s)$ introduced by the
quadrature rule $R$ also satisfies a linear Fredholm integral
equation of the second kind. The kernel is the same as in \mref{Fred0}
but the driving term is now $E[\lambda K x_R](s)$ instead of $y(s)$.
Taking the function norm of \mref{erreR} gives
\be
\|e_R\| \le \|\lambda K\|.\|e_R\| + \|E[\lambda K x_R]\| ,
\ee
thus,
\be
\|e_R\| \le \frac{\|E[\lambda K x_R]\|}{1 - |\lambda K\|}, 
\ee
provided $|\lambda K\| < 1$.

A more practical error formula can be derived in vector form. We rewrite
\mref{Fred0} using \mref{errEf}:
\be
x(s) = y(s) + \lambda \sum_{j=1}^N w_j K(s,t_j) x(t_j) + E[\lambda K
x](s) .
\mlab{Fred3}
\ee

Now subtract \mref{1.9} for $\mvec{x_R}$ from \mref{Fred3},
\be
x(s) - x_R(s) = \lambda \sum_{j=1}^N w_j K(s,t_j) (x(t_j)-x_R(t_j))
 + E[\lambda K x](s) ,
\ee
or
\be
e_R(s) = \lambda \sum_{j=1}^N w_j K(s,t_j) e_R(t_j)
 + E[\lambda K x](s) .
\ee

Evaluating this last equation at the nodes $t_i$ gives:
\be
e_R(t_i) = \lambda \sum_{j=1}^N w_j K(t_i,t_j) e_R(t_j)
 + E[\lambda K x](t_i) ,
\ee
or in vector notation,
\be
(I - \lambda K) {\bf e_R} = {\bf E}[\lambda K x] \,.
\mlab{err4}
\ee

This set of equation for $\mvec{e_R}$ has the same coefficients as
\mref{Fred2} for $\mvec{x_R}$ but with a different constant vector.
The solution to \mref{err4} is given by:
\be
{\bf e_R} = (I - \lambda K)^{-1} \, {\bf E}[\lambda K x] .
\ee

After taking the vector and matrix norm,
\be
\|{\bf e_R}\| = \|(I - \lambda K)^{-1}\|.\|{\bf E}[\lambda K x]\| ,
\mlab{erreR2}
\ee
which only requires that we can solve \mref{Fred2} and so is applicable
even in the case $\|\lambda K\| \ge 1$. Of course this is a formal result
where we note that the error term $\|{\bf E}[\lambda K x]\|$ is a function
of the unknown solution $x(s)$. To use this formula in practice one could
approximate $x(s)$ by $x_R(s)$ to evaluate the error term.

We can now combine the error estimates \mref{defeR} and \mref{2.1}
(neglecting precision errors) to
give a total error bound when approximating the solution $x(t_i)$ of
the original integral equation \mref{Fred0} at the grid points by the
approximated solution $({\bf x_S})_i$ of the system of equations
derived using the quadrature rule $R$. The error vector ${\bf e}$ is
defined as:
\be
{\bf e} = {\bf x} -{\bf x_S} .
\mlab{err}
\ee

This can be written as:
\be
{\bf e} = ({\bf x} - {\bf x_R}) + ({\bf x_R} - {\bf x_S}) 
= {\bf e_R} + {\bf e_S} .
\ee

Taking the norm:
\be
\|{\bf e}\| \le \|{\bf e_R}\| + \|{\bf e_S}\| ,
\ee
where both error terms have been bounded in \mref{erreR2} and
\mref{erren} (for the iterative procedure).

\section{Non-linear Fredholm equation of the second kind}
\label{NonlinFredholm}

We will use the method described in the previous section to build a
numerical solution for the {\it non-linear} Fredholm equation of the
second kind. The equation to solve is:
\be
x(s) = y(s) + \lambda\int_a^b K(s,t,x(s),x(t)) \: dt \, .
\mlab{nonlin}
\ee 

Applying the {\it Nystrom} method developed in the previous section,
we again will approximate the integral by some quadrature formula $R$,
replacing the integral equation \mref{nonlin} by the approximate equation:
\be
\hat{x}(s) = y(s) + \lambda \sum_{j=1}^N w_j K(s,t_j,\hat{x}(s),\hat{x}(t_j)) .
\mlab{nonlin1}
\ee

If we require that this equation should hold at the points $s=t_i$,
\mref{nonlin1} becomes a set of $N$~non-linear algebraic equations
with $N$ unknowns $x_i$ where $x_i=\hat{x}(t_i)$:
\be
x_i = y_i + \lambda \sum_{j=1}^N w_j K(t_i,t_j,x_i,x_j) .
\mlab{4.9}
\ee

In matrix form this equation can be written as:
\be
{\bf x} = {\bf y} + \lambda {\bf K}({\bf t}, {\bf x}) .
\mlab{5}
\ee

It is this system of non-linear algebraic equations we want to solve
numerically to approximate the solution of the integral equation
\mref{nonlin}. A straightforward way to try to solve such a set of
non-linear algebraic equations is to start from an initial guess ${\bf
x_0}$ and to define the {\it natural} iterative procedure:
\be
{\bf x_{n+1}} = {\bf y} + \lambda {\bf K}({\bf t}, {\bf x_n}) .
\mlab{itnonlin}
\ee

One can assume that if this procedure converges when
$n\rightarrow\infty$, it will converge to one of the solutions of
\mref{5}.  The achievement of convergence as well as its rate depend
on the behaviour of the kernel ${\bf K}({\bf t}, {\bf x})$.

From the previous description one sees that two major numerical
approximations are involved in solving the original equation. First we
have to make a proper choice of numerical quadrature formula and
second we truncate the iterative procedure \mref{itnonlin} after a
finite number of steps. It is important in order to assess the
approximate solution to have a good idea of the size of the errors
introduced by both approximations.  We will return to this in detail
in future sections.

To be able to discuss the size of the errors when solving \mref{5} we will
first introduce a suitable vector norm on our solution space (pp.~2-17 of
Ref.~\cite{Isaac}).  In general the {\it p-norm} of a vector is defined as,
\be
\|\mvec{x}\|_p \equiv  \l(\sum_j |x_i|^p\r)^{1/p} .
\mlab{normp}
\ee

Some of the most frequently used vector norms derived from
\mref{normp} are:
\ba
\|\mvec{x}\|_1 &\equiv&  \sum_i |x_i| ,\\
\|\mvec{x}\|_2 &\equiv&  \sqrt{\sum_i |x_i|^2} ,\\
\|\mvec{x}\|_\infty &\equiv& \max_i |x_i| .
\ea
The norm $\norm{\mvec{x}}_2$ is called the {\it Euclidean norm}, while the
norm $\norm{\mvec{x}}_\infty$ is called the {\it maximum norm}.

To investigate the error on the solution to our problem we will use
the maximum norm, defining,
\be
\|\mvec{x}\| \equiv \|\mvec{x}\|_\infty \equiv \max_i |x_i| .
\ee
The maximum norm is quite interesting because it makes sure that no
deviation in any point is larger than $\|\mvec{x}\|_\infty$. To
discuss the error bounds on the solution we also need the {\it matrix
norm} induced by the corresponding vector norm. The matrix norm
induced by the maximum norm is,
\be
\norm{A}_\infty \equiv \max_i \sum_j |A_{ij}| ,
\ee
i.e., the {\it maximum absolute row sum} of the matrix, which is
straightforward to compute.

The matrix norm $\norm{A}_1$ is given by the {\it maximum absolute
column} sum of the matrix:
\be
\norm{A}_1 \equiv \max_j \sum_i |A_{ij}| ,
\ee
while the matrix norm induced by the {\it Euclidean norm} is,
\be
\norm{A}_2 \equiv \sqrt{\rho(A^\dagger A)} ,
\ee
where $A^\dagger$ is the hermitian conjugate of A and $\rho(M)$ is the
spectral radius of the matrix $M$ defined by 
\be
\rho(M) \equiv \max_s |\lambda_s(M)|,
\ee  
where $\lambda_s(M)$ denotes the eigenvalues of $M$. From this it is
obvious that the Euclidean norm of a matrix is not easily computable
and therefore we opt for the maximum norm in our treatment.

The induced matrix norm ensures that the norm of the vector
$A\mvec{x}$ satisfies:
\be
\norm{A\mvec{x}} \le \norm{A}.\norm{\mvec{x}} .
\ee 

The error $\mvec{e_n}$ on the approximate solution $\mvec{x_n}$ to the
exact solution $\mvec{x}$ is,
\be
\|\mvec{e_n}\| = \|\mvec{x}-\mvec{x_n}\| 
= \max_i\l|x_i-(x_n)_i\r| .
\ee 

We will use this vector norm to compute the distance between two
successive iterations of \mref{itnonlin}. In practice we will
terminate the iterative procedure when a certain criterion is
fulfilled. Normally we will require that 2 successive iterations are
no more distant than a tolerance $\tol$ from each other,
\be
\|\mvec{x_{n+1}} - \mvec{x_n}\| \le \tol .
\mlab{tol}
\ee

We must be careful however if we want to translate $\|\mvec{x_{n+1}} -
\mvec{x_n}\|$ to the error $\norm{\mvec{e_n}}$ on the exact solution of
the vector equation \mref{5}. Can we find a relation between those two
quantities~?

In the case of a linear Fredholm equation of the second kind,
\mref{erren} gave us a bound on the error,
\be
\|{\bf e_n}\| \le \frac{\|{\bf x_{n+1}} - {\bf x_n}\|}{1-\|\lambda K\|} .
\ee
This expression is very useful because it relates the error on the
solution to the distance between two successive iterations. Can we
derive an analogous expression in the non-linear case~?

We can write:
\be
\mvec{e_n}-\mvec{e_{n+1}} = (\mvec{x}-\mvec{x_n}) - (\mvec{x}-\mvec{x_{n+1}}) 
= \mvec{x_{n+1}} - \mvec{x_n} ,
\ee
or
\be
\mvec{e_n} = \mvec{e_{n+1}} + (\mvec{x_{n+1}} - \mvec{x_n}) .
\ee

Taking the norm,
\be
\|\mvec{e_n}\| \le \|\mvec{e_{n+1}}\| + \|\mvec{x_{n+1}} -
\mvec{x_n}\| .
\mlab{err10}
\ee

Let us write the set of non-linear equations as:
\be
\mvec{x} = \mvec{g}(\mvec{x}),
\mlab{eq100}
\ee
or
\be
\hspace{3cm}x_i = g_i(x_1,\ldots,x_N) , \hspace{1cm} i=1,\ldots,N.
\ee

The iterative procedure is:
\be
\mvec{x_{n+1}} = \mvec{g}(\mvec{x_n}) .
\mlab{it10}
\ee

Subtracting \mref{it10} from \mref{eq100}, we have
\be
\mvec{x}-\mvec{x_{n+1}} = \mvec{g}(\mvec{x})-\mvec{g}(\mvec{x_n}),
\ee
or
\be
\mvec{e_{n+1}} = \mvec{g}(\mvec{x})-\mvec{g}(\mvec{x_n}).
\mlab{err101}
\ee

We now make a Taylor expansion of $\mvec{g}(\mvec{x})$ around 
$\mvec{g}(\mvec{x_n})$:
\be
\hspace{1cm}g_i(\mvec{x}) = g_i(\mvec{x_n}) 
+ \sum_{j=1}^N \frac{\partial g_i}{\partial x_j}(\mvec{x_n}).(x_j-x_{n,j})
 + \Order(\mvec{x}-\mvec{x_n})^2, \hspace{1cm} i=1,\ldots,N ,
\ee
or in vector notation:
\be
\mvec{g}(\mvec{x}) = \mvec{g}(\mvec{x_n}) 
+ \frac{\partial \mvec{g}}{\partial\mvec{x}}(\mvec{x_n}).(\mvec{x}-\mvec{x_n})
 + \Order(\mvec{x}-\mvec{x_n})^2 .
\ee

Thus,
\be
\mvec{g}(\mvec{x}) - \mvec{g}(\mvec{x_n}) = 
\frac{\partial \mvec{g}}{\partial\mvec{x}}(\mvec{x_n}).\mvec{e_n}
 + \Order(\mvec{e_n})^2 . 
\mlab{Taylor10}
\ee

Substitute \mref{Taylor10} in \mref{err101},
\be
\mvec{e_{n+1}} = 
\frac{\partial \mvec{g}}{\partial\mvec{x}}(\mvec{x_n}).\mvec{e_n}
 + \Order(\mvec{e_n})^2 .
\ee

After taking the norm,
\ba
\|\mvec{e_{n+1}}\| &\le& 
\l\|\frac{\partial \mvec{g}}{\partial\mvec{x}}(\mvec{x_n}).\mvec{e_n}\r\|
 + \|\Order(\mvec{e_n})^2\| \nn\\
&\le&\l\|\frac{\partial \mvec{g}}{\partial\mvec{x}}(\mvec{x_n})\r\|
.\|\mvec{e_n}\|
 + \|\Order(\mvec{e_n})^2\| .
\mlab{err102}
\ea

Substituting \mref{err102} in \mref{err10},
\be
\|\mvec{e_n}\| \le \|\mvec{x_{n+1}} - \mvec{x_n}\|
+ \l\|\frac{\partial \mvec{g}}{\partial\mvec{x}}(\mvec{x_n})\r\|.\|\mvec{e_n}\|
 + \|\Order(\mvec{e_n})^2\|.
\ee

Thus, provided 
$\l\|\frac{\partial\mvec{g}}{\partial\mvec{x}}(\mvec{x_n})\r\| < 1$,
\be
\|\mvec{e_n}\| \le \frac{\|\mvec{x_{n+1}} - \mvec{x_n}\|
 + \|\Order(\mvec{e_n})^2\|}
{1 - 
\l\|\frac{\partial \mvec{g}}{\partial\mvec{x}}(\mvec{x_n})\r\|}.
\mlab{err103}
\ee

This again relates the error $\|\mvec{e_n}\|$ to the distance between two
successive iterations $\|\mvec{x_{n+1}} - \mvec{x_n}\|$. We can neglect
the terms of $\Order(\mvec{e_n})^2$ if $\mvec{x_n}$ is sufficiently
close to the solution $\mvec{x}$.

\section{Schwinger-Dyson equations}

\subsection{The 3 coupled equations}

In this section we will formulate the integral equations for which we have
to develop the numerical formalism. There are 3 coupled non-linear integral
equations describing 3 unknown functions: the dynamical fermion mass $\S$,
the wavefunction renormalization $\F$ and the photon renormalization
function $\G$.

We recall the integral equations, Eqs.~(\oref{189}, \oref{190},
\oref{104}), derived in Euclidean space with the bare vertex approximation.
In the Landau gauge ($\xi=0$) with zero bare mass ($m_0=0$), these are:
\ba
\frac{\S(x)}{\F(x)} &=& \frac{3\alpha}{2\pi^2} \int dy \, 
\frac{y\F(y)\S(y)}{y+\S^2(y)} \int d\theta \, \sin^2\theta
\, \frac{\G(z)}{z} \mlab{S0} \\
\frac{1}{\F(x)} &=& 1 - \frac{\alpha}{2\pi^2 x} \int dy \, 
\frac{y\F(y)}{y+\S^2(y)} \mlab{F0} \\
&& \times\int d\theta \, \sin^2\theta \, \G(z) \, 
\l[\frac{2 x y \sin^2\theta}{z^2} - \frac{3 \sqrt{xy} \cos\theta}{z} \r] \nn \\
\frac{1}{\G(x)} &=& 1 + \frac{4 N_f \alpha}{3\pi^2 x} \int dy 
\frac{y\F(y)}{y+\S^2(y)} \mlab{G0} \\
&& \times \int d\theta \, \sin^2\theta \, \frac{\F(z)}{z+\S^2(z)}
\l[y(1-4\cos^2\theta) + 3\sqrt{xy}\cos\theta\r] , \nn 
\ea
where $z \equiv x+y-2\sqrt{xy}\cos\theta$.

If we look at these integral equations, one of the striking features is
that the integrals involved are 2-dimensional integrals, the radial
and the angular integrals both involve the unknown functions. This is
different from quenched QED, where $\G(x)=1$ implies that the angular
integrals in the $\F$ and $\S$ equations can be performed
analytically, leaving 2 coupled 1-dimensional integral equations to
solve. The fact that the problem of unquenched QED involves
2-dimensional integrals is a major problem because it is very computer
time consuming.

Symbolically we can write the equations as:
\be
\left\{\renewcommand{\arraystretch}{2}
\begin{array}{r@{\quad}c@{\quad}l}
\D \frac{\S}{\F} &=& f_1[\S,\F,\G] \\
\D \frac{1}{\F} &=& f_2[\S,\F,\G] \\
\D \frac{1}{\G} &=& f_3[\S,\F,\G], 
\end{array}
\right.
\ee
or
\ba
\frac{\S(x)}{\F(x)} &=& \int dy \, \int d\theta \, 
K_1\l(y,\S(y),\F(y),\theta,z,\G(z)\r) \mlab{S1}\\
\frac{1}{\F(x)} &=& 1 + \int dy \, \int d\theta \, 
K_2\l(x,y,\S(y),\F(y),\theta,z,\G(z)\r) \mlab{F1}\\
\frac{1}{\G(x)} &=& 1+ \int dy \, \int d\theta \, 
K_3\l(x,y,\S(y),\F(y),\theta,z,\S(z),\F(z)\r). \mlab{G1}
\ea

A straightforward method is to discretize the problem. This means we
will solve the problem for a finite number of function values $\S_i$,
$\F_i$ and $\G_i$. In practice these function values are taken at
momenta $x_i$ which are the nodes of the quadrature rule used to
approximate the integrals. Because of the expected behaviour of the
unknown functions and the integration kernels we choose the grid
points on a logarithmic scale in momentum squared. For numerical
purposes we introduce an infrared cutoff $\kappa$ as well as an
ultraviolet cutoff $\Lambda$. We define a grid of $N+1$ equidistant
points $t_i$, $i=0, \ldots, N$:
\be
t_i = \logten(\kappa^2) + \frac{i}{N} 
\l(\logten(\Lambda^2)-\logten(\kappa^2)\r) .
\ee

Discretizing the functions at the integration nodes implies that the
function values needed in the numerical approximation of the radial
integrals are exactly the tabulated function values. We are thus
looking for the function values $\S_i$, $\F_i$ and $\G_i$
satisfying,
\ba
\frac{\S_i}{\F_i} &=& \sum_j w_j \, \sum_k w_k \, 
K_1\l(x_j,\S_j,\F_j,\theta_k, z_k,\G_k \r) 
\mlab{S1.1}\\
\frac{1}{\F_i} &=& 1 + \sum_j w_j \, \sum_k w_k \, 
K_2\l(x_i,x_j,\S_j,\F_j,\theta_k,z_k,\G_k \r)
\mlab{F1.1} \\
\frac{1}{\G_i} &=& 1 + \sum_j w_j \, \sum_k w_k \, 
K_3\l(x_i,x_j,\S_j,\F_j,\theta_k,z_k,\S_k,\F_k \r),
\mlab{G1.1} 
\ea
where $z_k = x_i+x_j-2\sqrt{x_i x_j}\cos\theta_k$. 

Unfortunately the angular integrals of the equations will use function
values which are not tabulated. In Eqs.~(\oref{S1.1}, \oref{F1.1}) we
need the function values for the photon renormalization function $\G$
at the momentum $z_k$, while analogously in
\mref{G1.1} we need the function values of $\S$ and $\F$ at the momentum
$z_k$. Whatever numerical quadrature formula one uses for the angular
integrals, we always need the unknown functions at values which are
not tabulated. These untabulated function values can be estimated by
interpolating the functions $\S$, $\F$ and $\G$ between two tabulated
values. The simplest interpolation scheme will be to use linear
interpolation on the logarithm of momentum squared:
\be
f(z) = f(x) +
\frac{\logten z-\logten x}{\logten y-\logten x}\l[f(y)-f(x)\r] .
\ee

\subsection{Simplified approach: the $\S$ equation}
\label{S-equation}

To develop the numerical program we will first introduce a number of
approximations to simplify the problem. As we are working in the
Landau gauge with the bare vertex we approximate $\F$ by
$\F(x)\equiv 1$. This is motivated by the results of quenched QED
and can even be found in unquenched QED after introducing the
LAK(Landau-Abrikhozov-Khalatnikov)-approximation $\G(z) =
\G(\max(y,x))$. In both cases $F(x)$ will be equal to one. A
further approximation consists in replacing the full vacuum
polarization by its 1-loop perturbative value instead of solving the
photon Schwinger-Dyson equation,
\be
\Pi(z) = \frac{N_f\alpha}{3\pi} \l(\ln\frac{\Lambda^2}{z} + C\r),
\ee
where C is a renormalization constant. If we choose $C=0$ such that
$\Pi(\Lambda^2)=0$ the photon renormalization function $\G$ becomes,
\be
\G(z) = \l(1+\frac{N_f\alpha}{3\pi}\ln\frac{\Lambda^2}{z}\r)^{-1},
\ee
with $\G(\Lambda^2)=1$.

The coupled set of integral equations, Eqs.~(\oref{S0}, \oref{F0},
\oref{G0}), now simplifies to a single non-linear integral equation for
the dynamical fermion mass $\S$:
\be
\boxeq{
\S(x) = \frac{3\alpha}{2\pi^2} \int
dy\,\frac{y\S(y)}{y+\S^2(y)}
\int d\theta\,
\frac{\sin^2\theta}{z(1+\frac{N_f\alpha}{3\pi}\ln\frac{\Lambda^2}{z})}\;.}
\mlab{S2}
\ee

To develop the numerical method we could as well start from the
$\S$-equation in quenched QED, however in the 1-loop approximation the
problem is more realistic because the angular integrals cannot be
computed analytically as will also be the case in the complete
treatment of the coupled system of integral equations.

We now face two problems: firstly how do we choose the quadrature rules to
compute the radial and angular integrals, secondly how do we find a
solution for $\S(x)$ once the quadrature rules have been introduced~?

To make a sensible choice of integration rule we have to study the
behaviour of the integration kernel using some assumption for the
unknown function. From previous studies of quenched QED we know that
the integration nodes are best chosen on a logarithmic scale in
momentum squared. Therefore we will change variables in \mref{S2}:
\ba
t &=& \logten y \\
dt &=& \frac{dy}{y\ln10} .
\ea

Substituting this in \mref{S2},
\be
\S(x) = \frac{3\alpha\ln10}{2\pi^2} \int_{-\infty}^\infty
dt\,\frac{y^2\S(y)}{y+\S^2(y)}
\int d\theta\,
\frac{\sin^2\theta}{z(1+\frac{N_f\alpha}{3\pi}\ln\frac{\Lambda^2}{z})}
, \mlab{S3}
\ee
where $y=10^t$.

To compute the integrals numerically we will introduce an ultraviolet
cutoff $\Lambda^2$ and an infrared cutoff $\kappa^2$ on the radial
integration. The ultraviolet cutoff is introduced to regularize the
integrals while the infrared cutoff only serves numerical
purposes. When introducing the infrared cutoff one has to ensure that
either the neglected part of the integrals, i.e. $\int_0^{\kappa^2}$
is negligible, or else one must evaluate analytically the contribution
of the lower part of the integral and add it to the numerical
integral.  As we will see later we will choose the infrared cutoff
$\kappa^2$ so that the infrared part of the integral is negligible and
\mref{S3} can be replaced by:
\be
\S(x) = \frac{3\alpha\ln10}{2\pi^2} \int_{\logten\kappa^2}^
{\logten\Lambda^2}
dt\,\frac{y^2\S(y)}{y+\S^2(y)}
\int d\theta\,
\frac{\sin^2\theta}{z(1+\frac{N_f\alpha}{3\pi}\ln\frac{\Lambda^2}{z})}. 
\mlab{S4}
\ee

We now introduce the quadrature rules to approximate the integrals
numerically, \mref{S4} is replaced by the approximate equation:
\be
\S(x) = \frac{3\alpha\ln10}{2\pi^2} \sum_{j=0}^{N} w_j
\frac{x_j^2\S(x_j)}{x_j+\S^2(x_j)}
\sum_{k=0}^M w'_k 
\frac{\sin^2\theta_k}{z_k(1+\frac{N_f\alpha}{3\pi}
\ln\frac{\Lambda^2}{z_k})}
, \mlab{S5}
\ee
$w_j$, $w'_k$ are the weights of the quadrature rules $R$ and $R'$
(which can be different) used respectively to compute the radial and
angular integrals. The photon momentum is
$z_k=x+x_j-2\sqrt{x x_j}\cos\theta_k$. A simple choice of integration
rule could be a closed formula with $N+1$ equidistant nodes such as
a composite Newton-Cotes formula \cite{Isaac}. The nodes are then
given by:
\be
t_i = \logten\kappa^2 + \frac{i}{N} 
\l(\logten\Lambda^2-\logten\kappa^2\r) ,
\ee
with corresponding momenta squared,
\be
x_i = k_i^2 = 10^{t_i}.
\ee

How do we find a solution $\S(x)$ of \mref{S5}~? One possible solution
method is the {\it collocation method} where one only requires the equation
\mref{S5} to hold at the integration nodes $t_i$,
\be
\S_i = \frac{3\alpha\ln10}{2\pi^2} \sum_{j=0}^{N} w_j
\frac{x_j^2\S_j}{x_j+\S_j^2}
\sum_{k=0}^M w'_k 
\frac{\sin^2\theta_k}{z_k(1+\frac{N_f\alpha}{3\pi}
\ln\frac{\Lambda^2}{z_k})}
, \hspace{0.5cm} i=0,\ldots,N , \mlab{S6}
\ee
where we denote $\S_i=\S(x_i)$ and $z_k=x_i+x_j-2\sqrt{x_i x_j}
\cos\theta_k$.

This set of equations is self-consistent and only involves the function
values of $\S$ at the integration points; we do not need any information
about $\S$ at any other point in momentum space. \mref{S6} is in fact just
the application of the Nystrom method for non-linear Fredholm equations
\mref{4.9} to the $\S$-equation \mref{S4}. If we succeed in solving this
set of equations, our knowledge about the function $\S(x)$ will
completely reside in the knowledge of the function at a finite number of
points; $\S(x)$ has been discretized.

\mref{S6} is a set of $(N+1)$ non-linear algebraic equations for the
$(N+1)$ unknowns $\S_i$. An evident method to solve \mref{S6} would be
to use the natural iterative procedure proposed in
\mref{itnonlin}. We start from an initial guess $(\S_0)_i$,
$i=0,\ldots,N$, and define the iterative procedure:
\be
\l(\S_{n+1}\r)_i = \frac{3\alpha\ln10}{2\pi^2} \sum_{j=0}^{N} w_j
\frac{x_j^2\l(\S_n\r)_j}{x_j+\l(\S_n\r)_j^2}
\sum_{k=0}^M w'_k 
\frac{\sin^2\theta_k}{z_k(1+\frac{N_f\alpha}{3\pi}
\ln\frac{\Lambda^2}{z_k})}
, \hspace{0.5cm} i=0,\ldots,N . \mlab{S7}
\ee

Does this iterative procedure converge~? If it converges, what is the
error on the approximate solution if we truncate the procedure after
$n$ steps~?

If we use the empirical approach, implementing the iterative procedure
in a computer program and turning the handle, we observe that for
sufficiently large coupling $\alpha$ and for a suitably chosen
starting guess $\mvec{\S_0}$ the iterations tend to converge to a
non-trivial solution but at an extremely low rate. For example, if we
require that $\norm{\mvec{\S_{n+1}}-\mvec{\S_n}} \le \tol$, with
$\tol=\S(0)/1000$, the natural iteration scheme requires several
thousands iteration steps to reach the convergence criterion.

Furthermore, it seems very difficult to obtain a reasonable error
estimate on the truncated solution $\mvec{\S_n}$ as an approximate
solution to the solution $\mvec{\S}$ of \mref{S6}. To get an idea of
the accuracy of the approximation $\mvec{\S_n}$ we now decrease the
value of the tolerance to $\tol'=\tol/10$ and continue the iterative
procedure till a solution $\mvec{\S_{n'}}$ has been found which
satisfies the new tolerance condition. For the iterative method to be
reliable we expect the new approximate solution $\mvec{\S_{n'}}$ not
to be much more distant from $\mvec{\S_n}$ than $\tol$ as the
approximation $\mvec{\S_n}$ was found by imposing the tolerance
$\tol$. In reality this seems not to be fulfilled, the results of the
numerical program show that the difference
$\|\mvec{\S_{n'}}-\mvec{\S_n}\|$ is much larger than $\tol$.  This
means that we cannot rely on the approximation $\mvec{\S_n}$ to be an
approximate solution to $\mvec{\S}$ of \mref{S6} with an accuracy of
$\Order(\tol)$ and that we have no definite error estimate of the
solution $\mvec{\S_n}$.

To understand this feature we will investigate if the error formula
\mref{err103} can be used on the system of non-linear equations
\mref{S6}. From \mref{S6} we can formulate our system of non-linear
equations in a general form:
\be
\S_i = \sum_j K_{ij} ,
\mlab{eq150}
\ee
where the kernel can be written as:
\be
K_{ij} = C_{ij} \frac{\S_j}{x_j + \S_j^2} .
\mlab{eq155}
\ee

The iterative procedure to solve this will be:
\be
(\S_{n+1})_i = \sum_j (K_n)_{ij} 
= \sum_j C_{ij} \frac{(\S_n)_j}{x_j + (\S_n)_j^2}.
\mlab{eq151}
\ee

The error discussion in Section~\ref{NonlinFredholm} led to an error
bound \mref{err103} on the error $\mvec{e_n}$, after truncation of the
iterative procedure \mref{eq151}, which can be rewritten in the
current notation as,
\be
\|\mvec{e_n}\| \le \frac{\|\mvec{\S_{n+1}} - \mvec{\S_n}\|
 + \|\Order(\mvec{e_n})^2\|}
{1 - 
\l\|\frac{\partial \mvec{K}}{\partial\mvec{\S}}(\mvec{\S_n})\r\|},
\mlab{err104}
\ee
provided 
$\l\|\frac{\partial\mvec{K}}{\partial\mvec{\S}}(\mvec{\S_n})\r\| < 1$,
where $K_i = \sum_j K_{ij}$.

Taking the partial derivative of \mref{eq155} with respect to $\S_k$, we
obtain
\be
\parder{K_{ij}}{\S_k} = C_{ij} \l[\frac{x_j-\S_j^2}{(x_j +
\S_j^2)^2}\r] \delta_{jk} .
\mlab{eq156}
\ee

Thus,
\be
\l(\frac{\partial\mvec{K}}{\partial\mvec{\S}}\r)_{ij}
= \l(\frac{\partial K_i}{\partial \S_j}\r)
= \sum_k \l(\parder{K_{ik}}{\S_j}\r)
= C_{ij} \l[\frac{x_j-\S_j^2}{(x_j + \S_j^2)^2}\r] .
\mlab{156.1}
\ee

For \mref{err104} to be valid we know that
\be
\l\|\frac{\partial\mvec{K}}{\partial\mvec{\S}}(\mvec{\S_n})\r\| 
= \max_i \sum_j \l|\frac{\partial K_i}{\partial \S_j}(\mvec{\S_n})\r|
< 1,
\mlab{156.2}
\ee
or,
\be
\max_i \sum_j \l|C_{ij}
\l[\frac{x_j-(\S_n)_j^2}{(x_j + (\S_n)_j^2)^2}\r] \r| < 1.
\mlab{156.3}
\ee

From the numerical results we learn that
$\l\|\frac{\partial\mvec{K}}{\partial\mvec{\S}}(\mvec{\S_n})\r\|
\approx 3$
and thus condition \mref{156.2} is not satisfied by the kernel of the
fermion equation and therefore the error bound \mref{err104} cannot
be used in this case. This is the reason why the rate of convergence
is so slow and the error estimate so unreliable.

\section{Newton's iterative method}
\label{Newton}

Let us now consider an alternative method to improve the convergence rate
of the iterative procedure (pp.~109-119 of Ref.~\cite{Isaac}). Consider a
general system of non-linear algebraic equations
\be
\mvec{f(\mvec{x})} = \mvec{0} .
\mlab{203}
\ee

The most natural way to solve this set of equations iteratively
consists in rewriting \mref{203} as
\be
\mvec{x} = \mvec{g(\mvec{x})} , 
\mlab{203.1}
\ee
with
\be
\mvec{g(\mvec{x})} \equiv \mvec{x} - \mvec{f(\mvec{x})}
\mlab{203.2}
\ee
and then to define the iterative procedure
\be
\mvec{x_{n+1}} = \mvec{g(\mvec{x_n})} =  \mvec{x_n} + \mvec{f(\mvec{x_n})}.
\mlab{203.3}
\ee

However, as we saw in the previous section, this iterative procedure
does not always converge and if it converges the convergence rate can
be very slow and the error estimate unreliable.

In order to define an alternative iterative procedure to \mref{203.3}
we now replace \mref{203.2} by: 
\be
\mvec{g(\mvec{x})} \equiv \mvec{x} - A(\mvec{x})\mvec{f(\mvec{x})} ,
\mlab{205}
\ee
where $A(\mvec{x})$ is a square matrix of order $N+1$. If
$A(\mvec{x})$ is non-singular, \mref{203.1} and \mref{203}
will have the same solutions.

The simplest choice for $A(\mvec{x})$ is a constant non-singular
matrix,
\be
A(\mvec{x}) = A .
\mlab{206}
\ee

Next we define a matrix $J(\mvec{x})$ by:
\be
J(\mvec{x}) \equiv \l(\parder{f_i(\mvec{x})}{x_j}\r) 
\mlab{207}
\ee
the determinant of which is the Jacobian of the function $f_i(\mvec{x})$.

We also define a matrix $G(\mvec{x})$:
\be
G(\mvec{x}) \equiv \l(\parder{g_i(\mvec{x})}{x_j}\r) . 
\mlab{208}
\ee
 
Substituting Eqs.~(\oref{205}, \oref{206}, \oref{207}) in \mref{208}
gives,
\be
G(\mvec{x}) = I - A J(\mvec{x}) .
\mlab{209}
\ee

We now define an iterative procedure to solve \mref{203.1}:
\be
\mvec{x_{n+1}} = \mvec{g(\mvec{x_n})},
\mlab{210}
\ee
or using \mref{205},
\be
\mvec{x_{n+1}} = \mvec{x_n} - A \,\mvec{f(\mvec{x_n})} .
\mlab{211}
\ee

One can prove \mref{211} will converge, for $\mvec{x_0}$ sufficiently
close to a solution $\mvec{\tilde{x}}$ of \mref{203}, if the elements in
the matrix $G(\mvec{x})$ of
\mref{209} are sufficiently small. This could be realized in the case
that $J(\mvec{\tilde{x}})$ is non-singular and taking the constant
matrix $A$ to be approximately the inverse of
$J(\mvec{\tilde{x}})$. This naturally suggests a modification where we
replace the constant matrix $A$ from \mref{206} by the choice
\be
A(\mvec{x}) \equiv J^{-1}(\mvec{x}) \,.
\mlab{212}
\ee
The iterative procedure constructed with the matrix of \mref{212} is
called {\it Newton's method}.

Substituting \mref{212} in \mref{211} gives the following iterative
procedure equation:
\be
\mvec{x_{n+1}} = \mvec{x_n} - J^{-1}(\mvec{x_n})
\:\mvec{f(\mvec{x_n})} .
\mlab{213}
\ee

Although this suggests that we have to invert a matrix of order $N+1$
at each iteration step, we can transform the procedure from
\mref{213} so that we only have to solve a linear system of order
$N+1$ at each iteration step. Multiply both sides of \mref{213} with
$J(\mvec{x_n})$,
\be
J(\mvec{x_n})\, \l(\mvec{x_n}-\mvec{x_{n+1}}\r) =
\mvec{f(\mvec{x_n})} .
\mlab{214}
\ee

If we define 
\be
\mvec{\Delta_{n+1}} = \mvec{x_n}-\mvec{x_{n+1}}
\mlab{214.1}
\ee
then \mref{214} is a linear system of order $N+1$ to be solved for the
vector $\mvec{\Delta_{n+1}}$,
\be
\boxeq{
J(\mvec{x_n})\, \mvec{\Delta_{n+1}} = \mvec{f(\mvec{x_n})} .
}
\mlab{215}
\ee

From $\mvec{x_n}$ and the solution $\mvec{\Delta_{n+1}}$ of \mref{215}
we derive the next approximation $\mvec{x_{n+1}}$ using \mref{214.1}.

One can generally show that provided
\be
G(\mvec{\tilde{x}}) \equiv \l(\parder{g_i(\mvec{\tilde{x}})}{x_j}\r) 
= 0 , \qquad i,j = 0,\ldots,N .
\mlab{215.1}
\ee
there is a radius $\rho$ for which the iteration procedure
$\mvec{x_{n+1}} = \mvec{g}(\mvec{x_n})$ converges quadratically to the
solution $\mvec{\tilde{x}}$ of \mref{203} for any starting guess
$\mvec{x_0}$ satisfying:
\be
\norm{\mvec{x_0}-\mvec{\tilde{x}}} \le \rho .
\mlab{225}
\ee

We will now show that Newton's method satisfies \mref{215.1} so that
the method converges quadratically to a solution $\mvec{\tilde{x}}$ of
\mref{203} provided the starting guess $\mvec{x_0}$ is sufficiently
close to $\mvec{\tilde{x}}$.

From \mrefb{205}{212} the $j^{th}$ column of $G$ is given by,
\ba
\parder{\mvec{g(\mvec{x})}}{x_j} &=& \parder{\mvec{x}}{x_j} 
- \parder{}{x_j}\l[J^{-1}(\mvec{x})\,\mvec{f}(\mvec{x}) \r] \nn \\
&=&\parder{\mvec{x}}{x_j} -
J^{-1}(\mvec{x})\,\parder{\mvec{f}(\mvec{x})}{x_j}
- \parder{J^{-1}(\mvec{x})}{x_j}\,\mvec{f}(\mvec{x}) .
\mlab{221}
\ea

Setting $\mvec{x}=\mvec{\tilde{x}}$ in \mref{221} and recalling that
the solution $\mvec{\tilde{x}}$ satisfies $\mvec{f}(\mvec{\tilde{x}}) =
\mvec{0}$ and $J = \partial f_i/\partial x_j$ we get
\be
G(\mvec{\tilde{x}}) = I -
J^{-1}(\mvec{\tilde{x}})\,J(\mvec{\tilde{x}})
- \parder{J^{-1}(\mvec{\tilde{x}})}{\mvec{x}}\,\mvec{0} = 0 ,
\mlab{222}
\ee 
provided the matrix ${\partial J^{-1}(\mvec{\tilde{x}})}/{\partial x_j}$ in
\mref{221} exists. To determine ${\partial J^{-1}(\mvec{x})}/{\partial
x_j}$ we compute:
\be
\parder{(J^{-1}J)}{x_j} = J^{-1} \parder{J}{x_j} +
\parder{J^{-1}}{x_j} J
\mlab{223.0}
\ee
but also,
\be
\parder{(J^{-1}J)}{x_j} = \parder{I}{x_j} = 0 .
\mlab{223}
\ee

This means,
\be
\parder{J^{-1}(\mvec{x})}{x_j} = - J^{-1}(\mvec{x})
\parder{J(\mvec{x})}{x_j} J^{-1}(\mvec{x}) .
\mlab{224}
\ee

Thus, \mref{222} will be satisfied and the Newton method will be
quadratically convergent if $\mvec{f}(\mvec{x})$ has two derivatives
and $J(\mvec{x})$ is non-singular at the root $\mvec{\tilde{x}}$.

Furthermore one could also show that provided the starting guess
$\mvec{x_0}$ is close enough to $\mvec{\tilde{x}}$ the error on the
approximate solution is bound by:
\be
\norm{\mvec{e_{n+1}}} \le \norm{\mvec{x_{n+1}} - \mvec{x_n}}.
\mlab{226}
\ee

Although this bound is hugely overestimated it is very useful for
practical purposes as we will now explain. Indeed, this bound tells us
that the distance between the approximation $\mvec{x_{n+1}}$ and the
exact solution cannot exceed the distance between the solutions of
the last iteration and that of the previous one. 

Let us now apply Newton's method to the system of non-linear equations
\mref{S6} for the dynamically generated fermion mass. The equations
can be written symbolically as:
\be
\S_i = \sum_j K_{ij} , \qquad i=0,\ldots,N ,
\mlab{200}
\ee
with kernel
\be
K_{ij} = C_{ij} \frac{\S_j}{x_j + \S_j^2} , \qquad i,j=0,\ldots,N.
\mlab{200.1}
\ee

This can be written in the form of \mref{203} by defining
$\mvec{f(\mvec{\S})}$ as,
\be
f_i(\mvec{\S}) = \S_i - \sum_j K_{ij} = 0 , \qquad i=0,\ldots,N.
\mlab{216}
\ee

Using \mref{216} we derive the matrix $J$ from \mref{207}
\be
J_{ij} \equiv \l(\parder{f_i}{\S_j}\r) 
= \delta_{ij} - \sum_k \parder{K_{ik}}{\S_j} , \qquad i,j=0,\ldots,N .
\mlab{217}
\ee

The derivative of the kernel \mref{200.1} with respect to $\S_k$ is,
\be
\parder{K_{ij}}{\S_k} =  
C_{ij} \frac{x_j-\S_j^2}{(x_j + \S_j^2)^2} \delta_{jk}, 
\qquad i,j,k=0,\ldots,N.
\mlab{218.2}
\ee

Substituting \mref{218.2} in \mref{217} yields,
\be
J_{ij} = \delta_{ij} - C_{ij} \frac{x_j-\S_j^2}{(x_j + \S_j^2)^2} , 
\qquad i,j=0,\ldots,N . 
\mlab{218.21}
\ee

Substituting Eqs.~(\oref{216}, \oref{200.1}, \oref{218.21}) in
\mref{215} yields:
\be
\sum_j \l(\delta_{ij} - C_{ij} \frac{x_j-(\S_n)_j^2}
{(x_j + (\S_n)_j^2)^2}\r) \: (\Delta_{n+1})_j = (\S_n)_i - \sum_j
C_{ij} \frac{(\S_n)_j}{x_j + (\S_n)_j^2} , \qquad i=0,\ldots,N .
\mlab{218.3}
\ee

For each iteration we have to solve \mref{218.3} for
$\mvec{\Delta_{n+1}}$. Then from this solution we compute a new
approximation $\mvec{\S_{n+1}}$ with,
\be
\mvec{\S_{n+1}} = \mvec{\S_n} - \mvec{\Delta_{n+1}} .
\mlab{219}
\ee

From the numerical results we can say that the implementation of
Newton's method has given a tremendous improvement as well in
convergence rate (number of iteration steps needed to satisfy
$\norm{\mvec{\S^{n+1}}-\mvec{\S^n}} \le \tol$) as in reliability of
the error estimate on the approximate solution (see \mref{226}).

The required accuracy is achieved in less than 10 steps. Although each
step requires the solution of a linear system of order $N+1$ we
observe an important decrease of the computer time needed to find the
approximate solution satisfying the convergence criterion. Another
consequence of the quadratic convergence is that the distance between
two successive iteration decrease very rapidly, often as,
\be 
\norm{\mvec{\S_{n+1}} - \mvec{\S_n}} \approx 
\frac{\norm{\mvec{\S_n} - \mvec{\S_{n-1}}}}{10} .
\mlab{227}
\ee

From the numerical results it is clear that terminating the iterative
procedure when two successive iterations are closer than a tolerance
$\tol$ ensures that the exact solution is within $\tol$ of the last
iteration and surely even much closer than that. 

A straightforward check of the reliability of Newton's method compared
to the natural iterative procedure \mref{eq151} can be performed by varying
the starting guess $\mvec{\S_0}$. The convergence of the natural
iterative procedure is very sensitive to the starting guess: it only
converges (although very slowly) if the starting guess is larger than
but close to the exact solution; it will diverge if the starting guess
is too large, i.e. much larger that the exact solution, and it will
converge to the trivial solution $\mvec{\S}=\mvec{0}$ (which is always
a solution to the equation) as soon as the starting guess is
taken smaller than the exact solution. Even when the method does
converge to the non-trivial solution we observe that for varying
$\mvec{\S_0}$ the natural iteration scheme gives very different
approximate solutions $\mvec{\S_n}$ satisfying the convergence
criterion \mref{tol} for a fixed $\tol$. This is due to the fact that
the error $\mvec{e_n}$ on the approximate solutions in the natural
iterative procedure is much larger than the required tolerance $\tol$
between two successive iterations. We could in fact use this
information to get some better estimate of the error on the
approximate solution, by comparing the solutions reached from
different initial guesses $\mvec{\S_0}$.

In contrast to this, the Newton method performs exceptionally well.
Its convergence is almost always guaranteed, independent of the
starting guess. Only if the starting guess is chosen very far from the
exact solution will it just need a couple more iteration steps to
reach the solution and if the starting guess is chosen too close to
zero, the method will converge to the trivial solution
$\mvec{\S}=\mvec{0}$.  In the case of convergence to the non-trivial
solution the method is completely independent of the starting guess
$\mvec{\S_0}$: the approximate solutions $\mvec{\S_n}$ satisfying
\mref{tol} are all equal within this tolerance and even much
closer than that.

The iterative procedure \mref{216} can be extended in a
straightforward manner to a system of two or more coupled equations
with two or more unknown functions. As an example we take the case of
the coupling of the integral equations for the fermion wavefunction $\F$
and for the dynamical fermion mass $\S$,
\ba
f_{1,i}(\mvec{\S},\mvecF) &=& 0 , \qquad i=0,\ldots,N, \\
f_{2,i}(\mvec{\S},\mvecF) &=& 0 , \qquad i=0,\ldots,N.
\mlab{228}
\ea

This system can be written in the form of \mref{203},
\be
\mvec{f}(\mvec{x}) = {\mvec{f_1}(\mvec{x}) \choose
\mvec{f_2}(\mvec{x})} 
= \l( \begin{array}{c}
f_{1,0}(\mvec{x})\\
\vdots\\
f_{1,N}(\mvec{x})\\
f_{2,0}(\mvec{x})\\
\vdots\\
f_{2,N}(\mvec{x})
\end{array}\r) = \mvec{0}
\mlab{229}
\ee
with
\be
\mvec{x} = {\mvec{\S} \choose \mvecF} = 
\l( \begin{array}{c}
\S_0\\
\vdots\\
\S_N\\
\F_0\\
\vdots\\
\F_N
\end{array}\r). 
\mlab{230}
\ee

\mref{229} can be solved using Newton's iterative procedure \mref{215},
\be
J(\mvec{x_n})\, \mvec{\Delta_{n+1}} = \mvec{f(\mvec{x_n})} ,
\mlab{231}
\ee
where
\ba
J(\mvec{x}) &=& \l( \parder{f_i(\mvec{x})}{x_j} \r), \hspace{1cm} i,j =
0\dots 2N+1 \\
&=& \l( \renewcommand{\arraystretch}{2}
\begin{array} {c@{\quad}c}
\D\parder{f_{1,i}(\mvec{\S},\mvecF)}{\S_j} &
\D\parder{f_{1,i}(\mvec{\S},\mvecF)}{\F_j} \\
\D\parder{f_{2,i}(\mvec{\S},\mvecF)}{\S_j} &
\D\parder{f_{2,i}(\mvec{\S},\mvecF)}{\F_j} 
\end{array} \r), \hspace{1cm} i,j = 0,\ldots,N .\\
\mlab{232}
\ea

Every iteration now requires the solution of a system of $2N+2$ linear
equations \mref{231} for the $2N+2$ unknown components of the vector
$\mvec{\Delta_{n+1}}$.  Successive iterations will yield new function
approximations for $\S_0,\ldots,\S_N,\F_0,\ldots,\F_N$ computed from:
\be
\mvec{x_{n+1}} = \mvec{x_n} - \mvec{\Delta_{n+1}} ,
\mlab{233}
\ee
or,
\be
{\mvec{\S_{n+1}} \choose \mvec{\mvecF_{n+1}}} = 
{\mvec{\S_{n}} \choose \mvec{\mvecF_{n}}} - \mvec{\Delta_{n+1}} .
\ee

\section{Numerical integration rules}

In the previous sections we replaced the original integral equation by
a system of non-linear algebraic equations using some integration rule
and derived a method to solve this set of equations. If we look at the
error $\mvec{e}$ on the approximate solution $\mvec{\S_n}$ with
respect to the exact solution $\S(x)$ of the original integral
equation we have to consider two error sources (neglecting precision
errors): the error $\mvec{e_R}$ due to the approximation of the
integrals by a finite sum using some quadrature rule $R$ and the error
$\mvec{\e_n}$ introduced by solving the system of non-linear equations
numerically,
\be
\norm{\mvec{e}} \le \norm{\mvec{e_R}} + \norm{\mvec{e_n}} .
\mlab{234}
\ee

As we saw in the previous section the error $\norm{\mvec{e_n}}$ is
well controlled if we use Newton's method, because of its
quadratic convergence. The only limitation on the accuracy in Newton's
method seems to come from the numerical precision of the computation
and the available computer time. Therefore, the major source of error
on the solution will be caused by the quadrature rule. It is important
to note that the situation is very critical when solving integrals as
part of an integral equation problem because the quadrature error gets
amplified quite dramatically in the final solution of the integral
equation as we will see later. Furthermore the choice of quadrature
rule is even more important as the integrals to solve are
two-dimensional and therefore the potential errors even bigger.

If we use a quadrature formula with $N+1$ grid points to approximate
the radial integrals we will end up with a system of $N+1$ non-linear
algebraic equations to solve. To reduce the computing time we want to
get maximum accuracy with a minimum of grid points.

The best-known quadrature formulae are probably the {\it Newton-Cotes}
formulae, using equidistant grid points. To approximate the
integral:
\be
\int_a^b f(x) dx \;,
\mlab{235}
\ee
the most frequently used Newton-Cotes formulae, with their corresponding
error-term are:\vspace{2mm}
\ba
\parbox[b]{14cm}{\mbox{the midpoint rule:}\\[2mm]
\hspace*{1cm}
\mbox{$\D\int_{x_0}^{x_2} f(x) dx = 2h f_1 + \frac{h^3}{3} f^{(2)}(\xi) 
, \qquad x_0 < \xi < x_2$}}  \mlab{238}\\[5mm]
\parbox[b]{14cm}{\mbox{the trapezoidal rule:}\\[2mm]
\hspace*{1cm}
\mbox{$\D\int_{x_0}^{x_1} f(x) dx = \frac{h}{2} (f_0+f_1) 
- \frac{h^3}{12} f^{(2)}(\xi) 
, \qquad x_0 < \xi < x_1$}}   \mlab{239}\\[5mm]
\parbox[b]{14cm}{\mbox{Simpson's rule:}\\[2mm]
\hspace*{1cm}
\mbox{$\D\int_{x_0}^{x_2} f(x) dx = \frac{h}{3} (f_0+4f_1+f_2) 
- \frac{h^5}{90} f^{(4)}(\xi)
, \qquad x_0 < \xi < x_2$}}  \mlab{240}\\[5mm]
\parbox[b]{14cm}{\mbox{3/8-rule:}\\[2mm]
\hspace*{1cm}
\mbox{$\D\int_{x_0}^{x_3} f(x) dx = \frac{3h}{8} (f_0+3f_1+3f_2+f_3) 
- \frac{3h^5}{80} f^{(4)}(\xi)
, \qquad x_0 < \xi < x_3 $}}  \mlab{241}
\ea
where we define $f_j \equiv f(x_j)$ and
\be
x_j = x_0 + jh \;, \hspace{1cm} j = 0,\ldots,N ,
\mlab{236}
\ee
with
\be
x_0 = a , \qquad x_N = b , \qquad h = \frac{b-a}{N} \;.
\mlab{237}
\ee

To approximate the integral value to a good accuracy it will not be
sufficient to use an integration rule with 1, 2, 3 or 4 integration
points, we will normally need many more grid points. For this purpose we
could use the corresponding $(N+1)$-point Newton-Cotes formula.  In practice,
this is not useful because interpolation theory, upon which the
Newton-Cotes formulae are based, tells us that a very high order polynomial
does not in general approximate a function well at all.  Furthermore for $N
\ge 8$ the weights in the quadrature formula start to have different signs
so that the numerical precision of the calculation becomes a worry. A much
better method to increase the number of points is to use composite
integration rules. This consists in dividing the integration interval
$[a,b]$ in $m$ subintervals of size $H$,
\be
H \equiv \frac{b-a}{m} ,
\mlab{242}
\ee
rewriting the total integral as:
\be
\int_a^b f(x)dx = \sum_{j=0}^{m-1} \int_{y_j}^{y_{j+1}} f(x) dx ,
\mlab{243}
\ee
with
\be
y_j = a + jH , \qquad j= 0,\ldots,m \; .
\mlab{244}
\ee

We apply a low order basic Newton-Cotes formula with $n+1$ points on
each subinterval:
\be
\int_a^b f(x)dx = \sum_{j=0}^{m-1} \sum_{k=0}^{n} w_k f(y_j + kh)  \; ,
\mlab{245}
\ee
with the grid spacing $h$ defined as:
\be
h \equiv \frac{H}{n} = \frac{b-a}{mn} \; .
\mlab{247}
\ee

Substituting \mrefb{244}{247} in \mref{245} gives,
\be
\boxeq{
\int_a^b f(x)dx = \sum_{j=0}^{m-1} \sum_{k=0}^{n} w_k f(x_{jn+k})  \; ,
}
\mlab{245.1}
\ee
defining $x_k$ as:
\be
x_k = a + kh  , \qquad k=0,\ldots,mn \; .
\mlab{247.1}
\ee

The composite trapezoidal rule ($n=1$) is (using \mref{239}):
\be
\int_a^b f(x) dx = \sum_{j=0}^{m-1} \frac{h}{2} \l(f_j + f_{j+1}\r) ,
\mlab{248}
\ee
where $f_k = f(x_k)$, such that:
\be
\int_a^b f(x) dx = \frac{h}{2}\l( f_0 + 2\sum_{j=1}^{m-1} f_j +
f_m \r)  \;.
\mlab{250}
\ee
The error term on this rule is:
\be
E_{trap} = -\frac{b-a}{12} h^2 f^{(2)}(\xi) , \qquad a < \xi < b \; .
\mlab{250.1}
\ee

For $n=2$ we derive the composite Simpson's rule using \mref{240}:
\ba
\int_a^b f(x) dx &=& \sum_{j=0}^{m-1} \frac{h}{3}
\l(f_2j+4f_{2j+1}+f_{2j+2}\r) \nn\\
&=& \frac{h}{3} \l( f_0 + 2 \sum_{j=1}^{m-1} f_{2j} 
+ 4  \sum_{j=0}^{m-1} f_{2j+1} + f_{2m} \r) \;,
\mlab{251}
\ea
with error term:
\be
E_{Simp} = -\frac{b-a}{180} h^4 f^{(4)}(\xi)  , \qquad a < \xi < b \; .
\mlab{252}
\ee

The composite Simpson's rule requires an odd total number of grid
points. If for some reason the grid has an even number of points $N+1$
we can use the basic 3/8-rule of \mref{241} on the four first points:
\be
\int_{x_0}^{x_3} f(x) dx = \frac{3h}{8} \l( f_0 + 3 f_1 +
3 f_2 + f_3 \r) \;,
\mlab{253}
\ee
and use the composite Simpson's rule \mref{251} on the remaining
integral which has an odd number of points $N-2$. From
\mrefb{240}{241} we note that both basic rules are of comparable
accuracy so that the global accuracy of this mixed composite rule
will be comparable to \mref{252} .

\section{Implementation of the quadrature rule}
\label{quadrule}

\subsection{Estimate of computing time}

In this section we are going to apply the quadrature rules mentioned before
to the integral equation \mref{S4} describing the dynamical generation of
fermion mass in QED. We will at first use the trapezoidal rule and roughly
study the behaviour of the solution of the integral equation and the
computing time needed to find this solution using Newton's method from
Section~\ref{Newton} with an increasing number of grid points in the radial
integrals (with a fixed number of points in the angular integrals).
Throughout our study we will take an ultraviolet cutoff \fbox{$\Lambda^2 =
1\e10$}. In Table~\ref{Tab:1} we increase the number of grid panels,
$N_R$, from 100 to 1000 and tabulate the values of $\S(0)$, because it is
representative for the scale of the generated fermion mass, and the real
time (min:s) needed to compute the angular integrals. The other parameters
are chosen as $\alpha=2.086$ and the infrared cutoff $\kappa^2=0.1$.

\begin{table}[htbp]
\begin{center}
\begin{tabular}{|r|r|c|r|}
\hline
$N_R$ & $\S(0)$ & $\Delta t_\theta$ & $\Delta\S(0)$ \\
\hline
 100 & 112.74 & 0:04 &  \\
 200 &  91.90 & 0:15 & 20.84 \\
 300 &  88.78 & 0:32 &  3.12 \\
 400 &  87.88 & 0:55 &  0.90 \\
 500 &  87.50 & 1:27 &  0.38 \\
1000 &  87.10 & 5:16 &  0.40 \\
\hline
\end{tabular}
\caption{$\S(0)$ versus number of radial integration panels $N_R$ using
the trapezoidal rule. $\Delta t_\theta$ is the real time (min:s)
needed to compute the angular integrals. $\Delta \S(0)$ is the change
in $\S(0)$ when increasing the number of points. $\alpha=2.086$,
$\kappa^2=0.1$.}
\label{Tab:1}
\end{center}
\end{table}

Changing the number of radial points $N_R+1$ to $N'_R+1$ produces an
increase of $[(N'_R+1)/(N_R+1)]^2$ angular integrations to be computed as
there are $N_R+1$ equations each with $N_R+1$ radial points for which to
compute angular integrals. We note from Table~\ref{Tab:1} that indeed the
time $\Delta t_\theta$ increases as $[(N'_R+1)/(N_R+1)]^2$. $\Delta
t_\theta$ only gives part of the computer time needed by the program. Once
the angular integrals have been calculated, we have to solve the system of
$N_R+1$ non-linear equations. This will be done using Newton's iterative
procedure,
\mref{218.3}, involving the solution of a system of $N_R+1$ linear equations
at each iteration step. Because of the large size of these linear systems
they will be solved numerically. The solution time of this procedure
increases with the increasing number of equations.  Although the computing
time needed to solve the system of linear equations is small compared to
$\Delta t_\theta$ for reasonable $N_R$ ($< 500$), it becomes quite large
for larger $N_R$. For $N_R=1000$, each iteration step of Newton's method
takes about 2 minutes. If we start from a good initial guess for
$\mvec{\S^0}$ the iterative procedure will converge after 5 iterations; the
total computing time will then approximately be $\Delta t(N_R=1000) \approx
5 + (5 \times 2) \approx 15$ minutes. This is quite long considering the
relative simplicity of the integral equation. From $\Delta\S(0)$ in
Table~\ref{Tab:1} we see that $\S(0)$ improves significantly for $N_R$ up
to 500. The stagnation when $N_R \to 1000$ probably means that the accuracy
of the angular integrals only permits a global relative accuracy of about
0.45\% for the final solution.

\subsection{Influence of infrared cutoff}

\begin{figure}[htbp]
\begin{center}
\mbox{\epsfig{file=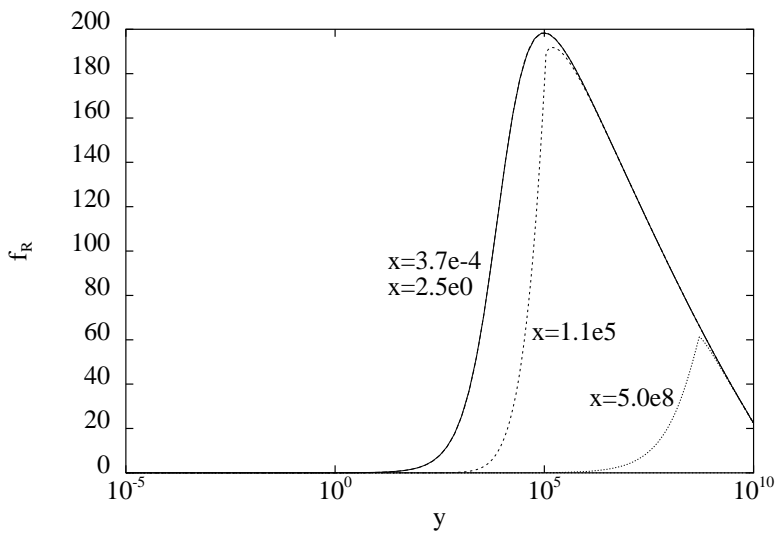,angle=-90,height=8cm}}
\end{center}
\vspace{-5mm}
\caption{Radial integrand $f_R(y)$ for $\alpha=2.086$ as a function 
of $y$ for various values of external fermion momentum $x=$
3.7e-4, 2.5e0, 1.1e5 and 5.0e8.}
\label{Fig:radint}
\end{figure}

In Section~\ref{S-equation} we have mentioned that the introduction
of an infrared cutoff $\kappa^2$ for numerical purposes requires that
we either have an analytic evaluation of the truncated infrared part
of the integral or that $\kappa^2$ should be chosen so that this part
is negligible. In this section we will look at the influence of the
variation of the infrared cutoff $\kappa^2$ on the numerical results
of the calculation. If we plot the radial integrand as in
Fig.~\ref{Fig:radint} we see that this integrand decreases rapidly
for momenta below the scale of the generated fermion mass ($y <
\S^2(0)$).  

Consequently, we do not expect any significant contribution
from that part of the integral. This means the choice of $\kappa^2$ is
dependent on the generated fermion mass and thus on the coupling for
which we solve the integral equation. If we fix $\kappa^2$ at some
value, and vary the coupling $\alpha$ we will only get reliable
results for couplings down to $\alpha_{min}$ for which the generated
fermion mass is larger than $\kappa$. In practice we only expect
to be able to find accurate solutions down to $\S(0)
\approx \Order(1)$ (taking $\Lambda^2 = 1\e10$) because of the limitations
imposed by the numerical precision of the calculation. In Table~\ref{Tab:2}
we show $\S(0)$, varying $\kappa^2$ from $1\ten{4}$ to $1\ten{-5}$ for
$\alpha=2.086$.  When changing the value of the infrared cutoff $\kappa^2$,
we accordingly modify the number of integration panels $N_R$ in order to
have the same grid spacing in every case:
\be
h \equiv \frac{\logten\Lambda^2-\logten\kappa^2}{N_R} =
\frac{1}{N_R}\logten\frac{\Lambda^2}{\kappa^2} = \frac{1}{30} \;.
\mlab{254}
\ee

\begin{table}[htbp]
\begin{center}
\begin{tabular}{|r|r|r|c|}
\hline
$\kappa^2$ & $N_R$ & $\S(0)$\hspace{2mm} & $\Delta t_\theta$ \\
\hline
1e4 & 180 & 69.0990 & 0:11 \\
1e3 & 210 & 87.0883 & 0:12 \\
100 & 240 & 88.2892 & 0:19 \\
10 & 270 & 88.3980 & 0:23 \\
1 & 300 & 88.4083 & 0:29 \\
0.1 & 330 & 88.4093 & 0:42 \\
0.01 & 360 & 88.4094 & 0:46 \\
1e-5 & 450 & 88.4094 & 1:05 \\
\hline
\end{tabular}
\caption{$\S(0)$ versus infrared cutoff $\kappa^2$ for $\alpha=2.086$ 
using the trapezoidal rule. The number of radial integration panels
$N_R$ is chosen to have a fixed grid spacing $h=\frac{1}{30}$. $\Delta
t_\theta$ is the real time (min:s) needed to compute the angular
integrals.  }
\label{Tab:2}
\end{center}
\end{table}

From Table~\ref{Tab:2} we see that indeed taking the infrared cutoff
$\kappa^2 < \Order(\frac{\S^2(0)}{100})$ is sufficient to achieve an
accuracy of $\approx 0.1\%$. For $\alpha=2.086$ a suitable choice
could be $\kappa^2 \sim 100$. If we are to investigate smaller values
of the coupling, closer to its critical value, we will have to choose
a smaller value of $\kappa^2$.

\subsection{Influence of grid spacing, kink in the integrand}

Having fixed the infrared cutoff $\kappa^2$, we will now turn our
attention to the influence of the grid spacing $h$, which is inversely
proportional to the number of grid panels, $N$. We performed the
calculation using the composite trapezoidal rule and the composite
Simpson's rule. From textbooks it is well known that Simpson's rule
generally yields better results than the trapezoidal rule because the
convergence of the composite Simpson's rule is proportional to $1/N^4$,
while the convergence of the trapezoidal rule is proportional to
$1/N^2$. With convergence of a quadrature rule we mean the convergence
of the finite sum to the exact integral value when $N\to\infty$. The
results from Table~\ref{Tab:3} were computed for $\alpha=2.086$ with
$\kappa^2=100$.
\begin{table}[htbp]
\begin{center}
\begin{tabular}{|r|r|r|r|r|r|}
\hline
$N_R$ & $1/h$ & $\S_{trap}(0)$ & $\S_{Simp}(0)$ & $\Delta t_\theta$ &
$\Delta t_{iter}$ \\
\hline
80 & 10 & 107.387 & 172.565 & 0:03  & \\
120 & 15 & 94.640 & 124.963 & 0:05 & \\
160 & 20 & 90.773 & 108.192 & 0:09 & \\
200 & 25 & 89.115 & 100.280 & 0:13 & \\
240 & 30 & 88.289 & 96.072 & 0:18 & \\
280 & 35 & 87.844 & 93.555 & 0:26 & \\
320 & 40 & 87.572 & 91.932 & 0:34 & \\
360 & 45 & 87.397 & 90.840 & 0:39 & \\
400 & 50 & 87.276 & 90.056 & 0:51 & \\
600 & 75 & 87.027 & 88.249 &  &  \\
800 & 100 & 86.959 & 87.642 &  & 9:00 \\
1000 & 125 & 86.933 & 87.367 &  & 16:00 \\
2000 & 250 & 86.908 &  &  & 2:35:25 \\
5000 & 625 & crash &  &  & \\
\hline
\end{tabular}
\caption{$\S(0)$ versus number of radial integration panels $N_R$ and
grid density $1/h$ using the trapezoidal rule and Simpson's
rule. $\Delta t_\theta$ is the real time (min:s) needed to compute the
angular integrals, $\Delta t_{iter}$ (h:min:s) is the total real
time. $\alpha=2.086$, $\kappa^2=100$.}
\label{Tab:3}
\end{center}
\end{table}

The run with $N_R=5000$ crashed because of memory allocation problems
in solving the linear system of equations. From Table~\ref{Tab:3} we
see that Simpson's rule gives worse results than the trapezoidal rule.
This is quite puzzling as it has a higher degree of precision.
However, for the error formulae to be applicable, the integrand has to
be sufficiently smooth. If not, the accuracy of Simpson's rule can be
just as good or bad as the one from the trapezoidal rule. 

If we look at the radial integrand it becomes clear why this happens.
In the quenched case, where the angular integrals can be computed
analytically, a typical angular integration will yield:
\be
I_\theta \equiv \int_0^\pi d\theta \; \frac{\sin^2\theta}{z}
= \frac{\pi}{2}\l[\frac{\theta(x-y)}{x} +
\frac{\theta(y-x)}{y} \r] \; .
\mlab{255}
\ee

The various angular integrals all have this characteristic feature:
\be
I_\theta \sim a_{<}(y,x) \,\theta(x-y) + a_{>}(y,x) \,\theta(y-x) \; .
\mlab{256}
\ee

\begin{figure}[htbp]
\begin{center}
\mbox{\epsfig{file=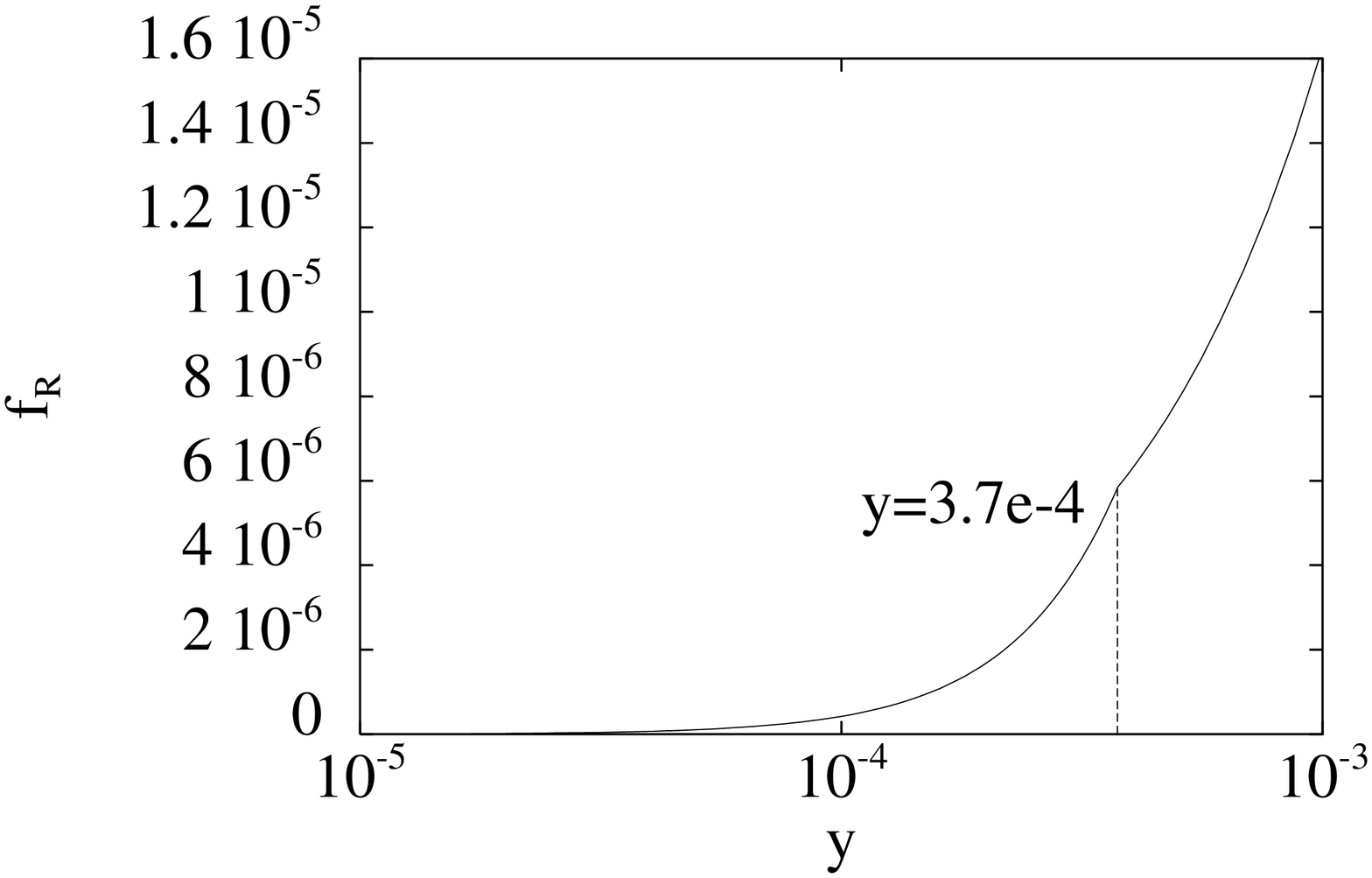,angle=-90,height=5.5cm}
\epsfig{file=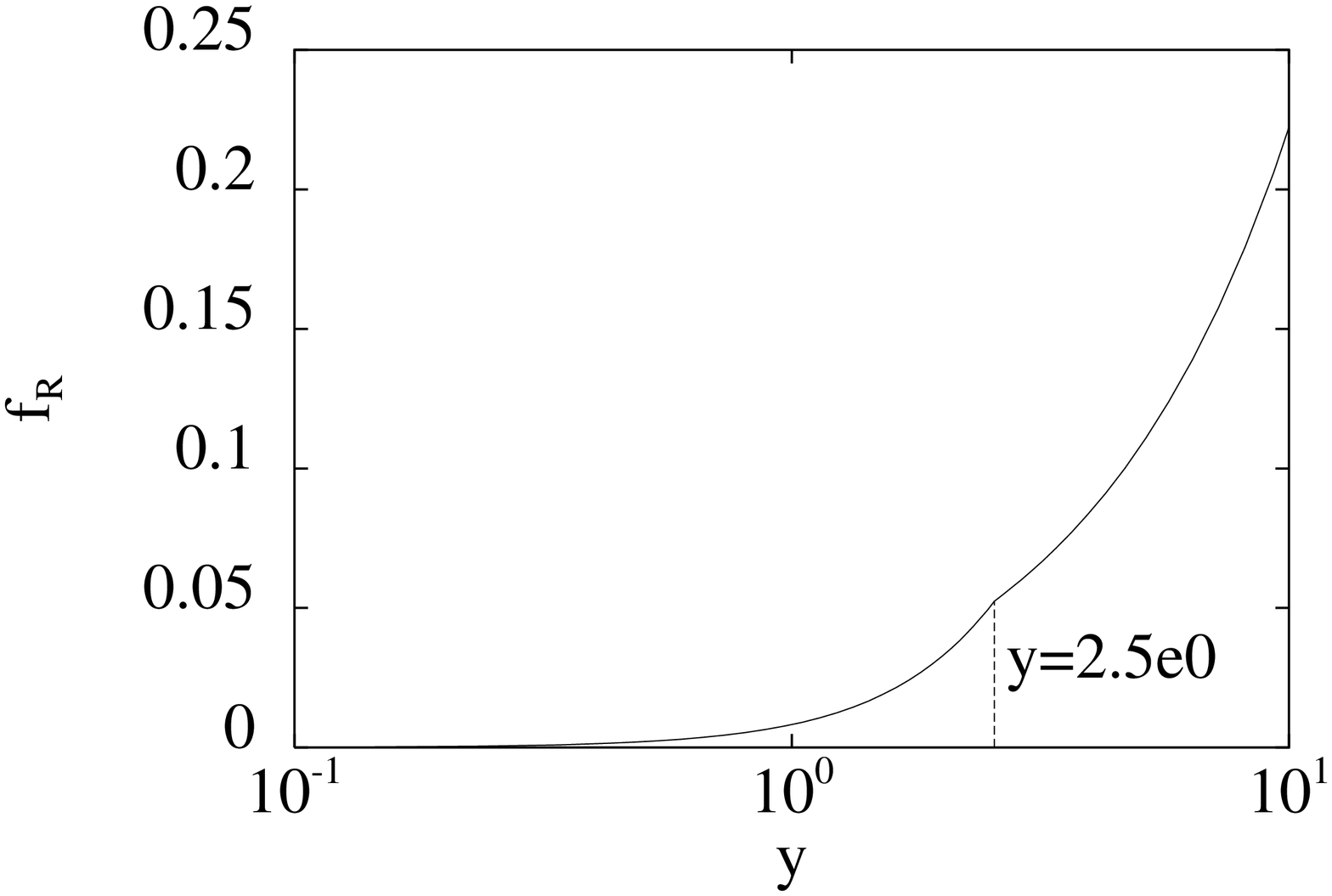,angle=-90,height=5.5cm}}
\mbox{\epsfig{file=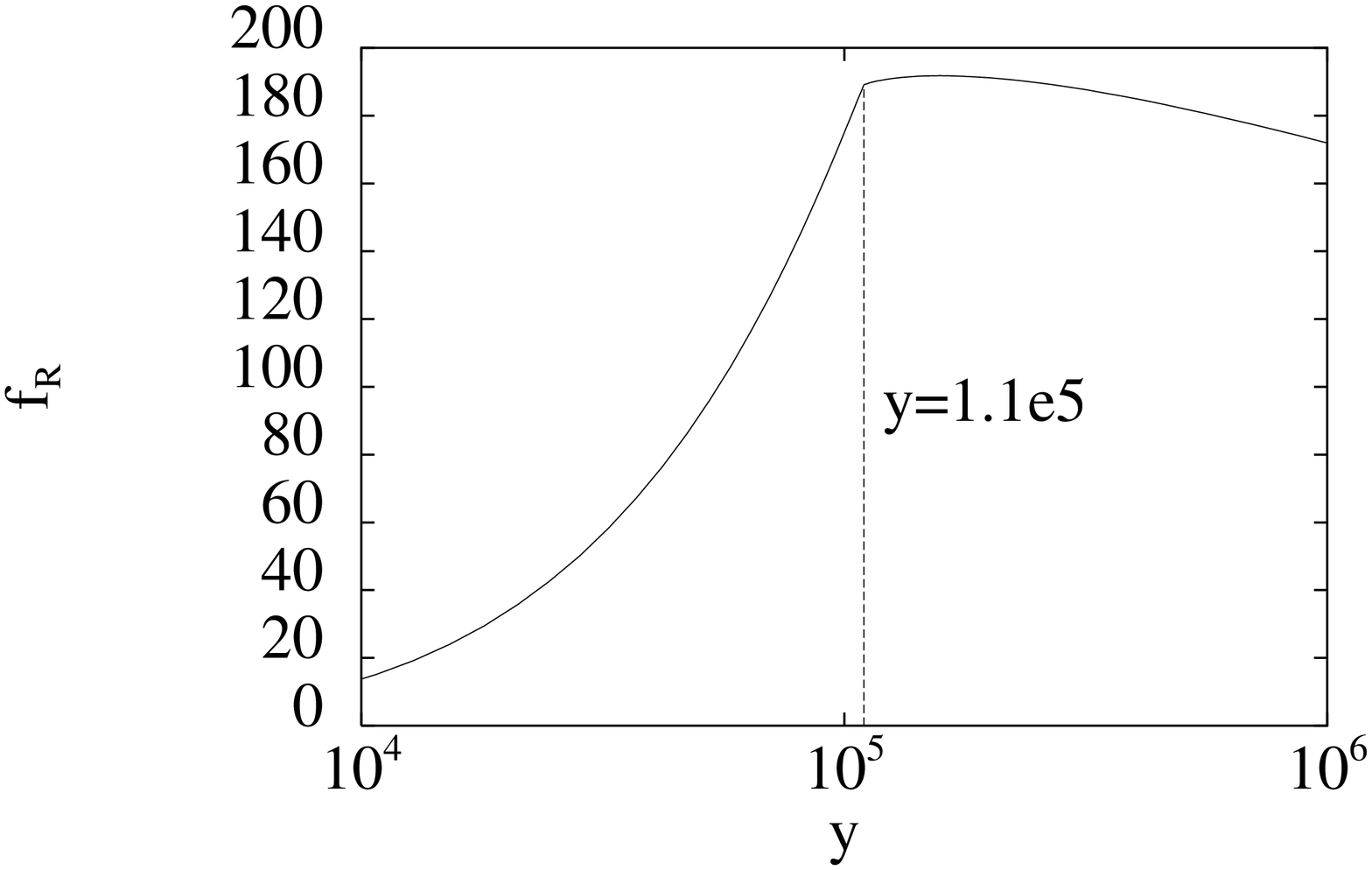,angle=-90,height=5.5cm}
\epsfig{file=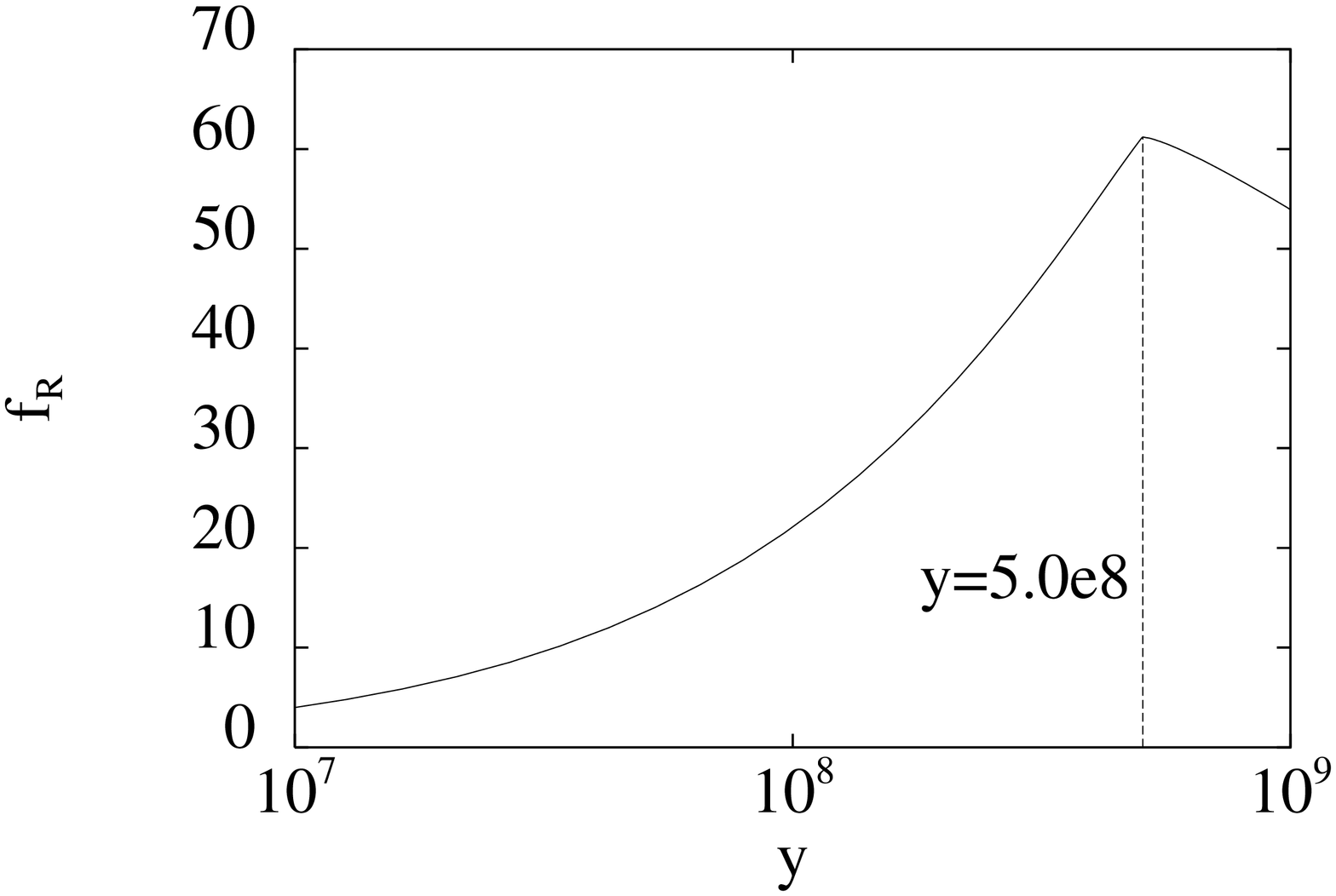,angle=-90,height=5.5cm}}
\end{center}
\vspace{-5mm}
\caption{Zoomed views of the kink at $y=x$ in the radial integrand 
$f_R(y)$ for various values of external fermion momentum $x=$
3.7e-4, 2.5e0, 1.1e5 and 5.0e8, for $\alpha=2.086$}
\label{Fig:radint.zoom}
\end{figure}

In the unquenched case the angular integrals are solved numerically
because of the function $\G(z)$ appearing in the angular integrals
of the fermion equation.  Still the shape of \mref{256} will remain
valid and this will cause a kink in the kernel of the radial integral
at $y = x$. This can be seen in Figs.~\ref{Fig:radint},
\ref{Fig:radint.zoom} where we plot the radial integrand for a number of
values of external momentum $x$. Fig.~\ref{Fig:radint.zoom} shows
enlargements of the radial integrands in the neighbourhood of the kink
at $y=x$. Although the kernel is continuous, it is not smooth as
its first derivative has a discontinuity. This implies that the error
formulae on the integration rules, Eqs.~(\oref{238}-\oref{241}, \oref{250.1},
\oref{252}), are not applicable. According to \mref{250.1} the
composite trapezoidal rule has an error decreasing as $h^2$ if it has a
continuous 2$^{nd}$ derivative, while from \mref{252} Simpson's rule goes as
$h^4$ if it has a continuous 4$^{th}$ derivative. Because of the
discontinuity in the 1$^{st}$ derivative of the integrand, no higher degree
rule applied on the interval $[a,b]$ will be able to give us a better
result than the trapezoidal rule.

Even so, we see from Table~\ref{Tab:3} that, for an equal number
of grid points, the results of the trapezoidal rule are better
than those from Simpson's rule. The reason for this is that Simpson's
rule uses three points on each subinterval, while the trapezoidal rule
only uses two. If we consider the $i^{th}$ equation from the system,
\mref{S6}, the radial integrand will have a kink at $x_j=x_i$. If $x_i$
is an endpoint of a subinterval the integrands over all the individual
subintervals are smooth and Simpson's rule should behave according
to its error formula \mref{252}. In contrast, if $x_i$ is a midpoint
of a subinterval then the integrand over the subinterval
$[x_{i-1},x_{i+1}]$ is not smooth, unlike over the other subintervals,
so that the integration rule will generate a considerable error. The
trapezoidal rule always has the kink as an endpoint of a subinterval
and so its error formula is always applicable. The numerics of this
phenomenon have been checked by comparing the values of the total
integrals for various values of external momentum $x_i$ using the
trapezoidal rule and Simpson's rule. The integrals with a kink in the
middle of a subinterval, i.e. for odd $i$, definitely yield worse
values with Simpson's rule than with the trapezoidal rule. Because the
integrals are the building blocks of the integral equation, the error
on each individual integral propagates into the final solution of the
integral equation. The accuracy of this solution will only be as good
as the worst integral evaluation. Therefore the trapezoidal rule will
yield a better solution of the integral equation than Simpson's
rule. If we are not satisfied with the results computed with the
trapezoidal rule (slow convergence when the number of grid points
$N_R+1$ is increased) and would like to use a higher degree rule
efficiently, we will have to handle the kink in the radial integrand
in an appropriate way.

The evident way to take care of the kink in the radial integrand is to
split the integration range into two separate integrals:
\ba
\S_i &=& \int_{\kappa^2}^{\Lambda^2} dy \; K(x_i, y) \nn \\
&=& \int_{\kappa^2}^{x_i} dy \; K(x_i, y) 
+ \int_{x_i}^{\Lambda^2} dy \; K(x_i, y)  \; , \qquad i=0,\ldots,N \;,
\mlab{257}
\ea
and approximate each of the integrals by an appropriate integration
rule.

Each of these two subintegrals now has an integrand which is smooth
over the integration interval. The accuracy of the numerical
integration should now respect the theoretical error formula. 

When implementing the composite integration rules we have to avoid two
pitfalls. Firstly, if the total number of panels (number of grid
points minus one) is not a multiple of the number of panels of the
basic rule, we have to combine different basic rules preferably of
comparable accuracy. Secondly, because of the kink in the radial
integrand, the composite rule must have the kink as an endpoint of one
of its subintervals if we want to achieve the accuracy predicted by
the theoretical error formula. The importance of avoiding these
pitfalls will now be demonstrated by considering the numerical
integration of functions behaving in a way similar to \mref{S2} but
for which the exact integral value can be calculated
analytically. This will allow us to compare the numerical and
analytical results.

\subsection{Smooth toy kernel}

To study the construction of the composite formula in the absence of
any kink in the integrand, we will replace the non-smooth radial
integrand by a function behaving in a similar way but without any
kink:
\be
f_R(x,y) = \frac{My}{y+M^2} \frac{2}{y+x} \;.
\mlab{258}
\ee   

The integral of this function,
\be
I(x) = \int_{\kappa^2}^{\Lambda^2} dy \; f_R(x,y) 
= \int_{\kappa^2}^{\Lambda^2} dy \; \frac{My}{y+M^2} \frac{2}{y+x} \;,
\mlab{259.0}
\ee
 is readily computed analytically,
\be
I(x) = \l\{
\begin{array}{l}
\D \frac{2M}{M^2-x}\l[M^2\ln\frac{\Lambda^2+M^2}{\kappa^2+M^2} 
-x\ln\frac{\Lambda^2+x}{\kappa^2+x}\r] \;, \qquad x \neq M^2 \\
\D 2M\l[\frac{M^2}{\Lambda^2+M^2} - \frac{M^2}{\kappa^2+M^2} +
\ln\frac{\Lambda^2+M^2}{\kappa^2+M^2}\r] \;, \qquad x = M^2 \;.
\end{array}
\r.
\mlab{259}
\ee

For $M=100$, $\kappa^2=100$ and $\Lambda^2=1\e10$, we have $I(x=1\e4) =
2563.093$, for instance.

We now apply various composite $(n+1)$-point Newton-Cotes rules to this
integral for $x=1\e4$ and show the results in Table~\ref{Tab:4}.
\begin{table}[htbp]
\begin{center}
\begin{tabular}{|r|c|c|c|c|c|c|c|}
\hline
$N_R$ & $E_1$ & $E_2$ & $E_3$ & $E_4$ & $E_5$ & $E_6$ & $E_7$ \\
\hline
 100 & 4.2e-08 & 3.7e-10 & 1.8e-08 & 1.7e-11 & 7.4e-11 & 7.3e-11 & 3.6e-10 \\ 
 200 & 1.1e-08 & 2.3e-11 & 6.4e-11 & 2.7e-13 & 5.8e-13 & 9.9e-12 & 2.5e-13 \\
 300 & 4.7e-09 & 4.6e-12 & 1.0e-11 & 2.4e-14 & 5.2e-14 & 2.2e-16 & 5.6e-16 \\
 400 & 2.6e-09 & 1.5e-12 & 2.5e-10 & 4.4e-15 & 9.4e-15 & 1.8e-15 & 2.5e-10 \\
 500 & 1.7e-09 & 5.9e-13 & 1.5e-12 & 8.9e-16 & 1.4e-15 & 9.2e-14 & 3.2e-13 \\
 600 & 1.2e-09 & 2.9e-13 & 6.4e-13 & 5.6e-16 & 1.1e-15 & 2.2e-16 & 4.4e-16 \\
 700 & 8.7e-10 & 1.6e-13 & 4.6e-11 & 0       & 5.6e-16 & 0       & 6.7e-16 \\
 800 & 6.6e-10 & 9.1e-14 & 2.1e-13 & 4.4e-16 & 0       & 8.4e-15 & 8.9e-15 \\
 900 & 5.2e-10 & 5.7e-14 & 1.3e-13 & 3.3e-16 & 6.7e-16 & 0       & 2.2e-16 \\
1000 & 4.2e-10 & 3.7e-14 & 1.6e-11 & 6.7e-16 & 6.7e-16 & 4.4e-16 & 2.2e-16 \\
\hline
\end{tabular}
\caption{Relative error $E_n=\l|(I_{num}-I_{exact})/
I_{exact}\r|$ from the numerical calculation of $I(x)$ of
\protect{\mref{259.0}} for $x=1\e4$ using composite $(n+1)$-point 
Newton-Cotes formulae with $n=1,\ldots,7$ for increasing total number of
grid panels $N_R$.}
\label{Tab:4}
\end{center}
\end{table}

\def\dop{D}
Comparing the results from Table~\ref{Tab:4} shows that for a fixed total
number of grid panels $N_R$, the higher degree rules perform significantly
better than the lower ones (except for $n=3,6,7$). The degree of precision
$\dop$ of a quadrature rule is defined such that all polynomials of degree
at most equal to the degree of precision are integrated exactly by the
quadrature formula. The degree of precision of the various rules is given
in Table~\ref{Tab:5}.
\begin{table}[htbp]
\begin{center}
\begin{tabular}{|c||c|c|c|c|c|c|c|c|}
\hline
$n$ & 1 & 2 & 3 & 4 & 5 & 6 & 7 \\
\hline
$\dop$ & 1 & 3 & 3 & 5 & 5 & 7 & 7 \\
\hline
\end{tabular}
\end{center}
\caption{Degree of precision $\dop$ of the $(n+1)$-point Newton-Cotes 
formulae with $n=1,\ldots,7$.}
\label{Tab:5}
\end{table}

Using a rule with a higher degree of precision seems to yield significantly
improved results for the integral evaluation, till the maximum accuracy
of about $1\e{-16}$ imposed by the use of double precision arithmetics
has been reached.

Also from Table~\ref{Tab:4}, we see that increasing the total number of
grid panels from $N_R$ to $N'_R$, using the same basic rule, seems to yield
the expected $(N'_R/N_R)^{(\dop+1)}$ improvement in accuracy, again till
the maximum accuracy is reached. This is not true when the number of panels
in the basic NC-rule is $n=3,6,7$. To construct a composite formula using a
single basic NC-rule, the total number of panels in the integration
interval must be a multiple of the number of panels of the basic
NC-rule. For $n=1,2,4,5$ all the total number of panels $N_R$ considered in
Table~\ref{Tab:4} are indeed multiples of the number of panels of their
basic NC-rule. This is not so for $n=3,6,7$. For $n=3$ this will be
satisfied for $N_R=300,600,900$. For other values of $N_R$ we have to
adapt the composite rule by taking as many $n$-panel rules as will fit in
$N_R$ panels and use an $n'$-panel rule on the remaining interval as shown
in Table~\ref{Tab:6}. For example, the $n=3$ case with 400 radial panels
will be composed of 133 3-panel or 3/8-rules and one trapezoidal rule.
\begin{table}[htbp]
\begin{center}
\begin{tabular}{|r|c|c|c|}
\hline
$N_R$ & $n=3$ & $n=6$ & $n=7$ \\
\hline
 100 &  33\maal 3+1 &  16\maal 6+4 &  14\maal 7+2  \\
 200 &  66\maal 3+2 &  33\maal 6+2 &  28\maal 7+4 \\
 300 & 100\maal 3   &  50\maal 6   &  42\maal 7+6 \\
 400 & 133\maal 3+1 &  66\maal 6+4 &  57\maal 7+1 \\
 500 & 166\maal 3+2 &  83\maal 6+2 &  71\maal 7+3 \\
 600 & 200\maal 3   & 100\maal 6   &  85\maal 7+5 \\
 700 & 233\maal 3+1 & 116\maal 6+4 & 100\maal 7   \\
 800 & 266\maal 3+2 & 133\maal 6+2 & 114\maal 7+2 \\
 900 & 300\maal 3   & 150\maal 6   & 128\maal 7+4 \\
1000 & 333\maal 3+1 & 166\maal 6+4 & 142\maal 7+6 \\
\hline
\end{tabular}
\caption{Structure of the composite $(n+1)$-point NC-rules for $n=3,6,7$ for a
total number of grid panels $N_R$, written as $m$ times an $n$-panel rule
complemented by one $n'$-panel rule.}
\label{Tab:6}
\end{center}
\end{table}

We now look back at the $n=3$ results of Table~\ref{Tab:4} using
Table~\ref{Tab:6}. For $N_R=300,600,900$, the composite rule can be wholly
constructed with basic 3/8-rules. For $N_R=200,500,800$, the composite
3/8-rule has to be complemented by one Simpson's rule. Because Simpson's
rule and the 3/8-rule have comparable accuracy this does not affect the
global accuracy of the composite rule. However, for $N_R=100,400,700,1000$,
the composite 3/8-rule has to be complemented with one trapezoidal rule
yielding significantly worse results. From the error term in
\mrefb{239}{241} one can prove theoretically that this mixed composite rule
behaves as $1/N^3$, while the pure composite trapezoidal rule goes as
$1/N^2$ and the pure Simpson's and 3/8-rule go as $1/N^4$.  The results of
Table~\ref{Tab:4} for $N_R=100,400,700,1000$ have indeed a $1/N^3$
convergence rate. We see an analogous pattern for $n=6,7$. We can deduce
from Tables~\ref{Tab:4}, \ref{Tab:6} that the error on the integral
evaluation is determined by the least accurate of the subrules used even if
it is only used once in the total evaluation. Further demonstration of this
can be found in Table~\ref{Tab:7} where we alter the total number of grid
panels in the cases $n=3,6,7$ to make it a multiple of the number of panels
of the basic rules. It is clear that for $n=3$ the error formula is now
well respected and that the error is comparable to that of Simpson's rule
($n=2$) from Table~\ref{Tab:4}, as it should be for composite rules of equal
degree of precision. For $n=6,7$ the improvement of the accuracy with
increasing total number of grid points seems to respect the error formula
although the maximum accuracy of
\Order(1e-16) is reached very rapidly.
    
\begin{table}[htbp]
\begin{center}
\begin{tabular}{|r|c||r|c||r|c|}
\hline
$N_R$ & $E_3$ & $N_R$ & $E_6$ & $N_R$ & $E_7$ \\
\hline
 99 & 8.5e-10 &  97 & 2.3e-10 &  99 & 1.0e-09 \\
198 & 5.4e-11 & 193 & 5.0e-15 & 197 & 8.2e-15 \\  
297 & 1.1e-11 & 289 & 2.2e-16 & 295 & 0       \\  
396 & 3.4e-12 & 385 & 6.7e-16 & 393 & 2.2e-16 \\  
495 & 1.4e-12 & 481 & 4.4e-16 & 491 & 0       \\  
594 & 6.7e-13 & 577 & 5.6e-16 & 589 & 8.9e-16 \\  
693 & 3.6e-13 & 673 & 4.4e-16 & 687 & 3.3e-16 \\  
792 & 2.1e-13 & 769 & 5.6e-16 & 785 & 4.4e-16 \\  
891 & 1.3e-13 & 865 & 3.3e-16 & 883 & 2.2e-16 \\  
990 & 8.7e-14 & 961 & 8.9e-16 & 981 & 2.2e-16 \\  
\hline
\end{tabular}
\caption{Relative error $E_n=\l|(I_{num}-I_{exact})/
I_{exact}\r|$ from the numerical calculation of $I(x)$ of
\protect{\mref{259.0}} for $x=1\e4$, adapting the number of panels $N_R$ to
use pure composite $(n+1)$-point Newton-Cotes formulae with $n=3,6,7$ for
increasing total number of grid panels $N_R$.}
\label{Tab:7}
\end{center}
\end{table}

\subsection{Toy kernel with kink}

We will now make an analogous study using a simplified kernel which
has a kink in the integration region:
\be
f_R(x,y) = \frac{My}{y+M^2} \frac{1}{\max(x,y)} \;.
\mlab{260}
\ee

The integration of this function, 
\be
I(x) =  \int_{\kappa^2}^{\Lambda^2} dy \; f_R(x,y) 
= \int_{\kappa^2}^{\Lambda^2} dy \; \frac{My}{y+M^2}
\frac{1}{\max(x,y)}
\mlab{261}
\ee
can be performed analytically yielding:
\be
I(x) = M - \frac{M\kappa^2}{x} - \frac{M^3}{x}\ln\frac{x+M^2}{\kappa^2+M^2} +
         M\ln\frac{\Lambda^2+M^2}{x+M^2} \;.
\mlab{261.1}
\ee 

For $M=100$, $\kappa^2=100$ and $\Lambda^2=1\e10$, we compute
$I(x=1\e4)=1342.917$. The numerical results computed with the composite
NC-formulae are tabulated in Table~\ref{Tab:8}.

\begin{table}[htbp]
\begin{center}
\begin{tabular}{|r|r|c|c|c|c|c|c|c|}
\hline
$N_R$ & $i_{kink}$ & $E_1$ & $E_2$ & $E_3$ & $E_4$ &
$E_5$ & $E_6$ & $E_7$ \\
\hline
 100 &  25 & 1.1e-04 & 2.1e-04 & 1.4e-06 & 2.3e-04 & 5.3e-09 & 4.3e-04 & 7.1e-05 \\
 200 &  50 & 2.6e-05 & 3.7e-08 & 8.4e-08 & 1.4e-05 & 9.8e-11 & 1.1e-11 & 4.6e-05 \\
 300 &  75 & 1.2e-05 & 2.3e-05 & 1.7e-08 & 2.7e-05 & 8.8e-12 & 4.8e-05 & 2.8e-05 \\
 400 & 100 & 6.6e-06 & 2.3e-09 & 5.5e-09 & 7.4e-13 & 1.6e-12 & 7.1e-15 & 1.5e-05 \\
 500 & 125 & 4.2e-06 & 8.4e-06 & 2.1e-09 & 9.5e-06 & 4.2e-13 & 1.7e-05 & 7.9e-06 \\
 600 & 150 & 2.9e-06 & 4.6e-10 & 1.0e-09 & 1.6e-06 & 1.4e-13 & 2.2e-16 & 2.0e-06 \\
 700 & 175 & 2.1e-06 & 4.3e-06 & 6.0e-10 & 4.9e-06 & 5.5e-14 & 8.8e-06 & 0       \\
 800 & 200 & 1.6e-06 & 1.5e-10 & 3.3e-10 & 1.1e-14 & 2.5e-14 & 8.4e-15 & 1.1e-06 \\
 900 & 225 & 1.3e-06 & 2.6e-06 & 2.0e-10 & 2.9e-06 & 1.2e-14 & 5.3e-06 & 2.4e-06 \\
1000 & 250 & 1.1e-06 & 6.0e-11 & 1.5e-10 & 5.6e-07 & 7.6e-15 & 4.4e-16 & 2.5e-06 \\
\hline
\end{tabular}
\caption{Relative error $E_n=\l|(I_{num}-I_{exact})/
I_{exact}\r|$ from the numerical calculation of $I(x)$ of
\protect{\mref{261}} for $x=1\e4$ using composite $(n+1)$-point 
Newton-Cotes formulae with $n=1,\ldots,7$ for
increasing total number of grid panels $N_R$. $i_{kink}$ gives the position
of the kink within the $N_R+1$ points of the grid, $i \in [0,N_R]$.}
\label{Tab:8}
\end{center}
\end{table}

From Table~\ref{Tab:8} we see that increasing the total number of
integration panels $N_R$ in the composite trapezoidal rule gives the
improvement expected from \mref{250.1}. However, for Simpson's rule ($n=2$)
this is not so.  When $N_R$ is such that the index of the kink is odd, the
integral evaluation is clearly worse than when it is even. This reflects
the fact that an odd index means that the kink is {\it not} an endpoint of
a basic Simpson's rule. The accuracy, in this case, is comparable to the
one achieved with the trapezoidal rule. The 3/8-rule ($n=3$) behaves in a
better way than Simpson's rule. As we saw in Table~\ref{Tab:6} the
composite 3/8-rule is a mixed one. In our exercise we use the first $n'$
($n'<3$) panels to apply one $n'$-point NC-rule, the remaining points are
used to apply a pure composite 3/8-rule. One can check that this
implies that the kink will always be an endpoint of a basic 3/8-rule for any
$N_R$ of Table~\ref{Tab:8}. For $n=5$, it is obvious that the kink will be
an endpoint as the index of the kink is a multiple of 5, which is also the
number of panels in the basic NC-rule.  For $n=4,6,7$ the kink will only be
an endpoint for some values of $N_R$, hence the erratic behaviour of the
computed integral value. To improve on the previous calculation we will now
split the integral as suggested in \mref{257}. The results are shown in
Table~\ref{Tab:9}.
\begin{table}[htbp]
\begin{center}
\begin{tabular}{|r|r|c|c|c|c|c|c|c|}
\hline
$N_R$ & $i_{kink}$ & $E_1$ & $E_2$ & $E_3$ & $E_4$ &
$E_5$ & $E_6$ & $E_7$ \\
\hline
 100 &  25 & 1.1e-04 & 6.1e-07 & 1.4e-06 & 2.1e-08 & 5.3e-09 & 1.9e-08 & 1.1e-09 \\
 200 &  50 & 2.6e-05 & 3.7e-08 & 8.4e-08 & 5.6e-11 & 9.8e-11 & 1.1e-11 & 2.1e-09 \\
 300 &  75 & 1.2e-05 & 7.9e-09 & 1.7e-08 & 7.1e-12 & 8.8e-12 & 4.5e-12 & 1.3e-12 \\
 400 & 100 & 6.6e-06 & 2.3e-09 & 5.5e-09 & 7.4e-13 & 1.6e-12 & 7.1e-15 & 3.0e-13 \\
 500 & 125 & 4.2e-06 & 1.1e-09 & 2.1e-09 & 1.3e-10 & 4.2e-13 & 1.1e-15 & 1.1e-15 \\
 600 & 150 & 2.9e-06 & 4.6e-10 & 1.0e-09 & 1.0e-13 & 1.4e-13 & 2.2e-16 & 1.3e-13 \\
 700 & 175 & 2.1e-06 & 2.9e-10 & 6.0e-10 & 3.1e-14 & 5.6e-14 & 4.5e-11 & 2.2e-16 \\
 800 & 200 & 1.6e-06 & 1.5e-10 & 3.3e-10 & 1.2e-14 & 2.5e-14 & 8.2e-15 & 3.3e-16 \\
 900 & 225 & 1.3e-06 & 1.1e-10 & 2.0e-10 & 2.1e-11 & 1.2e-14 & 1.7e-14 & 2.1e-11 \\
1000 & 250 & 1.1e-06 & 6.0e-11 & 1.5e-10 & 7.1e-15 & 7.1e-15 & 6.7e-16 & 3.9e-14 \\
\hline
\end{tabular}
\caption{Relative error $E_n=\l|(I_{num}-I_{exact})/
I_{exact}\r|$ from the numerical calculation of $I(x)$ of
\protect{\mref{261}} for $x=1\e4$, splitting the integral and using 
composite $(n+1)$-point Newton-Cotes formulae with 
$n=1,\ldots,7$ for increasing total number of grid panels
$N_R$. $i_{kink}$ gives the position of the kink within the grid with
$N_R+1$ points, $i \in [0,N_R]$.}
\label{Tab:9}
\end{center}
\end{table}

As expected the results for $n=1,3,5$ are the same as in
Table~\ref{Tab:8}. For Simpson's rule ($n=2$) the results are now
significantly better and comparable to the 3/8-rule for the various
values of $N_R$. For the other NC-rules, $n=4,6,7$, although the results
have improved because the kink is an endpoint in every case, the
behaviour is not consistent for increasing $N_R$. The explanation for
this can be found in the composition of the various {\it mixed}
composite rules, mixing the main $n$-panel rule with one rule of lower
degree of precision.

To improve on those integral evaluations we can modify the number of
integration points such that, after splitting the integral in two
subintegrals at the kink, the number of panels in both integrals is a
multiple of the number of panels of the basic rule. The new results
are shown in Table~\ref{Tab:9.1}. The results for $n=1,5$ are not
shown as they are the same as in Table~\ref{Tab:9}. All the composite
rules now have the accuracy predicted by their error formula. 

\begin{table}[htbp]
\begin{center}
\begin{tabular}{|r|r|c|c|c|c||r|r|c|}
\hline
$N_R$ & $i_{kink}$ & $E_2$ & $E_3$ & $E_4$ & $E_6$
& $N_R$ & $i_{kink}$ & $E_7$ \\
\hline
 96 &  24 & 7.0e-07 & 1.6e-06 & 3.4e-09 & 6.4e-10 &  85 &  21 & 4.7e-09 \\
192 &  48 & 4.4e-08 & 9.8e-08 & 5.9e-11 & 1.9e-12 & 169 &  42 & 1.2e-11 \\
288 &  72 & 8.6e-09 & 1.9e-08 & 5.3e-12 & 7.1e-14 & 253 &  63 & 4.5e-13 \\ 
384 &  96 & 2.7e-09 & 6.2e-09 & 9.5e-13 & 7.5e-15 & 337 &  84 & 4.4e-14 \\ 
480 & 120 & 1.1e-09 & 2.5e-09 & 2.5e-13 & 4.4e-16 & 421 & 105 & 7.5e-15 \\ 
576 & 144 & 5.4e-10 & 1.2e-09 & 8.5e-14 & 0       & 505 & 126 & 2.0e-15 \\ 
672 & 168 & 2.9e-10 & 6.6e-10 & 3.3e-14 & 4.4e-16 & 589 & 147 & 4.4e-16 \\ 
768 & 192 & 1.7e-10 & 3.8e-10 & 1.5e-14 & 4.4e-16 & 673 & 168 & 5.6e-16 \\ 
864 & 216 & 1.1e-10 & 2.4e-10 & 7.8e-15 & 3.3e-16 & 757 & 189 & 2.2e-16 \\ 
960 & 240 & 7.0e-11 & 1.6e-10 & 4.0e-15 & 1.1e-15 & 841 & 210 & 3.3e-16 \\ 
\hline
\end{tabular}
\caption{Relative error $E_n=\l|(I_{num}-I_{exact})/
I_{exact}\r|$ from the numerical calculation of $I(x)$ of
\protect{\mref{261}} for $x=1\e4$, splitting the integral and adapting
the number of panels $N_R$ to use pure composite $(n+1)$-point Newton-Cotes
formulae with $n=2,3,4,6,7$ for increasing total number of grid panels
$N_R$. $i_{kink}$ gives the position of the kink within the grid with
$N_R+1$ points, $i \in [0,N_R]$.}
\label{Tab:9.1}
\end{center}
\end{table}

From the previous discussion it is clear that, even for an integrand
with a kink, it is far more advantageous to use a quadrature rule with
a higher degree of precision, as the 6-panel or 7-panel rules, than one of
lower degree, for an equal total number of integration points. 

\subsection{Split Simpson's rule and the integral equation}

Of course the quadrature rules are only building blocks of the
integral equation and we must keep in mind how these rules are used in
the global solution scheme of the integral equations. The various
tables in the previous discussion were all derived for one value of
external momentum, $x=1\e4$. Because this also coincides with the kink
in the radial integrand, some conclusions drawn from these tables rely
specifically on this value or rather on its index in the vector of
integration points. If we consider the system of non-linear equations,
\mref{S6}, instead of just an individual integral, we note that the 
external momentum $x_i$ takes on values corresponding to the momenta
of the radial integration nodes, $i=0,\ldots,N_R$.  This means that
it may be difficult to satisfy the requirements needed to obtain an
optimal accuracy, as derived from the previous discussion, for all of
them at the same time, as we will now clarify.

When we split the integration interval in two subintervals to avoid
the kink in the integrand at the value $x_i, i=0,\ldots,N_R$, we
will have $i$ panels in the lower interval and $N_R-i$ panels in the
upper interval. We will then apply some composite integration rule to
each of these subintegrals. Unfortunately, we now encounter a problem
due to the use of the collocation method to solve the integral
equation. In the collocation method, we use the same fixed set of
integration points for {\it all} the radial integrals, independently
of the external momentum, as we only know the function values $\S_i$
and thus the values of the integrands at a fixed number of momenta
$x_i$. We are not able to choose the number of points in the various
integrals according to the external momentum and the position of the
kink, such that the number of panels is a multiple of that from the
basic rules as suggested by the results of Table~\ref{Tab:9.1}. The
collocation method {\it forces} us to use non-optimal {\it mixed}
composite rules.

Let us show this in the following example. We want to apply the
composite Simpson's rule to evaluate the integrals. Let us take
$N_R=100$ and vary the position of the kink corresponding to the
external momentum in \mref{S6}. Table~\ref{Tab:10} shows the number of
radial integration panels $N_{R1}$, $N_{R2}$ in each integral after we
have split the total integral in two at the kink $x_i$. It also shows
the composition of the integration rule if we use the composite
Simpson's rule, complemented with one 3/8-rule or trapezoidal rule when
needed.
\begin{table}[htbp]
\begin{center}
\begin{tabular}{|c|c|c|c|c|}
\hline
$x_i$ & $N_{R1}$ & $N_{R2}$ & Rule 1 & Rule 2  \\
\hline
$x_0$  & 0 & 100  & -           & Simpson's \\
$x_1$  & 1 &  99  & Trapezoidal & Simpson's+3/8 \\
$x_2$  & 2 &  98  & Simpson's     & Simpson's \\
$x_3$  & 3 &  97  & Simpson's+3/8 & Simpson's+3/8 \\
$\vdots$ & & & &\\
$x_{\mbox{even}}$ & even & even   & Simpson's     & Simpson's \\
$x_{\mbox{odd}}$ & odd   & odd & Simpson's+3/8 & Simpson's+3/8 \\  
$\vdots$ & & & &\\
$x_{97}$ &  97 &  3 & Simpson's+3/8 & Simpson's+3/8 \\
$x_{98}$ &  98 &  2 & Simpson's     & Simpson's \\
$x_{99}$ & 99 &   1 & Simpson's+3/8 & Trapezoidal \\
$x_{100}$ & 100 & 0 & Simpson's     & - \\
\hline
\end{tabular}
\end{center}
\caption{Number of radial integration panels $N_{R1}$, $N_{R2}$ and 
structure of the mixed composite Simpson's rule in each integral after
splitting the total integral in two at the kink $x_i$ ($N_R=100$).}
\label{Tab:10}
\end{table}

From Table~\ref{Tab:10} we see that varying $x_i$ leads to different
mixed composite rules to be used. Even if we try to combine rules with
comparable accuracy, this is never possible for $x_i=x_1$ or
$x_{N_R-1}$ where a trapezoidal rule is always involved. This reduces
the accuracy to about the same level as the one achieved with the pure
trapezoidal rule. Even using composite rules formed with a basic rule
of higher degree of precision does not help because, depending on the
position of the kink, the composite rule will have to be complemented
with rules of lower degree of precision.  There is nothing we can do
about this as long as we use the {\it collocation} method. In a later
section, when we will introduce the {\it polynomial expansion} of the
unknown functions, this will be cured in an elegant way.

We now apply the splitting of the integral to the original kernel of
\mref{S2} using the split Simpson's rule described in
Table~\ref{Tab:10}. The results are shown in Table~\ref{Tab:11}. For
comparison we also tabulate the results for the pure composite trapezoidal
and Simpson's rule (without splitting the integral).

\begin{table}[htbp]
\begin{center}
\begin{tabular}{|r|r|c|c|c|}
\hline
$N_R$ & $1/h$ & $\S(0)$ & $\S(0)$ & $\S(0)$ \\
&  & Split Simp & trapez & Simpson's \\
\hline
 32 &  4   &    17.444 &    246.741 &   587.681 \\
 64 &  8   &    72.730 &    121.505 &   219.550  \\
128 & 16   &    83.790 &     93.557 &   120.363 \\
256 & 32   &    86.236 &     88.075 &    94.905 \\
512 & 64   &    86.768 &     87.093 &    88.778 \\
\hline
\end{tabular}
\caption{$\S(0)$ versus number of radial integration panels $N_R$ and
grid density $1/h$ using the split Simpson's rule and the pure composite
trapezoidal and Simpson's rule for $\alpha=2.086$,
$\kappa^2=100$.}
\label{Tab:11}
\end{center}
\end{table}

For equal values of $N_R$, the result of the split Simpson's rule is
better than the results achieved with the other methods, given that the
correct answer for $\S(0) \approx 87.009$.

\subsection{Heuristic improvement of the split Simpson's rule}

\def\I{I}
We now make an interesting observation starting from the error
formulae on the integral evaluation. Recall the error formula,
\mref{250.1}, for the composite trapezoidal rule with grid spacing~$h$,
\be
\I - I_{h} = E_{trap} = -\frac{b-a}{12} h^2 f^{(2)}(\xi) ,
\qquad a < \xi < b \; ,
\mlab{262}
\ee
where $I$ is the exact integral value and $I_h$ represents the approximate
value of the integral computed with a composite trapezoidal rule with grid
spacing $h$. If we perform two independent integral evaluations with grid
spacings $h_1$ and $h_2$, and divide their respective errors using
\mref{262} we get:
\be
\frac{\I - I_{h_1}}{\I - I_{h_2}} =
\frac{h_1^2 f^{(2)}(\xi_1)}{h_2^2 f^{(2)}(\xi_2)} \; .
\mlab{263}
\ee

\def\hoh{\l(h_2/h_1\r)^2}
If the second derivative of $f(x)$ varies slowly we can put
$f^{(2)}(\xi_1) \approx f^{(2)}(\xi_2)$ and \mref{263} becomes,
\be
\frac{\I - I_{h_1}}{\I - I_{h_2}} \approx
\l(\frac{h_1}{h_2}\r)^2 \; ,
\mlab{264}
\ee
and thus,
\be
\I \approx \frac{\hoh I_{h_1} - I_{h_2}}{\hoh-1} \;.
\mlab{265}
\ee

If we take $h_1=h$ and $h_2=2h$, this expression gives,
\be
\I \approx \frac{4I_h -I_{2h}}{3} \; .
\mlab{266}
\ee

This last equations tells us that we can get an improved integral
evaluation, if we know the evaluations of the integral with some number
of panels and for half this number of panels. If we apply this to an
integral evaluation using three points, $x_0$, $x_1$ and $x_2$, this
yields,
\bann
\I &\approx& \frac{1}{3}\l[(4\frac{h}{2}(f_0+2f_1+f_2) -
h(f_0+f_2)\r] \\ 
&\approx& \frac{h}{3}(f_0+4f_1+f_2) \;,
\eann
which is exactly Simpson's rule \mref{240}. Thus, using the
trapezoidal rules with $N$ and $N/2$ panels we can construct some rule
with a higher degree of precision, i.e. Simpson's rule.

From the error formula \mref{262} we can also compute,
\bann
\frac{I_{h_2} - I_{h_1}}{I_{h_3} - I_{h_2}} &=&
\frac{(I_{h_2}-\I) - (I_{h_1}-\I)}
{(I_{h_3}-\I) - (I_{h_2}-\I)} \\ &=& \frac{h_2^2 f^{(2)}(\xi_2) -
h_1^2 f^{(2)}(\xi_1)} {h_3^2 f^{(2)}(\xi_3) - h_2^2 f^{(2)}(\xi_2)} \\
&\approx& \frac{h_2^2 - h_1^2}{h_3^2-h_2^2} \;,
\mlab{266.1}
\eann
which allows us to estimate the improvement of the integral evaluation
with successive doubling of the number of panels. If we take $h_3=h/2$,
$h_2=h$, $h_1=2h$, then,
\be
\frac{I_h - I_{2h}}{I_{h/2} - I_h} \approx 4 \;.
\mlab{266.2}
\ee

Although the previous relations were derived for integral evaluations,
we can check if the numerical solution of the integral equation follow
some analogous relations. From the trapezoidal results of
Table~\ref{Tab:11} for $\S(0)$ we indeed find that,
\be
\frac{\S_0(\mbox{trap},h)-\S_0(\mbox{trap},2h)}
{\S_0(\mbox{trap},h/2)-\S_0(\mbox{trap},h)} \approx 4 \;,
\mlab{266.3}
\ee
which means that \mref{266.2} gets propagated from the integral
evaluations to the solution of the integral equation.

For curiosity we can also check if a relation analogous to \mref{266}
can be derived from Table~\ref{Tab:11}, using the final solution of
the integral equation rather than the individual integral values.  We
rather surprisingly see that:
\be
\S_0(\mbox{split},h) \approx \frac{4\S_0(\mbox{trap},h)-
\S_0(\mbox{trap},h/2)}{3} \; .
\mlab{266.4}
\ee

This tells us that the results of the split Simpson's rule can be
approximated by combining the results from the trapezoidal rule for the
same number of panels and for half this number of panels.  
\mrefb{266.3}{266.4} seem to be a feature of the
$h^2$ accuracy of the composite trapezoidal rule.

In analogy with this, we now study the behaviour of Simpson's rule. 
Using the error formulae \mref{252},
\be
E_{Simp} = -\frac{b-a}{180} h^4 f^{(4)}(\xi)  , \qquad a < \xi < b \;
,
\ee
for two different evaluations using grid spacings $h_1$ and $h_2$. If the
fourth derivative of $f$ varies slowly, we can write,
\be
\frac{\I - I_{h_1}}{\I - I_{h_1}} \approx
\l(\frac{h_1}{h_2}\r)^4 \; ,
\mlab{267}
\ee
and thus,
\def\hoh4{\l(h_2/h_1\r)^4}
\be
\I \approx \frac{\hoh4 I_{h_1} - I_{h_2}}{\hoh4-1} \;.
\mlab{268}
\ee

Using this for $h_1=h$ and $h_2=2h_1$ yields,
\be
\I \approx \frac{16I_h -I_{2h}}{15} \; .
\mlab{269}
\ee

Analogously to \mref{266.1}, we have,
\be
\frac{I_{h_2} - I_{h_1}}{I_{h_3} - I_{h_2}} =
\approx \frac{h_2^4 - h_1^4}{h_3^4-h_2^4} \;,
\mlab{269.1}
\ee
and for $h_3=h/2$, $h_2=h$, $h_1=2h$,
\be
\frac{I_h - I_{2h}}{I_{h/2} - I_h} \approx 16 \;.
\mlab{269.2}
\ee

Nevertheless, Table~\ref{Tab:11} shows that the rate of convergence of
the split Simpson's rule, when $N_R$ is increased does not follow
\mref{269.2} but tends to be the same
as for the trapezoidal rule,
\be
\frac{\S_0(\mbox{split},h)-\S_0(\mbox{split},2h)}
{\S_0(\mbox{split},h/2)-\S_0(\mbox{split},h)} \approx 4 \;.
\mlab{269.3}
\ee

This is probably because there are always two trapezoidal rules
involved in the calculation using the split rules and the error
propagation keeps the global degree of precision down to that of the
trapezoidal rule.

The previous observations can be used to construct a heuristic method
to improve the split Simpson's solution on the integral equation.
Eqs.~(\oref{266.3}, \oref{266.4}, \oref{269.3}) suggest the following
improvement:
\be
\S_0(\mbox{Improved split},h) \approx \frac{4\S_0(\mbox{split},h)-
\S_0(\mbox{split},h/2)}{3} \; .
\mlab{269.4}
\ee

The application of \mref{269.4} to the results of Table~\ref{Tab:11}
are tabulated in Table~\ref{Tab:12}.

\begin{table}[htbp]
\begin{center}
\begin{tabular}{|r|r|c|c|}
\hline
$N_R$ & $1/h$ & $\S(0)$ & $\S(0)$ \\
&  & Split Simp & Improved split Simp \\
\hline
 32 &  4   &    17.444 &  \\
 64 &  8   &    72.730 &  91.159 \\
128 & 16   &    83.790 &  87.477 \\
256 & 32   &    86.236 &  87.051 \\
512 & 64   &    86.768 &  86.945 \\
\hline
\end{tabular}
\caption{$\S(0)$ versus number of radial integration panels $N_R$ and
grid density $1/h$ using the split Simpson's rule (from
Table~\protect{\ref{Tab:11}}) 
and the improved split Simpson's rule of
\protect{\mref{269.4}} for $\alpha=2.086$, $\kappa^2=100$.
}
\label{Tab:12}
\end{center}
\end{table}

\section{Critical coupling in the 1-loop approximation to $\Pi$}
\label{1loop}

\begin{figure}[htbp]
\begin{center}
\mbox{\epsfig{file=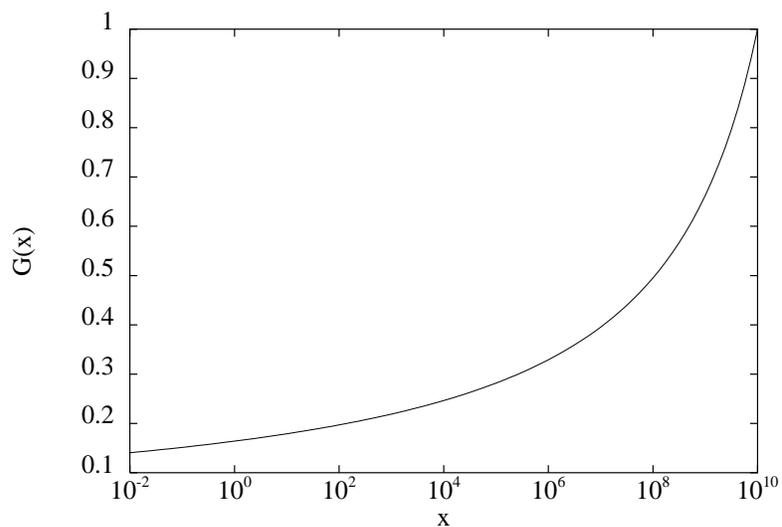,angle=-90,height=8cm}}
\end{center}
\caption{1-loop photon renormalization function $\G(x)$ as a
function of the photon momentum $x$ for
$\alpha=2.086$. $\kappa^2=0.01$, $\Lambda^2=1\e10$.}
\label{Fig:G-1loop}
\end{figure}

\begin{figure}[htbp]
\begin{center}
\mbox{\epsfig{file=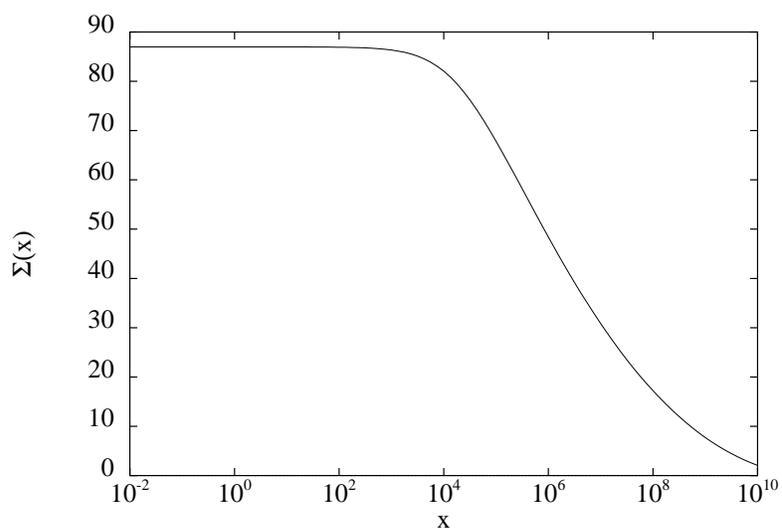,angle=-90,height=8cm}}
\end{center}
\caption{Dynamical mass function $\S(x)$ as a function of the 
fermion momentum $x$ for $\alpha=2.086$ in the 1-loop approximation
to $\G$. $\kappa^2=0.01$, $\Lambda^2=1\e10$.}
\label{Fig:S-1loop}
\end{figure}

We will now apply the previously discussed method to the integral
equation, \mref{S2}, describing the dynamical fermion mass generation in
QED, in the 1-loop approximation to the vacuum polarization. In
Fig.~\ref{Fig:G-1loop} we plot the 1-loop behaviour of $\G(x)$ used
as input in \mref{S2}.
The $\S$-equation is solved for various values of $\alpha$ using the
improved split Simpson's rule and Newton's iterative method. In
Fig.~\ref{Fig:S-1loop} we show a typical plot of the dynamical mass
function $\S(x)$ for $\alpha=2.086$, $\kappa^2=0.01$ and
$\Lambda^2=1\e10$.

\begin{figure}[htbp]
\begin{center}
\mbox{\epsfig{file=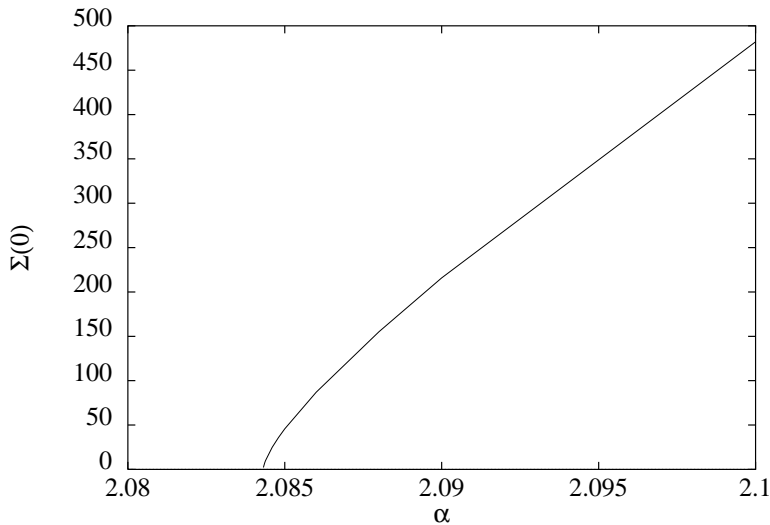,angle=-90,height=8cm}}
\end{center}
\caption{Dynamically generated mass $\S(0)$ versus coupling $\alpha$
in the 1-loop approximation to $\G$. $\kappa^2=0.01$,
$\Lambda^2=1\e10$.}
\label{Fig:M-1loop}
\end{figure}

In Fig.~\ref{Fig:M-1loop} we plot the evolution of $\S(0)$, which
is representative for the scale of the dynamically generated fermion
mass, versus the coupling strength $\alpha$. For small $\alpha$ there
is no fermion mass generation. At a certain value of the coupling,
called the critical coupling, $\alpha_c$, fermion mass generation sets
in. The generated fermion mass increases further with increasing values
of $\alpha$. To pin down the value of $\alpha_c$ numerically, we start
from some large value of the coupling and decrease it till the mass
generation disappears. From this we find \fbox{$\alpha_c = 2.08432$}.

In this chapter we have set up the numerical framework to solve non-linear
integral equations. We have applied this to a single integral equation for
the dynamical mass $\S$, corresponding to a specific truncation of the
Schwinger-Dyson equations describing fermion mass generation in QED.  In
the next chapter we will relax some of the simplifications introduced in
this chapter and will investigate the behaviour of the coupled set of
integral equations for $\S$ and $\G$.

\chapter{Solving the coupled ($\S$, $\G$)-system: first attempt}
\label{Paper2}

\section{Numerical method to solve the coupled ($\S$, $\G$)-system}
\label{Sec:NumethSG}

In this chapter we are going to extend the study started in the previous
chapter by including the photon equation in our procedure instead of
approximating the vacuum polarization by its 1-loop approximation.

We recall the integral equations, Eqs.~(\oref{189}, \oref{190},
\oref{1105}), derived with the bare vertex approximation. In the Landau
gauge and with zero bare mass these equations are:
\ba
\frac{\S(x)}{\F(x)} &=& \frac{3\alpha}{2\pi^2} \int dy \, 
\frac{y\F(y)\S(y)}{y+\S^2(y)} \int d\theta \, \sin^2\theta
\, \frac{\G(z)}{z}  \mlab{700.0}\\
\frac{1}{\F(x)} &=& 1 + \frac{\alpha}{2\pi^2 x} \int dy \, 
\frac{y\F(y)}{y+\S^2(y)} 
\mlab{700.1}\\
&& \times \int d\theta \, \sin^2\theta \, \G(z) \,
\l(\frac{3 \sqrt{xy} \cos\theta}{z} 
- \frac{2 x y \sin^2\theta}{z^2} \r)  \nn \\
\frac{1}{\G(x)} &=& 1 + \frac{2 N_f \alpha}{3\pi^2 x} \int dy
\frac{y\F(y)}{y+\S^2(y)} 
\int d\theta \, \sin^2\theta \, \frac{\F(z)}{z+\S^2(z)} \mlab{700.2} \\
&& \times  
\l[ (n-2)y - 2ny\cos^2\theta + (n+2)\sqrt{xy}\cos\theta 
+ (n-4)\S(y)\S(z) \r] \nn
\ea
where $z \equiv x+y-2\sqrt{xy}\cos\theta$.

Although we will generally set $n=4$ in \mref{700.2} throughout this work
to avoid the quadratic divergence in the vacuum polarization integral, as
explained in Section~\ref{Sec:Coupeq}, we will use an alternative procedure
in this chapter, taking $n=0$, which corresponds to the operator
$\Proj_{\mu\nu} = g_{\mu\nu}$ in \mref{197}, in order to investigate the
results obtained by Kondo, Mino and Nakatani in Ref.~\cite{Kondo92}.
Setting $n=0$ in \mref{700.2} yields:
\be
\frac{1}{\G(x)} = 1 - \frac{4 N_f \alpha}{3\pi^2 x} \int dy
\frac{y\F(y)}{y+\S^2(y)} 
\int d\theta \, \sin^2\theta \, \frac{\F(z)}{z+\S^2(z)}
\l[ y - \sqrt{xy}\cos\theta + 2\S(y)\S(z) \r] \,. \mlab{700.3}
\ee

The vacuum polarization integral in \mref{700.3} contains a quadratic
divergence which can be removed explicitly by imposing~:
\be
\lim_{x \rightarrow 0} \;\frac{x}{\G(x)} = 0 \, ,
\mlab{limit}
\ee
to ensure a massless photon. If we write the photon renormalization
function as~:
\be
\G(x) = \frac{1}{1 + \Pi(x)} \, ,
\mlab{G-Pi}
\ee
\mref{limit} can be satisfied by defining a renormalized vacuum polarization
$\Pi(x)$~:
\be
x \tilde{\Pi}(x) = x \Pi(x) - \lim_{x \rightarrow 0} x \Pi(x) \, .
\mlab{vacpolrenorm}
\ee

This is the procedure adopted by Kondo et al.~\cite{Kondo92}. They
solve numerically the coupled set of integral equations for the
dynamical fermion mass $\S(x)$ and the photon renormalization
function $\G(x)$ in the case of zero bare mass, $m_0 \equiv 0$. The
calculations are performed in the Landau gauge ($\xi=0$) with the bare
vertex approximation, i.e.  $\Gamma^\mu(k,p) \equiv \gamma^\mu$. As a
further approximation they decouple the $\F$-equation by putting
$\F(x) \equiv 1$. While the quadratic divergence in the vacuum polarization
is removed by imposing \mref{vacpolrenorm}, the fact that the
Ward-Takahashi identity is not satisfied, when dynamical mass is
generated, makes the results procedure dependent.  The main
improvement with respect to Section~\ref{S-equation} is that we now
determine the photon renormalization function $\G(x)$ using the
photon Schwinger-Dyson equation instead of using the 1-loop perturbative
result.
 
The coupled integral equations for $\S$ and $\G$ obtained using these
approximations, in Euclidean space and introducing an ultraviolet cutoff
$\Lambda^2$ on the radial integrals, are given by~:
\be
\S(x) = \frac{3\alpha}{2\pi^2} \int_0^{\Lambda^2} dy \, 
\frac{y\S(y)}{y+\S^2(y)} \int_0^\pi d\theta \, \sin^2\theta
\, \frac{\G(z)}{z}
\mlab{700}
\ee
and, performing the subtraction, \mref{vacpolrenorm}, on \mref{700.3}:
\ba
\frac{1}{\G(x)} &=& 1 - \frac{4N_f\alpha}{3\pi^2 x}
\int_0^{\Lambda^2} dy \frac{y}{y+\S^2(y)} 
\int_0^\pi d\theta \sin^2\theta \mlab{702} \\
&&\hspace{3cm}\times \left\{ \frac{y - \sqrt{xy}\cos\theta +
2\S(y)\S(z)}{z+\S^2(z)} -
\frac{y+2\S^2(y)}{y+\S^2(y)} \right\} \, .\nonumber
\ea

Recall that in QED the momentum dependence of the coupling comes
wholly from the photon renormalization function, so solutions for
$\G(x)$ give the running of the coupling. Kondo et al.\ solve this
coupled set of non-linear integral equations, Eqs.~(\oref{700},
\oref{702}), for $N_f=1$ and find a symmetry breaking phase for
$\alpha$ greater than some critical coupling $\alpha_c
\approx 2.084$.

We now describe how to solve the coupled set of integral equations
\mrefb{700}{702} for $\S(x)$ and $\G(x)$. As in
the previous chapter we will replace the integral equations by a set
of non-linear algebraic equations (see Section~\ref{S-equation}). For
the purpose of numerical integration we introduce an ultraviolet
cutoff $\Lambda^2$ and an infrared cutoff $\kappa^2$ and change
variables to the logarithm of momentum squared, $t = \logten y$. We
then evaluate the integrals by some quadrature rule and consider the
resulting equations only for external momenta equal to the integration
nodes,
\ba
\S_i &=& \frac{3\alpha\ln10}{2\pi^2} \sum_{j=0}^{N} w_j \, 
\frac{x_j^2\S_j}{x_j+\S^2_j} \sum_{k=0}^M w'_k \, \sin^2\theta_k
\, \frac{\G(z_k)}{z_k}
\mlab{702.1} \\
\frac{1}{\G_i} &=& 1 - \frac{4N_f\alpha\ln10}{3\pi^2 x_i}
\sum_{j=0}^{N} w_j \, \frac{x_j^2}{x_j+\S^2_j} 
\sum_{k=0}^M w'_k \, \sin^2\theta_k \mlab{702.2} \\
&&\hspace{3cm}\times \left\{ \frac{x_j - \sqrt{x_i x_j}\cos\theta_k +
2\S_j\S(z_k)}{z_k+\S^2(z_k)} -
\frac{x_j+2\S^2_j}{x_j+\S^2_j} \right\} \, ,\nonumber
\ea
where the equidistant logarithmic nodes are distributed as,  
\be
t_i = \logten\kappa^2 + \frac{i}{N} 
\l(\logten\Lambda^2-\logten\kappa^2\r) , \qquad i = 0,\ldots,N .
\mlab{702.3}
\ee
The corresponding momenta squared of the external particle and the
radial integration nodes are
\be
x_i = 10^{t_i} , \qquad i = 0,\ldots,N .
\mlab{702.4}
\ee

The angular integration nodes are 
\be
\theta_k = \frac{k\pi}{M} , \qquad  k = 0,\ldots,M ,
\ee
such that the momenta squared of the angular integration nodes are
given by
\be
z_k=x_i+x_j-2\sqrt{x_i x_j}\cos\theta_k , \qquad k = 0,\ldots,M .
\mlab{702.5}
\ee

The unknowns of the system of non-linear algebraic equations are
the function values at the radial integration nodes,
\be
\parbox{3cm}{
\bann
\S_i = \S(x_i) \\
\G_i = \G(x_i)
\eann
} 
\parbox{3cm}{\[ i = 0,\ldots,N. \]}
\mlab{702.6}
\ee

However, the collocation method cannot yet be applied to
\mrefb{702.1}{702.2} as the equations do not only refer to the unknown
functions at the radial integration nodes $x_i$. The angular parts of
\mrefb{702.1}{702.2} contain the function values $\S(z_k)$ and
$\G(z_k)$ where the momentum $z_k$ defined in \mref{702.5} is not one
of the quadrature nodes $x_i$ as it also depends on the angle between
the external momentum and the internal momentum. Therefore $\S(z_k)$
and $\G(z_k)$ are not one of the components $\S_i$, $\G_i$ of
\mref{702.6} which are the solution vectors of the problem and the
collocation method cannot be applied directly. To compute the angular
parts of
\mrefb{702.1}{702.2} we have to interpolate the values of $\S(z_k)$
and $\G(z_k)$.  A straightforward choice is to perform a linear
interpolation on the logarithmic scale between the function values at
the surrounding integration nodes $x_i$ and $x_{i+1}$, where $z_k \in
[x_i,x_{i+1}]$ :
\be
\parbox{10cm}{\bann
\S(z_k) &=& \S_i + \frac{\logten z_k - \logten x_i}{\logten x_{i+1} - \logten
x_i} \l( \S_{i+1} - \S_i \r) \\[5pt]
\G(z_k) &=& \G_i + \frac{\logten z_k - \logten x_i}{\logten x_{i+1} - \logten
x_i} \l( \G_{i+1} - \G_i \r) \,.
\eann }
\mlab{702.7}
\ee

After using these interpolation rules for $\S(z_k)$ and $\G(z_k)$ in
\mrefb{702.1}{702.2} the system of non-linear equations now only
depends on the function values $\S_i$ and $\G_i$ of \mref{702.6} so
that the collocation method can be applied.

\label{FullNewton}
We then have to solve this system of non-linear equations using some
appropriate numerical technique. From the discussion in the previous
chapter it is clear that Newton's iterative procedure is the right
choice for this. However the full implementation of this method on the
system of equations is very tedious and requires a large amount of
computing time and memory allocation. A major consumption of computer
time will come from the computation of the angular integrals in
\mrefb{702.1}{702.2}. Because the kernels of the angular integrals
depend on the unknown functions, the angular integrals have to be
recalculated for {\it each} iteration in Newton's method using the new
approximations for $\S$ and $\G$.  Furthermore, Newton's method
requires the partial derivatives of the non-linear equations with
respect to the function values $\S_i$ and $\G_i$. As these are present
in the radial and angular integrals, the computation of the
derivatives will use a huge amount of computer time and memory
allocation. Because these resources are limited we will settle for
some compromise.

\begin{figure}[htbp]
\begin{center}
\vspace*{-2.5cm}\hspace*{-1cm}
\begin{picture}(160,200) 
\lijndikte
\put(60,170){\framebox(40,10){Initial guess: $\S_0$, $\G_0$}}
\put(80,170){\vector(0,-1){10}}
\put(40,150){\framebox(80,10){Compute angular integrals for $\S$-equation}}
\put(80,150){\vector(0,-1){10}}
\put(40,120){\framebox(80,20){\parbox{80mm}
{\begin{center}Initialize Newton's method for $\S$-equation\\
Set: $\S_{n+1,0} = \S_n$ \end{center}}}}
\put(80,120){\vector(0,-1){10}}
\put(20,100){\framebox(120,10){Solve linear equations to compute 
             $\S_{n+1,m+1} = f_1(\S_{n+1,m}, \G_n)$}}
\put(80,100){\vector(0,-1){10}}
\put(79,78){
\usebox{\ifbox}
\makebox(0,0){\parbox{60mm}
{\begin{center}Convergence\\$\norm{\S_{n+1,m+1}-\S_{n+1,m}}$ \\ $< \tol$ ?
\end{center}}}}
\put(104,78){\vector(1,0){20}}
\put(108,80){\makebox(0,0)[bl]{no}}
\put(124,73){\framebox(30,10){Increment $m$}}
\put(150,83){\line(0,1){22}}
\put(150,105){\vector(-1,0){10}}
\put(80,66){\vector(0,-1){10}}
\put(82,62){\makebox(0,0)[bl]{yes}}
\put(60,46){\framebox(40,10){Set: $\S_{n+1} = \S_{n+1,m+1}$}}
\put(80,46){\vector(0,-1){10}}
\put(50,26){\framebox(60,10){Compute $\G_{n+1} = f_2(\S_{n+1})$}}
\put(80,26){\vector(0,-1){10}}
\put(79,4){
\usebox{\ifbox}
\makebox(0,0){\parbox{60mm}
{\begin{center}Convergence\\$\norm{\S_{n+1}-\S_{n}} < \tol$ ? 
\end{center}}}}
\put(104,4){\vector(1,0){20}}
\put(108,6){\makebox(0,0)[bl]{no}}
\put(124,-1){\framebox(30,10){Increment $n$}}
\put(154,4){\line(1,0){15}}
\put(169,4){\line(0,1){151}}
\put(169,155){\vector(-1,0){49}}
\put(80,-8){\vector(0,-1){10}}
\put(82,-13){\makebox(0,0)[bl]{yes}}
\put(80,-23){\circle{10}\makebox(0,0){OK}}
%
\put(10,59){\framebox(150,86)[tr]
{\parbox{4cm}{\begin{center}Newton's method\\for $\S$-equation\end{center}}}}
\end{picture}
\end{center}
\vspace{2.5cm}
\caption{Program flow to solve the coupled ($\S$,$\G$)-system.}
\label{Fig:flow-SG}
\end{figure}
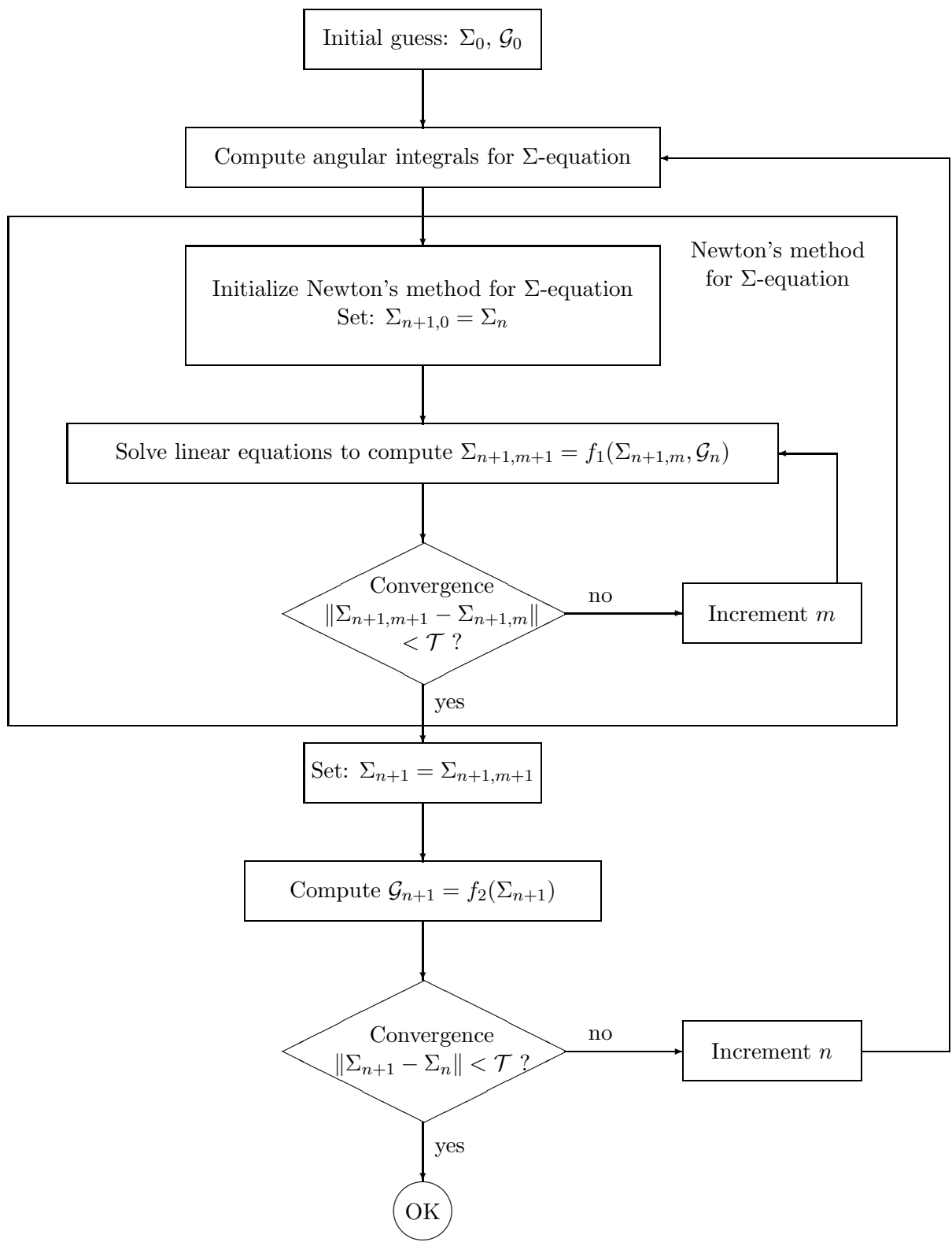

We therefore introduce a hybrid method between Newton's method and the
natural iterative procedure based on the observation that the kernel of the
angular integrals in the $\S$-equation, \mref{702.1}, is a function of
$\G(z)$ but is independent of $\S$, while the kernel of the angular
integrals of the $\G$-equation, \mref{702.2}, is a function of $\S(z)$ and
has no dependency on $\G$. In this hybrid method we apply Newton's method
on the $\S$-equation for a given $\G$, but the coupling between the $\S$
and $\G$ equations is solved using a global natural iterative procedure. We
now give more details about the program flow of this method shown in
Fig.~\ref{Fig:flow-SG}.

Let us start from some initial guess $\mvec{\S_{0}}$ and $\mvec{\bG_{0}}$
for the unknown vectors of function values at the quadrature points
$x_i$. This could for example be the 1-loop perturbative approximation for
$\G$ and some arbitrary, realistic function for the dynamical mass $\S$. We
now describe how to derive new approximations ($\mvec{\S_{n+1}}$,
$\mvec{\bG_{n+1}}$) starting from the current approximations
($\mvec{\S_{n}}$, $\mvec{\bG_{n}}$). We first compute the angular integrals
$(\Theta_{n})_{ij}$,
\be
(\Theta_n)_{ij} = \sum_{k=0}^M w'_k \, \sin^2\theta_k
\, \frac{\G_n(z_k)}{z_k} ,
\ee
of the $\S$-equation, \mref{702.1}, using $\mvec{\bG_n}$ and the
interpolation rule \mref{702.7}.

Then, \mref{702.1} becomes:
\be
(\S_{n+1})_i - \frac{3\alpha\ln10}{2\pi^2} \sum_{j=0}^{N} w_j \, 
\frac{x_j^2(\S_{n+1})_j(\Theta_n)_{ij}}{x_j+(\S_{n+1})^2_j} = 0 , 
\qquad i=0,\ldots,N \;.
\mlab{703}
\ee

\mref{703} describes a system of non-linear algebraic equations
determining the solution vector $\mvec{\S_{n+1}}$ computed from
$\mvec{\bG_n}$. This equation is very similar to \mref{S6} and can be
solved using Newton's iterative method. The iterative method starts
from an initial guess $\mvec{\S_{n+1,0}}$, for which $\mvec{\S_n}$
seems an obvious but in no way necessary choice. At each iteration
step the method requires the solution of a linear set of equations,
\mref{218.3}, to compute $\mvec{\S_{n+1,m+1}}$ from the previous
solution $\mvec{\S_{n+1,m}}$.  Because we only improve $\S$ in this
part of the calculation, the angular integrals $\Theta$ remain
unchanged throughout Newton's method. The iterations of the Newton
method will be repeated till two successive approximations
$\mvec{\S_{n+1,m+1}}$ and $\mvec{\S_{n+1,m}}$ are sufficiently close,
this approximation will be identified as $\mvec{\S_{n+1}}$ .

Once the Newton method has converged, the function $\mvec{\S_{n+1}}$
is used to compute a new approximation to $\mvec{\bG_{n+1}}$ using the
photon equation, \mref{702.2}, and the interpolation rule,
\mref{702.7}. Note that the integral in this equation does only depend
on $\S$ and so we need not apply any iterative procedure to compute
$\mvec{\bG_{n+1}}$, all we have to do is evaluate the double sum in
\mref{702.2}, corresponding to the two dimensional integrals of
\mref{702}. This provides the end-point of one global iteration where the new
approximations ($\mvec{\S_{n+1}}$, $\mvec{\bG_{n+1}}$) has been constructed
from the previous approximation ($\mvec{\S_{n}}$, $\mvec{\bG_{n}}$). The
whole procedure is iterated till the new solutions for $\S$ and $\G$
satisfy a global convergence criterion.

As before the part computed using Newton's method is solved very
efficiently. The coupling of the $\G$-iteration on the other hand
slows down the whole procedure as the angular integrals for both
equations have to be recalculated for every main
iteration. Fortunately the $\G$-equation seems to converge relatively
rapidly in this hybrid iteration scheme, i.e. after a few iterations, 
so that the overall computing time remains reasonable.

In the next section we will show the results obtained with this method
and discuss how the photon quadratic divergence, which is easily
removed theoretically, could effectively be cancelled numerically.

\section{Numerical cancellation of the photon quadratic divergence}

\begin{figure}[htbp]
\begin{center}
\mbox{\epsfig{file=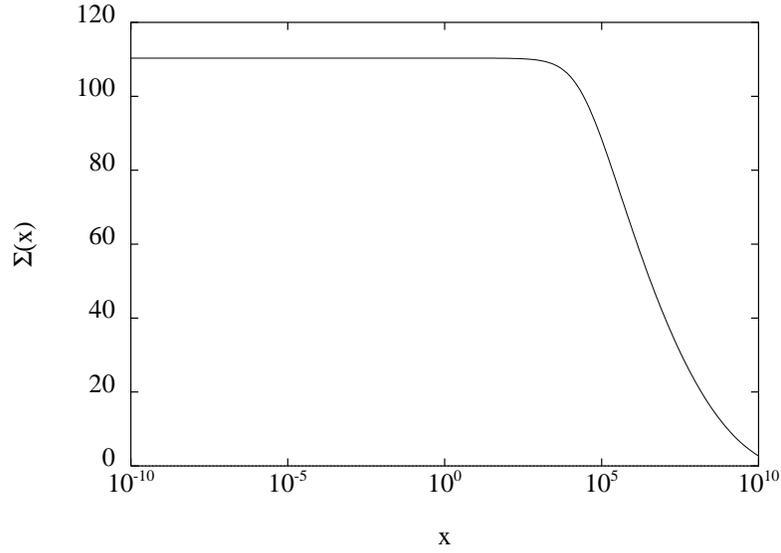,height=8cm,angle=-90}}
\end{center}
\vspace{-0.5cm}
\caption{Dynamical mass function $\S(x)$, as a function of momentum
$x$ for $N_f=1$ and $\alpha=2.086$ as calculated in a
self-consistent way as in Ref.~\protect{\cite{Kondo92}} ($\Lambda^2 = 1\e10$).}
\label{fig:figB}
\end{figure}

\begin{figure}[htbp]
\begin{center}
\mbox{\epsfig{file=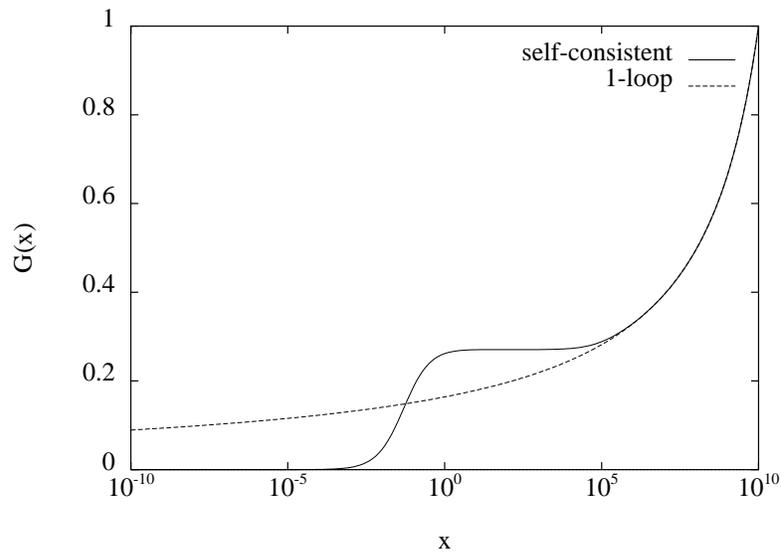,height=8cm,angle=-90}}
\end{center}
\vspace{-0.5cm}
\caption{Photon renormalization function $\G(x)$, as a function of momentum
$x$ for $N_f=1$ and $\alpha=2.086$ as calculated in a
self-consistent way as in Ref.~\protect{\cite{Kondo92}} and in 1-loop
approximation ($\Lambda^2 = 1\e10$).}
\label{fig:figG}
\end{figure}

In this section we apply the previously developed method to determine the
critical coupling and study the behaviour of the photon renormalization
function of the coupled ($\S$, $\G$)-system.  A numerical solution to this
problem has also been recently presented in Ref.~\cite{Kondo92} by Kondo,
Mino and Nakatani. As in Ref.~\cite{Bloch95}, we discuss the peculiar
behaviour they find for the photon renormalization function $\G$ at
intermediate low momentum.  For $N_f=1$ we find a symmetry breaking phase
for $\alpha$ greater than some critical coupling $\alpha_c
\approx 2.084$.  In Figs.~\ref{fig:figB}, \ref{fig:figG} we display the
results for a value of $\alpha = 2.086$, close to its critical value. The
dynamical mass function, $\S(x)$, is illustrated in Fig.~\ref{fig:figB}.

Fig.~\ref{fig:figG} shows the photon renormalization function,
$\G(x)$, found from the solution of the coupled ($\S$, $\G$)-system
and this is compared with its 1-loop approximation. To allow the
comparison with the 1-loop result the vacuum polarization is
renormalized such that $\tilde{\Pi}(\Lambda^2)=0$. 

One observes that at high momenta the self-consistent $\G(x)$ follows the
1-loop result very nicely. For decreasing momenta the effect of the
dynamically generated mass comes into play and the value of $\G(x)$, and
hence that of the running coupling, seems to stabilize for a while, as one
could expect. Then, surprisingly, at some lower momentum there is a sudden
fall in $\G(x)$, which drops below the 1-loop value and almost vanishes
completely. This is a rather strange behaviour for the running coupling at
low momenta. This decrease corresponds to the vacuum polarization integral
of \mref{702} becoming large.  We will show that this sharp decrease is an
artefact of the method used by Kondo et al.~\cite{Kondo92} to remove the
quadratic divergence in the vacuum polarization. We also discuss how this
can be avoided in numerical studies of the Schwinger-Dyson equations.

To solve the problem numerically we have to make additional
assumptions about the ultraviolet behaviour of $\S(x)$ and
$\G(x)$. These arise from the need to handle loop momenta beyond the
ultraviolet (UV) cutoff.  If in \mrefb{700}{702} $0 \le x, y
\le \Lambda^2$, then the momentum in the angular integration, 
$z = x + y - 2\sqrt{xy}\cos\theta$, will lie in the interval $0 \le z
\le 4\Lambda^2$. Therefore, the angular integrals need values of
$\S$ and $\G$ at momenta above the UV-cutoff, i.e. outside the {\it
physical momentum region}. Therefore one will have to extrapolate $\S$ and
$\G$ outside this region. In their work, Kondo et al.~\cite{Kondo92}
define~:
\ba
\Sigma(x > \Lambda^2) &\equiv& 0 \mlab{extrapS} \\
\Pi(x > \Lambda^2) &\equiv& 0  \; \Rightarrow \;  \G(x > \Lambda^2)
\equiv  1 \mlab{extrapG} \, .
\ea

Both dynamical mass and vacuum polarization vanish above the UV-cutoff
and the theory then behaves as a free theory.  Although this
assumption seems reasonable, \mref{extrapS} introduces a jump
discontinuity in the dynamical mass function at $x = \Lambda^2$
because $\S(\Lambda^2) \ne 0$ for $\alpha > \alpha_c$ (see
Fig.~\ref{fig:figB}), while \mref{extrapG} introduces a relatively
sharp kink in the photon renormalization function at that point (see
Fig.~\ref{fig:figG}).

A more detailed investigation shows that the step in the photon
renormalization function found by Kondo et al.\ is an artefact of the
way they renormalize the quadratic divergence in the vacuum
polarization integral, \mref{702}, combined with the presence of the
jump discontinuity in the dynamical mass function, \mref{extrapS}, as
we now explain.

From the angular integrand of the $\G$-equation, \mref{702}~, we define
$f_\theta$ as~:
\be
f_\theta =  
\frac{y - \sqrt{xy}\cos\theta + 2\S(y)\S(z)}{z+\S^2(z)} -
\frac{y+2\S^2(y)}{y+\S^2(y)} \, .
\mlab{angint}
\ee

Both terms in \mref{angint} cancel exactly at $x = 0$ to remove the
quadratic singularity. Of course the description of the real world has to
be such that the approximate cancellation of the quadratically divergent
terms at low $x$ becomes exact at $x = 0$ in a continuous way.

To investigate this, we now look analytically at the behaviour of
$f_\theta$ at low $x$, for some arbitrary value of $y$ and $\theta$. We can
write $z$ as:
\be
z = y + \deltay
\mlab{p}
\ee
where we define:
\be
\deltay =  x - 2\sqrt{xy}\cos\theta \, ,
\ee
and we know $\deltay$ is small if $x$ is small. Furthermore, we also write
$\S(z)$ as:
\be
\S(z) = \S(y) + \delta\S \, .
\mlab{deltaS}
\ee

Substituting Eqs.~\nref{p}, \nref{deltaS} in the expression for the angular
integrand $f_\theta$, \mref{angint}, yields:
\be
f_\theta \approx \frac{y - \sqrt{xy}\cos\theta + 2\S^2(y) + 2\dS\S(y)}
{y + \S^2(y) + \deltay + 2\dS\S(y) + \dS^2}
- \frac{y + 2\S^2(y)}{y + \S^2(y)} \, .
\ee

Performing a Taylor expansion of the denominator of the first term, we
get (neglecting terms of $\Order(\deltay)^2, \dS^2$):
\ba
f_\theta &\approx& \frac{1}{y + \S^2(y)}\left[ - \sqrt{xy}\cos\theta 
+ 2\dS\S(y) - \l(\deltay+2\dS\S(y)\r) \frac{y - \sqrt{xy}\cos\theta 
+ 2\S^2(y)}{y + \S^2(y)} \right]
\nonumber \\
&\approx& \Order(x, \sqrt{xy}\cos\theta, \dS) \, . \mlab{ftheta1}
\ea

If $\S(z)$ is smooth, we can make a Taylor expansion of $\S(z)$
around $\S(y)$:
\be
\S(z) = \S(y) + \deltay \: \S'(y) + \Order(\deltay)^2 \,,
\mlab{Taylor}
\ee
and $\dS$ of \mref{deltaS} is,
\be
\dS = \deltay \: \S'(y) + \Order(\deltay)^2 \, .
\ee

In this case, \mref{ftheta1} becomes,
\be
f_\theta \approx \Order(x,\sqrt{xy}\cos\theta)
\mlab{ftheta1.1}
\ee
and it is clear that the angular integrand $f_\theta$ is continuous for all
$\theta \in [0,\pi]$ and goes to zero in a continuous way when $x$ goes to
zero. From \mref{ftheta1} we see that the same argument holds even when
$\S$ is continuous, but not necessarily smooth, at $z=\Lambda^2$.

Now let us look at the angular integrand $f_\theta$ in the
approximation of Kondo et al.~\cite{Kondo92} when $x$ is small but $y$
is very large, indeed larger than $y_0 = (\Lambda - \sqrt{x})^2$.
Then, for values of $\theta$ greater than $\theta_0(y) =
\arccos((x + y - \Lambda^2)/2\sqrt{xy})$ we will have
$z > \Lambda^2$. If we now use Kondo et al.'s extrapolation,
\mref{extrapS}, then $\S(z > \Lambda^2) = 0$ and the angular 
integrand \mref{angint}, now becomes~:
\be
f_\theta =
\frac{y - \sqrt{xy}\cos\theta}{z} -
\frac{y+2\S^2(y)}{y+\S^2(y)} ,\qquad  \mbox{for} \quad z > \Lambda^2 .
\ee

For small $x$, and $z > \Lambda^2$ (corresponding to $y>y_0$ and
$\theta>\theta_0(y)$) we have,
\ba
f_\theta &\approx& -\frac{\S^2(y)}{y + \S^2(y)} + 
{\cal O}(x,\sqrt{xy}\cos\theta) \, ,
\mlab{ftheta2}
\ea
while for small $x$ and $z \le \Lambda^2$ we still have the expected
behaviour of \mref{ftheta1.1},
\ba
f_\theta &\approx& {\cal O}(x,\sqrt{xy}\cos\theta) \, .
\mlab{ftheta2.1}
\ea

From \mrefb{ftheta2}{ftheta2.1} we see that as soon as $x$ deviates
from zero, the angular integrands for $y>y_0$ contain a jump
discontinuity at $\theta = \theta_0(y)$, and part of the angular
integrand will not vanish continuously when $x \rightarrow 0$. In fact
the angular integral $I_\theta$ will receive an extra contribution
$\delta I_\theta$ when $y$ is larger than $y_0 = (\Lambda -
\sqrt{x})^2$~:
\ba
\delta I_\theta(y>y_0) &=& -\frac{\S^2(y)}{y + \S^2(y)}
\int_{\theta_0(y)}^{\pi} d\theta \, \sin^2\theta 
\nonumber \\
&=& -\frac{\S^2(y)}{y + \S^2(y)}\,\left(\frac{\pi}{2} -
\frac{\theta_0(y)}{2} + \frac{\sin 2\theta_0(y)}{4}\right)\, .
\mlab{dItheta}
\ea

Substituting \mref{dItheta} in \mref{702} we see that the vacuum
polarization receives an extra contribution $\delta\Pi(x)$~:
\be
\delta\Pi(x) =
\frac{4N_f\alpha}{3\pi^2 x}\int_{y_0}^{\Lambda^2} dy \,
\frac{y\S^2(y)}{(y+\S^2(y))^2}\left(\frac{\pi}{2} -
\frac{\theta_0(y)}{2} + \frac{\sin2\theta_0(y)}{4}\right) \, .
\mlab{extracon}
\ee

Writing $\sqrt{y}=\Lambda + \sqrt{x}\cos{\psi}$, so that $\theta_0
\simeq \psi$ for $x \ll \Lambda^2$, we have, using the mean value
theorem~:
\be
\delta\Pi(x) \simeq
\frac{8N_f\alpha}{3\pi^2}\frac{\Lambda^3\S^2(\Lambda^2)}
{\sqrt{x}(\Lambda^2+\S^2(\Lambda^2))^2}
\int_{\pi/2}^{\pi} \, d\psi \sin{\psi} \left(\frac{\pi}{2} - \frac{\psi}{2} 
+ \frac{\sin2\psi}{4}\right) \, ,
\ee

so that~:
\be
\delta\Pi(x) \simeq
\frac{8N_f\alpha}{9\pi^2}\frac{\S^2(\Lambda^2)}{\sqrt{x}\Lambda} \, .
\mlab{deltavp}
\ee

Because of the $1/\sqrt{x}$ this change in $\Pi(x)$ would be
noticeable at very small values of $x$. However, this analytic
calculation does not explain the sharp decrease of $\G(x)$ at
intermediate low momenta we and Kondo et al.~\cite{Kondo92} find 
(see Fig.~\ref{fig:figG}).

To understand why this happens we have to consider how the numerical
program computes the extra contribution, \mref{extracon}, to the vacuum
polarization integral. As shown in \mref{702.2}, the integrals of
\mref{702} are approximated by a finite sum of integrand values at 
momenta uniformly spread on a logarithmic scale. For small $x$, the extra
contribution is entirely concentrated at the uppermost momentum region of
the radial integral with $y \in [y_0,\Lambda^2]$. There the numerical
integration program will have only one grid point $x_i$ (\mref{702.4})
situated in the interval $[y_0,\Lambda^2]$ for any realistic grid
distribution. This point will lie at $x_N=\Lambda^2$ if we use a closed
(N+1)-point quadrature formula. Therefore the integral will be approximated
by the value of the integrand at $\Lambda^2$ times a weight factor
$W(\Lambda^2) = w\Lambda^2$ ($w$ is ${\cal O}(1)$)~:
\be
\delta\Pi(x) \approx
\frac{4N_f\alpha}{3\pi^2}
\frac{W(\Lambda^2)\Lambda^2\S^2(\Lambda^2)}
{x(\Lambda^2+\S^2(\Lambda^2))^2}\left(\frac{\pi}{2}
- \frac{\theta_0(\Lambda^2)}{2} +
\frac{\sin2\theta_0(\Lambda^2)}{4}\right) \, .
\ee

For small $x$ we have $\theta_0(\Lambda^2)\approx\pi/2$ and the
extra contribution to the vacuum polarization will be~:
\be
\delta\Pi(x) \approx \frac{N_f\alpha
w}{3\pi}\frac{\S^2(\Lambda^2)}{x} \, .
\mlab{deltavpnum}
\ee

This will effectively add a huge correction to the vacuum polarization
at low $x$. This has been extensively checked numerically and shown to
be completely responsible for the sudden decrease in the photon
renormalization function $\G(x)$ at low momenta. To reproduce our
previous analytic result of \mref{deltavp} numerically, the
integration grid would have to be tuned unnaturally fine to include
more points in the region $[y_0,\Lambda^2]$. Without such tuning one
has the result of \mref{deltavpnum}. Then $x\Pi(x)$ does not vanish
continuously as $x \rightarrow 0$. Instead, for $x > 0$, $x\Pi(x) \approx
N_f\alpha w \S^2(\Lambda^2)/3\pi$ and so as soon as $x$ is non-zero
the cancellation of the quadratic divergence disappears suddenly and
not gradually as the physical world requires. 
\mref{deltavpnum} tells us that the step in $\G$ (see
Fig.~\ref{fig:figG}), is due to an unsuccessful numerical
cancellation of the quadratic divergence in the vacuum polarization
integral $\Pi(x)$.
It is significant for the sensitivity of the problem that, against all
expectations, the high momentum behaviour of $\Sigma(x)$, where its
value is quite small, plays such a major role in the behaviour of
$\G(x)$ at low $x$. This will even become more apparent in the
following discussion.

It is natural to expect that the function $\S$ from the physical world will
be smooth. To improve on the discontinuous extrapolation rule
\mref{extrapS}, we can replace it 
by the following simple extrapolation rule~:
\be
\S(x > \Lambda^2) = \S(\Lambda^2) \, \frac{\Lambda^2}{x} \, .
\mlab{extrapS2}
\ee

\begin{figure}[htbp]
\begin{center}
\mbox{\epsfig{file=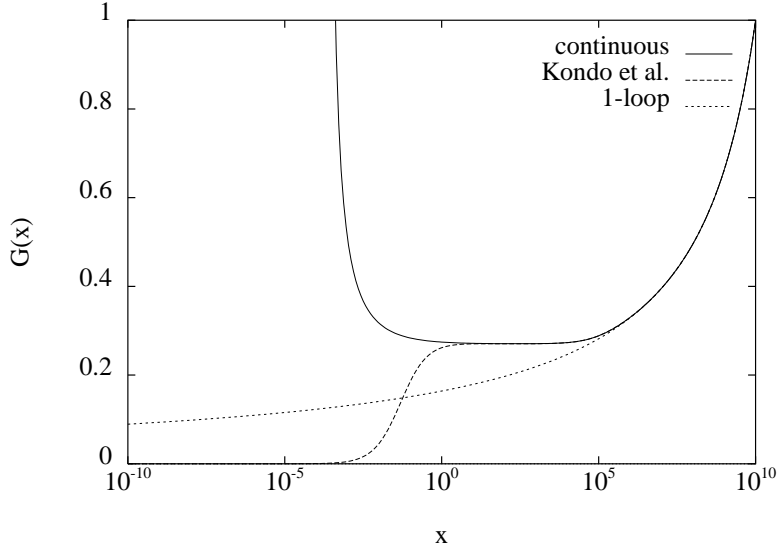,height=8cm,angle=-90}}
\end{center}
\vspace{-0.5cm}
\caption{Photon renormalization function $\G(x)$, as a function of 
momentum $x$ for $N_f=1$ and $\alpha=2.086$ as calculated in a
self-consistent way with a continuous extrapolation for $\S$, 
with the jump discontinuity in $\S$ as in Ref.~\protect{\cite{Kondo92}} 
and in 1-loop approximation ($\Lambda^2 = 1\e10$).}
\label{fig:figG3}
\end{figure}

This will get rid of the jump discontinuity in the dynamical mass
function, leaving instead a very slight kink. Although $\S(x)$ is not
yet smooth at $x=\Lambda^2$ it now is continuous.
When solving the integral equations using this extrapolation rule, the
step in the photon renormalization function at intermediate low
momenta disappears, only to be replaced by a singularity as can be
seen in Fig.~\ref{fig:figG3}.

However, the new singularity in $\G$ is not as worrying as it may seem
at first sight. If we recall \mref{G-Pi},
\[
\G(x) = \frac{1}{1 + \Pi(x)} \,,
\]
we see that the singularity in $\G$ corresponds to $\Pi(x) \to -1$.
A closer numerical investigation shows that this is due to the
inadequacy of the interpolations, \mref{702.7}, to compute the angular
integrals in \mref{702.2}. The functions $\S(x)$ and $\G(x)$
constructed with these interpolation rules are in fact piecewise
linear polynomials (on logarithmic scale) with interpolation points
$x_i$. Although these functions are continuous, they are not smooth
and this leads to cancellation mismatches in the angular integrals of
the $\G$-equation and thus to unphysical singularities in $\G$. 

This points the way to a possible solution of this problem: we want
smooth approximations to the functions in order to get a realistic,
physical answer to the problem. 

To study the validity of this statement without completely modifying
the numerical program straight away, we just add one more step at the
very end of the previous calculation.
There we used the collocation method to
construct the system of non-linear equations, \mrefb{702.1}{702.2},
enhanced with the interpolation rules, \mref{702.7}, and extrapolation rules,
\mrefb{extrapS2}{extrapG}. This system of equations was then solved to
determine the unknown function values $\S_i$ and $\G_i$ at the quadrature
nodes of the radial integrals.

Starting from this solution vector $\S_i$ we now construct a smooth
polynomial approximation $\tilde{\S}(x)$ using, for instance,
Chebyshev polynomials to replace the piecewise linear construction
achieved previously with the interpolation rule \mref{702.7}. The
Chebyshev approximation to $\S(x)$ can be written as,
\be
\tilde{\S}(x) = \sum_{j=0}^{N-1} a_j T_j(x) .
\mlab{polynom}
\ee

As will be shown in Section~\ref{Chebapprox}, the coefficients $a_j$ can be
easily determined provided we know the function values $\S(y_j)$ at the $N$
distinct roots $y_j$ of the Chebyshev polynomial $T_N(x)$ of degree~$N$.
The function values $\S(y_j)$, needed to determine $a_j$, can be
approximated by applying the interpolation rule \mref{702.7} on the
solution vector $\S_i$.  The polynomial approximation $\tilde{\S}(x)$
coincides with the interpolated, piecewise linear, dynamical mass function
$\S(x)$ at the $N$ points $y_j$,
\be
\tilde{\S}(y_j) = \S(y_j), \qquad j=1,\ldots,N.
\mlab{polynom.1}
\ee

Note that the interpolation points $y_j$ of the new smoothed function
$\tilde{\S}(x)$ do not coincide with the original collocation points $x_i$
of the collocation method, and thus, $\tilde{\S}(x_i)
\ne \S_i$.

\begin{figure}[htbp]
\begin{center}
\mbox{\epsfig{file=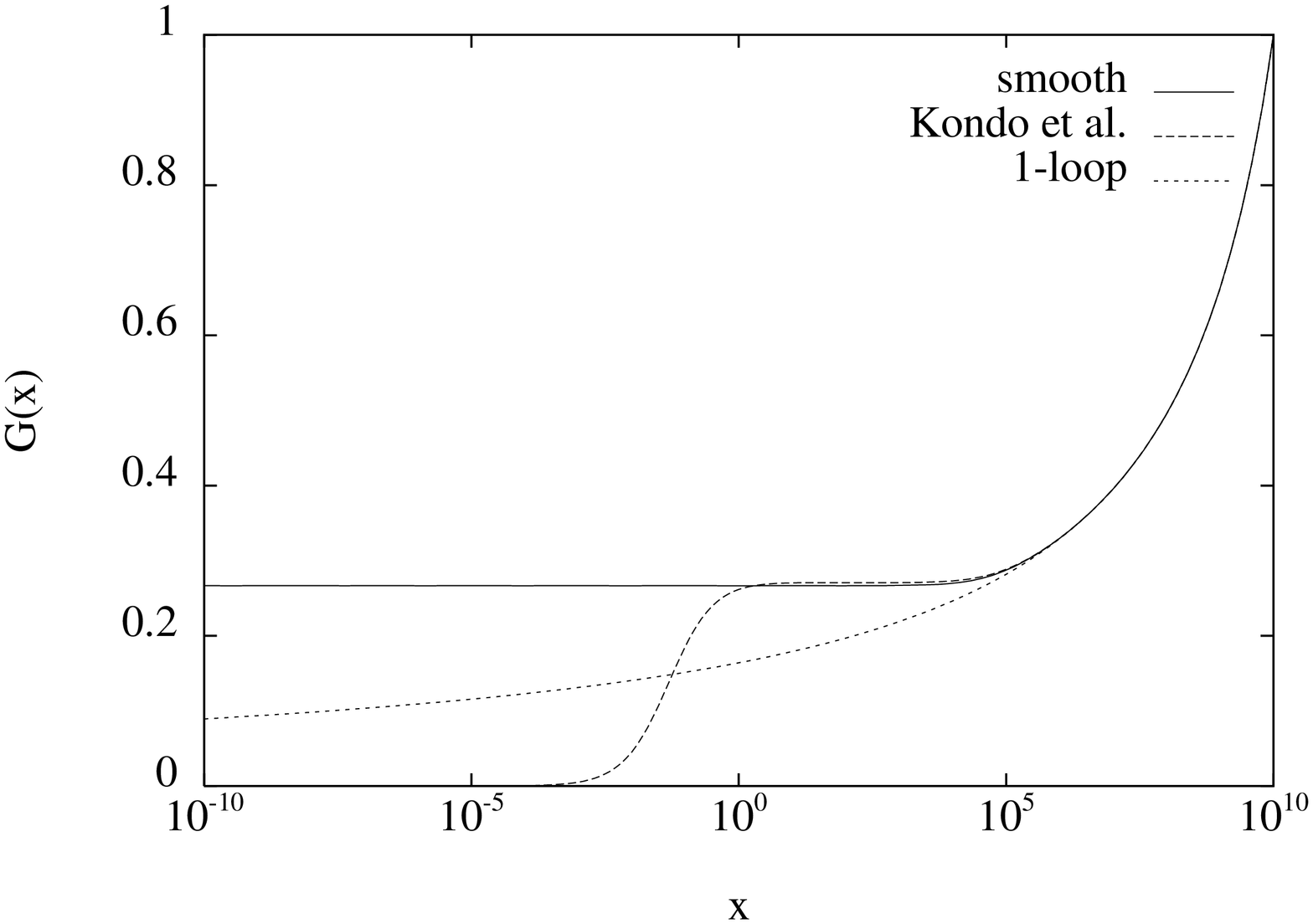,height=8cm,angle=-90}}
\end{center}
\vspace{-0.5cm}
\caption{Photon renormalization function $\G(x)$, as a function of 
momentum $x$ for $N_f=1$ and $\alpha=2.086$ as calculated in a
self-consistent way with a smoothened approximation to $\S$, 
with the jump discontinuity in $\S$ as in Ref.~\protect{\cite{Kondo92}} 
and in 1-loop approximation ($\Lambda^2 = 1\e10$).}
\label{fig:figG2}
\end{figure}

We now use the smooth function $\tilde{\S}(x)$ of \mref{polynom} to compute
numerically the integral evaluations in \mref{702.2}, and determine $\G$. As
we see in Fig.~\ref{fig:figG2}, the smoothing of $\S$ has the desired
effect on the behaviour of $\G$ . The singularity disappears and is
replaced by a flat line down into the infrared. This agrees with our
physical intuition about the behaviour of the running of the coupling, when
fermion mass is generated.

In this chapter we have seen that the proper numerical cancellation
of the quadratic divergence in the vacuum polarization requires the
dynamical mass function $\S(x)$ to be smooth. This ensures that the
cancellation of the quadratic divergence takes place smoothly as $x
\rightarrow 0$.

From the previous discussion we conclude that the collocation method,
where the unknowns of the problem are the function values at the
radial integration nodes, has definite drawbacks. Because the unknown
functions are also present in the angular integrals, we have to
complement the method with some appropriate interpolation and
extrapolation rules. The function $\S$ constructed with these rules
will not be smooth and therefore $\G$ will behave unphysically.
Furthermore, as mentioned in the previous chapter, the use of a
single fixed set of radial integration points in the collocation
method, and the kink in the radial integrand forcing us to split the
integral in two, reduces the accuracy of the integration rules.

To avoid these problems it is therefore preferable to search for
smooth solutions for the dynamical mass function $\S(x)$, the
fermion wavefunction renormalization $\F(x)$ and the photon
renormalization function $\G(x)$. In the next chapter we are going
to develop the formalism to approximate the unknown functions by a
smooth, polynomial expansion instead of discretizing the function at the
radial integration points.

\chapter{Chebyshev expansion method}
\label{Cheby}

In the previous chapters we solved the integral equations using the
collocation method. This method discretizes the unknown functions at the
nodes of the quadrature rule used to evaluate the integrals. However we
gathered enough evidence supporting the need to develop an alternative
procedure where these functions are smoothly approximated, for example by
the use of some polynomial expansion. For various reasons one of the
favoured polynomial approximations of functions is the expansion in
Chebyshev polynomials.

\section{Chebyshev polynomials}

The Chebyshev polynomial of degree $n$ is denoted $T_n(x)$, and is
given by the explicit formula~\cite{Numrecep},
\be
T_n(x) = \cos(n \arccos x) \; .
\mlab{800}
\ee 

Although this looks trigonometric at first glance, the use of
trigonometric expressions in \mref{800} gives the following polynomial
forms,
\be
\renewcommand{\arraystretch}{1.3}
\begin{array}{ccl}
T_0(x) &=& 1 \\
T_1(x) &=& x \\
T_2(x) &=& 2x^2 - 1 \\
T_3(x) &=& 4x^3 - 3x \\
T_4(x) &=& 8x^4 - 8x^2 + 1 \\
\vdots
\end{array}
\mlab{800.1}
\ee

In general one can derive the following recursion relation:
\be
T_{n+1}(x) = 2x T_n(x) - T_{n-1}(x) , \qquad n \ge 1.
\mlab{801}
\ee

\begin{figure}[htbp]
\begin{center}
\mbox{\epsfig{file=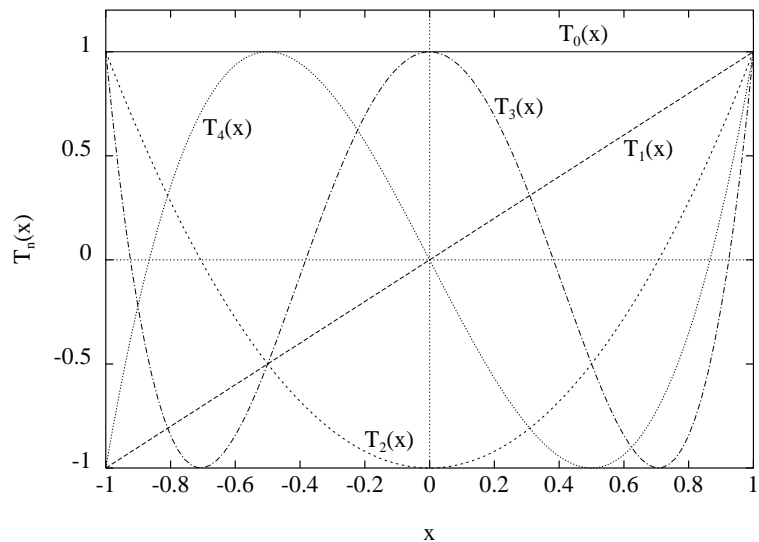,height=8cm,angle=-90}}
\end{center}
\vspace{-0.5cm}
\caption{Chebyshev polynomials $T_n(x)$ for $n$=0,\ldots,4.}
\label{Fig:Cheby}
\end{figure}

We plot the first few Chebyshev polynomials $T_n(x)$, $n$=0,\ldots,4
in Fig.~\ref{Fig:Cheby}.

The polynomial $T_n(x)$ has $n$ zeros in the interval [-1,1] at
\be
x = \cos \l(\frac{(k-1/2)\pi}{n}\r), \qquad k=1,\ldots,n.
\mlab{803}
\ee

$T_n(x)$ also has $n+1$ extrema in [-1,1] located at
\be
x = \cos \l( \frac{k\pi}{n}\r) , \qquad k=0,\ldots,n.
\mlab{804}
\ee

All the minima have a value $T_n(x)=-1$, while the maxima all have a
value $T_n(x)=1$.

The Chebyshev polynomials are orthogonal in the interval $[-1,1]$ over a
weight $\sqrt{1-x^2}$,
\be
\int_{-1}^1 dx \, \frac{T_i(x) T_j(x)}{\sqrt{1-x^2}} = \l\{
\renewcommand{\arraystretch}{1.1}
\begin{array}{l@{\qquad}l}
\D 0 & i \ne j \\ 
\D \pi/2 & i=j \ne 0 \\
\D \pi & i=j=0
\end{array} \r. \,.
\mlab{802}
\ee

In addition to the continuous orthogonality relation \mref{802}, the
Chebyshev polynomials also satisfy a discrete orthogonality
relation. If $x_k$ are the $n$ zeros of $T_n(x)$ given by \mref{803},
$k=1,\ldots,n$, and if $i,j < n$, then
\be
\sum_{k=1}^n \, T_i(x_k) T_j(x_k) = \l\{
\renewcommand{\arraystretch}{1.1}
\begin{array}{l@{\qquad}l}
\D 0 & i \ne j \\ 
\D n/2 & i=j \ne 0 \\
\D n & i=j=0
\end{array} \r. \,.
\mlab{805}
\ee

\section{Chebyshev approximation}
\label{Chebapprox}

We now want to determine the coefficients $c_j$ of the polynomial
approximation to an arbitrary function $f(x)$, 
\be
f(x) \approx \sum_{j=0}^{N-1}{'} c_j T_j(x) 
\equiv \sum_{j=0}^{N-1} c_j T_j(x) - \frac{c_0}{2}, 
\mlab{806}
\ee
such that the approximation becomes {\it exact} at the N zeros of $T_N(x)$.

For these zeros we then have
\be
f(x_k) = \sum_{j=0}^{N-1}{'} c_j T_j(x_k) , \qquad k=1,\ldots,N.
\mlab{807}
\ee

Multiply both sides with $T_i(x_k)$ where $i<N$ and sum over all zeros
of $T_N(x)$:
\be
\sum_{k=1}^N T_i(x_k) f(x_k) = \sum_{j=0}^{N-1}{'} c_j 
\sum_{k=1}^N T_i(x_k) T_j(x_k) .
\mlab{808}
\ee

Using the orthogonality relation \mref{805} yields:
\be
\sum_{k=1}^N T_i(x_k) f(x_k) = \frac{N}{2} c_i .
\mlab{809}
\ee

The coefficients $c_j$ of \mref{806} are
\be
c_j = \frac{2}{N} \sum_{k=1}^N T_j(x_k) f(x_k) .
\mlab{810}
\ee

If we substitute the expression \nref{803} for the zeros of $T_N(x)$
this becomes:
\be
c_j = \frac{2}{N} \sum_{k=1}^N T_j\l[\cos \l(\frac{(k-1/2)\pi}{N}\r)\r] 
f\l[\cos \l(\frac{(k-1/2)\pi}{N}\r)\r] .
\mlab{811}
\ee

Substituting the definition \mref{800} for the Chebyshev polynomial
$T_j(x)$ the coefficients can be computed as
\be
c_j = \frac{2}{N} \sum_{k=1}^N \cos\l(\frac{j(k-1/2)\pi}{N}\r) 
f\l[\cos \l(\frac{(k-1/2)\pi}{N}\r)\r] .
\mlab{812}
\ee

The Chebyshev expansion is often used because the error
generated by replacing the function by its expansion is smeared out
over the complete interval.

\section{Evaluation of Chebyshev approximation}

To evaluate a Chebyshev approximation of a function using a set of
Chebyshev coefficients $c_j$, we could use the recurrence relation
\mref{801} to evaluate the successive values of $T_j(x)$ and then sum
up these contributions multiplied by their respective
coefficient. However, there is a more efficient way to evaluate a sum
of polynomials using {\it Clenshaw's recurrence formula}. 

Suppose we want to evaluate the polynomial sum
\be
f(x) \equiv \sum_{j=0}^N c_j F_j(x) ,
\mlab{817}
\ee
where the polynomials $F_j(x)$ obey a recurrence relation of the kind,
\be
F_{n+1}(x) = \alpha(n,x) F_n(x) + \beta(n,x) F_{n-1}(x).
\mlab{818}
\ee  

Define the quantities $d_j$ by the following recurrence relation:
\be
d_j = \alpha(j,x) d_{j+1} + \beta(j+1,x) d_{j+2} + c_j , 
\qquad j=N,N-1,\ldots,1
\mlab{819}
\ee
where $d_{N+2} = d_{N+1} = 0$.
 
Then {\it Clenshaw's recurrence formula} to compute $f(x)$ defined in
\mref{817} is
\be
f(x) = \beta(1,x) F_0(x) d_2 + F_1(x) d_1 + F_0(x) c_0 \,.
\mlab{820}
\ee

If we apply Clenshaw's formula to the Chebyshev polynomials obeying
the recurrence relation \mref{801}, the function approximation
\mref{806} is given by
\ba
d_{N+1} &=& d_N = 0 \nn\\
d_j &=& 2x d_{j+1} - d_{j+2} + c_j \;,\qquad j=N-1,N-2,\ldots,1 \nn\\ 
f(x) &=& x d_1 - d_2 + \frac{c_0}{2} .
\mlab{820.1}
\ea

The Chebyshev polynomials define a polynomial approximation over the
interval [-1,1]. To approximate a function $f(x)$ over an arbitrary
interval [a,b] we introduce a change of variable
\be
s \equiv \frac{x-\frac{1}{2}(b+a)}{\frac{1}{2}(b-a)},
\mlab{813}
\ee  
so that,
\be
x\in[a,b] \mapsto s\in[-1,1] .
\mlab{814}
\ee

The Chebyshev approximation will now be
\be
f(x) \approx \sum_{j=0}^{N-1}{'} c_j T_j(s) ,
\mlab{815}
\ee
where $x$ is mapped into $s$ using \mref{813}.

\section{Chebyshev expansions for $\S$, $\F$ and $\G$.}

For the specific case of the numerical solution of the Schwinger-Dyson
equations, we have already pointed out in the previous chapters that a
convenient variable to perform the numerical integrations is $t =
\logten x$. Therefore we will consider $\S(x)$, $\F(x)$ and $\G(x)$
as functions of $t$, defined over the interval
$t \in [\logten\kappa^2, \logten\Lambda^2]$. According to \mref{813} the
Chebyshev polynomials, used to construct the Chebyshev expansions,
will be written as function of the new variable $s$ defined as
\be
s = \frac{\logten x-\frac{1}{2}(\logten\Lambda^2+\logten\kappa^2)}
{\frac{1}{2}(\logten\Lambda^2-\logten\kappa^2)},
\mlab{821}
\ee
or
\be
s \equiv \frac{\logten(x/\Lambda\kappa)}{\logten(\Lambda/\kappa)}.
\mlab{822}
\ee

We will define the Chebyshev expansions of the unknown functions as:
\ba
\S(x) \equiv \sum_{j=0}^{\NS-1}{'} a_j T_j(s) \mlab{823} \\ 
\F(x) \equiv \sum_{j=0}^{\NF-1}{'} b_j T_j(s) \mlab{824} \\
\G(x) \equiv \sum_{j=0}^{\NG-1}{'} c_j T_j(s) \mlab{825} ,
\ea
where $s$ is defined by \mref{822}.

In principle the number of Chebyshev polynomials $\NS, \NF, \NG$ used
to approximate the functions will be chosen so that the error on the
three functions is comparable.

We now mention some of the advantages of using the Chebyshev expansion
to approximate the unknown functions. First of all, it guarantees the
smoothness of the solutions and in doing so it should also ensure the
correct cancellation of the quadratical divergence in the vacuum
polarization integral. Related to this is the fact that the Chebyshev
expansions are extremely useful to handle the two-dimensional
integrals because the function values can be computed at any value of
$x\in[a,b]$. There is no need for any complementary interpolation
method anymore. Furthermore, because we can compute the function
values at any point it allows us to use whatever quadrature rule we
want, we are not bound anymore to use the same set of equidistant
integration points for all the individual equations in the system of
non-linear equations. We can now freely choose a different, optimal
set of points for each integration.

\section{$\S$-equation and Chebyshev expansion}
\label{ChebNewton}

We will now use the Chebyshev expansion for $\S$ to construct an
alternative method for solving the integral equation to replace the
previously used collocation method.

Let us recall the $\S$-equation \mref{S4},
\be
\S(x) = \frac{3\alpha\ln10}{2\pi^2} \int_{\logten\kappa^2}^
{\logten\Lambda^2}
dt\,\frac{y^2\S(y)}{y+\S^2(y)}
\int d\theta\,
\frac{\sin^2\theta}{z(1+\frac{N_f\alpha}{3\pi}\ln\frac{\Lambda^2}{z})},
\quad x\in [\kappa^2,\Lambda^2]
\mlab{826}
\ee
where $y = 10^t$ and $z=x+y-2\sqrt{xy}\cos\theta$.

We will now look for an approximate solution $\S(x)$ to \mref{826}
which can be written as a Chebyshev expansion,
\be
\S(x) = \sum_{j=0}^{\NS-1}{'} a_j T_j\l(s(x)\r)
\mlab{827}
\ee
where $x\in[\kappa^2,\Lambda^2]$ and, from \mref{822},
\be
s(x) = \frac{\logten(x/\Lambda\kappa)}{\logten(\Lambda/\kappa)}.
\mlab{828}
\ee

The integral equation \mref{827} contains $\NS$ unknown Chebyshev
coefficients $a_j$. To determine these coefficients we need at least
$\NS$ constraints. These constraints are obviously found by imposing
that \mref{827} should be satisfied at $M$ different values of $x$
(where $M>=\NS$). 
\be
\S_i = \frac{3\alpha\ln10}{2\pi^2} \int_{\logten\kappa^2}^
{\logten\Lambda^2}
dt\,\frac{y^2\S(y)}{y+\S^2(y)}
\int d\theta\,
\frac{\sin^2\theta}{z(1+\frac{N_f\alpha}{3\pi}\ln\frac{\Lambda^2}{z})},
\quad i=1,\ldots,M,
\mlab{829}
\ee
where $\S_i=\S(x_i)$ using the Chebyshev expansion \mref{827}.

If $M=\NS$, \mref{829} is a system of $\NS$ non-linear equations with
$\NS$ unknowns. If $M>\NS$ the system of equations will be
overconstrained and the coefficients can be determined by {\it
minimizing} the error between the right and left hand sides of the
complete system of $M$ equations.  Such a minimization procedure is
quite tedious for a non-linear problem and does not have any advantage
compared to solving the system of equations when $M=\NS$~\cite{Baker}.
In practice we will choose the $\NS$ external momenta to be located at
the $\NS$ zeros of the Chebyshev polynomial $T_{\NS}$,
\be
s_i = \cos \l(\frac{(i-1/2)\pi}{\NS}\r), \qquad i=1,\ldots,\NS,
\mlab{829.1}
\ee
and from \mref{828} the external momenta $x_i$ are given by ,
\be
x_i =\Lambda\kappa \l(\frac{\Lambda}{\kappa}\r)^{s_i}.
\mlab{829.2}
\ee

In contrast to the use of the expansion method to solve linear integral
equations, the non-linearity of \mref{829}
does not allow us to take the expansion coefficients $a_j$ out of
the integrals. To make further progress in the numerical solution of
\mref{829}, we have to approximate the integrals of \mref{829} by some 
suitable quadrature rules. The quadrature rule $R_i$, the number of
integration nodes and the position of the nodes can vary depending on
the external momentum $x_i$. The actual choice of the quadrature rule
will be discussed in a later section. \mref{829} can now be written as:
\be
\S_i = \frac{3\alpha\ln10}{2\pi^2} \sum_{k=0}^{(N_R)_i} w_{ik} 
\,\frac{y_{ik}^2\S_{ik}}{y_{ik}+\S^2_{ik}}
\sum_{\ell=0}^{N_\theta} w'_\ell 
\frac{\sin^2\theta_\ell}
{z_\ell(1+\frac{N_f\alpha}{3\pi}\ln\frac{\Lambda^2}{z_\ell})},
\quad i=1,\ldots,\NS ,
\mlab{830}
\ee
where $\S_{ik}=\S(y_{ik})$, $y_{ik}=10^{t_{ik}}$, $t_{ik}$ are the
$(N_R)_i+1$ integration nodes and $w_{ik}$ the weights corresponding to
the integration rule $R_i$. The photon momentum in the angular part is
given by $z_\ell=x_i+y_{ik}-2\sqrt{x_i y_{ik}}\cos\theta_\ell$.

The angular part of \mref{830} is independent of the unknown
function $\S$. We define:
\be
\Theta_{ik} \equiv \sum_{\ell=0}^{N_\theta} w'_\ell 
\frac{\sin^2\theta_\ell}
{z_\ell(1+\frac{N_f\alpha}{3\pi}\ln\frac{\Lambda^2}{z_\ell})} \;.
\mlab{830.1}
\ee

Substituting \mref{830.1} in \mref{830} yields:
\be
\S_i = \frac{3\alpha\ln10}{2\pi^2} \sum_{k=0}^{(N_R)_i} w_{ik} 
\,\frac{y_{ik}^2\Theta_{ik}\S_{ik}}{y_{ik}+\S^2_{ik}} ,
\quad i=1,\ldots,\NS .
\mlab{830.2}
\ee
 
\mrefb{830.2}{827} form a system of $\NS$ non-linear
algebraic equations, where the $\NS$ Chebyshev coefficients are the
unknowns. To solve this system of equations, we will again use Newton's
method, developed in Section~\ref{Newton}. We apply Eqs.~(\oref{215},
\oref{214.1}, \oref{207}) to \mrefb{830.2}{827}.

Newton's iterative method will provide successive approximations
$\mvec{a_n}$ to the vector of Chebyshev coefficients $\mvec{a}$ solving
\mref{830.2}. Each iteration step requires the solution of a linear set of
equations:
\be
J(\mvec{a_n})\, \mvec{\Delta_{n+1}} = \mvec{f(\mvec{a_n})} .
\mlab{831}
\ee

Once the solution $\mvec{\Delta_{n+1}}$ of \mref{831} has been computed,
the new approximation $\mvec{a_{n+1}}$ is determined from
\be
\mvec{a_{n+1}} = \mvec{a_n} - \mvec{\Delta_{n+1}} .
\mlab{832}
\ee

To construct the system of linear equations, \mref{831}, we rewrite
\mref{830.2} as
\be
f_i(\mvec{a}) \equiv \S_i - \frac{3\alpha\ln10}{2\pi^2}
\sum_{k=0}^{(N_R)_i} w_{ik} 
\,\frac{y_{ik}^2\Theta_{ik}\S_{ik}}{y_{ik}+\S^2_{ik}} = 0,
\quad i=1,\ldots,\NS ,
\mlab{833}
\ee
where $\S_i$, $\S_{ik}$ are functions of $\mvec{a}$. The matrix of
derivatives $J$ is defined as
\be
J_{ij}(\mvec{a}) \equiv \parder{f_i(\mvec{a})}{a_j}.
\mlab{834}
\ee

Substituting \mref{833} in \mref{834} yields
\ba
J_{ij}(\mvec{a}) &=& \parder{}{a_j}
\l( \S_i - \frac{3\alpha\ln10}{2\pi^2}
\sum_{k=0}^{(N_R)_i} w_{ik} 
\,\frac{y_{ik}^2\Theta_{ik}\S_{ik}}{y_{ik}+\S^2_{ik}}\r) \nn\\
&=& \parder{\S_i}{a_j} - \parder{}{a_j}\l(\frac{3\alpha\ln10}{2\pi^2}
\sum_{k=0}^{(N_R)_i} w_{ik} 
\,\frac{y_{ik}^2\Theta_{ik}\S_{ik}}{y_{ik}+\S^2_{ik}}\r) \,.
\mlab{835}
\ea

From the Chebyshev expansion \mref{827} and the definition \mref{806} we
know that
\be
\parder{\S_i}{a_j} = T_j\l(s_i\r) - \frac{1}{2}\delta_{j0} \equiv \Tp_j(s_i).
\mlab{836}
\ee

Applying the chain rule and substituting \mref{836} in \mref{835} gives,
\ba
J_{ij}(\mvec{a}) &=& \Tp_j\l(s_i\r) 
- \frac{3\alpha\ln10}{2\pi^2}
\sum_{k=0}^{(N_R)_i} w_{ik} \, \parder{\S_{ik}}{a_j}\parder{}{\S_{ik}}
\l(\frac{y_{ik}^2\Theta_{ik}\S_{ik}}{y_{ik}+\S^2_{ik}}\r) \nn\\
&=& \Tp_j\l(s_i\r)
- \frac{3\alpha\ln10}{2\pi^2}
\sum_{k=0}^{(N_R)_i} w_{ik} \, \Tp_j\l(r_{ik}\r)
\frac{y_{ik}^2\Theta_{ik}(y_{ik}-\S^2_{ik})}{(y_{ik}+\S^2_{ik})^2},
\mlab{837}
\ea
where $r_{ik}$ maps $y_{ik}$ on the interval [-1,1] using \mref{828}.

After substitution of \mrefb{833}{837} in \mref{831} the linear system of
algebraic equation to be solved at each iteration step in Newton's method
is
\ba
\sum_{j=0}^{\NS-1}\l[\Tp_j\l(s_i\r)
- \frac{3\alpha\ln10}{2\pi^2}
\sum_{k=0}^{(N_R)_i} w_{ik} \, \Tp_j\l(r_{ik}\r)
\frac{y_{ik}^2\Theta_{ik}(y_{ik}-\S^2_{ik})}{(y_{ik}+\S^2_{ik})^2}
\r]_{\mvec{a}=\mvec{a_n}}(\mvec{\Delta_{n+1}})_j  \hspace{1.8cm} \nn\\
&& \hspace{-11.5cm} = \l[\S_i - \frac{3\alpha\ln10}{2\pi^2}
\sum_{k=0}^{(N_R)_i} w_{ik} 
\,\frac{y_{ik}^2\Theta_{ik}\S_{ik}}{y_{ik}+\S^2_{ik}}\r]_{\mvec{a}=\mvec{a_n}} ,
 \qquad i=1,\ldots,\NS.
\mlab{838}
\ea

\section{Splitting the integral}

To evaluate the integrals of \mref{829} we have to introduce some
appropriate quadrature rule. Following the discussion in
Section~\ref{quadrule} we know that in order to preserve the accuracy of
rules with a high degree of precision, the integrand has to be sufficiently
smooth. Because of the kink in the radial integrand it is therefore
necessary to split the radial integrals in two parts:
\be
\S_i = \int_{\logten\kappa^2}^{\logten x_i} dt \; K(x_i, y) 
+ \int_{\logten x_i}^{\logten\Lambda^2} dt \; K(x_i, y)  \; ,
\qquad i=1,\ldots,\NS \;,
\mlab{839}
\ee
where the total radial integrand is
\be
K(x, y) = \frac{3\alpha\ln10}{2\pi^2}
\frac{y^2\S(y)}{y+\S^2(y)}\,\Theta(x,y)
\mlab{840}
\ee

and the angular integral $\Theta(x,y)$ is defined as:
\be
\Theta(x,y) = \int d\theta\,
\frac{\sin^2\theta}{z(1+\frac{N_f\alpha}{3\pi}\ln\frac{\Lambda^2}{z})},
\mlab{840.1}
\ee
with $z=x+y-2\sqrt{xy}\cos\theta$.

We then choose a suitable quadrature rule to evaluate both integrals in
\mref{839}. The resulting system of non-linear algebraic equations is still
given by \mref{830.2} and will be solved using the method described in the
previous section. 

\section{Gaussian quadrature}

The quadrature rule in \mref{830} can again be chosen to be a composite
Newton-Cotes rule with equidistant points as in Section~\ref{quadrule},
but, because of the polynomial expansion of $\S(x)$ we are now free to use
other methods.

In the Newton-Cotes formulae the integration nodes are equidistant and the
weights are determined to maximize the degree of precision of the
integration rule. In a more general class of integration rules, we
determine not only the weights of the function values at the different
integration nodes, but also the location of these nodes such that the
degree of precision becomes maximal. This allows us to achieve a higher
degree of precision than the Newton-Cotes rules with an equal number of
integration points. Such methods are known as the {\it Gaussian}
integration rules~\cite{Isaac,Numrecep}. The n-point Gaussian quadrature
evaluates the integral
\be
\int_a^b w(x) f(x)\,dx = \sum_{j=1}^n w_j f(x_j) + E_n\{f\}.
\mlab{841}
\ee
such that its degree of precision is $2n-1$.

One can show in general that the quadrature formula, \mref{841}, has degree
of precision at most $2n-1$. This maximum degree of precision is attained
iff the $n$ nodes $x_j$ are the zeros of $p_n(x)$, the n$^{th}$ {\it
orthogonal polynomial} with respect to the weight $w(x)$ over $[a,b]$.

Orthogonal polynomials with respect to a specified weight function
$w(x)$ over [a,b] obey the relation:
\be
\int_a^b w(x) p_i(x) p_j(x) \, dx = 0 \qquad \mbox{if} \quad i \ne j.
\mlab{842}
\ee

The Gaussian quadrature with weight $w(x)\equiv 1$ over the interval
$[-1,1]$ is known as the {\it Gauss-Legendre} quadrature rule, which can be
written as
\be
\int_{-1}^1 f(x)\,dx = \sum_{j=1}^n w_j f(x_j) + E_n\{f\}.
\mlab{842.1}
\ee

The orthogonal polynomials with respect to the weight $w(x)\equiv 1$ over
[-1,1] are the {\it Legendre} polynomials $P_n(x)$. They can be
built by imposing the orthogonality relation
\be
\int_{-1}^1 P_i(x) P_j(x) \, dx = 0 \qquad \mbox{if} \quad i \ne j
\mlab{842.2}
\ee
and are normalized by 
\be
\int_{-1}^1 P_n^2(x) \, dx = \frac{2}{2n+1}.
\mlab{842.3}
\ee

The Legendre polynomials can be computed with the help of Rodrigues' formula:
\be
P_n(x) = \frac{1}{2^n n!}\frac{d^n}{dx^n}(x^2-1)^n,
\mlab{843}
\ee
or by using the recurrence relation,
\be
(n+1)P_{n+1}(x) = (2n+1) x P_n(x) - n P_{n-1}(x).
\ee

The abscissas of the quadrature, \mref{842.1}, are the roots of the Legendre
polynomial $P_n(x)$.  The coefficients of the Gauss-Legendre quadrature
formula, \mref{842.1}, over the interval [-1,1] are given by:
\be
w_j = \frac{2}{(1-x^2)[P_n'(x_j)]^2}.
\mlab{844}
\ee

One can prove that the coefficients $w_j$ in the Gauss-Legendre quadrature
formulae are always positive. This is important for the numerical accuracy
of the method because roundoff errors are not generally magnified in this
case.

The error term of the Gauss-Legendre quadrature over [-1,1] is:
\be
E_n\{f\} = \frac{2^{2n+1}(n!)^4}{(2n+1)[(2n)!]^3} f^{(2n)}(\xi),
\qquad -1<\xi<1.
\mlab{845}
\ee

To compute an integral over an arbitrary interval, the Gauss-Legendre will
be adapted as
\be
\int_a^b f(y)\,dy = \frac{b-a}{2}\sum_{j=1}^n w_j \;
f\l(\frac{b+a}{2} + \frac{b-a}{2}x_j\r) + E_n\{f\}.
\mlab{846}
\ee
where $w_j$ and $x_j$ are the weights and nodes of the Gauss-Legendre
quadrature over the interval [-1,1], \mref{842.1}.

The error term is now
\be
E_n\{f\} = \frac{(b-a)^{2n+1}(n!)^4}{(2n+1)[(2n)!]^3} f^{(2n)}(\xi),
\qquad a<\xi<b.
\mlab{846.1}
\ee

\section{Gaussian quadrature and the integral equations}

If we look back at the solution method used to solve the integral
equations, we can ask ourselves if we could have used the Gaussian
quadrature to evaluate the integrals in the collocation method. We have
indeed tried this method, but the results obtained were much worse than
those obtained with the Newton-Cotes formula.  Indeed, if we use a Gaussian
formula with $N$ nodes $x_j$, we will construct a system of non-linear
equations where the unknowns are the function values $\S_j$ at the
integration nodes, which are now unequally spaced. Although this problem is
solvable in the same way as before the accuracy obtained is rather poor
because the high degree of precision of the Gaussian rule requires the
integrand to be sufficiently smooth. This condition is obviously not
satisfied as the integrand has a kink.  Although we encountered the same
problem when we used the composite Newton-Cotes formulae when the degree of
precision was higher than that of the trapezoidal rule, we were able to
improve the accuracy by splitting the integration region in two at the
kink, so that each of the two integrations has a smooth integrand.
Unfortunately we cannot apply this method to the Gaussian quadrature
because the integration nodes are unequally spaced. If we want to apply
Gaussian quadratures to the equation with external momentum $x_i$, we
choose $N_1$ Gaussian nodes in the interval $[\kappa^2,x_i]$ and $N_2$
Gaussian nodes in the interval $[x_i,\Lambda^2]$. Therefore we will have a
set of $N_1+N_2$ integration nodes $y_j$, being the roots of the Legendre
polynomials $P_{N_1}(x)$ and $P_{N_2}(x)$. The reason why the collocation
method fails is that the position of the $N_1+N_2$ nodes changes with that
of the kink: the integration nodes will be different for each external
momentum and the collocation method is not applicable.

However, when we introduce the Chebyshev expansion for the unknown
functions the situation is completely different. After splitting the integral
at $y=x$ we can use a Gaussian quadrature with any number of nodes $N_1,
N_2$ on the intervals because the integrand can be computed at any point in
the interval $[\kappa^2,\Lambda^2]$. The integral equation, \mref{839}, for
$\S$ will be replaced by the system of non-linear equations
\be
\S_i = \sum_{j=1}^{(N_1)_i} w_{1ij} \; K(x_i, y_{1ij}) 
+ \sum_{j=1}^{(N_2)_i} w_{2ij} \; K(x_i, y_{2ij})  \; ,
\qquad i=1,\ldots,\NS \;, 
\mlab{847}
\ee
where the nodes $y_{1ij}, y_{2ij}$ and the weights $w_{1ij}, w_{2ij}$ are
defined according to \mref{846} and $K(x,y)$ is defined in \mref{840}.
Remember that as before it is the variable $t\equiv\logten y$, where $t \in
[\logten\kappa^2,\logten\Lambda^2]$, which is mapped on the interval [-1,1]
to apply the Gauss-Legendre quadrature.

Concatenating the two arrays of node locations $y_{1ij}$ and $y_{2ij}$ into
one array $y_{ij}$ and the weight arrays $w_{1ij}$ and $w_{2ij}$ into
$w_{ij}$, we can rewrite \mref{847} as
\be
\S_i = \sum_{j=1}^{(N_R)_i} w_{ij} \; K(x_i, y_{ij}),
\qquad i=1,\ldots,\NS \;,
\mlab{848}
\ee
where $(N_R)_i=(N_1)_i+(N_2)_i$.

The system of equations, \mref{848}, is similar to the system, \mref{830.2},
for which Newton's method was developed in Section~\ref{ChebNewton}. 
Therefore \mref{848} will be solved by Newton's iteration method,
\mref{838}.

Although we also considered the use of a two-dimensional adaptive
integration method, we did not retain this method. Its advantage is that it
only computes function values at positions which depend on the behaviour of
the integrand, minimizing the number of function evaluations. Furthermore,
the integration routine returns an integral value satisfying a requested
minimum accuracy. However, the method is not efficient to evaluate
integrals as part of an integral equation. The variable location of the
function values to be evaluated does not allow us to compute parts of the
integrands beforehand and to store them for multiple, future use. Moreover,
the main problem resides in the use of Newton's method, which
requires the knowledge of the derivatives of the integral with respect to
the Chebyshev coefficients.  The use of an adaptive method makes it
extremely hard and inefficient to compute these derivatives. It seems
therefore that a higher order method with a priori determined integration
nodes and weights, as in the Gaussian quadrature, is the best choice of
integral evaluation for the solution of integral equations with the
Chebyshev expansion method.

In the next chapter we will apply the Chebyshev expansion method to solve
the coupled integral equations of unquenched QED in the bare vertex
approximation for various approximations to the ($\S$, $\F$,
$\G$)-system of integral equations.

\chapter{Numerical results with Chebyshev expansion method}
\label{Chap:Numres}

In this chapter we will apply the Chebyshev expansion method to determine
the critical coupling above which fermion mass is generated dynamically in
unquenched QED in the bare vertex approximation. We will consider various
approximations to the coupled ($\S$, $\F$, $\G$)-system. First we will
decouple the $\G$ equation by using the 1-loop approximation to the vacuum
polarization. In a next section we will revisit the coupled ($\S$,
$\G$)-system which was discussed previously in Chapter~\ref{Paper2} and was
the motivation to introduce the Chebyshev expansion method. Finally we will
treat the complete system of coupled integral equations for $\S$, $\F$ and
$\G$.

\section{The 1-loop approximation}
\label{Sec:1loop}

We first simplify the ($\S$, $\F$, $\G$)-system of coupled integral
equations by approximating the vacuum polarization by its 1-loop result.
The $\G$-equation then decouples from the coupled ($\S$,~$\F$)-system
describing the dynamical generation of fermion mass. We recall
Eqs.~(\oref{S0}, \oref{F0}),
\ba
\frac{\S(x)}{\F(x)} &=& \frac{3\alpha}{2\pi^2} \int dy \, 
\frac{y\F(y)\S(y)}{y+\S^2(y)} \int d\theta \, \sin^2\theta
\, \frac{\G(z)}{z} \mlab{850}\\
\frac{1}{\F(x)} &=& 1 + \frac{\alpha}{2\pi^2 x} \int dy \, 
\frac{y\F(y)}{y+\S^2(y)} 
\mlab{851}\\
&& \times \int d\theta \, \sin^2\theta \, \G(z) \,
\l(\frac{3 \sqrt{xy} \cos\theta}{z} 
- \frac{2 x y \sin^2\theta}{z^2} \r) \nn
\ea 
where $z \equiv x+y-2\sqrt{xy}\cos\theta$ and the 1-loop approximation to
$\G(z)$ is given by:
\be
\G(z) = \frac{1}{1+\frac{N_f\alpha}{3\pi}\ln\frac{\Lambda^2}{z}}.
\mlab{852}
\ee

As discussed in Chapter~\ref{Sec:UnqQED} many other authors have used the
same approximation to determine the critical coupling of unquenched QED.
We will consider the three main variants encountered in the literature.
Firstly, we consider the LAK-approximation to remove the angular dependence
of the vacuum polarization; this automatically yields $\F(x)=1$ and leaves
us the $\S$-equation alone to solve.  In another approximation we keep the
full angular dependence in the vacuum polarization but approximate $\F(x)
\equiv 1$, which should be reasonable in the Landau gauge, and solve the
$\S$-equation.  Finally we will solve the coupled ($\S$, $\F$)-system in
the 1-loop approximation to the vacuum polarization.

\subsection{The LAK-approximation}
\label{Sec:LAK}

An often used variant to the 1-loop calculation is often referred to as the
LAK-approximation (in analogy to Landau, Abrikosov and
Khalatnikov~\cite{LAK56}) to the vacuum polarization:
\be
\Pi(z) = \Pi(\max(x,y)), 
\mlab{853}
\ee
where $z = x+y-2\sqrt{xy}\cos\theta$. This approximation has been often
introduced to allow the angular integrals to be computed analytically
\cite{Kondo91,Gusynin}. Furthermore, the angular integral of the
$\F$-equation, \mref{851}, vanishes in the Landau gauge when introducing
the LAK-approximation and $\F(x)=1$.

The mass equation, \mref{850}, now becomes 
\be
\S(x) = \frac{3\alpha}{2\pi^2} \int dy \, 
\frac{y\S(y)}{y+\S^2(y)}\G(\max(x,y)) \int d\theta \, 
\frac{\sin^2\theta}{z} .
 \mlab{854}
\ee

The angular integral can be computed analytically and is given in
Appendix~\ref{App:angint}. Substituting \mref{A1} into \mref{854} gives
\be
\S(x) = \frac{3\alpha}{4\pi} \int dy \, 
\frac{y\S(y)}{y+\S^2(y)}\frac{\G(\max(x,y))}{\max(x,y)}.
 \mlab{856}
\ee
with $\G(z)$ given by \mref{852}.

\mref{856} will be solved following the solution pattern for the Chebyshev
expansion method developed in Chapter~\ref{Cheby}. Change the integration
variable from $y$ to $t=\logten y$. Then, split the integral in two at
$y=x$, where the radial integrand obviously has a kink.  Consequently,
replace the integrals by a quadrature formula using a Gauss-Legendre
quadrature with $(N_1)_i=(N_2)_i=120$ nodes on every radial integral. The
resulting system of non-linear equations is:
\be
\hspace{2cm}\S_i = \frac{3\alpha\ln10}{4\pi} \sum_{j=1}^{(N_R)_i} w_{ij} 
\,\frac{y_{ij}^2\S_{ij}}{y_{ij}+\S^2_{ij}}
\frac{\G(\max(x_i,y_{ij}))}{\max(x_i,y_{ij})} \quad ,
\qquad i=1,\ldots,\NS .
\mlab{850.4}
\ee

This system can then be solved with a Newton's method analogous to the one
described in Section~\ref{ChebNewton} to determine the coefficients $a_j$,
$j=1,\ldots,\NS$ of the Chebyshev expansion for $\S(x)$. In practice we
choose $\NS=50$, such that the error due to the approximation of $\S(x)$ by
a Chebyshev expansion is negligible.

\begin{figure}[htbp]
\begin{center}
\mbox{\epsfig{file=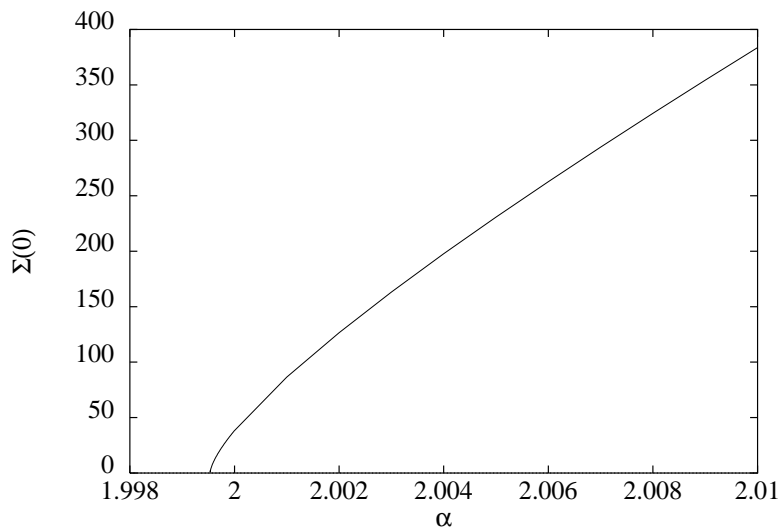,height=8cm,angle=-90}}
\end{center}
\vspace{-0.5cm}
\caption{Generated fermion mass $\S(0)$ versus coupling $\alpha$ for
$N_f=1$ in the 1-loop LAK approximation to $\Pi$.}
\label{Fig:fmg-1loop-LAK}
\end{figure}

\begin{figure}[htbp]
\begin{center}
\mbox{\epsfig{file=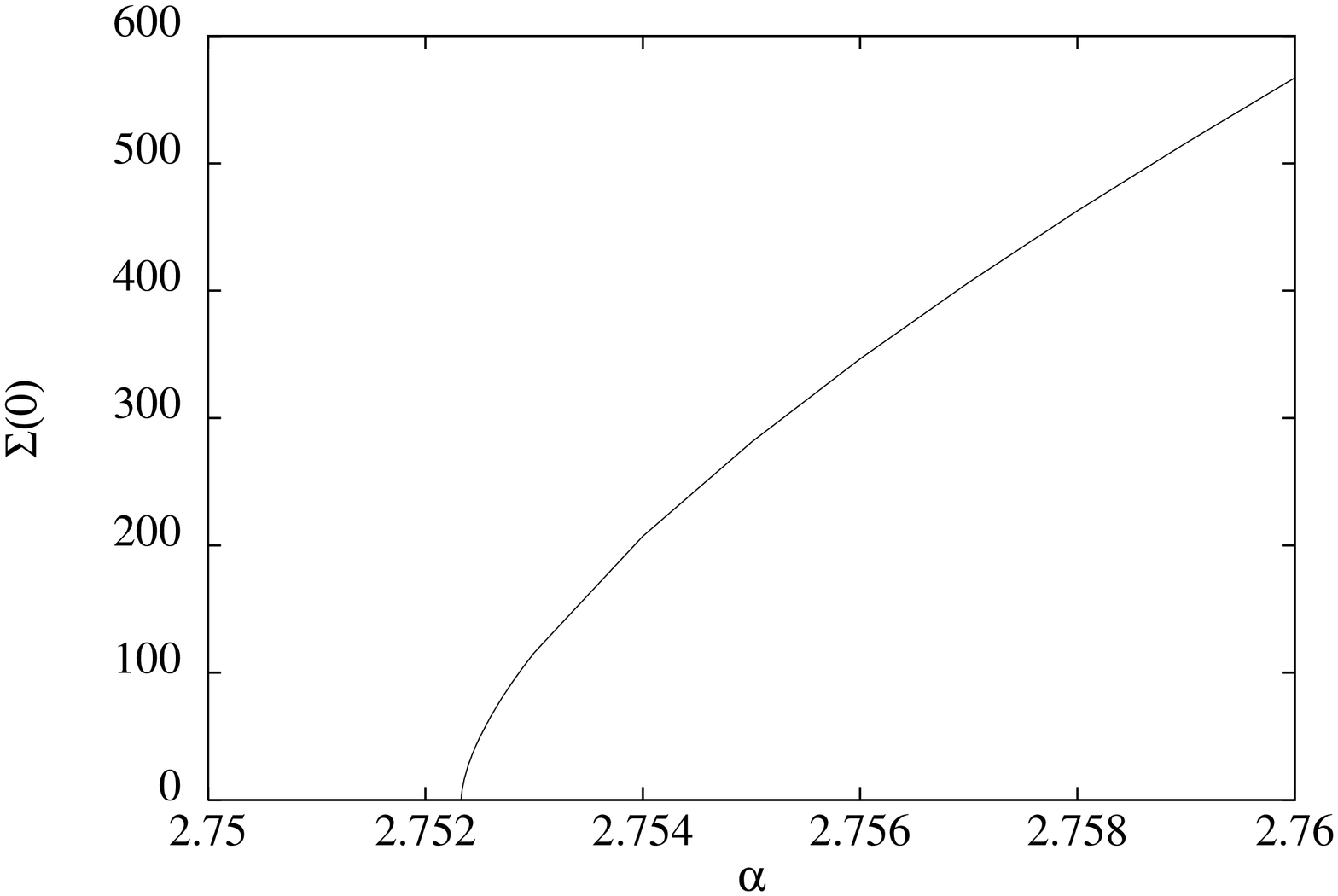,height=8cm,angle=-90}}
\end{center}
\vspace{-0.5cm}
\caption{Generated fermion mass $\S(0)$ versus coupling $\alpha$ for
$N_f=2$ in the 1-loop LAK approximation to $\Pi$.}
\label{Fig:fmg-1loop-LAK-N2}
\end{figure}

The numerical results of \mref{850.4} are summarized in
Figs.~\ref{Fig:fmg-1loop-LAK} for $N_f=1$ and
Fig.~\ref{Fig:fmg-1loop-LAK-N2} for $N_f=2$ where we show the evolution of
the generated fermion mass with changing coupling $\alpha$. The critical
couplings are \fbox{$\alpha_c(N_f=1)=1.99953$} and
\fbox{$\alpha_c(N_f=2)=2.75233$}.

\subsection{The $\F\equiv 1$ approximation}
\label{Cheb-S-eq}

In this section we approximate the system of equations,
\mrefb{850}{851}, by setting $\F(x)\equiv 1$, which is thought to be a
good approximation in the Landau gauge, and solve the remaining
$\S$-equation which is:
\be
\S(x) = \frac{3\alpha}{2\pi^2} \int dy \, 
\frac{y\S(y)}{y+\S^2(y)} \int d\theta \, 
\frac{\sin^2\theta}{z\l(1+\frac{N_f\alpha}{3\pi}\ln\frac{\Lambda^2}{z}\r)} \;.
\mlab{850.1}
\ee

The numerical method to solve this equation has been derived in
Chapter~\ref{Cheby}. We saw in \mref{848} that the Chebyshev expansion
method requires us to solve the following system of non-linear algebraic
equations for the Chebyshev coefficients $a$ of the Chebyshev expansion for
$\S(x)$:
\be
\hspace{2cm}\S_i = \frac{3\alpha\ln10}{2\pi^2} \sum_{j=1}^{(N_R)_i} w_{ij} 
\,\frac{y_{ij}^2\S_{ij}}{y_{ij}+\S^2_{ij}} \Theta(x_i,y_{ij}) \quad ,
\qquad i=1,\ldots,\NS .
\mlab{850.2}
\ee
with
\be
\hspace{-2cm}\Theta(x,y) = \int d\theta\,
\frac{\sin^2\theta}{z(1+\frac{N_f\alpha}{3\pi}\ln\frac{\Lambda^2}{z})} \;.
\mlab{850.3}
\ee

To solve \mref{850.2}, we first choose a set of
values $(N_1)_i,(N_2)_i$, $i=1,\ldots,\NS$, fixing the Gaussian quadrature
rule to be used on each single radial integral. In practice we opted for
$(N_1)_i = (N_2)_i = 120$ nodes on each interval $[\logten\kappa^2,\logten
x_i]$ and $[\logten x_i,\logten\Lambda^2]$ for all $i$ to yield sufficient
accuracy.  For each rule we then compute and store the corresponding
locations and weights of the integration nodes.

Then, the angular integrals $\Theta(x_i,y_{ij})$, \mref{850.3}, are computed
for $i=1,\ldots,\NS$ and $j=1,\ldots,(N_1)_i+(N_2)_i$, using some
appropriate quadrature formula. We will use a Gaussian quadrature rule with
$(N_\theta)_{ij}$ nodes to evaluate the angular integrals. In practice we
will choose the same number of nodes $N_\theta$ for all the angular
integrals. The angular integrals are evaluated by
\be
\Theta(x_i,y_{ij}) = \sum_{k=0}^{N_\theta} w_k
\frac{\sin^2\theta_k}{z_k(1+\frac{N_f\alpha}{3\pi}\ln\frac{\Lambda^2}{z_k})},
\mlab{849}
\ee
where $z_k=x_i+y_{ij}-2\sqrt{x_i y_{ij}}\cos\theta_k$. The locations
$\theta_k$ and weights $w_k$ of the Gaussian quadrature are determined by
\mref{846} for a Gaussian quadrature with $N_\theta$ points over the
interval $[0,\pi]$. In practice we take $N_\theta=32$ to give us sufficient
accuracy.

After the angular integrals have been computed and stored, we apply
Newton's method to \mref{850.2} as described in Section~\ref{ChebNewton} to
find the solution vector of Chebyshev coefficients $a_j$, $j=1,\ldots,\NS$
defining the Chebyshev approximation to $\S(x)$.  As in
Section~\ref{Sec:LAK} we will again take $\NS=50$.

\begin{figure}[htbp]
\begin{center}
\mbox{\epsfig{file=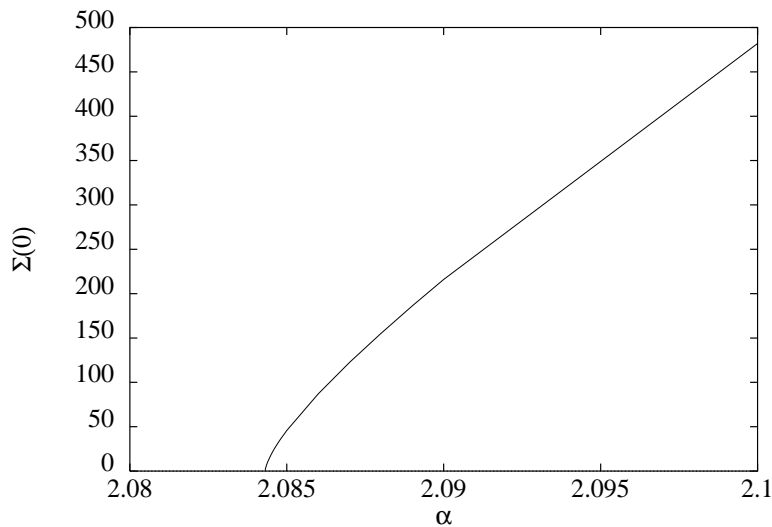,height=8cm,angle=-90}}
\end{center}
\vspace{-0.5cm}
\caption{Generated fermion mass $\S(0)$ versus coupling $\alpha$ for
$N_f=1$ in the 1-loop approximation to $\Pi$ and $\F\equiv 1$.}
\label{Fig:fmg-1loop}
\end{figure}

We now summarize the main results computed from \mref{850.2} with the above
described method.  The results are quite similar to the ones obtained with
the collocation method in Section~\ref{1loop}. We show the evolution of the
generated fermion mass with changing coupling $\alpha$ in
Fig.~\ref{Fig:fmg-1loop} for one flavour, $N_f=1$. The critical coupling is
\fbox{$\alpha_c(N_f=1)=2.08431$}, which is in total agreement with the results
obtained with the collocation method and the improved Simpson's
rule. However, the results show that we need much fewer integration nodes
using the Gauss-Legendre quadrature than we do using Simpson's rule to
obtain equal accuracy.

\begin{figure}[htbp]
\begin{center}
\mbox{\epsfig{file=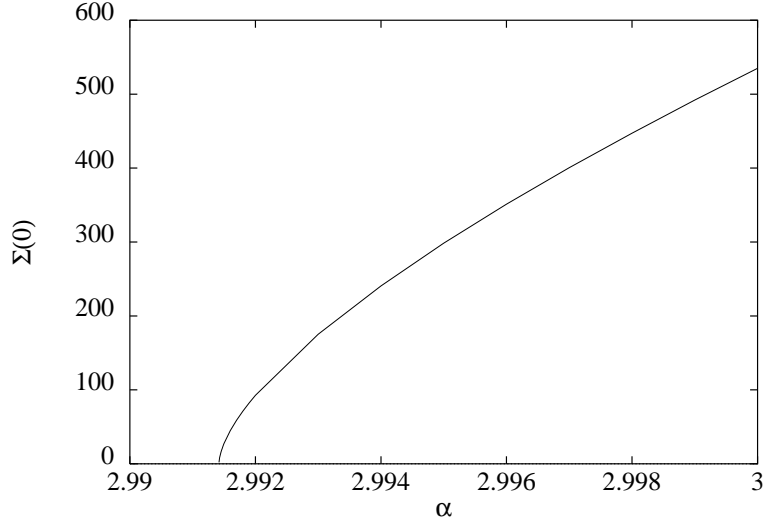,height=8cm,angle=-90}}
\end{center}
\vspace{-0.5cm}
\caption{Generated fermion mass $\S(0)$ versus coupling $\alpha$ for
$N_f=2$ in the 1-loop approximation to $\Pi$ and $\F\equiv 1$.}
\label{Fig:fmg-1loop-N2}
\end{figure}

We performed a similar calculation for two flavours, $N_f=2$. The evolution
of the generated fermion mass is shown in Fig.~\ref{Fig:fmg-1loop-N2}. The
critical coupling is \fbox{$\alpha_c(N_f=2)=2.99142$}.

\subsection{The coupled ($\S$, $\F$)-system}
\label{Sec:BF}

A further improvement on the calculation in the 1-loop approximation to the
vacuum polarization is to solve the coupled system of fermion equations
(\oref{850}, \oref{851}) for $\S$ and $\F$. We have seen in
Section~\ref{Newton} how such a coupled system can be solved using Newton's
method.  

We are looking for approximate solutions to Eqs.~(\oref{850}, \oref{851}),
which can be written as the following Chebyshev expansions:
\ba
\S(x) \equiv \sum_{j=0}^{\NS-1}{'} a_j T_j(s(x)) \mlab{861} \\ 
\F(x) \equiv \sum_{j=0}^{\NF-1}{'} b_j T_j(s(x)) \mlab{862}
\ea
where $s(x)$ satisfies \mref{822} and where the sum $\sum'$ is defined in
\mref{806}.

To solve the problem numerically we will again go through the following
steps. Change the integration variable from $y$ to $t=\logten y$. Then,
select $\NS$ external momenta where we impose that \mref{850} has to be
satisfied and $\NF$ external momenta where \mref{851} has to be
satisfied. We then split the radial integrals in two at $x=y$ to avoid
integrating numerically over the kink. Consequently we introduce the
quadrature rules to evaluate the integrals. We will again use the
Gauss-Legendre quadrature to solve the radial and angular integrals. We now
evaluate and store the angular integrals,
\ba
\Theta_\S(x_i,y_{ij}) &=& \sum_{k=1}^{N_\theta} w_k \, \sin^2\theta_k
\, \frac{\G(z_k)}{z_k} \mlab{863} \\
\Theta_\F(x_i,y_{ij}) &=& \sum_{k=1}^{N_\theta} w_k \, \sin^2\theta_k \, 
\G(z_k) \, \l(\frac{3 \sqrt{x_i y_{ij}} \cos\theta_k}{z_k} 
- \frac{2 x_i y_{ij} \sin^2\theta_k}{z_k^2} \r) \mlab{864}
\ea
where $z_k=x_i+y_{ij}-2\sqrt{x_i y_{ij}}\cos\theta_k$ and $\G(z)$ is
defined by its 1-loop approximation \mref{852}.

By doing so we are left with a system of $\NS+\NF$ non-linear equations to
determine the $\NS$ Chebyshev coefficients $a_j$, defining the Chebyshev
expansion of $\S(x)$, and the $\NF$ Chebyshev coefficients $b_j$, defining
the Chebyshev expansion of $\F(x)$:
\ba
f_{1,i} &\equiv& \frac{\S(x_i)}{\F(x_i)} -
\frac{3\alpha\ln10}{2\pi^2}
\sum_{j=1}^{(N_R)_i} w_{ij} \, 
\frac{y_{ij}^2\F(y_{ij})\S(y_{ij})}{y_{ij}+\S^2(y_{ij})} 
\Theta_\S(x_i,y_{ij}) = 0, \quad i=1,\ldots,\NS \hspace*{5mm} \mlab{865}\\
f_{2,i} &\equiv& \frac{1}{\F(x_i)} - 1 -
\frac{\alpha\ln10}{2\pi^2 x_i}
\sum_{j=1}^{(N_R)_i} w_{ij} \,
\frac{y_{ij}^2\F(y_{ij})}{y_{ij}+\S^2(y_{ij})} \Theta_\F(x_i,y_{ij}) = 0, 
\quad i=1,\ldots,\NF \mlab{866}
\ea
where $(N_R)_i=(N_1)_i+(N_2)_i$ is the total number of nodes of the
two Gauss-Legendre rules used to compute the split radial integrals.
This system of non-linear equations will be solved with Newton's iterative
method. Each iteration step requires the solution of the following system
of linear equations,
\be
J(\mvec{a_n}, \mvec{b_n})\, \mvec{\Delta_{n+1}} = 
\mvec{f}(\mvec{a_n},\mvec{b_n}) ,
\mlab{867}
\ee

which can be written out as:
\ba
\sum_{j=0}^{\NS-1}\parder{f_{1,i}(\mvec{a_n},\mvec{b_n})}{a_j}(\Delta_{a,n+1})_j
+\sum_{j=0}^{\NF-1}
\parder{f_{1,i}(\mvec{a_n},\mvec{b_n})}{b_j}(\Delta_{b,n+1})_j
&=& f_{1,i}(\mvec{a_n},\mvec{b_n}), \quad i=1,\ldots,\NS \nn \\[-5pt]
\mlab{868} \\[-5pt] 
\sum_{j=0}^{\NS-1}\parder{f_{2,i}(\mvec{a_n},\mvec{b_n})}{a_j}(\Delta_{a,n+1})_j
+\sum_{j=0}^{\NF-1}
\parder{f_{2,i}(\mvec{a_n},\mvec{b_n})}{b_j}(\Delta_{b,n+1})_j
&=& f_{2,i}(\mvec{a_n},\mvec{b_n}), \quad i=1,\ldots,\NF \,.\nn
\ea

The partial derivatives in \mref{868} are computed from \mrefb{865}{866}
using the expression \nref{836}:
\ba
\parder{f_{1,i}(\mvec{a},\mvec{b})}{a_j} &=& \parder{}{a_j}\l(
\frac{\S(x_i)}{\F(x_i)} - \frac{3\alpha\ln10}{2\pi^2}
\sum_{k=1}^{(N_R)_i} w_{ik} \, 
\frac{y_{ik}^2\F(y_{ik})\S(y_{ik})}{y_{ik}+\S^2(y_{ik})} 
\Theta_\S(x_i,y_{ik})\r) \nn \\
&=& \frac{\Tp_j\l(s_i\r)}{\F(x_i)}  - \frac{3\alpha\ln10}{2\pi^2}
\sum_{k=1}^{(N_R)_i} w_{ik} \, 
\frac{y_{ik}^2\F(y_{ik})\l(y_{ik}-\S^2(y_{ik})\r)\Tp_j\l(r_{ik}\r)}
{\l(y_{ik}+\S^2(y_{ik})\r)^2} 
\Theta_\S(x_i,y_{ik}) \hspace{1cm} \mlab{869} \\[2mm]
\parder{f_{1,i}(\mvec{a},\mvec{b})}{b_j} &=& \parder{}{b_j}\l(
\frac{\S(x_i)}{\F(x_i)} - \frac{3\alpha\ln10}{2\pi^2}
\sum_{k=1}^{(N_R)_i} w_{ik} \, 
\frac{y_{ik}^2\F(y_{ik})\S(y_{ik})}{y_{ik}+\S^2(y_{ik})} 
\Theta_\S(x_i,y_{ik})\r) \nn \\
&=& -\frac{\S(x_i)\Tp_j\l(s_i\r)}{\F^2(x_i)} - \frac{3\alpha\ln10}{2\pi^2}
\sum_{k=1}^{(N_R)_i} w_{ik} \, 
\frac{y_{ik}^2\Tp_j\l(r_{ik}\r)\S(y_{ik})}{y_{ik}+\S^2(y_{ik})} 
\Theta_\S(x_i,y_{ik}) \mlab{870} \\[2mm]
\parder{f_{2,i}(\mvec{a},\mvec{b})}{a_j} &=& \parder{}{a_j}\l(
\frac{1}{\F(x_i)} - 1 -
\frac{\alpha\ln10}{2\pi^2 x_i}
\sum_{k=1}^{(N_R)_i} w_{ik} \,
\frac{y_{ik}^2\F(y_{ik})}{y_{ik}+\S^2(y_{ik})} \Theta_\F(x_i,y_{ik})\r)\nn\\
&=& \frac{\alpha\ln10}{2\pi^2 x_i}
\sum_{k=1}^{(N_R)_i} w_{ik} \,
\frac{2y_{ik}^2\F(y_{ik})\S(y_{ik})\Tp_j(r_{ik})}{\l(y_{ik}+\S^2(y_{ik})\r)^2}
\Theta_\F(x_i,y_{ik}) \mlab{871} \\[2mm]
\parder{f_{2,i}(\mvec{a},\mvec{b})}{b_j} &=& \parder{}{b_j}\l(
\frac{1}{\F(x_i)} - 1 -
\frac{\alpha\ln10}{2\pi^2 x_i}
\sum_{k=1}^{(N_R)_i} w_{ik} \,
\frac{y_{ik}^2\F(y_{ik})}{y_{ik}+\S^2(y_{ik})} \Theta_\F(x_i,y_{ik})\r)\nn\\
&=&-\frac{\Tp_j\l(s_i\r)}{\F^2(x_i)} -
\frac{\alpha\ln10}{2\pi^2 x_i}
\sum_{k=1}^{(N_R)_i} w_{ik} \,
\frac{y_{ik}^2\Tp_j(r_{ik})}{y_{ik}+\S^2(y_{ik})} \Theta_\F(x_i,y_{ik}) \;.
\mlab{872} 
\ea

We now substitute Eqs.~(\oref{865}, \oref{866}, \oref{869}-\oref{872}) in
the system of linear equations \mref{868} and solve it for
$(\mvec{\Delta_{a,n+1}}, \mvec{\Delta_{b,n+1}})$. Then, the new
approximations to the Chebyshev coefficients are computed by:
\ba
\mvec{a_{n+1}} &=& \mvec{a_n} - \mvec{\Delta_{a,n+1}} \nn \\[-12pt]
\mlab{873}\\[-12pt]
\mvec{b_{n+1}} &=& \mvec{b_n} - \mvec{\Delta_{b,n+1}} \nn .
\ea

\begin{figure}[htbp]
\begin{center}
\mbox{\epsfig{file=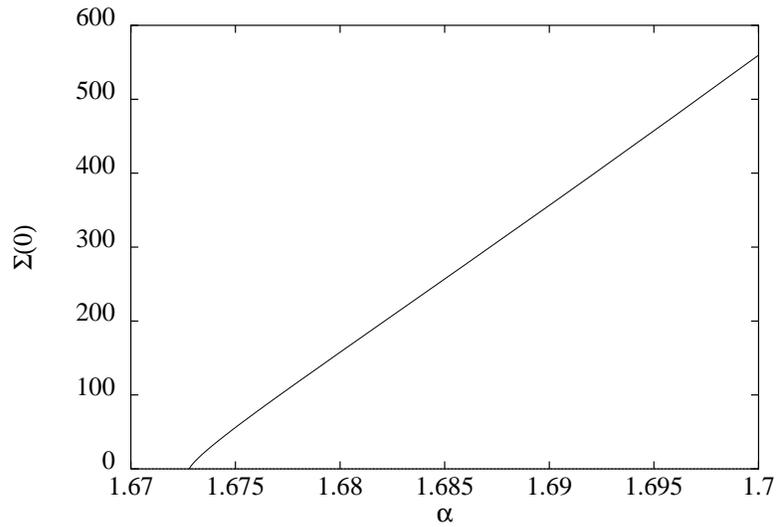,height=8cm,angle=-90}}
\end{center}
\vspace{-0.5cm}
\caption{Generated fermion mass $\S(0)$ versus coupling $\alpha$ for
the coupled $(\S, \F)$-system, for $N_f=1$ in the 1-loop approximation to
$\Pi$.}
\label{Fig:fmg-1loop-BF}
\end{figure}

In the program implementing this we choose $\NS=\NF=50$. The Gauss-Legendre
quadratures are performed with $N_\theta=32$ nodes for the angular
integrations and $(N_1)_i=(N_2)_i=120$ nodes for the split radial
integrations.
The evolution of the generated mass for the coupled $(\S, \F)$-system for
$N_f=1$ is shown in Fig.~\ref{Fig:fmg-1loop-BF}. The value of the critical
coupling is \fbox{$\alpha_c(N_f=1)=1.67280$}.

\begin{figure}[htbp]
\begin{center}
\mbox{\epsfig{file=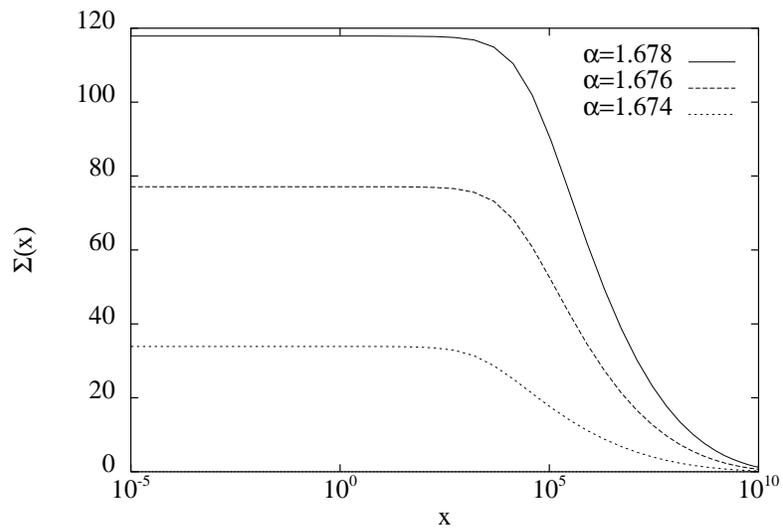,height=8cm,angle=-90}}
\end{center}
\vspace{-0.5cm}
\caption{Dynamical fermion mass $\S(x)$ versus momentum squared $x$ for
the coupled $(\S, \F)$-system, for $N_f=1$ in the 1-loop approximation to
$\Pi$, for $\alpha=1.678, 1.676, 1.674$.}
\label{Fig:B-1loop-BF}
\end{figure}

\begin{figure}[htbp]
\begin{center}
\mbox{\epsfig{file=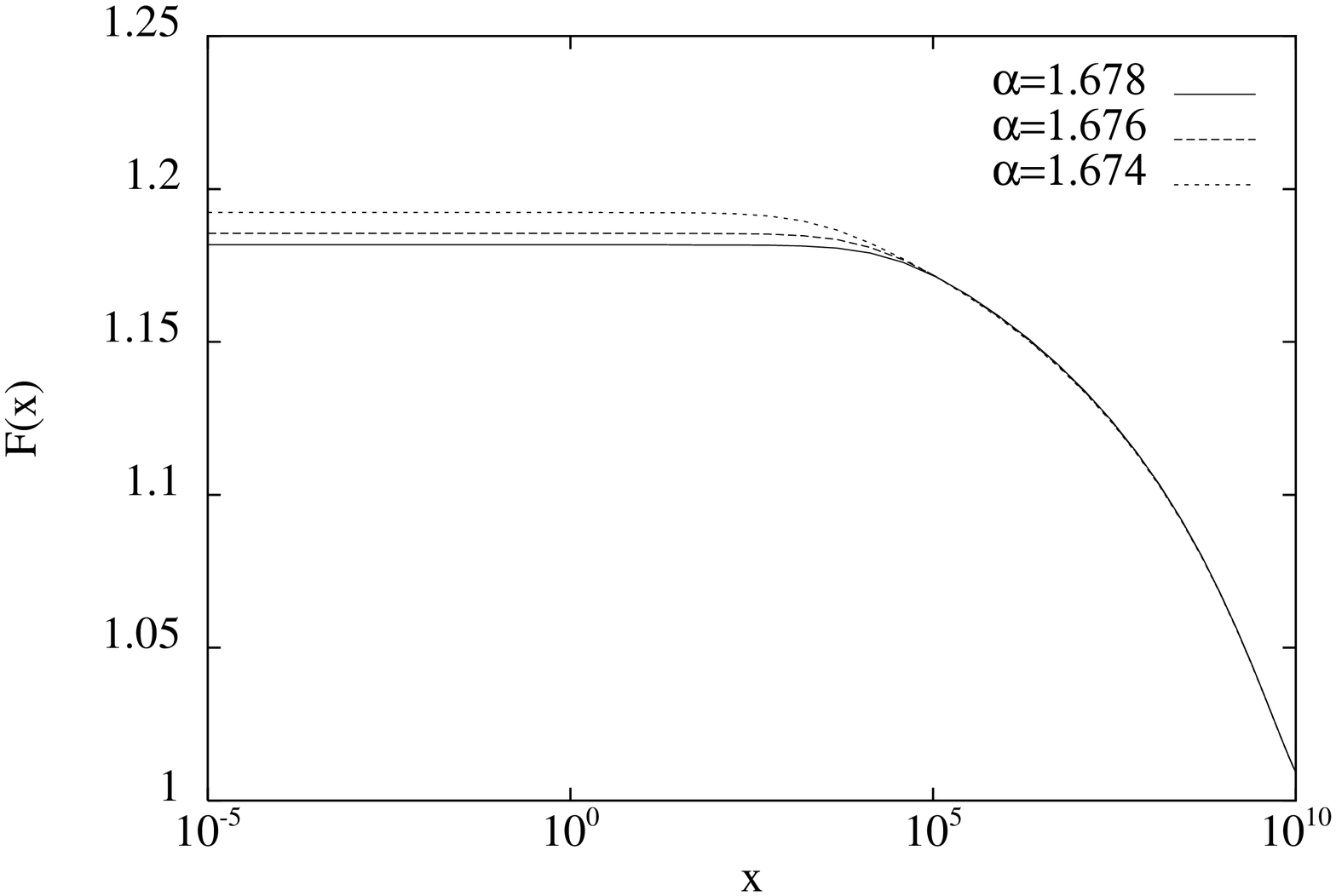,height=8cm,angle=-90}}
\end{center}
\vspace{-0.5cm}
\caption{Fermion wavefunction renormalization $\F(x)$ versus momentum
squared $x$ for the coupled $(\S, \F)$-system, for $N_f=1$ in the 1-loop
approximation to $\Pi$, for $\alpha=1.678, 1.676, 1.674$.}
\label{Fig:F-1loop-BF}
\end{figure}

Typical plots of the dynamical mass function $\S(x)$ and the fermion
wavefunction renormalization $\F(x)$ are shown in Fig.~\ref{Fig:B-1loop-BF}
and Fig.~\ref{Fig:F-1loop-BF} for $\alpha=1.678, 1.676, 1.674$.

\begin{figure}[htbp]
\begin{center}
\mbox{\epsfig{file=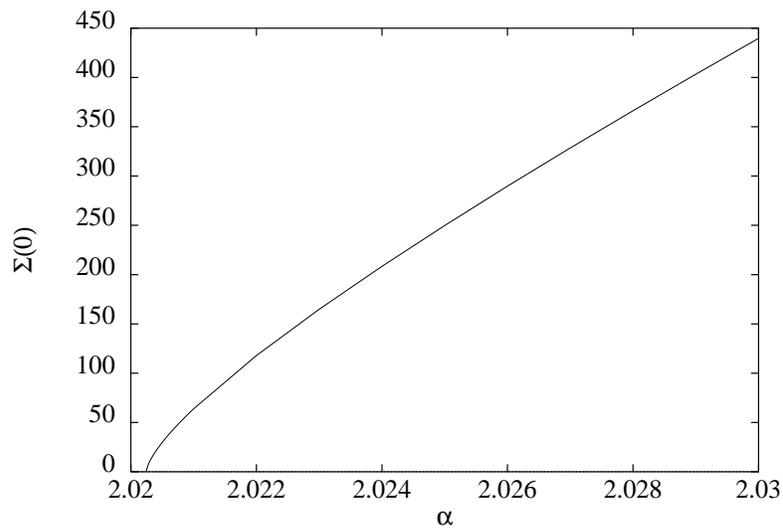,height=8cm,angle=-90}}
\end{center}
\vspace{-0.5cm}
\caption{Generated fermion mass $\S(0)$ versus coupling $\alpha$ for
the coupled $(\S, \F)$-system, for $N_f=2$ in the 1-loop approximation to
$\Pi$.}
\label{Fig:fmg-1loop-BF-N2}
\end{figure}

Fig.~\ref{Fig:fmg-1loop-BF-N2} shows the generated fermion mass versus
coupling for $N_f=2$. The value of the critical coupling is
\fbox{$\alpha_c(N_f=2)=2.02025$}.

\clearpage
\subsection{Summary}
\label{Sec:sum-1loop}

\vspace{-2mm}
In Table~\ref{Tab:alphac-1loop} we compare the various results obtained for
the critical coupling $\alpha_c$ in the 1-loop approximation to the vacuum
polarization in the Landau gauge. It is clear that the inclusion of the
$\F$-equation affects the value of the critical coupling. We note a
decrease of $\alpha_c$ by about 20\% for $N_f=1$ and by more than 30\% for
$N_f=2$.\vspace{-3mm}

\begin{table}[htbp]
\begin{center}
\begin{tabular}{|c|c|c|}
\hline
Approximation & $\alpha_c(N_f=1)$ & $\alpha_c(N_f=2)$ \\
\hline
LAK          & 1.99953 & 2.75233 \\
$\F\equiv 1$ & 2.08431 & 2.99142 \\
$(\S,\F)$    & 1.67280 & 2.02025 \\
\hline
\end{tabular}
\caption{Critical coupling $\alpha_c$ for $N_f=1$ and $N_f=2$ for various
approximations to the ($\S$,~$\F$)-system in the 1-loop approximation to
the vacuum polarization.}
\label{Tab:alphac-1loop}
\end{center}
\end{table}

\vspace{-5mm}
In Table~\ref{Tab:comp_1loop} we compare our results with those found in
the literature as discussed in Chapter~\ref{Sec:UnqQED} for the
LAK-approximation and in the $\F\equiv 1$ approximation.  From
Table~\ref{Tab:comp_1loop}.A we see that in the LAK-approximation all the
analytical and numerical calculations agree extremely well. The largest
deviation for $N_f=1$ is found in the analytical calculation
Ref.~\cite{Gusynin} and is only about 2\%.  For the $\F\equiv 1$
approximation only numerical work has been done as the angular integrals
cannot be computed analytically. The deviation between previously published
work and our calculation is at most 0.5\% for $N_f=1$ and almost 6\% for
$N_f=2$ as can be seen in Table~\ref{Tab:comp_1loop}.B. For the coupled
($\S$, $\F$)-system no evaluation of the critical coupling has been found
in the literature. There is only a qualitative assessment in
Ref.~\cite{Kondo92b} to verify that the approximation $\F\equiv 1$ is
justified. However, as we noted above, we found in our numerical
calculation that the critical coupling does change considerably by
including the corrections to $\F$ in the calculation.

\vspace{-2mm}\begin{table}[htbp]
\begin{center}
\parbox[t]{0.8cm}{~\\(A)}\begin{tabular}[t]{|c|l|l|}
\hline
Ref. & $\alpha_c(N_f=1)$ & $\alpha_c(N_f=2)$ \\
\hline
JCRB & 1.99953 & 2.75233 \\
\cite{Kondo91} & 1.99972 & 2.71482 \\
\cite{Gusynin} & 1.95 & \\
\cite{Kondo92b} & 1.9989 & 2.7517 \\
\cite{Oli90} & 1.999534163 &\\
\cite{Kondo94} & 1.9995 & \\
\hline
\end{tabular}\hspace{1cm}
\parbox[t]{0.8cm}{~\\(B)}\begin{tabular}[t]{|c|l|l|}
\hline
Ref. & $\alpha_c(N_f=1)$ & $\alpha_c(N_f=2)$ \\
\hline
JCRB &  2.08431 & 2.99142\\
\cite{Kondo92b} & 2.0728 & 2.8209 \\ 
\cite{Kondo92} & 2.084 & \\
\hline
\end{tabular}
\caption{Comparison of our numerical results~(JCRB) with those found in the
literature for the critical coupling in the 1-loop approximation to the
vacuum polarization, for $N_f=1,2$: (A) in the LAK-approximation, (B) in
the $\F\equiv 1$-approximation.}
\label{Tab:comp_1loop}
\end{center}
\end{table}

\clearpage

\section{Coupled ($\S$, $\G$)-system: revisited}
\label{Chap:BG}

We now take a new look at the solution of the coupled ($\S$, $\G$)-system
which was discussed previously in Chapter~\ref{Paper2}. There we solved
this system of equations using the collocation method and found that we
encountered difficulties cancelling the photon quadratic divergence
properly. It was then suggested that with some smooth approximations to the
functions $\S$ and $\G$ we could prevent these problems.  In this section
we are going to investigate how the approximation of $\S$ and $\G$ by
Chebyshev expansions affects the numerical cancellation of the quadratic
divergence.  

In contrast to Chapter~\ref{Paper2}, we now use the conventional operator
$\Proj_{\mu\nu} = g_{\mu\nu}- 4 {q_\mu q_\nu}/{q^2}$ to derive the photon
equation, as motivated in Section~\ref{Sec:n=4}. Although the vacuum
polarization is then theoretically free of quadratic divergences, this does
not ensure that it will be automatically so numerically. Setting $\F
\equiv 1$, the coupled system of integral equations, \mrefb{189}{104}, in
the Landau gauge and with zero bare mass, becomes:
\ba
\S(x) &=& \frac{3\alpha}{2\pi^2} \int dy \, 
\frac{y\S(y)}{y+\S^2(y)} \int d\theta \, \sin^2\theta
\, \frac{\G(z)}{z}  \mlab{874.00}\\
\frac{1}{\G(x)} &=& 1 + \frac{4 N_f \alpha}{3\pi^2 x} \int dy 
\frac{y}{y+\S^2(y)}\int d\theta \, \sin^2\theta \, 
\frac{ y(1-4\cos^2\theta) + 3\sqrt{xy}\cos\theta}{z+\S^2(z)}  \;.
\mlab{874.01} 
\ea

In analogy to \mrefb{702.1}{702.2}, after having changed variables with
$t=\logten y$, we replace the integral equations by a system of non-linear
equations by introducing quadrature rules to evaluate the integrals
numerically:
\ba
\S(x_i) &=& \frac{3\alpha\ln10}{2\pi^2} \sum_{j=1}^{(N_R)_i} w_{ij} \, 
\frac{y_{ij}^2\S(y_{ij})}{y_{ij}+\S^2(y_{ij})} \Theta_\S(x_i,y_{ij}) ,
 \qquad i=1,\ldots,\NS
\mlab{874} \\
\frac{1}{\G(x_i)} &=& 1 + \frac{4N_f\alpha\ln10}{3\pi^2 x_i}
\sum_{j=1}^{(N_R)_i} w_{ij} \, \frac{y_{ij}^2}{y_{ij}+\S^2(y_{ij})} 
\Theta_\G(x_i,y_{ij})  , \qquad i=1,\ldots,\NG
\mlab{874.1} 
\ea
where
\ba
\Theta_\S(x_i,y_{ij}) &=& \sum_{k=1}^{N_\theta} w'_k \, \sin^2\theta_k
\, \frac{\G(z_k)}{z_k} 
\mlab{875}\\
\Theta_\G(x_i,y_{ij}) &=& \sum_{k=1}^{N_\theta} w'_k \, \sin^2\theta_k
\frac{ y_{ij}(1-4\cos^2\theta_k) + 3\sqrt{x_i y_{ij}}\cos\theta_k}
{z_k+\S^2(z_k)}
\mlab{875.1}
\ea
and $z_k = x_i + y_{ij} - 2\sqrt{x_i y_{ij}}\cos\theta_k$.

The unknowns of the system of equations are the coefficients $a_j$ and
$c_j$ of the Chebyshev expansions for $\S(x)$ and $\G(x)$:
\ba
\S(x) \equiv \sum_{j=0}^{\NS-1}{'} a_j T_j(s(x)) \mlab{876} \\ 
\G(x) \equiv \sum_{j=0}^{\NG-1}{'} c_j T_j(s(x)) \mlab{877} ,
\ea
where $s(x)$ is defined by \mref{822}.

Instead of being equidistant as in Chapter~\ref{Paper2} the logarithms of
the external momenta in \mrefb{874}{874.1}, mapped to the interval
$[-1,1]$, now correspond to the roots of the Chebyshev polynomials
$T_{\NS}(x)$ and $T_{\NG}(x)$ as shown in \mrefb{829.1}{829.2}.

The evaluation of the kernels in the radial and angular integrals is
straightforward as the functions $\S(x)$ and $\G(x)$ can be computed at any
point in the interval $[\logten\kappa^2,\logten\Lambda^2]$. 

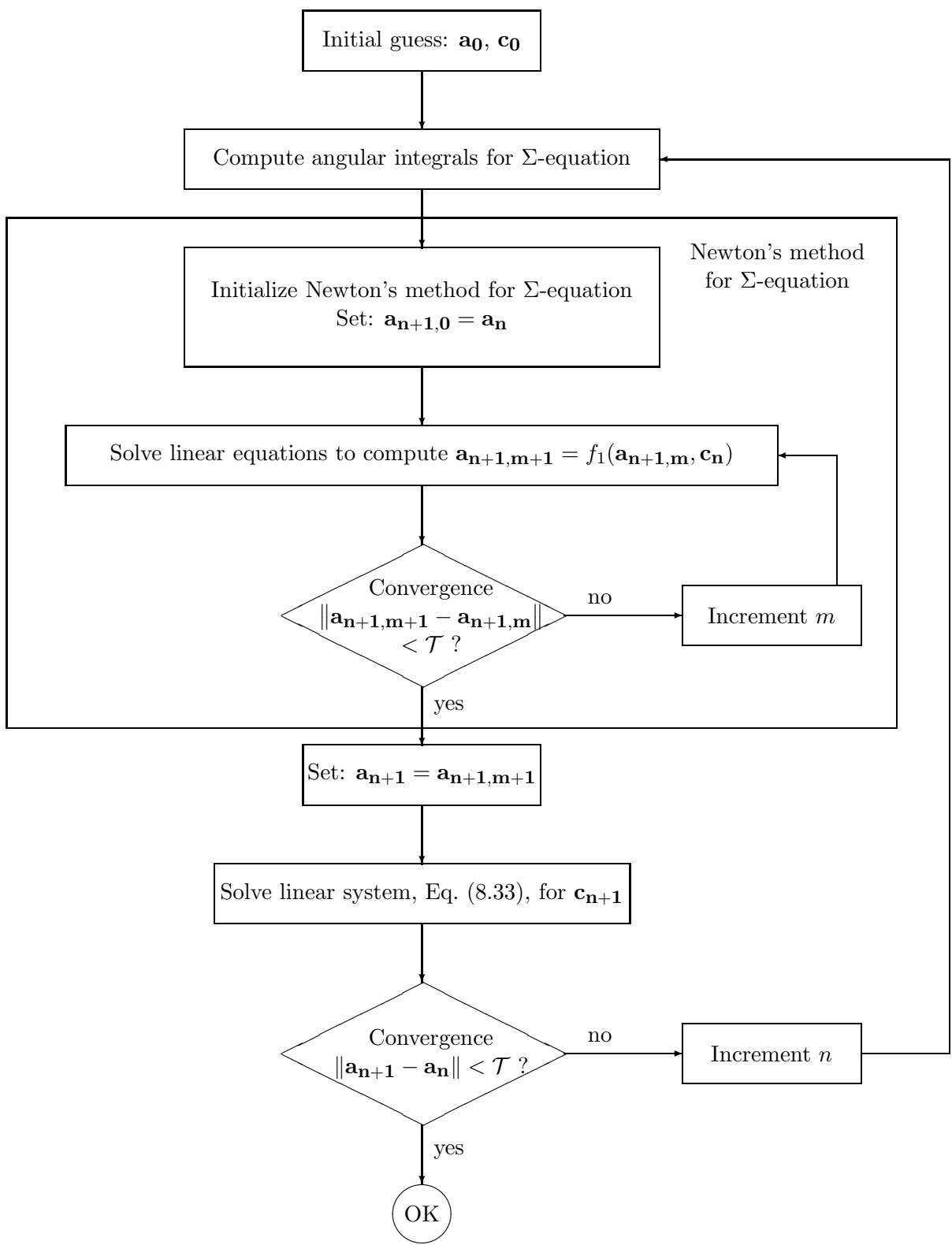
\begin{figure}[htbp]
\begin{center}
\vspace*{-2.5cm}\hspace*{-1cm}
\begin{picture}(160,200) 
\lijndikte
\put(60,170){\framebox(40,10){Initial guess: $\mvec{a_0}$, $\mvec{c_0}$}}
\put(80,170){\vector(0,-1){10}}
\put(40,150){\framebox(80,10){Compute angular integrals for $\S$-equation}}
\put(80,150){\vector(0,-1){10}}
\put(40,120){\framebox(80,20){\parbox{80mm}
{\begin{center}Initialize Newton's method for $\S$-equation\\
Set: $\mvec{a_{n+1,0}} = \mvec{a_n}$ \end{center}}}}
\put(80,120){\vector(0,-1){10}}
\put(20,100){\framebox(120,10){Solve linear equations to compute 
             $\mvec{a_{n+1,m+1}} = f_1(\mvec{a_{n+1,m}}, \mvec{c_n})$}}
\put(80,100){\vector(0,-1){10}}
\put(79,78){
\usebox{\ifbox}
\makebox(0,0){\parbox{60mm}
{\begin{center}Convergence\\$\norm{\mvec{a_{n+1,m+1}}-\mvec{a_{n+1,m}}}$ \\
$< \tol$ ?
\end{center}}}}
\put(104,78){\vector(1,0){20}}
\put(108,80){\makebox(0,0)[bl]{no}}
\put(124,73){\framebox(30,10){Increment $m$}}
\put(150,83){\line(0,1){22}}
\put(150,105){\vector(-1,0){10}}
\put(80,66){\vector(0,-1){10}}
\put(82,62){\makebox(0,0)[bl]{yes}}
\put(60,46){\framebox(40,10){Set: $\mvec{a_{n+1}} = \mvec{a_{n+1,m+1}}$}}
\put(80,46){\vector(0,-1){10}}
\put(45,26){\framebox(70,10){Solve linear system, \mref{878}, for
$\mvec{c_{n+1}}$}}
\put(80,26){\vector(0,-1){10}}
\put(79,4){
\usebox{\ifbox}
\makebox(0,0){\parbox{60mm}
{\begin{center}Convergence\\$\norm{\mvec{a_{n+1}}-\mvec{a_{n}}} < \tol$ ? 
\end{center}}}}
\put(104,4){\vector(1,0){20}}
\put(108,6){\makebox(0,0)[bl]{no}}
\put(124,-1){\framebox(30,10){Increment $n$}}
\put(154,4){\line(1,0){15}}
\put(169,4){\line(0,1){151}}
\put(169,155){\vector(-1,0){49}}
\put(80,-8){\vector(0,-1){10}}
\put(82,-13){\makebox(0,0)[bl]{yes}}
\put(80,-23){\circle{10}\makebox(0,0){OK}}
%
\put(10,59){\framebox(150,86)[tr]
{\parbox{4cm}{\begin{center}Newton's method\\for $\S$-equation\end{center}}}}
\end{picture}
\end{center}
\vspace{2.5cm}
\caption{Program flow to solve the coupled ($\S$, $\G$)-system using the
Chebyshev expansion method.}
\label{Fig:flow-SG-Cheby}
\end{figure}

In analogy with the discussion in Chapter~\ref{Paper2}, the coupled system of
non-linear algebraic equations will be solved by the method described by
the program flow, Fig.~\ref{Fig:flow-SG-Cheby} which is similar to
Fig.~\ref{Fig:flow-SG}.

We start from an initial guess $\mvec{a_0}$ and $\mvec{c_0}$ to the
Chebyshev coefficients. To derive the approximation $\mvec{a_{n+1}}$,
$\mvec{c_{n+1}}$ from $\mvec{a_{n}}$, $\mvec{c_{n}}$ we apply the following
procedure. Keeping the coefficients $\mvec{c_n}$ fixed, we compute the
angular integrals, \mref{875}, of the $\S$-equation using the Chebyshev
expansion, \mref{877}, for $\G$. After substituting the values of these
angular integrals in \mref{874}, this equation will represent a set of
non-linear algebraic equations for the unknown $\mvec{a_{n+1}}$, which is
analogous to \mref{830.2}. This set of equations can be solved by applying
Newton's iterative procedure to the Chebyshev expansion method as described
in Section~\ref{ChebNewton}. This will involve successive solutions of
linear systems of equations.

We then compute the angular integrals, \mref{875.1}, of the $\G$-equation
using the Chebyshev expansion for $\S(x)$ with coefficients
$\mvec{a_{n+1}}$. Then, these angular integrals are substituted into the
$\G$-equation, \mref{874.1}. Taking the reciprocal of this equation and
substituting the Chebyshev expansion, \mref{877}, for $\G(x)$ yields:
\be
\sum_{j=0}^{\NG-1}{'} c_j T_j(s(x_i)) = 
\l[ 1 + \frac{4N_f\alpha\ln10}{3\pi^2 x_i}
\sum_{j=1}^{(N_R)_i} w_{ij} \, \frac{y_{ij}^2}{y_{ij}+\S^2(y_{ij})} 
\Theta_\G(x_i,y_{ij}) \r]^{-1} , \qquad i=1,\ldots,\NG.
\mlab{878} 
\ee 

This is a system of $N_\G$ linear equations for the $N_\G$ Chebyshev
coefficients $c_j$ with known right hand sides which can easily be solved
numerically. In this way we have constructed a new set of Chebyshev
coefficients $\mvec{a_{n+1}}$, $\mvec{c_{n+1}}$. We repeat the whole
procedure till convergence has been reached.

We now show the main results obtained with this method. As before the
numbers of Chebyshev polynomials in the expansions are $N_\S=N_\G=50$,
while the number of radial integration nodes for the Gauss-Legendre rule
are taken to be $(N_R)_i=(N_1)_i+(N_2)_i$ with $(N_1)_i=(N_2)_i=120$ and
for the angular integrals $N_\theta=32$.

\begin{figure}[htbp]
\begin{center}
\mbox{\epsfig{file=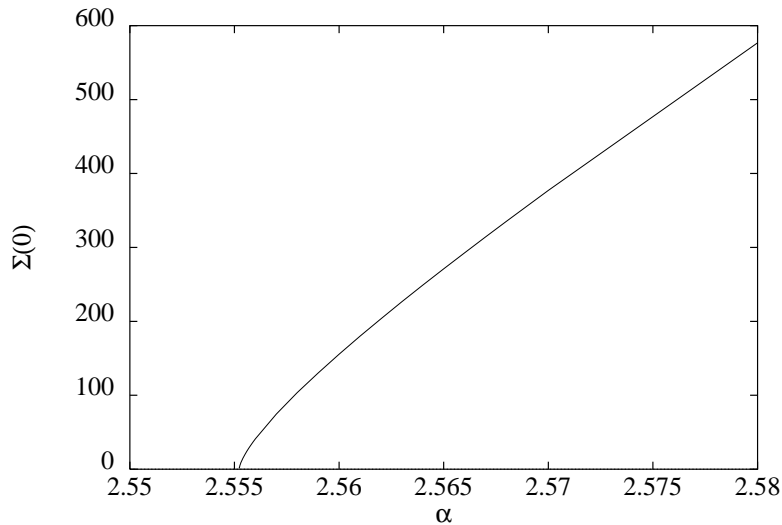,height=8cm,angle=-90}}
\end{center}
\vspace{-0.5cm}
\caption{Generated fermion mass $\S(0)$ versus coupling $\alpha$ 
for the coupled ($\S$, $\G$)-system, for $N_f=1$.}
\label{Fig:fmg-BG}
\end{figure}

\begin{figure}[htbp]
\begin{center}
\mbox{\epsfig{file=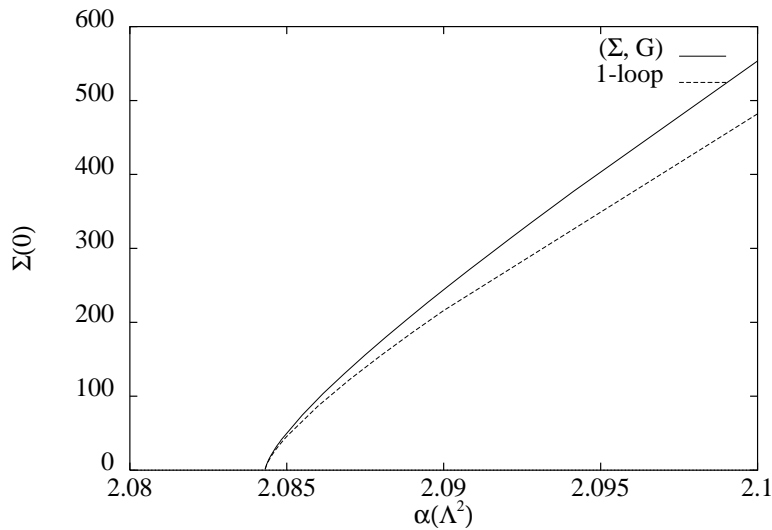,height=8cm,angle=-90}}
\end{center}
\vspace{-0.5cm}
\caption{Generated fermion mass $\S(0)$ versus running coupling 
$\alpha(\Lambda^2)$ for the coupled $(\S, \G)$-system and in the 1-loop
approximation to $\G$, for $N_f=1$.}
\label{Fig:fmg-BG-rc}
\end{figure}

The evolution of the generated fermion mass $\S(0)$ versus the coupling
$\alpha$ is shown in Fig.~\ref{Fig:fmg-BG}.
The critical coupling is found to be $\alpha_c(N_f=1)=2.55523$. Although
this could seem in contradiction with the results of Chapter~\ref{Paper2}
where $\alpha_c(N_f=1)\approx 2.084$, this is only apparently so. In
Chapter~\ref{Paper2} we followed the treatment of Kondo et
al.~\cite{Kondo92} and renormalized the coupling such that
$\alpha(\Lambda^2)=\alpha$. As the renormalization was not performed
consistently on all the quantities under consideration we will leave the
coupling unrenormalized in the current method. However, it is clear from
the study of dynamical fermion mass generation in quenched QED that
the scale of the generated mass depends on the strength of the coupling,
which is constant in that case. As we mentioned previously the running of
the coupling in unquenched QED is completely determined by the photon
renormalization function $\G$, and the running coupling can be written as
$\alpha(x) = \alpha \, \G(x)$. It is obvious from the $\S$-equation,
\mref{874.00}, that in unquenched QED the scale of the generated fermion 
mass will depend on the size and behaviour of the running coupling. In the
1-loop approximation to $\G$, Section~\ref{Sec:1loop}, the generated mass
scale is related to the value of $\alpha$, which is also equal to
$\alpha(\Lambda^2)$ as the 1-loop corrected $\G$ is there chosen to be
$\G(\Lambda^2)=1$. In order to compare the new calculations with those of
Chapter~\ref{Paper2} and Section~\ref{Cheb-S-eq} (where $\F\equiv 1$), we
plot $\S(0)$ versus $\alpha(\Lambda^2)=\alpha \,
\G(\Lambda^2)$ using the Chebyshev expansion method for the coupled ($\S$,
$\G$)-system and in the 1-loop approximation to $\G$ in
Fig.~\ref{Fig:fmg-BG-rc}.
Here we see that the critical coupling at the UV-cutoff for the ($\S$,
$\G$)-system is \fbox{$\alpha_c(\Lambda^2, N_f=1)=2.08431$}, which is
consistent with the calculation of Chapter~\ref{Paper2} and the results of
Kondo et al.~\cite{Kondo92} and moreover is identical to the value found
previously in the 1-loop approximation with $\F\equiv 1$ in
Section~\ref{Cheb-S-eq}. From Fig.~\ref{Fig:fmg-BG-rc} we see that the
generation of fermion mass starts at the same value
$\alpha(\Lambda^2)=2.08431$ for both approximations, but this mass evolves
differently for increasing coupling.  This is easy to explain as the
$\G$-equation,
\mref{874.01}, derived for $\F \equiv 1$, can be solved analytically 
at the critical point, where $\S(x)=0$. $\G(x)$ is then identical to its
1-loop approximation, hence the same value of the critical coupling. For
larger couplings the generated mass function will alter the behaviour of
$\G(x)$ and so the evolution of the generated mass scale will differ
between the ($\S$, $\G$)-system and the 1-loop approximation. In another
study of the coupled ($\S$, $\G$)-system with $\F\equiv 1$, Atkinson et
al.~\cite{Atk92} introduce the LAK-approximation on the vacuum polarization
and on the mass function to compute the angular integrals analytically,
and simplify some of the remaining integrals to derive a differential
equation which is then solved numerically. With these approximations they
find $\alpha_c=2.100286$, which only deviates about 0.8\% from our, more
accurate, solution.

\begin{figure}[htbp]
\begin{center}
\mbox{\epsfig{file=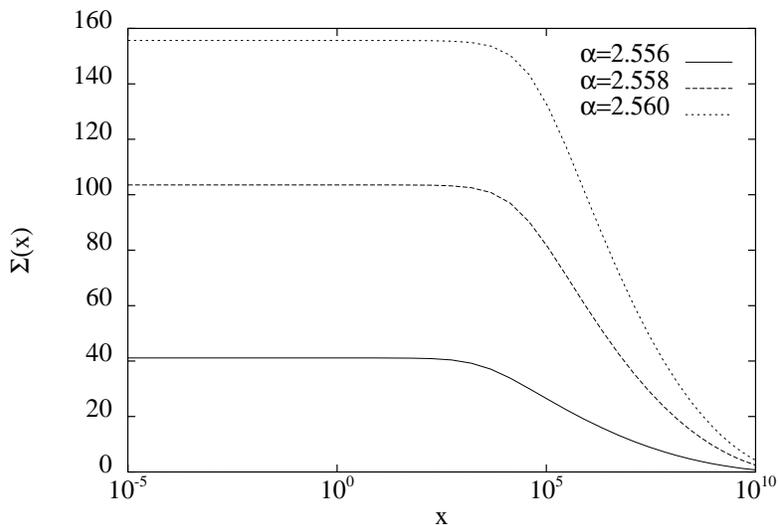,height=8cm,angle=-90}}
\end{center}
\vspace{-0.5cm}
\caption{Dynamical fermion mass $\S(x)$ versus momentum squared $x$ for
the coupled $(\S, \G)$-system, for $N_f=1$ and $\alpha=2.556, 2.558, 2.56$.}
\label{Fig:B-BG}
\end{figure}

\begin{figure}[htbp]
\begin{center}
\mbox{\epsfig{file=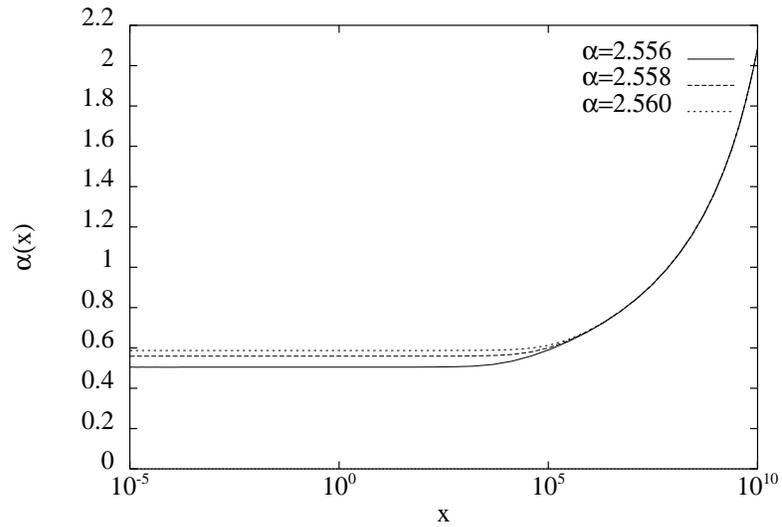,height=8cm,angle=-90}}
\end{center}
\vspace{-0.5cm}
\caption{Running coupling $\alpha(x)$ versus momentum squared $x$ for
the coupled $(\S, \G)$-system, for $N_f=1$ and $\alpha=2.556, 2.558, 2.56$.}
\label{Fig:rc-BG}
\end{figure}

Typical plots of the dynamical fermion mass $\S(x)$ for various values of
the bare coupling $\alpha$ are shown in Fig.~\ref{Fig:B-BG}.
The corresponding running couplings, $\alpha(x)=\alpha\G(x)$, are shown in
Fig.~\ref{Fig:rc-BG}. 
As expected from the discussion of Chapter~\ref{Paper2}, we indeed see that
any unphysical behaviour in the running coupling has now been removed: the
quadratic divergence has been cancelled properly.

\begin{figure}[htbp]
\begin{center}
\mbox{\epsfig{file=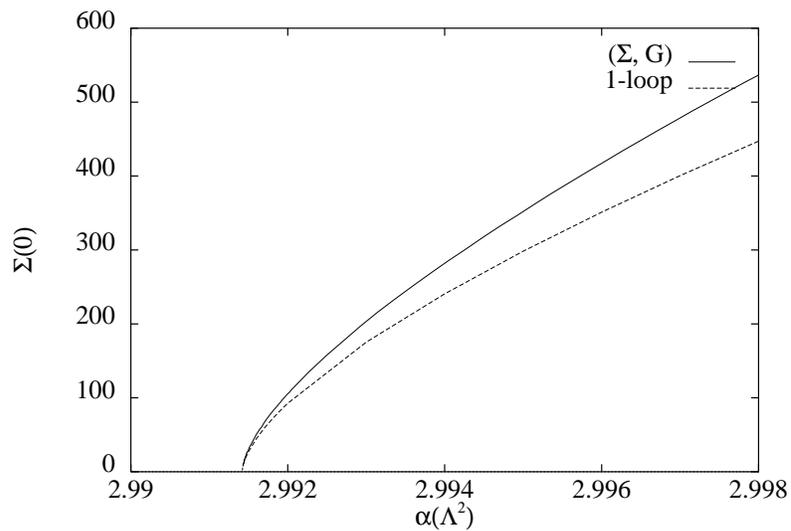,height=8cm,angle=-90}}
\end{center}
\vspace{-0.5cm}
\caption{Generated fermion mass $\S(0)$ versus running coupling 
$\alpha(\Lambda^2)$ for the coupled $(\S, \G)$-system and in the 1-loop
approximation to $\G$, for $N_f=2$.}
\label{Fig:fmg-BG-rc-N2}
\end{figure}

The results of fermion mass generation for two flavours ($N_f=2$) are shown
in Fig.~\ref{Fig:fmg-BG-rc-N2}. Here we show the evolution of $\S(0)$
versus the value of the running coupling at the UV-cutoff,
$\alpha(\Lambda^2)$ for the coupled $(\S, \G)$-system and in the 1-loop
approximation to $\G$. The critical coupling at the UV-cutoff is
\fbox{$\alpha_c(\Lambda^2, N_f=2)=2.99142$}.

\begin{figure}[htbp]
\begin{center}
\mbox{\epsfig{file=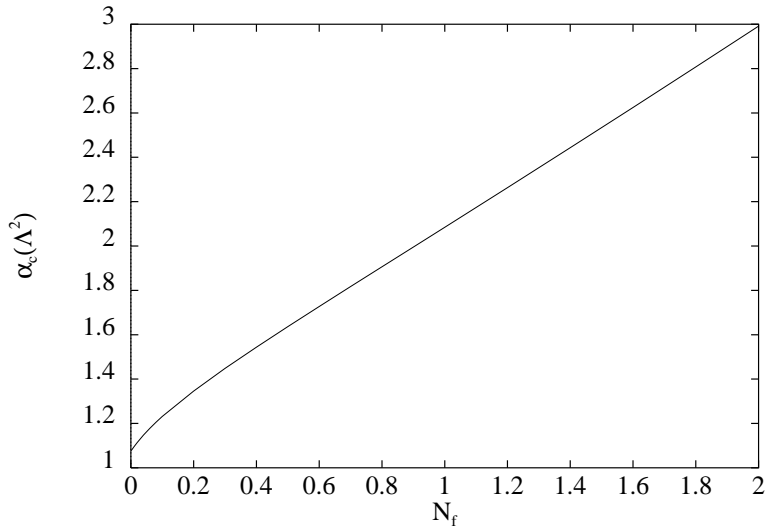,height=8cm,angle=-90}}
\end{center}
\vspace{-0.5cm}
\caption{Critical coupling $\alpha_c(\Lambda^2)$ versus number of flavours
$N_f$ for the coupled $(\S, \G)$-system.}
\label{Fig:critcoup-BG-rc}
\end{figure}

If we consider the number of flavours $N_f$ in the Schwinger-Dyson
equations as an arbitrary parameter, its values do not necessarily need to
be integer but can take non-integer values.  We compute the critical
coupling for various number of flavours, $0 < N_f\le 2$, and plot the
results in Fig.~\ref{Fig:critcoup-BG-rc}. It is reassuring to see the
smooth evolution of the critical coupling from $\alpha_c=\pi/3$ for
quenched QED ($N_f=0$) to the above mentioned values for unquenched QED
with $N_f=1,2$.

There is one more remark which has to be made. When we compare the current
method with that from Chapter~\ref{Paper2}, we note that the interpolation
problems have completely disappeared thanks to the use of the Chebyshev
expansions of $\S$ and $\G$. Nevertheless, one could argue that the need to
extrapolate remains. It is known that a polynomial expansion built on a
certain interval, here $[\kappa^2,\Lambda^2]$, can only be used reliably
for values of the argument in that interval. However we can check that for
values, which only lie slightly outside the interval, the function
approximations remain realistic. If not, one could always introduce some
continuous extrapolation as proposed in
\mref{extrapS2}. Furthermore, the use of the Gaussian quadrature formulae,
in contrast to that of Newton and Cotes, has eliminated any problem
produced by the extrapolation method. Because the Gaussian rules are open
rules, the endpoints of the integration interval are not integration nodes
and no extrapolation needs to be made for small values of the external
photon momentum, where the mismatch in the cancellation of the quadratic
divergence appeared in Chapter~\ref{Paper2}, and thus, this problem will
not occur with the current procedure.

In this section we have shown how the use of the Chebyshev expansion method
has enabled us to cancel the quadratic divergence of the vacuum
polarization integral properly. We have consistently solved the coupled
system of equations for the dynamical fermion mass $\S$ and the photon
renormalization function $\G$ and determined the critical coupling
$\alpha_c$ above which fermion mass is generated dynamically.

In the next section we will relax the condition on the fermion wavefunction
renormalization, $\F \equiv 1$, to improve the study further and will
consider the system of three coupled equations for $\S$, $\F$ and $\G$,
still in the bare vertex approximation.

\section{Coupled ($\S$, $\F$, $\G$)-system}
\label{BFG}

As a logical extension of the study presented in the previous section we
now consider the solution of the system of three coupled non-linear
integral equations for $\S$, $\F$ and $\G$.
To solve this problem we will use ideas developed in Sections~\ref{Chap:BG}
and \ref{Sec:BF}.

We recall the three integral equations describing $\S$, $\F$ and $\G$,
Eqs.~(\oref{189}, \oref{190}, \oref{104}), in the Landau gauge and with zero
bare mass:
\ba
\frac{\S(x)}{\F(x)} &=& \frac{3\alpha}{2\pi^2} \int dy \, 
\frac{y\F(y)\S(y)}{y+\S^2(y)} \int d\theta \, \sin^2\theta
\, \frac{\G(z)}{z} \mlab{900}\\
\frac{1}{\F(x)} &=& 1 + \frac{\alpha}{2\pi^2 x} \int dy \, 
\frac{y\F(y)}{y+\S^2(y)} 
\mlab{901}\\
&& \times \int d\theta \, \sin^2\theta \, \G(z) \,
\l(\frac{3 \sqrt{xy} \cos\theta}{z} 
- \frac{2 x y \sin^2\theta}{z^2} \r) \nn \\
\frac{1}{\G(x)} &=& 1 + \frac{4 N_f \alpha}{3\pi^2 x} \int dy 
\frac{y\F(y)}{y+\S^2(y)} 
\mlab{902}\\
&& \times \int d\theta \, \sin^2\theta \, \frac{\F(z)}{z+\S^2(z)}
\l[y(1-4\cos^2\theta) + 3\sqrt{xy}\cos\theta\r] \nn
\ea
where $z = x + y - 2\sqrt{xy}\cos\theta$.

As previously, we derive a system of non-linear algebraic equations in the
following way. Introduce an ultraviolet cutoff, $\Lambda^2$, and an
infrared cutoff, $\kappa^2$ and change variables from $y$ to $t \equiv
\logten y$. Then, consider the integral equations at respectively,
$N_\S$, $N_\F$ and $N_\G$ external momenta $x_i$, which are chosen to be
the roots of the Chebyshev polynomial of corresponding degree. Finally, 
replace the integrals by Gauss-Legendre quadrature rules. This yields:
\ba
\frac{\S(x_i)}{\F(x_i)} - \frac{3\alpha}{2\pi^2} 
\sum_{j=1}^{(N_R)_i} w_{ij} \, 
\frac{y_{ij}^2\F(y_{ij})\S(y_{ij})}{y_{ij}+\S^2(y_{ij})} \,
\Theta_{\S}(x_i,y_{ij}) &=& 0, \qquad i=1,\ldots,N_\S \mlab{903}\\
\frac{1}{\F(x_i)} - 1 - \frac{\alpha}{2\pi^2 x_i} 
\sum_{j=1}^{(N_R)_i} w_{ij} \, \frac{y_{ij}^2\F(y_{ij})}{y_{ij}+\S^2(y_{ij})} \,
\Theta_{\F}(x_i,y_{ij}) &=& 0 , \qquad i=1,\ldots,N_\F \mlab{904}\\
\frac{1}{\G(x_i)} - 1 - \frac{4 N_f \alpha}{3\pi^2 x_i}
\sum_{j=1}^{(N_R)_i} w_{ij} \, \frac{y_{ij}^2\F(y_{ij})}{y_{ij}+\S^2(y_{ij})} \,
\Theta_{\G}(x_i,y_{ij}) &=& 0, \qquad i=1,\ldots,N_\G \mlab{905}
\ea
with
\ba
\Theta_{\S}(x_i,y_{ij}) &=& \sum_{k=1}^{N_\theta} w'_k \, \sin^2\theta_k
\, \frac{\G(z_k)}{z_k} \mlab{906} \\
\Theta_{\F}(x_i,y_{ij}) &=& \sum_{k=1}^{N_\theta} w'_k \, \sin^2\theta_k \, 
\G(z_k) \, \l(\frac{3 \sqrt{x_i y_{ij}} \cos\theta_k}{z_k} 
- \frac{2 x_i y_{ij} \sin^2\theta_k}{z_k^2} \r) \mlab{907} \\
\Theta_{\G}(x_i,y_{ij}) &=& \sum_{k=1}^{N_\theta} w'_k \, \sin^2\theta_k \, 
\frac{\F(z_k)}{z_k+\S^2(z_k)}
\l[y_{ij}(1-4\cos^2\theta_k) + 3\sqrt{x_i y_{ij}}\cos\theta_k\r] \mlab{908}
\ea
where $z_k = x_i + y_{ij} - 2\sqrt{x_i y_{ij}}\cos\theta_k$. The
unknowns of the system of equations are the Chebyshev coefficients $a_j$,
$b_j$ and $c_j$ of the following expansions:
\ba
\S(x) \equiv \sum_{j=0}^{\NS-1}{'} a_j T_j(s(x)) \mlab{909} \\ 
\F(x) \equiv \sum_{j=0}^{\NF-1}{'} b_j T_j(s(x)) \mlab{910} \\
\G(x) \equiv \sum_{j=0}^{\NG-1}{'} c_j T_j(s(x)) \mlab{911}
\ea
where $s$ is defined by
\be
s(x) \equiv \frac{\logten(x/\Lambda\kappa)}{\logten(\Lambda/\kappa)} \,.
\mlab{912}
\ee

\begin{figure}[htbp]
\begin{center}
\vspace*{-2.5cm}\hspace*{-1cm}
\begin{picture}(160,200) 
\lijndikte
\put(55,170){\framebox(50,10){Initial guess: $\mvec{a_0}$, $\mvec{b_0}$, 
$\mvec{c_0}$}}
\put(80,170){\vector(0,-1){10}}
\put(40,150){\framebox(80,10){Compute angular integrals for 
($\S$, $\F$)-system}}
\put(80,150){\vector(0,-1){10}}
\put(40,120){\framebox(80,20){\parbox{80mm}
{\begin{center}Initialize Newton's method for ($\S$, $\F$)-system\\
Set: \quad $\mvec{a_{n+1,0}} = \mvec{a_n}$, 
\quad $\mvec{b_{n+1,0}} = \mvec{b_n}$
\end{center}}}}
\put(80,120){\vector(0,-1){10}}
\put(20,100){\framebox(120,10){Solve linear system to compute 
             $\mvec{a_{n+1,m+1}}$, $\mvec{b_{n+1,m+1}}$}}
\put(80,100){\vector(0,-1){10}}
\put(79,78){
\usebox{\ifbox}
\makebox(0,0){\parbox{60mm}
{\begin{center}
Has step $m+1$ \\ converged ? 
\end{center}}}}
\put(104,78){\vector(1,0){20}}
\put(108,80){\makebox(0,0)[bl]{no}}
\put(124,73){\framebox(30,10){Increment $m$}}
\put(150,83){\line(0,1){22}}
\put(150,105){\vector(-1,0){10}}
\put(80,66){\vector(0,-1){10}}
\put(82,62){\makebox(0,0)[bl]{yes}}
\put(40,46){\framebox(80,10){Set: $\mvec{a_{n+1}} = \mvec{a_{n+1,m+1}}$, 
$\mvec{b_{n+1}} = \mvec{b_{n+1,m+1}}$}}
\put(80,46){\vector(0,-1){10}}
\put(45,26){\framebox(70,10){Solve linear system, \mref{913}, for
$\mvec{c_{n+1}}$}}
\put(80,26){\vector(0,-1){10}}
\put(79,4){
\usebox{\ifbox}
\makebox(0,0){\parbox{60mm}
{\begin{center}
Has step $n+1$ \\ converged ? 
\end{center}}}}
%
\put(104,4){\vector(1,0){20}}
\put(108,6){\makebox(0,0)[bl]{no}}
\put(124,-1){\framebox(30,10){Increment $n$}}
\put(154,4){\line(1,0){15}}
\put(169,4){\line(0,1){151}}
\put(169,155){\vector(-1,0){49}}
\put(80,-8){\vector(0,-1){10}}
\put(82,-13){\makebox(0,0)[bl]{yes}}
\put(80,-23){\circle{10}\makebox(0,0){OK}}
%
\put(10,59){\framebox(150,86)[tr]
{\parbox{4cm}{\begin{center}Newton's method\\for 
($\S$, $\F$)-system\end{center}}}}
\end{picture}
\end{center}
\vspace{2.5cm}
\caption{Program flow to solve the coupled ($\S$, $\F$, $\G$)-system using the
Chebyshev expansion method.}
\label{Fig:flow-BFG}
\end{figure}
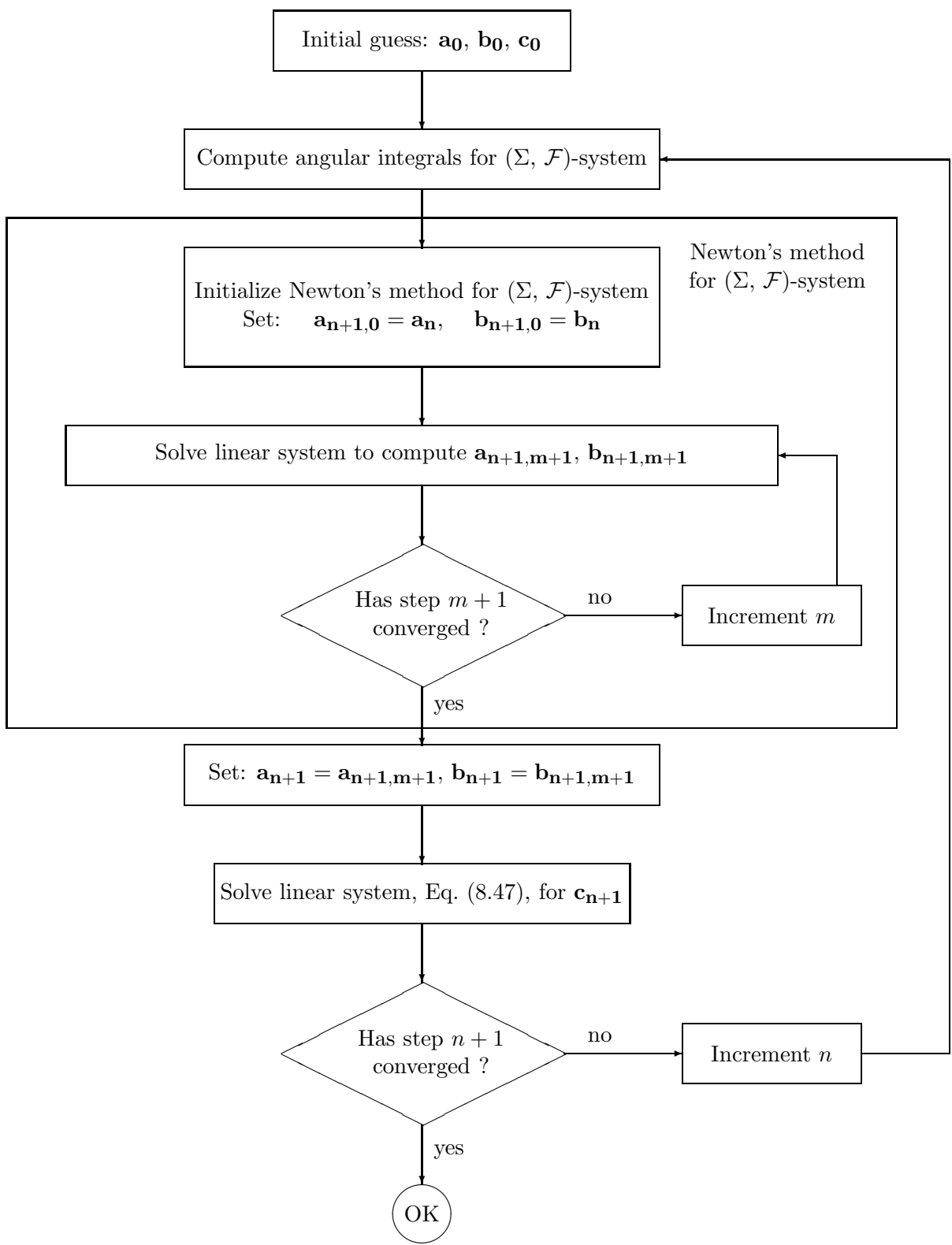

According to Section~\ref{Newton} the convergence rate of the numerical
method used to solve this system of equations will be quadratic if we use
Newton's method. However, as mentioned in Section~\ref{Sec:NumethSG}, it is
not convenient to implement Newton's method on the complete system of
equations. Newton's method is an iterative procedure where at each step a
system of linear equations has to be solved to derive new approximations
$(\mvec{a_{n+1}}, \mvec{b_{n+1}}, \mvec{c_{n+1}})$ from $(\mvec{a_{n}},
\mvec{b_{n}}, \mvec{c_{n}})$. This means that the angular integrals, 
Eqs.~(\oref{906}, \oref{907}, \oref{908}), have to be recalculated at each
iteration step. However, the main objection to this method is that Newton's
method requires the derivatives of the left hand sides of Eqs.~(\oref{903},
\oref{904}, \oref{905}) to be taken with respect of the Chebyshev
coefficients $a_j$, $b_j$ and $c_j$. The implementation of this method
would use a very large amount of memory space and running the program would
consume much computing time. Therefore, this procedure has not been
implemented here, although it remains an important objective for future
work, in order to enhance the consistency and accuracy of the method, if
some more powerful computer infrastructure is available.  In the meantime
we will use a hybrid method with the aim of retaining the advantages of the
quadratic convergence rate of Newton's method, while keeping the needs for
memory storage and computing time reasonably small. The program flow of
this method, which we explain in more detail below, is shown in
Fig.~\ref{Fig:flow-BFG}.

To start the program we introduce an initial guess for the Chebyshev
coefficients. These initial guesses can be the results of a previous
calculation or be derived from some reasonable choice for the unknown
functions (e.g. $\F(x)=1$, one-loop approximation for $\G(x)$). Starting
from these values we apply the following iterative procedure till
convergence is reached. We describe the $(n+1)^{th}$ iteration
step, supposing we know the $n^{th}$ approximations $\mvec{a_{n}}$,
$\mvec{b_{n}}$, $\mvec{c_{n}}$. From \mrefb{906}{907} we compute the
angular integrals for the $\S$- and $\F$-equations, using the Chebyshev
coefficients $\mvec{c_{n}}$ to compute the necessary $\G$-values with
\mref{911}. Now, \mrefb{903}{904} form an independent system of non-linear
equations determining the new approximations $\mvec{a_{n+1}}$,
$\mvec{b_{n+1}}$. This system of equations can be very efficiently solved
using Newton's iterative method, as shown in Section~\ref{Sec:BF}. Starting
from some initial values, which can be chosen to be $\mvec{a_{n}}$,
$\mvec{b_{n}}$, each iteration determines $\mvec{a_{n+1,m+1}}$,
$\mvec{b_{n+1,m+1}}$ from $\mvec{a_{n+1,m}}$, $\mvec{b_{n+1,m}}$ by solving
a system of linear equations till convergence is reached. The final iterate
gives the new approximations $\mvec{a_{n+1}}$, $\mvec{b_{n+1}}$. We now
substitute those Chebyshev coefficients in the expansions \mrefb{909}{910}
to compute the angular integrals \mref{908}. To determine the new values
$\mvec{c_{n+1}}$, we will take the reciprocal of \mref{905} and substitute
the expansion
\mref{911}, yielding:
\be
\sum_{j=0}^{\NG-1}{'} c_j T_j(s(x_i)) 
= \l[ 1 + \frac{4 N_f \alpha}{3\pi^2 x_i}
\sum_{j=1}^{(N_R)_i} w_{ij} \, \frac{y_{ij}^2\F(y_{ij})}{y_{ij}+\S^2(y_{ij})} \,
\Theta_{\G}(x_i,y_{ij}) \r]^{-1} , \qquad i=1,\ldots,N_\G, \mlab{913}
\ee
with $\S$ and $\F$ defined by the expansions \mrefb{909}{910} using the
coefficients $\mvec{a_{n+1}}$, $\mvec{b_{n+1}}$. \mref{913} represents a
linear system for the Chebyshev coefficients $\mvec{c_{n+1}}$, which can
easily be solved by standard numerical techniques. In this way we have
determined the new sets of Chebyshev coefficients $\mvec{a_{n+1}}$,
$\mvec{b_{n+1}}$ and $\mvec{c_{n+1}}$ from $\mvec{a_{n}}$, $\mvec{b_{n}}$
and $\mvec{c_{n}}$. The whole procedure is iterated till the convergence
criterion is satisfied. The final iterates are approximate solutions
$\mvec{a}$, $\mvec{b}$ and $\mvec{c}$ to the non-linear system,
Eqs.~(\oref{903}, \oref{904}, \oref{905}) within the required accuracy.

The results of the program are achieved by requesting a final relative
accuracy of 0.001. For the Chebyshev expansions, we take $\NS=\NF=\NG=50$.
The Gauss-Legendre integrations are performed with
$(N_R)_i=(N_1)_i+(N_2)_i$, where $(N_1)_i=(N_2)_i=120$, radial integration
points and $N_\theta=32$ angular integration points, which is sufficient to
obtain the above mentioned accuracy. The program needs between 3 and 8
global {\Large[}($\S$, $\F$), $\G${\Large]}-iterations to converge, while
each individual Newton's method to solve a ($\S$, $\F$)-system requires
between 2 and 8 iterations. The major part of the computing time involves
the recalculation of the angular integrals at each global iteration.

\begin{figure}[htbp]
\begin{center}
\mbox{\epsfig{file=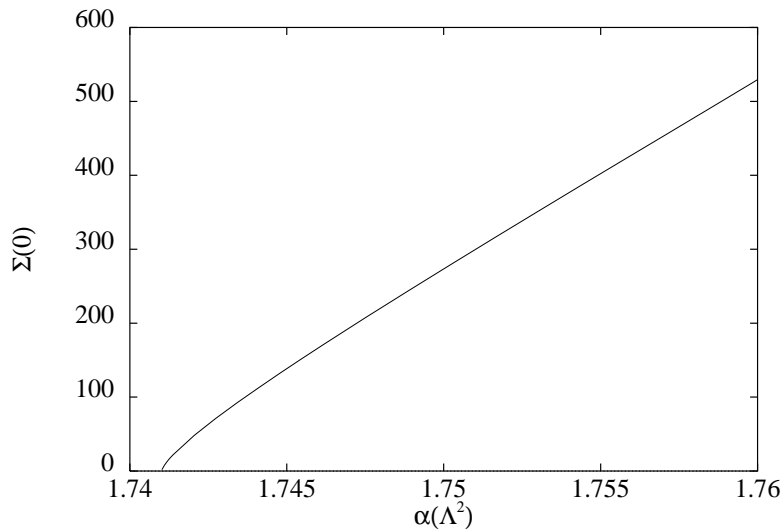,height=8cm,angle=-90}}
\end{center}
\vspace{-0.5cm}
\caption{Generated fermion mass $\S(0)$ versus running coupling 
$\alpha(\Lambda^2)$ for the coupled $(\S, \F, \G)$-system, for $N_f=1$.}
\label{Fig:fmg-BFG-rc}
\end{figure}

The evolution of the generated fermion mass $\S(0)$, is plotted versus
$\alpha(\Lambda^2)$, the value of the running coupling at the ultraviolet
cutoff, in Fig.~\ref{Fig:fmg-BFG-rc}. The critical coupling for the
($\S$,~$\F$,~$\G$)-system is \fbox{$\alpha_c(\Lambda^2, N_f=1) = 1.74102$}.

\begin{figure}[htbp]
\begin{center}
\mbox{\epsfig{file=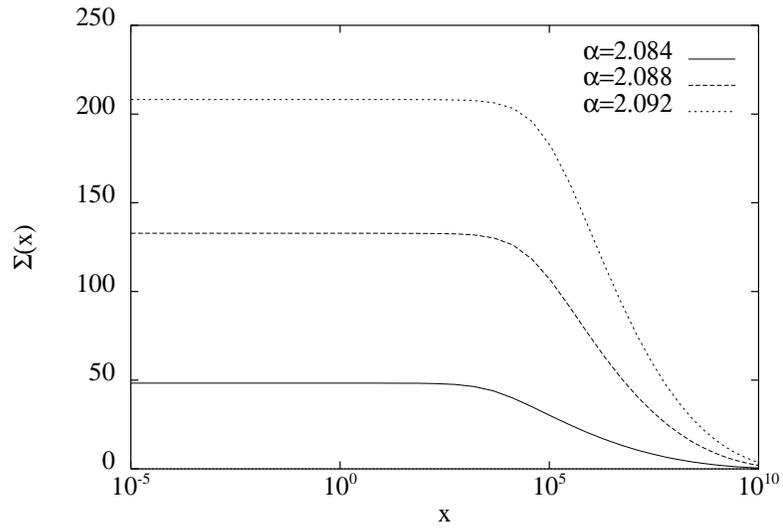,height=8cm,angle=-90}}
\end{center}
\vspace{-0.5cm}
\caption{Dynamical fermion mass $\S(x)$ versus momentum squared $x$ for
the coupled $(\S, \F, \G)$-system, for $N_f=1$ and $\alpha=2.084, 2.088,
2.092$.}
\label{Fig:B-BFG}
\end{figure}

\begin{figure}[htbp]
\begin{center}
\mbox{\epsfig{file=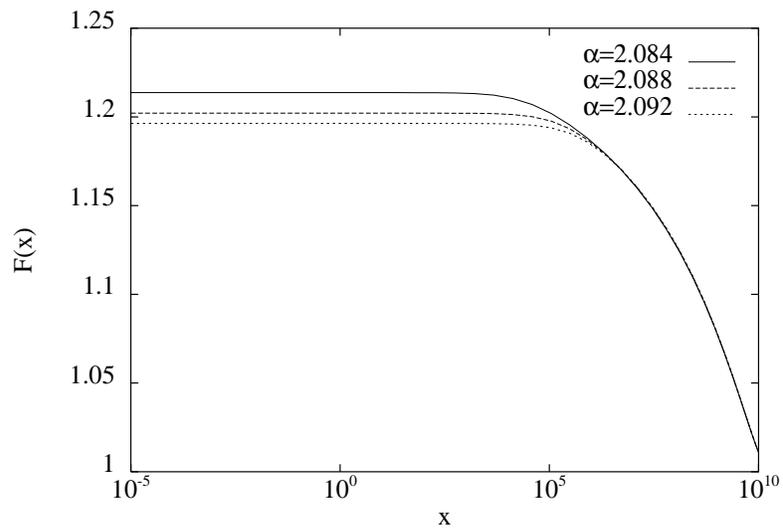,height=8cm,angle=-90}}
\end{center}
\vspace{-0.5cm}
\caption{Fermion wavefunction renormalization $\F(x)$ versus momentum 
squared $x$ for the coupled $(\S, \F, \G)$-system, for $N_f=1$ and
$\alpha=2.084, 2.088, 2.092$.}
\label{Fig:F-BFG}
\end{figure}

\begin{figure}[htbp]
\begin{center}
\mbox{\epsfig{file=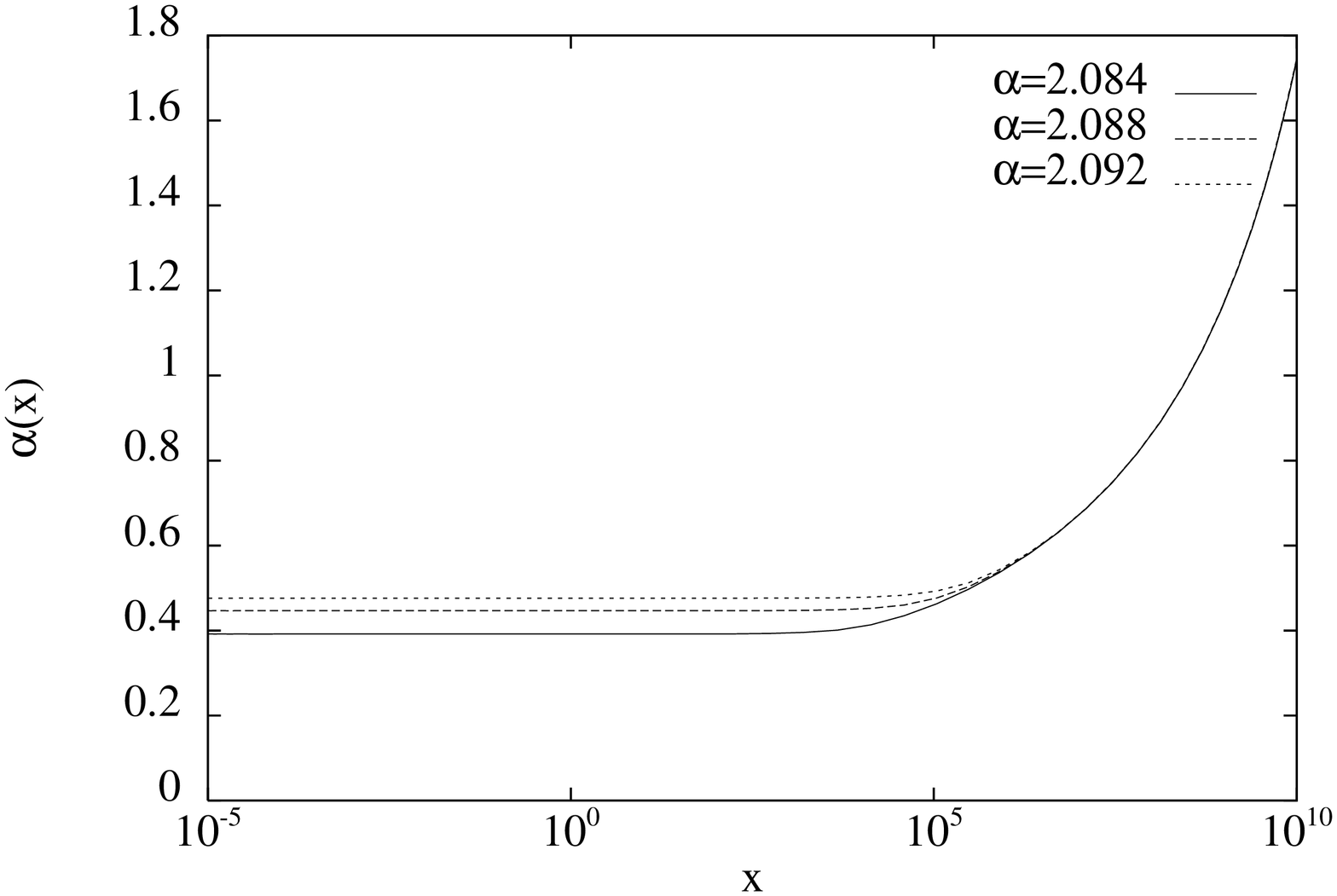,height=8cm,angle=-90}}
\end{center}
\vspace{-0.5cm}
\caption{Running coupling $\alpha(x)$ versus momentum squared $x$ for
the coupled $(\S, \F, \G)$-system, for $N_f=1$ and $\alpha=2.084, 2.088,
2.092$.}
\label{Fig:rc-BFG}
\end{figure}

Typical plots of $\S(x)$, $\F(x)$ and $\alpha(x)=\alpha \, \G(x)$ are shown
in Fig.~\ref{Fig:B-BFG}, Fig.~\ref{Fig:F-BFG} and  Fig.~\ref{Fig:rc-BFG}.

\begin{figure}[htbp]
\begin{center}
\mbox{\epsfig{file=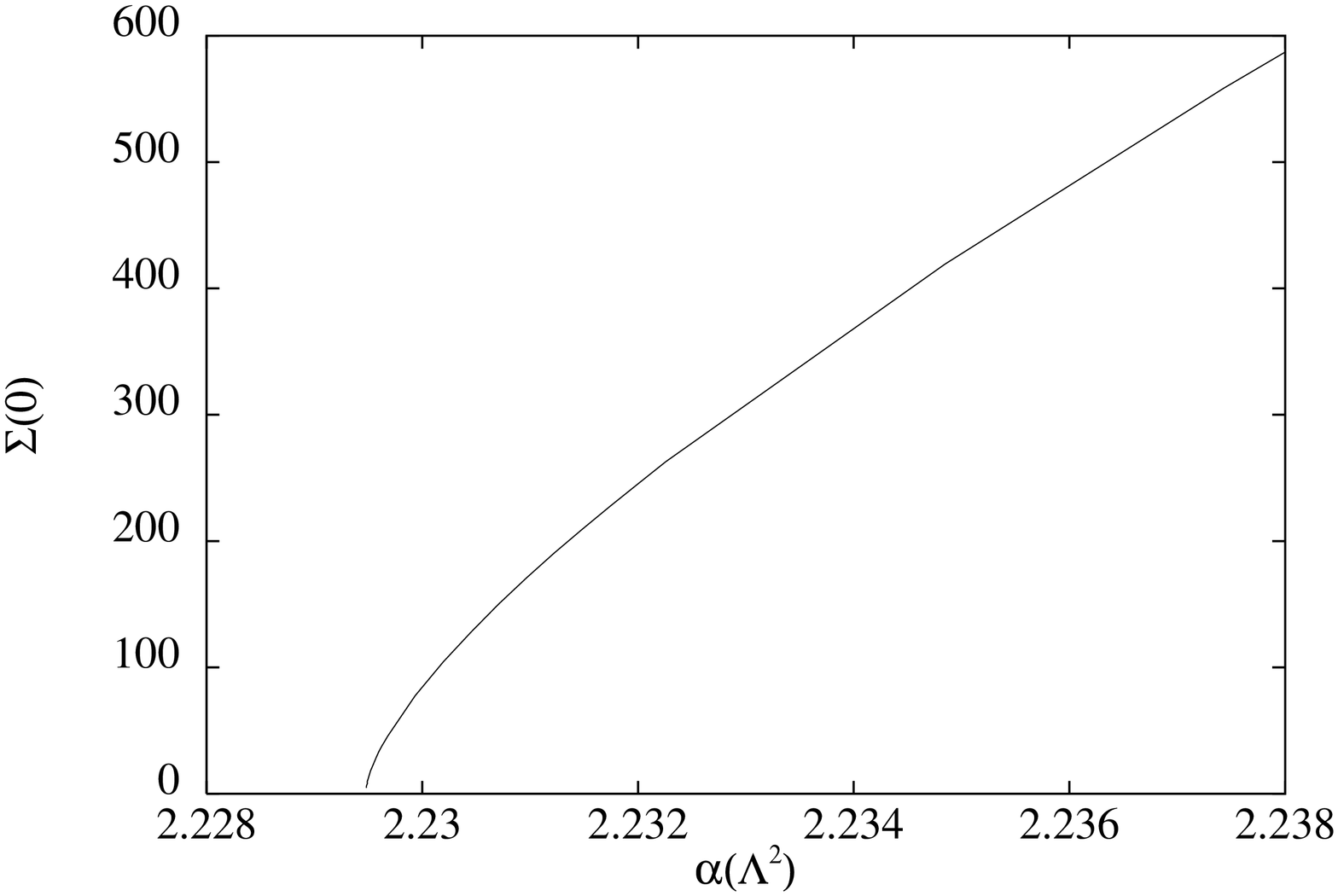,height=8cm,angle=-90}}
\end{center}
\vspace{-0.5cm}
\caption{Generated fermion mass $\S(0)$ versus running coupling 
$\alpha(\Lambda^2)$ for the coupled $(\S, \F, \G)$-system, for $N_f=2$.}
\label{Fig:fmg-BFG-rc-N2}
\end{figure}

For two flavours, $N_f=2$, the generated fermion mass $\S(0)$ versus the
running coupling $\alpha(\Lambda^2)$ is plotted in
Fig.~\ref{Fig:fmg-BFG-rc-N2}. The critical coupling is
\fbox{$\alpha_c(\Lambda^2,N_f=2) = 2.22948$}.

If we compare the results of the coupled $(\S, \F, \G)$-system with those
of the coupled $(\S, \F)$-system in the 1-loop approximation to $\G$ of
Section~\ref{Sec:BF}, we note that the critical coupling increases by
approximately 4\% for $N_f=1$ and 10\% for $N_f=2$ when we include the
$\G$-equation, \mref{902}, in the treatment instead of its 1-loop
approximation, \mref{852}.
On the other hand, if we compare these results with those computed for the
($\S$, $\G$)-system with $\F\equiv 1$ in Section~\ref{Chap:BG}, we note that
the consistent treatment of the $\F$-equation decreases $\alpha_c$ by 16\%
for $N_f=1$ and 25\% for $N_f=2$.

The only other calculation found in the literature treating the three
equations simultaneously is that of Rakow~\cite{Rak91}.  He finds a
critical coupling $\alpha_c=2.25$ for $N_f=1$, which is quite different
from our result with a deviation of almost 30\%. Unfortunately the lack of
details in that paper does not allow us to deduce where the difference
comes from. As explained previously we take the relevant critical coupling
to be $\alpha_c(\Lambda^2)=\alpha_c\G(\Lambda^2)$. We believe that only
then, calculations with different regularization procedures can be
compared. We are very confident about the accuracy of our calculation and
we can only guess that Rakow's result should be interpreted in some
different way to find agreement. Furthermore, his main result is that the
renormalized coupling $\alpha_r\to 0$ in the critical point which, as he
claims, would prove that the renormalized theory is trivial. However, the
renormalized coupling is strangely defined in the zero momentum point as
$\alpha_r = \alpha\G(0)\F^2(0)$. From plots analogous to
Fig.~\ref{Fig:rc-BFG} we see that for the critical point we have
$\alpha_c(0)=\alpha_c\G(0) \to 0$, as is the case in the massless 1-loop
approximation to $\G$, so that obviously $\alpha_r\to 0$. However, we do
not believe that the infrared behaviour of QED can explain its triviality,
but rather it is its ultraviolet behaviour which could, because of the
Landau pole. We do not think that Rakow's observation about his
renormalized coupling proves the triviality of QED. Intuitively we could
say that renormalizing the theory relates the overall evolution of the
coupling to the scales of the theory (bare mass and renormalization scale);
choosing to renormalize at zero momentum ($\mu^2=0$) in the critical point
(where $\S(x)=0$) and taking the continuum limit ($\Lambda\to\infty$) is in
contradiction with this as there are no finite scales available. This will
be different if we renormalize at some finite scale and only then will we
be able to discuss the triviality of the theory.

\section{Summary}

In this chapter we have seen that dynamical fermion mass generation does
occur in unquenched QED with bare vertex approximation for $N_f=1$ and
$N_f=2$.  In Table~\ref{Tab:alphac-bare-comp} we summarize the various
results obtained for the critical coupling for $N_f=1$ and $N_f=2$ from the
previous sections.

\def\cl#1{#1}
\begin{table}[htbp]
\begin{center}
\begin{tabular}{|c||c|c||c|c|}
\hline
System & $\G$ & $\F$ & $\alpha_c(N_f=1)$ & $\alpha_c(N_f=2)$ \\
\hline
$\S$ & 1-loop, LAK & $\F=1$ & 1.99953 & 2.75233 \\
$\S$ & 1-loop      & $\F\equiv 1$ & 2.08431 & 2.99142 \\
$(\S,\F)$ & 1-loop      & SD           & 1.67280 & 2.02025 \\
\hline
$(\S,\G)$ & SD          & $\F\equiv 1$ & 2.08431 & 2.99142 \\
\hline
$(\S,\F,\G)$ & SD          & SD           & 1.74102 & 2.22948 \\
\hline
\end{tabular}
\caption[Critical coupling $\alpha_c$ for $N_f=1$ and $N_f=2$, for various
approximations to the ($\S,\F,\G$)-system in bare vertex approximation.]
{Critical coupling $\alpha_c$ for $N_f=1$ and $N_f=2$ for various
approximations to the ($\S,\F,\G$)-system in the bare vertex
approximation. ({\small Column ``System'' states which coupled system of
equations was effectively solved. Columns ``$\G$'' and ``$\F$'' tell which
approximations were used for these functions, ``SD'' means that the
function is determined self-consistently by the coupled SD-equations in
``System'', ``LAK'' is the Landau-Abrikosov-Khalatnikov approximation of
Section~\ref{Sec:LAK}.})}
\label{Tab:alphac-bare-comp}
\end{center}
\end{table}

For the 1-loop approximation we compared our results with those found in
the literature in Section~\ref{Sec:sum-1loop}. We concluded that the
various results in the LAK-approximation and the $\F=1$ approximation agree
with each other within good accuracy and that our results are totally in
line with the most accurate ones. Furthermore we produced the first
critical coupling results for the coupled ($\S$, $\F$)-system. The
inclusion of self-energy corrections in $\F$ causes a decrease of
$\alpha_c$ by 20\% for $N_f=1$ and 30\% for $N_f=2$.

For the ($\S$, $\G$)-system with $\F\equiv 1$ we found the same 
critical coupling as Kondo et al.~\cite{Kondo92}. However, their
erroneous behaviour of $\G(x)$ at intermediate low energy has been
corrected thanks to the use of Chebyshev expansions for $\S$ and $\G$. The
results of Atkinson et al.~\cite{Atk92}, although less accurate because of
additional approximations, still agree very well with our calculation. We
also give the first results for $N_f=2$.

Finally, we have given a detailed description of the consistent and
accurate treatment of the complete ($\S$, $\F$, $\G$)-system in the bare
vertex approximation. From Table~\ref{Tab:alphac-bare-comp} we see that
replacing the 1-loop approximation to $\G$ by the consistent SD-treatment
of the $\G$-equation introduces an increase of $\alpha_c$ by 4\% for
$N_f=1$ and 10\% for $N_f=2$. On the other hand, if we consistently add the
$\F$-equation in the ($\S$, $\G$)-system we note a decrease of $\alpha_c$
by 16\% for $N_f=1$ and 25\% for $N_f=2$.  We have been unable to make a
useful quantitative comparison with Rakow's work~\cite{Rak91} as explained
in Section~\ref{BFG}.

Until now all the calculations have been made in the bare vertex
approximation. However, we know that the bare vertex violates the
Ward-Takahashi identity relating the QED vertex to the fermion propagator,
which is a direct consequence of the gauge invariance of the theory.  In
the next chapter we will examine the possibility of improving the vertex
Ansatz. This is the first time that the study of fermion mass generation in
unquenched QED will be taken beyond the bare vertex approximation.

\chapter{Improving the vertex Ansatz}
\label{Sec:Impvx}

In this chapter we are going to investigate the influence of the vertex
Ansatz on the dynamical generation of fermion mass. In the previous
chapters we have used the bare vertex approximation, $\Gamma^\mu(k,p) =
\gamma^\mu$. This vertex Ansatz has the advantage of being very simple and
therefore it makes the manipulation of the Schwinger-Dyson equations
easier.  However, this approximation does not satisfy the Ward-Takahashi
identity relating the QED vertex with the fermion propagator, which is a
consequence of the gauge invariance of the theory. Therefore, the bare
vertex approximation does not ensure that the physical quantities computed
with it are gauge invariant as they should be. The Ward-Takahashi identity
relating the QED vertex with the fermion propagator determines uniquely the
longitudinal part of the vertex~\cite{BC}, called Ball-Chiu
vertex. However, the transverse part of the vertex is still
arbitrary. Constraints on that part of the vertex can be imposed by
requiring the multiplicative renormalizability of the fermion and photon
propagator, the absence of kinematical singularities, the reproduction of
the perturbative results in the weak coupling limit, gauge invariance of
critical coupling,~...~\cite{CP90,Bashir94,Ayse}. In the next sections we
will investigate the generation of fermion mass using improved vertices and
highlight the problems which occur when doing so.

\clearpage
\section{$1/\F$-corrected vertex}

To investigate the influence of the vertex improvement on the dynamical
generation of fermion mass we will first introduce a {\it $1/\F$-corrected
vertex} defined as:
\be
\Gamma^\mu(k,p) = 
\frac{1}{2}\l[\frac{1}{\F(k^2)}+\frac{1}{\F(p^2)}\r]\gamma^\mu ,
\mlab{1009}
\ee 
which is just the first term of the Ball-Chiu vertex, \mref{1.1002}. The
motivation for this vertex Ansatz is that it introduces a wavefunction
renormalization dependence in the vertex. However, it avoids the numerical
difficulties which can occur with the complete Ball-Chiu vertex because of
the difference terms
\[
\frac{1}{k^2-p^2}\l[\frac{1}{\F(k^2)}-\frac{1}{\F(p^2)}\r]
\qquad , \qquad
\frac{1}{k^2-p^2}\l[\frac{\S(k^2)}{\F(k^2)}-\frac{\S(k^2)}{\F(p^2)}\r] \;.
\]

The coupled integral equations with this vertex Ansatz are easily derived
from Eqs.~(\oref{189}, \oref{190}, \oref{104}) for the bare vertex, as the
$1/\F$-vertex,
\mref{1009}, has the same Dirac structure as the bare vertex and merely
introduces a multiplicative factor in each integral where the full vertex
is replaced by the vertex Ansatz. The equations, with $m_0=0$ and in the
Landau gauge, are:
\ba
\frac{\S(x)}{\F(x)} &=& \frac{3\alpha}{2\pi^2} \int dy \; 
\frac{yA(y,x)\F(y)\S(y)}{\Ds{y}} \int d\theta \sin^2\theta \, 
\frac{\G(z)}{z}  \mlab{1010}\\[2mm]
\frac{1}{\F(x)} &=& 1 - \frac{\alpha}{2\pi^2 x} \int dy \;
\frac{yA(y,x)\F(y)}{y+\S^2(y)} 
\int d\theta \, \sin^2 \theta \, \G(z) 
\l[\frac{2yx\sin^2\theta}{z^2} - \frac{3\sqrt{yx}\cos\theta}{z}\r] 
\mlab{1011} \\[2mm]
\frac{1}{\G(x)} &=& 1 + \frac{4N_f\alpha}{3\pi^2 x} \int dy \; 
\frac{y\F(y)}{\Ds{y}} \int d\theta \, \sin^2\theta \,
\frac{A(y,z)\F(z)}{\Ds{z}} \bigg[y(1-4\cos^2\theta) 
+ 3\sqrt{yx}\cos\theta\bigg] \nn\\ \mlab{1012}\\[-13mm]\nn
\ea
where $\D A(y,x) = \frac{1}{2}\l[\frac{1}{\F(y)}+\frac{1}{\F(x)}\r]$ .

\begin{figure}[htbp]
\begin{center}
\mbox{\epsfig{file=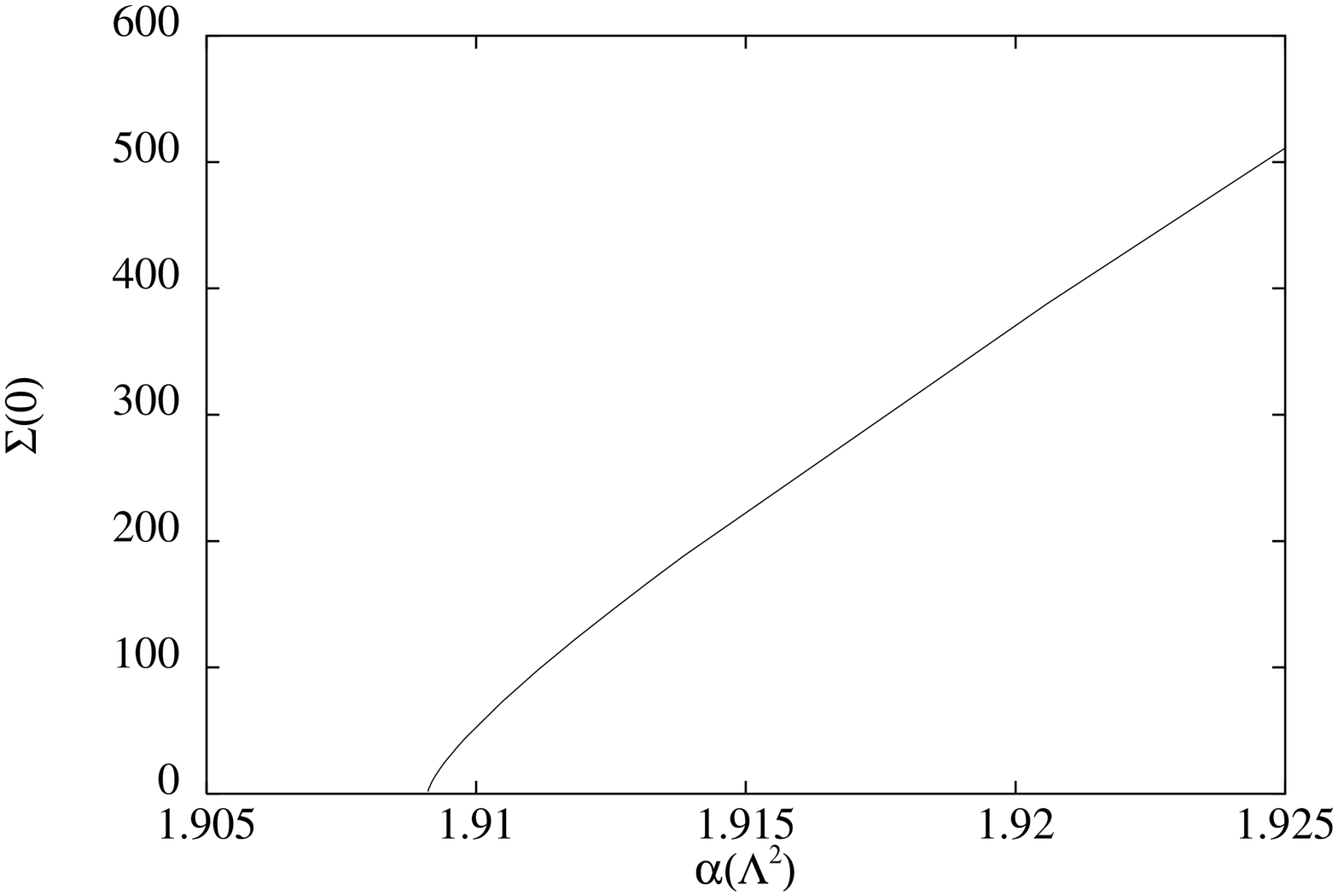,height=8cm,angle=-90}}
\end{center}
\vspace{-0.5cm}
\caption{Generated fermion mass $\S(0)$ versus running coupling 
$\alpha(\Lambda^2)$ for the coupled $(\S, \F, \G)$-system with the
$1/\F$-vertex Ansatz, for $N_f=1$.}
\label{Fig:fmg-BFG-vertex1-rc}
\end{figure}

\vspace{2mm}
We now solve this system of coupled integral equations using the same
method as in Section~\ref{BFG}. In Fig.~\ref{Fig:fmg-BFG-vertex1-rc} we
show the evolution of the generated fermion mass versus the running
coupling at the UV-cutoff, $\alpha(\Lambda^2)$, for $N_f=1$. The critical
coupling is \fbox{$\alpha_c(\Lambda^2, N_f=1) = 1.90911$}.

\begin{figure}[htbp]
\begin{center}
\mbox{\epsfig{file=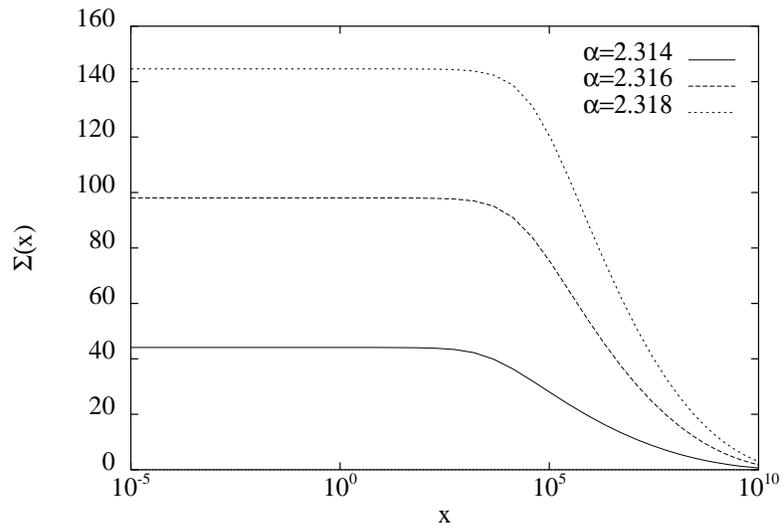,height=8cm,angle=-90}}
\end{center}
\vspace{-0.5cm}
\caption{Dynamical fermion mass $\S(x)$ versus momentum squared $x$ for
the coupled $(\S, \F, \G)$-system with the
$1/\F$-vertex Ansatz, for $N_f=1$ and $\alpha=2.314, 2.316, 2.318$.}
\label{Fig:B-BFG-vertex1}
\end{figure}

\begin{figure}[htbp]
\begin{center}
\mbox{\epsfig{file=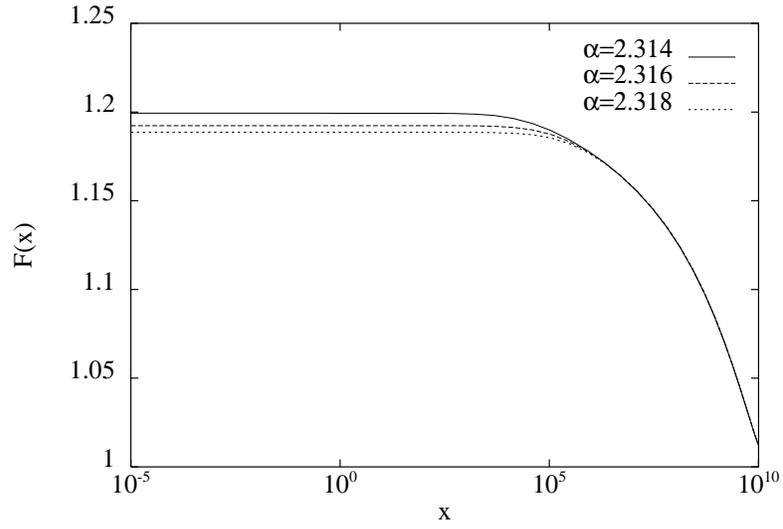,height=8cm,angle=-90}}
\end{center}
\vspace{-0.5cm}
\caption{Fermion wavefunction renormalization $\F(x)$ versus momentum 
squared $x$ for the coupled $(\S, \F, \G)$-system with the
$1/\F$-vertex Ansatz, for $N_f=1$ and
$\alpha=2.314, 2.316, 2.318$.}
\label{Fig:F-BFG-vertex1}
\end{figure}

\begin{figure}[htbp]
\begin{center}
\mbox{\epsfig{file=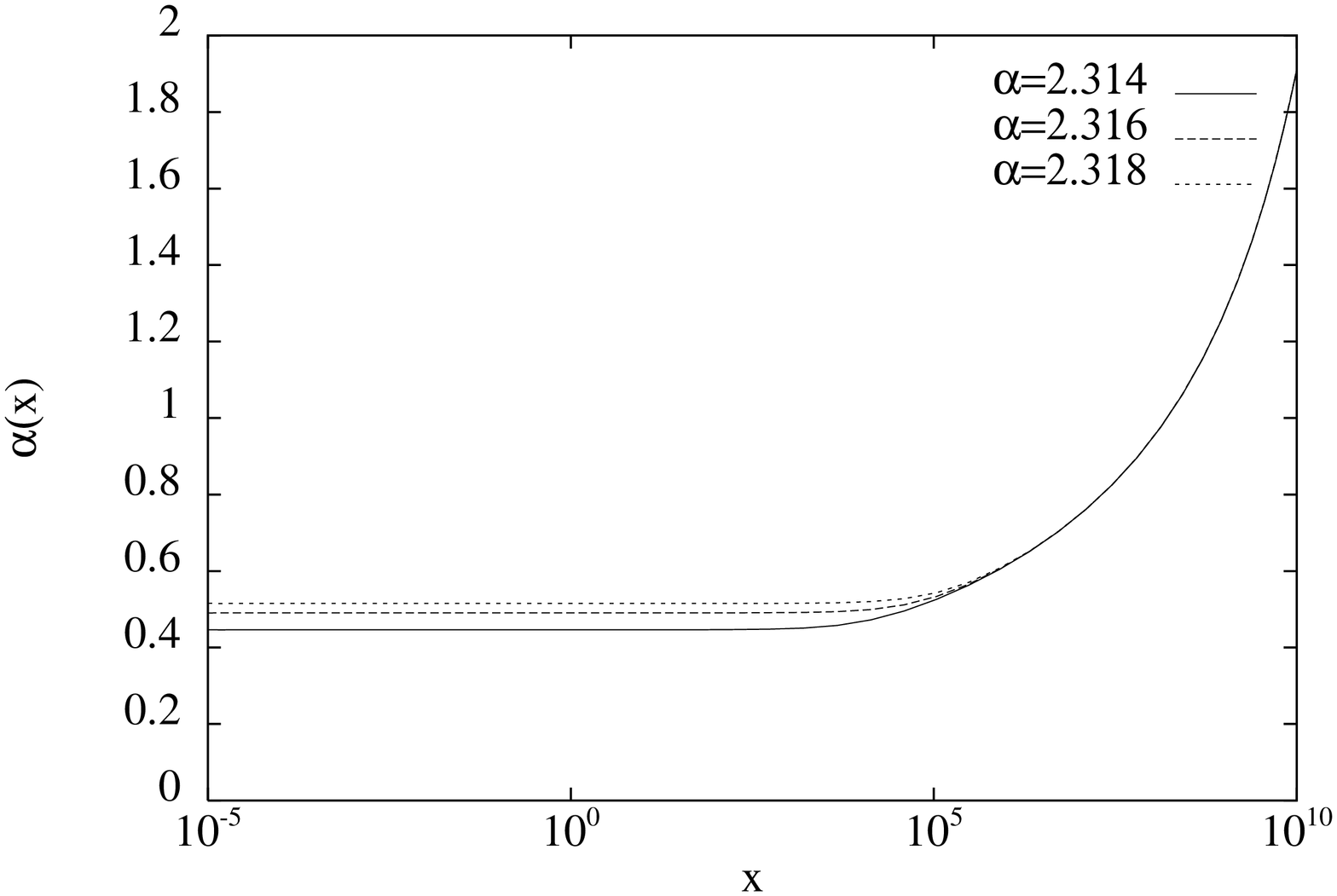,height=8cm,angle=-90}}
\end{center}
\vspace{-0.5cm}
\caption{Running coupling $\alpha(x)$ versus momentum squared $x$ for
the coupled $(\S, \F, \G)$-system with the
$1/\F$-vertex Ansatz, for $N_f=1$ and $\alpha=2.314, 2.316, 2.318$.}
\label{Fig:rc-BFG-vertex1}
\end{figure}

Typical plots of $\S(x)$, $\F(x)$ and $\alpha(x)=\alpha \, \G(x)$ are shown
in Fig.~\ref{Fig:B-BFG-vertex1}, Fig.~\ref{Fig:F-BFG-vertex1} and
Fig.~\ref{Fig:rc-BFG-vertex1}.

\begin{figure}[htbp]
\begin{center}
\mbox{\epsfig{file=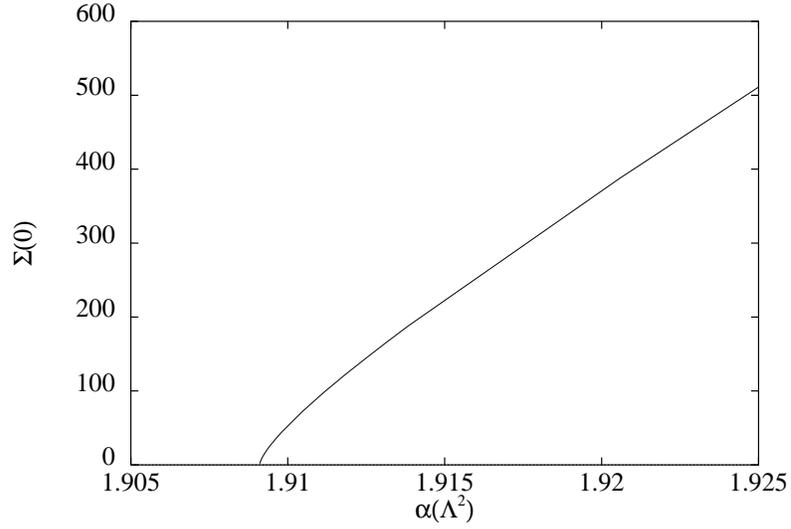,height=8cm,angle=-90}}
\end{center}
\vspace{-0.5cm}
\caption{Generated fermion mass $\S(0)$ versus running coupling 
$\alpha(\Lambda^2)$ for the coupled $(\S, \F, \G)$-system with the
$1/\F$-vertex Ansatz, for $N_f=2$.}
\label{Fig:fmg-BFG-rc-vertex1-N2}
\end{figure}

In Fig.~\ref{Fig:fmg-BFG-rc-vertex1-N2} we plot the generated fermion mass
$\S(0)$ versus the running coupling $\alpha(\Lambda^2)$ for $N_f=2$. The
critical coupling is \fbox{$\alpha_c(\Lambda^2,N_f=2) = 2.59578$}.

If we compare these results with those obtained with the bare vertex
approximation in Section~\ref{BFG} we note an increase of the critical
coupling, $\alpha_c(\Lambda^2)$ by about 10\% for $N_f=1$ and 16\% for
$N_f=2$. 

\section{Ball-Chiu vertex}

Next we will derive the results using the longitudinal or Ball-Chiu vertex,
which satisfies the Ward-Takahashi identity.  The Ball-Chiu vertex has been
introduced in \mref{1.1002} and is given by:\vspace{-5mm}
\ba
\Gamma^\mu_{L}(k,p) &=& 
\frac{1}{2}\l[\frac{1}{\F(k^2)}+\frac{1}{\F(p^2)}\r]\gamma^\mu
+ \frac{1}{2}\l[\frac{1}{\F(k^2)}-\frac{1}{\F(p^2)}\r]
\frac{(k+p)^\mu(\slash{k}+\slash{p})}{k^2-p^2} \mlab{1013.01} \\[1mm]
&& - \l[\frac{\S(k^2)}{\F(k^2)} - \frac{\S(p^2)}{\F(p^2)}\r]
\frac{(k+p)^\mu}{k^2-p^2} \nn .
\ea

The coupled integral equations are easily derived from Eqs.~(\oref{1.1006},
\oref{1.1007}, \oref{1.1008}) for the Curtis-Pennington vertex, by setting 
$\tau_6(y,x)=0$ to remove the transverse part of the vertex.  The
equations, with $m_0=0$ and in the Landau gauge, are:
\ba
\frac{\S(x)}{\F(x)} &=& \frac{\alpha}{2\pi^2} \int dy \, 
\frac{y\F(y)}{\Ds{y}} \int d\theta \sin^2\theta \, \G(z) \mlab{1013} \\ 
&& \hspace{10mm} \times 
\l\{  \frac{3A(y,x)\S(y)}{z} 
- \frac{\S(y)-\S(x)}{\F(x)(y-x)}\frac{2yx\sin^2\theta}{z^2} \r\} \nn\\[2mm]
\frac{1}{\F(x)} &=& 1 - \frac{\alpha}{2\pi^2 x} \int dy \,
\frac{y\F(y)}{y+\S^2(y)} \int d\theta \, \sin^2 \theta \, \G(z) \mlab{1014} \\
&& \times \l\{
A(y,x)\l[\frac{2yx\sin^2\theta}{z^2} - \frac{3\sqrt{yx}\cos\theta}{z}\r] 
+\l[B(y,x)(y+x)-C(y,x)\S(y)\r]\frac{2yx\sin^2\theta}{z^2} \r\} \nn\\[2mm]
\frac{1}{\G(x)} &=& 1 + \frac{2N_f\alpha}{3\pi^2 x} \int dy \, 
\frac{y\F(y)}{\Ds{y}} \int d\theta \, \sin^2\theta \,
\frac{\F(z)}{\Ds{z}} \mlab{1015}\\
&& \times \Bigg\{ 2A(y,z)\bigg[y(1-4\cos^2\theta) 
+ 3\sqrt{yx}\cos\theta\bigg] \nn\\
&& \hspace{5mm} + 
B(y,z) \bigg[\Big(y+z-2\S(y)\S(z)\Big)\,
\Big(2y(1-4\cos^2\theta)+3\sqrt{yx}\cos\theta\Big) \nn\\
&& \hspace{22mm} + 3(y-z)\Big(y-\S(y)\S(z)\Big)\bigg] \nn\\
&& \hspace{5mm} - C(y,z) 
\bigg[\Big(\S(y)+\S(z)\Big)\Big(2y(1-4\cos^2\theta)+3\sqrt{yx}\cos\theta\Big) 
+ 3(y-z)\S(y)\bigg] \Bigg\} \nn
\ea
where
\ba
\parbox{12cm}{
\vspace{-7mm}\bann
A(y,x) &=& \frac{1}{2}\l[\frac{1}{\F(y)}+\frac{1}{\F(x)}\r] \nn
\hspace{6cm}\\[2mm]
B(y,x) &=& \frac{1}{2(y-x)}\l[\frac{1}{\F(y)}-\frac{1}{\F(x)}\r] \nn\\[2mm]
C(y,x) &=& -\frac{1}{y-x}\l[\frac{\S(y)}{\F(y)}-\frac{\S(x)}{\F(x)}\r] \;.\nn
\eann}
\mlab{1016.14}
\ea

\subsection{Improper cancellation of quadratic divergences}
\def\x0{x_0=1.00856\ten{-05}}
\def\yN{y_N=9.96568\ten{+09}}

\begin{figure}[htbp]
\begin{center}
\mbox{\epsfig{file=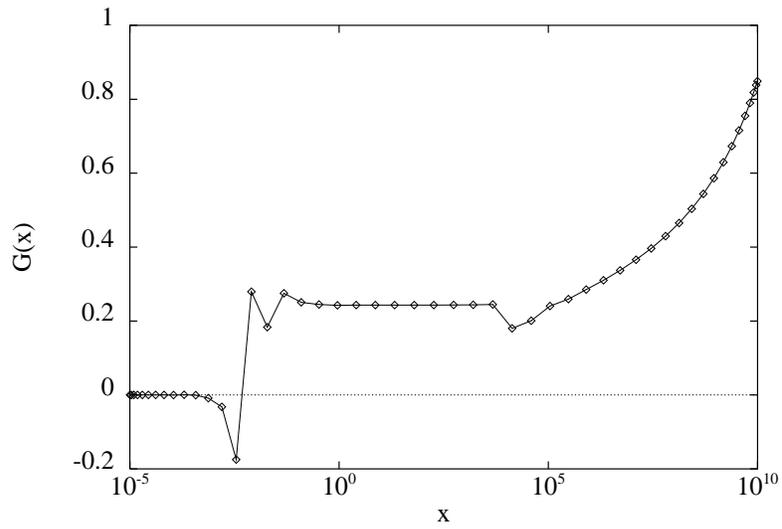,height=8cm,angle=-90}}
\end{center}
\vspace{-0.5cm}
\caption{Photon renormalization function $\G(x)$ versus momentum 
squared $x$ from the $\G$-equation with Ball-Chiu
vertex, for $\alpha=1.921$ and $N_f=1$.}
\label{Fig:G-nocancel}
\end{figure}

When solving this system of coupled integral equations with the method
described in Chapter~\ref{BFG} we encounter some serious new problems.
Although the solution of \mrefb{1013}{1014} does not seem to suffer by the
introduction of the Ball-Chiu vertex, the behaviour of the photon equation,
\mref{1015}, however, is somehow erratic. This can be seen in 
Fig.~\ref{Fig:G-nocancel} where we plotted the behaviour of $\G(x)$ for
$\alpha=1.921$ with realistic input functions $\S(x)$, $\F(x)$.
\def\filexx{figs/unquenched_QED/cheby/QuadDiv}

\begin{figure}[htbp]
\begin{center}
\mbox{\epsfig{file=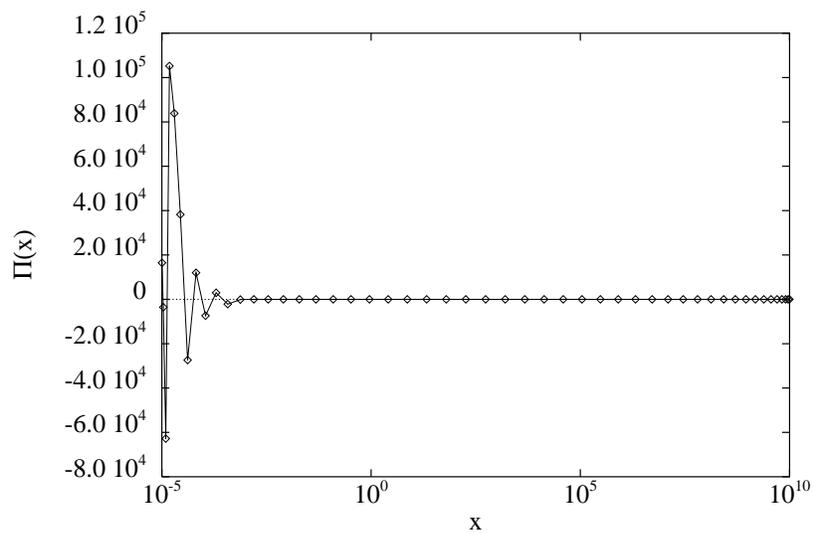,height=8cm,angle=-90}}
\end{center}
\vspace{-0.5cm}
\caption{Vacuum polarization $\Pi(x)$ versus momentum 
squared $x$ from the $\G$-equation with Ball-Chiu
vertex, for $\alpha=1.921$ and $N_f=1$.}
\label{Fig:vp-nocancel}
\end{figure}

To investigate the numerical cancellation of the quadratic divergence we
plot the vacuum polarization function $\Pi(x)$ of \mref{1015}
in Fig.~\ref{Fig:vp-nocancel}.
From the 1-loop perturbative results for the vacuum polarization we expect
the vacuum polarization to be roughly of the order of:
\be
\Pi(0) \simeq \frac{N_f\alpha}{3\pi}\ln\frac{\Lambda^2}{\S^2(0)} \approx 2.8
\mlab{expect}
\ee
for $\alpha=1.921$, $\Lambda=1\ten{5}$ and $\S(0)=100$.  The very large values
of the vacuum polarization at small values of $x$, in
Fig.~\ref{Fig:vp-nocancel}, clearly show that the quadratic divergence has
not been cancelled correctly. There seems to be a residual linear
divergence in the numerical solution. To examine this in more detail, we
will investigate the radial integrand, $K_R(x,y)$, of $\Pi(x)$ for small
values of $x$. We can write the vacuum polarization integral of \mref{1015}
as:
\be
\Pi(x) = \int dt \; K_R(x,y) 
\mlab{1016.1}
\ee
where $t=\logten y$ and the radial integrand $K_R(x,y)$ is defined by: 
\be
K_R(x,y) = \int d_\theta \; K_\theta(x,y,\theta) \;.
\mlab{1016.11}
\ee
In \mref{1016.11}, the angular integrand $K_\theta(x,y,\theta)$ is given
by:
\be
K_\theta(x,y,\theta) = \rho(x,y) f_\theta(x,y,\theta) \;.
\mlab{1016.12}
\ee
where the multiplicative factor $\rho(x,y)$, independent from $\theta$, is:
\be
\rho(x,y) = \frac{2 N_f \alpha\ln10}{3\pi^2 x} \frac{y^2\F(y)}{\Ds{y}}
\mlab{1016.13}
\ee
and the angular function $f_\theta$ from is defined as:\vspace{-3mm}
\ba
\parbox[t]{14.7cm}{\vspace{-8mm}\small\bann
\lefteqn{f_\theta(x,y,\theta) = \sin^2\theta \,
\frac{\F(z)}{\Ds{z}}
\Bigg\{ 2A(y,z)\bigg[y(1-4\cos^2\theta) 
+ 3\sqrt{yx}\cos\theta\bigg]} \\
&& 
+ B(y,z) \bigg[\Big(y+z-2\S(y)\S(z)\Big)\,
\Big(2y(1-4\cos^2\theta)+3\sqrt{yx}\cos\theta\Big) 
+ 3(y-z)\Big(y-\S(y)\S(z)\Big)\bigg] \nn\\
&& 
- C(y,z) 
\bigg[\Big(\S(y)+\S(z)\Big)\Big(2y(1-4\cos^2\theta)+3\sqrt{yx}\cos\theta\Big) 
+ 3(y-z)\S(y)\bigg] \Bigg\} \nn\\[2mm]
\lefteqn{\mbox{with} \quad z=y+x-2\sqrt{yx}\cos\theta \;.} \nn
\eann}
\mlab{1017}
\ea
 
\begin{figure}[htbp]
\begin{center}
\mbox{\epsfig{file=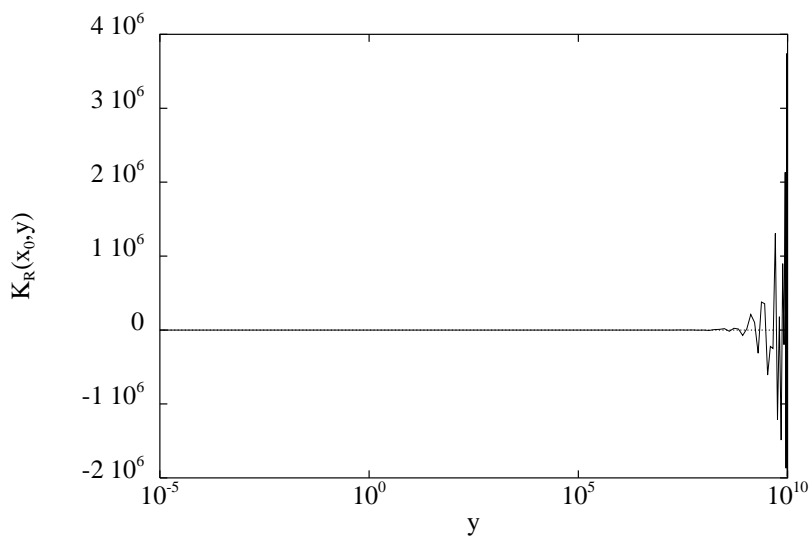,height=8cm,angle=-90}}
\end{center}
\vspace{-0.5cm}
\caption{Radial integrand $K_R(x_0,y)$, where $\x0$,
versus radial momentum squared $y$ from the $\G$-equation 
with Ball-Chiu vertex, for $\alpha=1.921$ and $N_f=1$.}
\label{Fig:radint-nocancel}
\end{figure}

In Fig.~\ref{Fig:radint-nocancel} we plot the radial integrand
$K_R(x_0,y)$, where $\x0$ is the smallest external momentum
value used in the numerical solution of Eqs.~(\oref{1013}, \oref{1014},
\oref{1015}). There, the integral value is $\Pi(x_0) = 16471.3$. This 
much too large value seems to be caused by the chaotic behaviour of the
radial integrand for large values of momentum. We will, therefore,
investigate where this behaviour originates from and examine $K_R(x_0,y)$
in more detail for large values of radial momentum $y$. For $\yN$, which is
the largest radial integration point for the external momentum $x_0$, the
radial integrand $K_R(x_0,y_N)= -1.82932\ten{+06}$.

\begin{figure}[htbp]
\begin{center}
\mbox{\epsfig{file=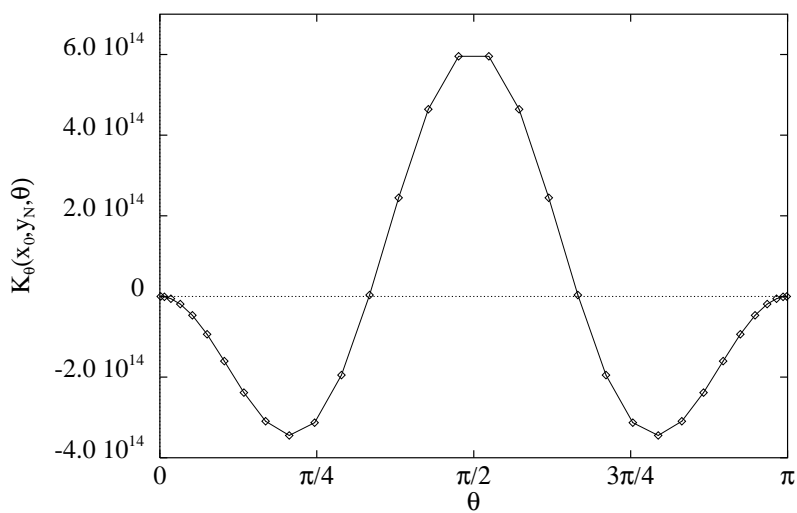,height=8cm,angle=-90}}
\end{center}
\vspace{-0.5cm}
\caption{Angular integrand $K_\theta(x_0,y_N,\theta)$ for
$\x0$, $\yN$, versus angle $\theta$ from the $\G$-equation 
with Ball-Chiu vertex, for $\alpha=1.921$ and $N_f=1$.}
\label{Fig:ti-nocancel}
\end{figure}

We now look at the behaviour of the angular integrand
$K_\theta(x_0,y_N,\theta)$ where $\x0$ and $\yN$. We note from
\mrefb{1016.12}{1016.13} that, for such small values of $x$ and large
values of $y$, accuracy problems in the angular function $f_\theta$ are
magnified enormously, here by a factor $\rho(x_0,y_N)\approx 3\ten{+14}$. We
plot the angular integrand $K_\theta(x_0,y_N,\theta)$ versus $\theta$ in
Fig.~\ref{Fig:ti-nocancel}.

For the quadratic divergence to cancel, the vacuum polarization integral
must satisfy:
\be
\lim_{x\to 0} x \Pi(x) = 0 .
\mlab{1019}
\ee

The cancellation of the quadratic divergence occurs if the terms
proportional to $1/x$ in the angular integrand, \mref{1016.12}, vanish as a
result of the angular integrals being equal to zero when $x\to 0$ . To
investigate if this is achieved numerically, we rewrite the angular function
$f_\theta$, \mref{1017}, as a sum of angular functions which all give
individual contributions to the vacuum polarization that are theoretically
free of quadratic divergences:
\be
f_\theta = \sin^2\theta \l(I_A + J_A + I_B + J_B + I_C + J_C\r) \;,
\mlab{1020}
\ee
where
\ba
I_A(x,y,\theta) &=& \frac{2 A(y,z) y  \F(z)}{\Ds{z}} \; (1 - 4\cos^2\theta)\nn\\
J_A(x,y,\theta) &=& \frac{6 A(y,z) \F(z)}{\Ds{z}} \sqrt{yx}\cos\theta\nn\\
I_B(x,y,\theta) &=& \frac{2yB(y,z)\F(z)}{\Ds{z}} 
\Big(y+z-2\S(y)\S(z)\Big) (1 - 4\cos^2\theta)\mlab{1023}\\
J_B(x,y,\theta) &=& \frac{3B(y,z)\F(z)}{\Ds{z}} \bigg[\Big(y+z-2\S(y)\S(z)\Big)
\sqrt{yx}\cos\theta + (y-z)\Big(y-\S(y)\S(z)\Big)\bigg]\nn\\
I_C(x,y,\theta) &=& -\frac{2y C(y,z)\F(z)}{\Ds{z}}\Big(\S(y)+\S(z)\Big) 
(1 - 4\cos^2\theta)\nn\\
J_C(x,y,\theta) &=& -\frac{C(y,z)\F(z)}{\Ds{z}}
\bigg[\Big(\S(y)+\S(z)\Big)3\sqrt{yx}\cos\theta 
+ 3(y-z)\S(y)\bigg]\nn \;.
\ea

We define the angular integrands $K_{\theta,i}$, $i=1,\ldots,6$, as:
\ba
K_{\theta,1}(x,y,\theta) &=& \rho(x,y) \sin^2\theta \, I_A(x,y,\theta) \nn\\
K_{\theta,2}(x,y,\theta) &=& \rho(x,y) \sin^2\theta \, J_A(x,y,\theta) \nn\\
K_{\theta,3}(x,y,\theta) &=& \rho(x,y) \sin^2\theta \, I_B(x,y,\theta) \mlab{1021}\\
K_{\theta,4}(x,y,\theta) &=& \rho(x,y) \sin^2\theta \, J_B(x,y,\theta) \nn\\
K_{\theta,5}(x,y,\theta) &=& \rho(x,y) \sin^2\theta \, I_C(x,y,\theta) \nn\\
K_{\theta,6}(x,y,\theta) &=& \rho(x,y) \sin^2\theta \, J_C(x,y,\theta) \nn
\ea
and the radial integrands $K_{R,i}$, after angular integration of
\mref{1021}, as:
\be
\hspace{3cm}
K_{R,i}(x,y) = \int d\theta \; K_{\theta,i}(x,y,\theta) \quad , \qquad
i=1,\ldots,6 \;.
\ee

The total radial kernel $K_R$ is given by:
\be
K_R = K_{R,1} + K_{R,2} + K_{R,3} + K_{R,4} + K_{R,5} + K_{R,6} \;.
\ee

Although the analytical cancellation of the quadratic divergence for $x\to
0$ is obvious, this is not ensured to happen numerically. We tabulate the
computed values of the individual radial kernels, $K_{R,i}(x_0,y_N)$, with
$\x0$ and $\yN$ in
Table~\ref{Tab:K_R-nocancel}. The main contribution to the radial integrand
comes from $K_{R,3}$.

\begin{table}[htbp]
\begin{center}
\begin{tabular}{|l|r|}
\hline
$K_{R,1}$ & $4.61709\ten{+00}$ \\ $K_{R,2}$ & $1.44340\ten{+00}$ \\
$K_{R,3}$ & $-1.82932\ten{+06}$ \\ $K_{R,4}$ & $7.64711\ten{-02}$ \\
$K_{R,5}$ & $-2.35535\ten{-02}$ \\ $K_{R,6}$ & $-3.40508\ten{-09}$ \\
\hline
\end{tabular}
\end{center}
\caption{Radial kernels $K_{R,i}(x_0,y_N)$, for $i=1,\ldots,6$, 
with $\x0$ and $\yN$.}
\label{Tab:K_R-nocancel}
\end{table}

\begin{figure}[htbp]
\begin{center}
\mbox{\epsfig{file=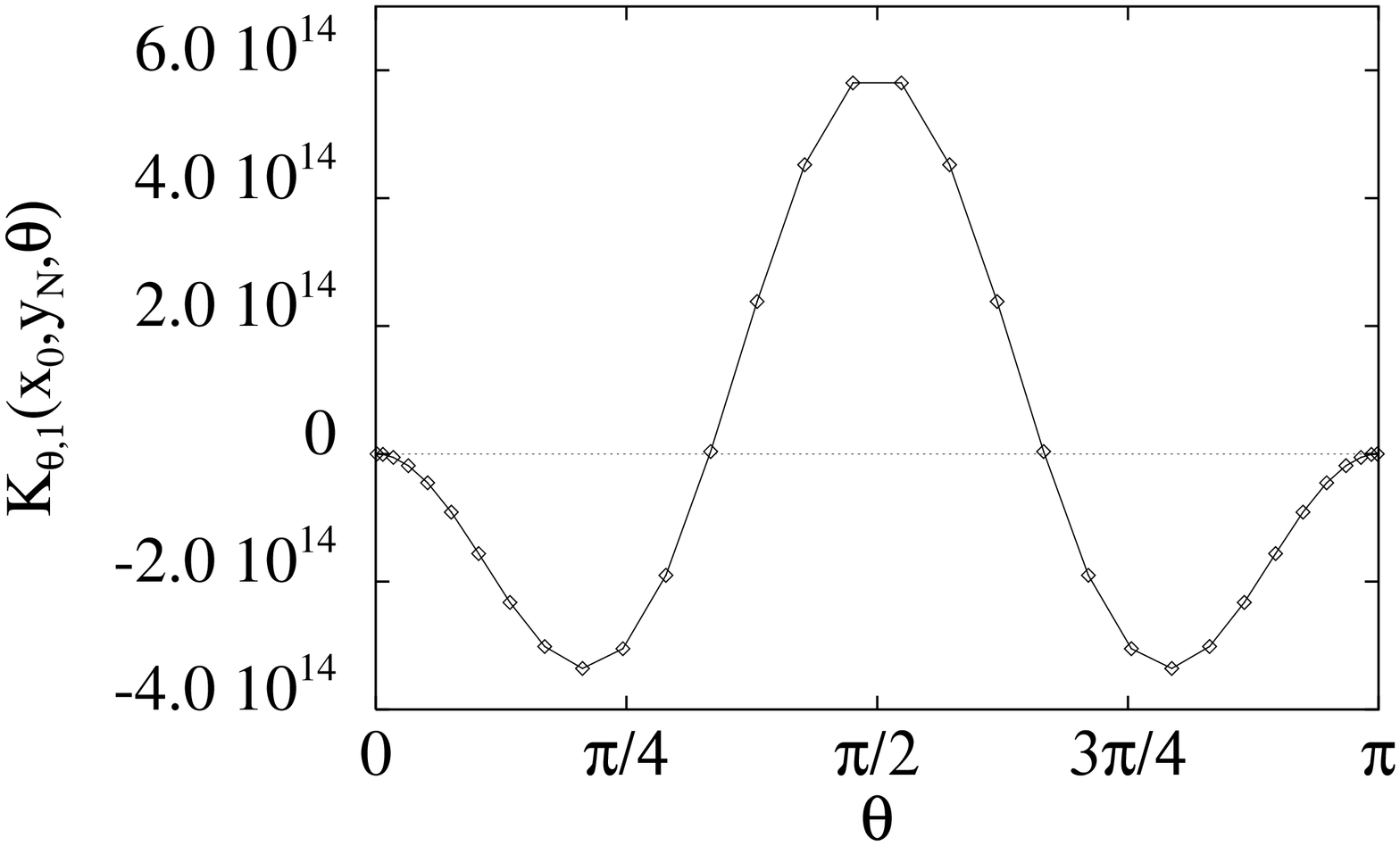,height=5.5cm,angle=-90}
\epsfig{file=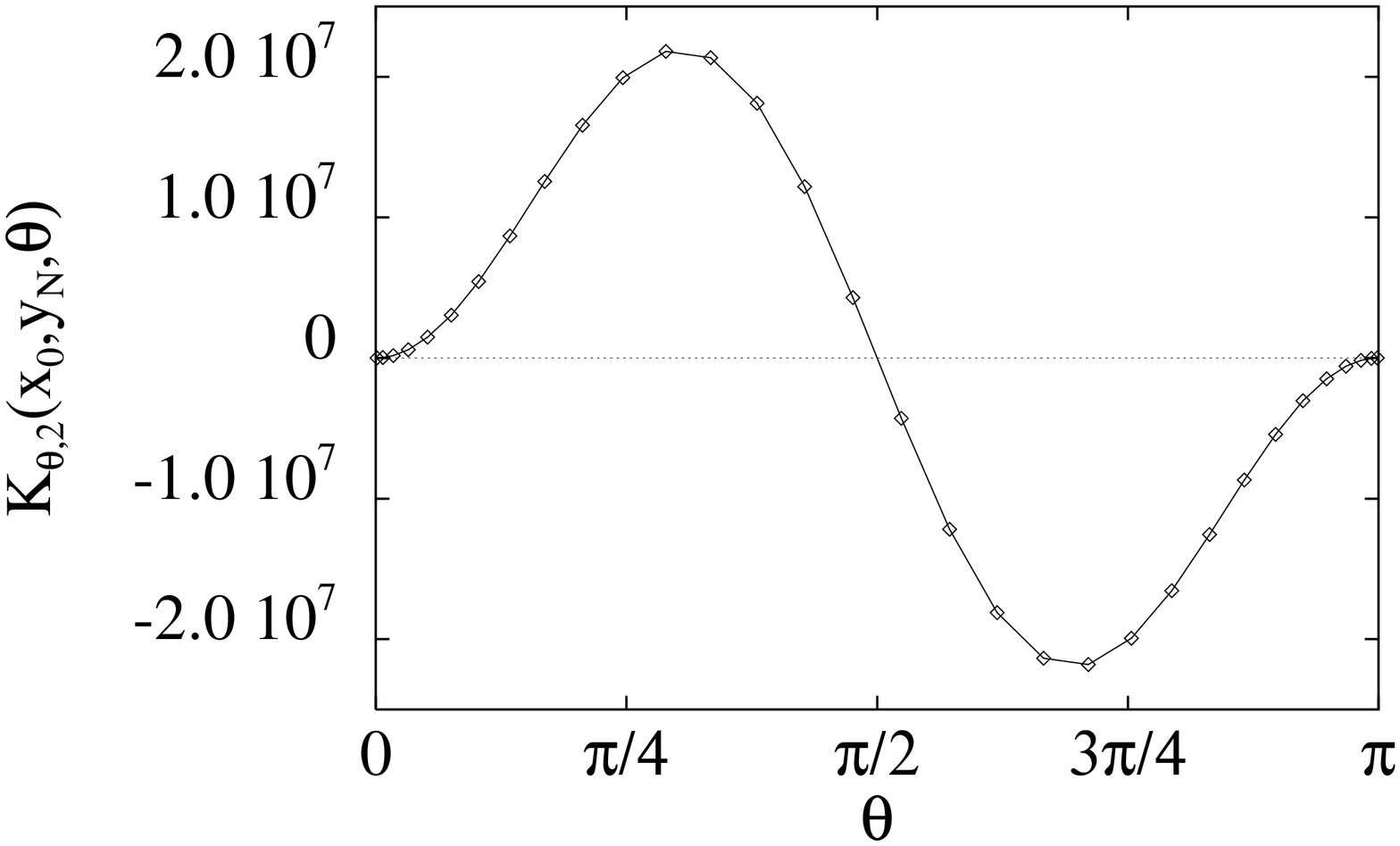,height=5.5cm,angle=-90}}
\mbox{\epsfig{file=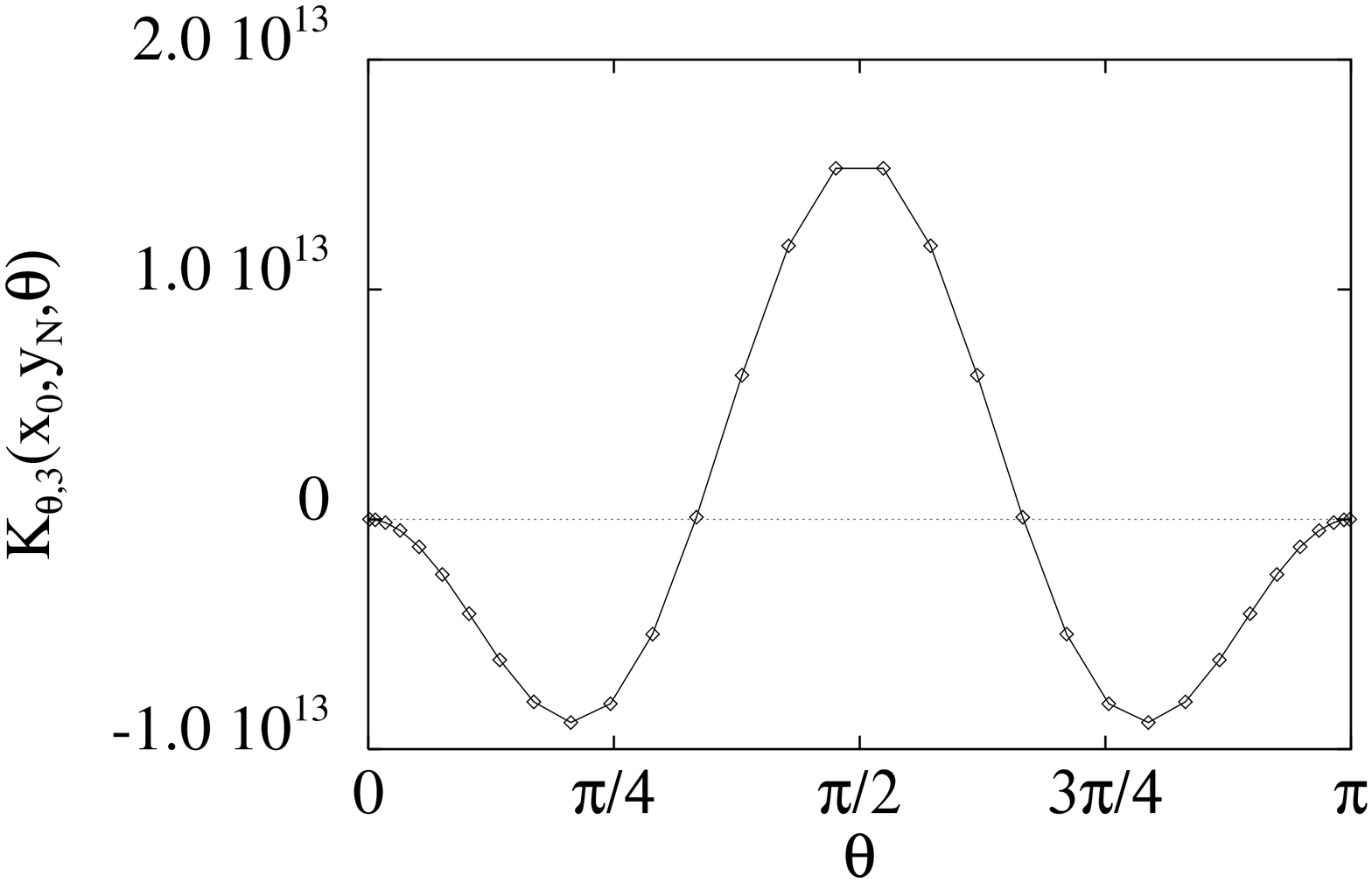,height=5.5cm,angle=-90}
\epsfig{file=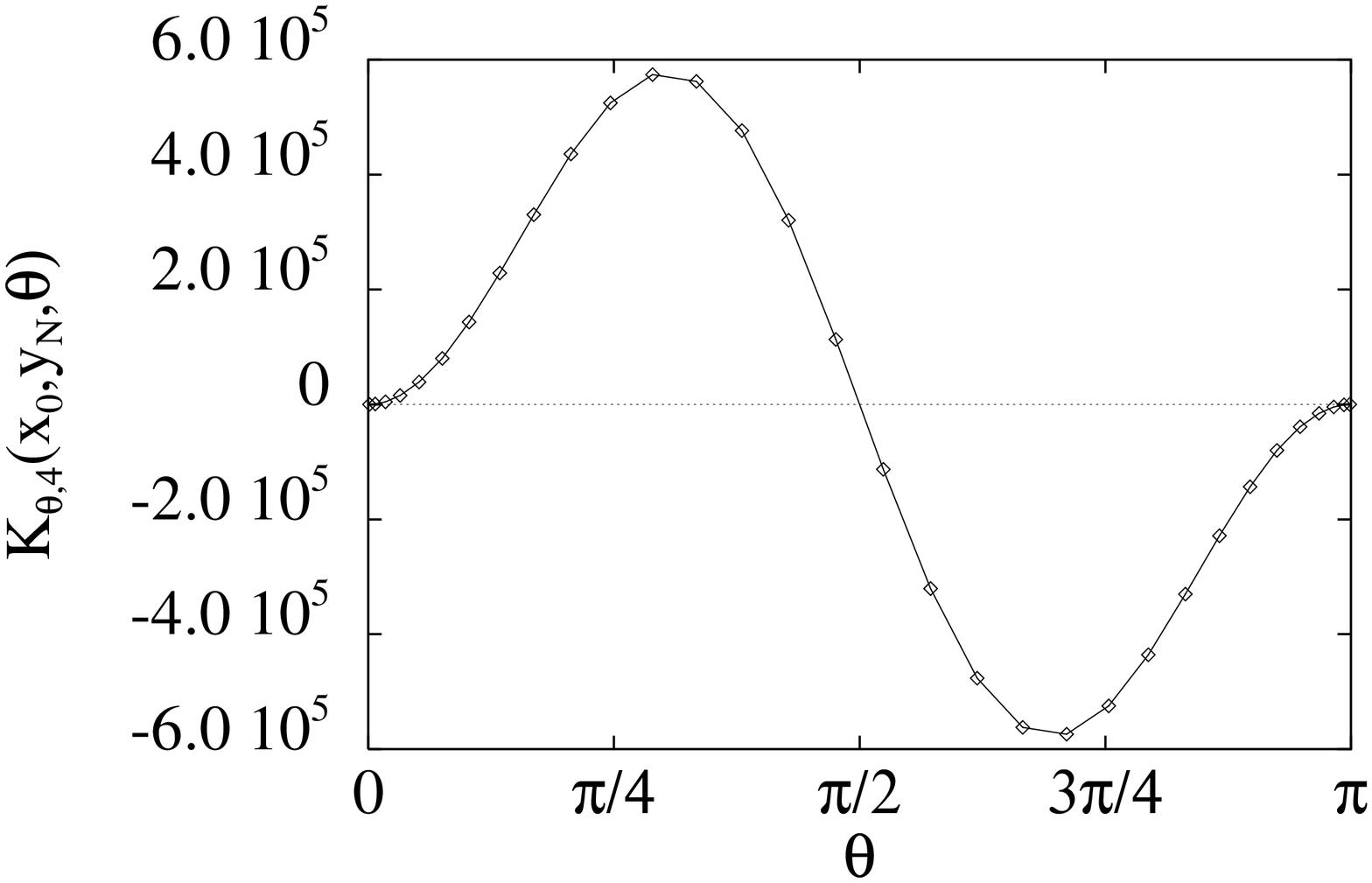,height=5.5cm,angle=-90}}
\mbox{\epsfig{file=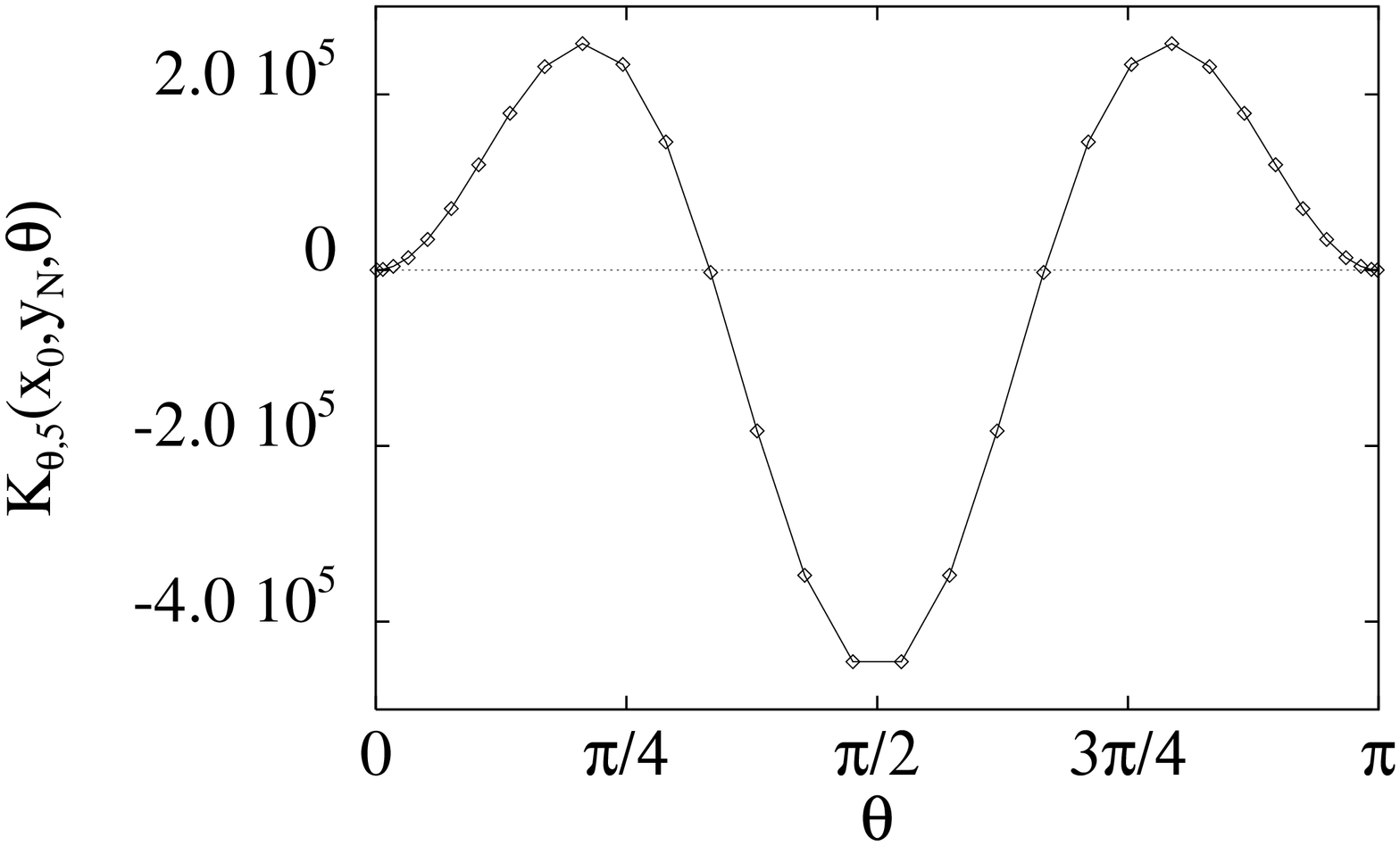,height=5.5cm,angle=-90}
\epsfig{file=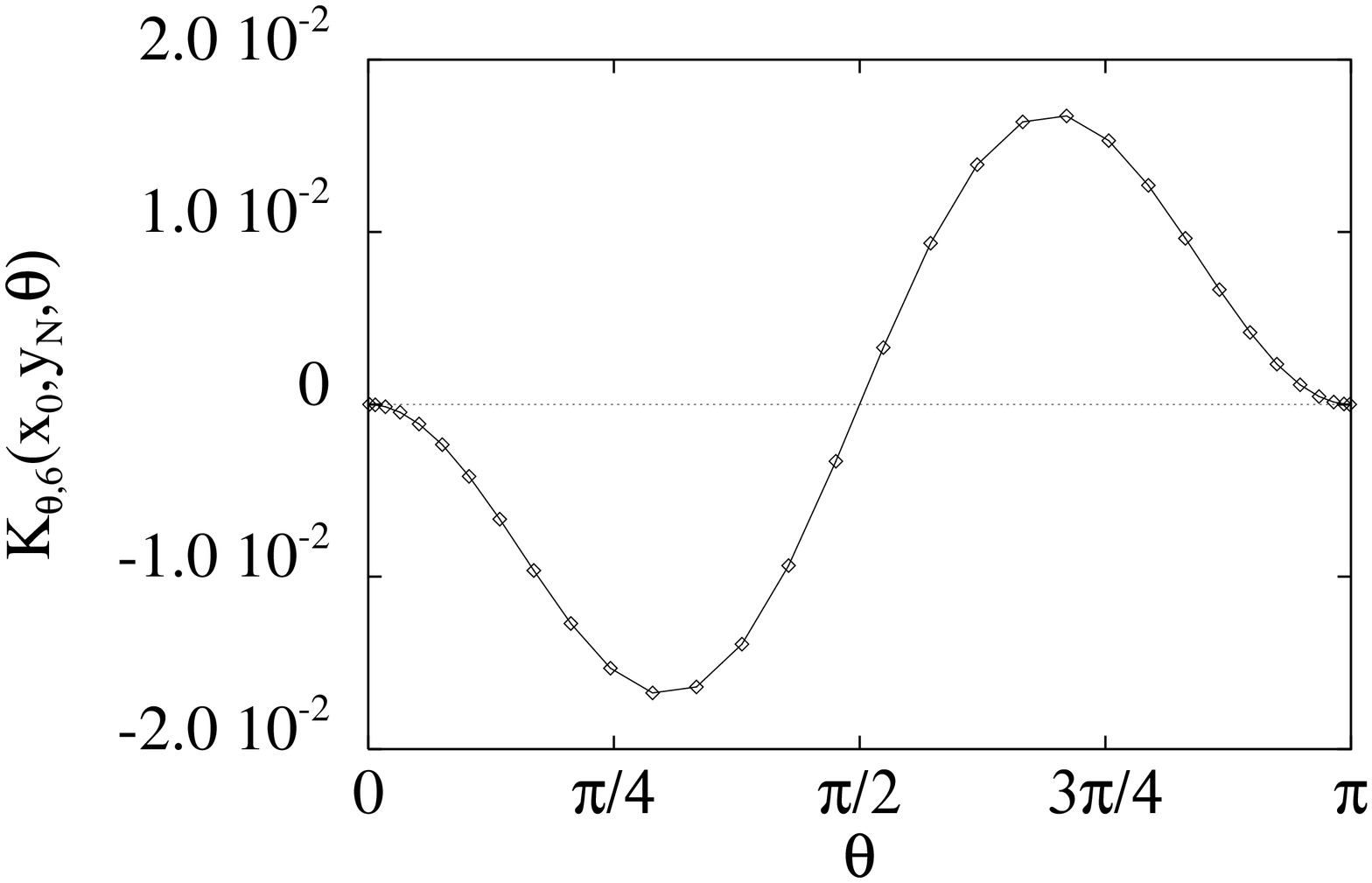,height=5.5cm,angle=-90}}
\end{center}
\vspace{-0.5cm}
\caption{Angular integrands $K_{\theta,i}(x_0,y_N,\theta)$, for $i=1,\ldots,6$,
with $\x0$, $\yN$, versus angle $\theta$ from the $\G$-equation 
with Ball-Chiu vertex, for $\alpha=1.921$ and $N_f=1$.}
\label{Fig:tii-nocancel}
\end{figure}

Plots of the various angular integrands $K_{\theta,i}(x_0,y_N,\theta)$,
\mref{1021}, are shown in Fig.~\ref{Fig:tii-nocancel}. From this figure
nothing suspicious can be detected. This is understandable, from
Table~\ref{Tab:K_R-nocancel}, as the main contribution to the radial
integrand comes from $K_{R,3}$ and is of $\Order(\oten{6})$, while the
angular integrand $K_{\theta,3}$ has a magnitude of
$\Order(\oten{13})$. The problem seems to be hidden as an undiscernible
noise in the much larger smooth envelope of the angular integrand.

\subsection{Small-$x$ expansion of angular functions}

To investigate the cancellation of quadratic and linear divergences we need
the terms of $K_{\theta,i}$ which are proportional to $1/x$ and
$1/\sqrt{x}$. Because of \mref{1021} and the $1/x$-proportionality of
$\rho(x,y)$, this corresponds to expanding the angular
functions, \mref{1023}, up to constant terms for the quadratic divergent
contributions and to $\sqrt{x}$ for the linear divergent
contributions. Therefore we look how the various angular functions,
\mref{1023}, depend on $x$ or $(z-y)=x-2\sqrt{xy}\cos\theta$ for small $x$.

To Taylor expand the angular functions, \mref{1023}, we first Taylor expand
their various components up to $\Order(z-y)$:
\ba
\F(z) &=& \F(y) + (z-y)\,\F'(y) + \Order(z-y)^2 \mlab{1024} \\[3mm]
\S(z) &=& \S(y) + (z-y)\,\S'(y)  + \Order(z-y)^2 \mlab{1025}\\[3mm]
\frac{1}{\Ds{z}} &=& 
\frac{1}{\Ds{y}} - (z-y)\l[\frac{1+2\S(y)\S'(y)}{(\Ds{y})^2}\r] + \Order(z-y)^2 
\mlab{1026}\\[5mm]
A(y,z) &=& \frac{1}{2}\l[\frac{1}{\F(z)}+\frac{1}{\F(y)}\r] \nn\\
&=& \frac{1}{2}\l\{\frac{1}{\F(y)}+(z-y)\l[\frac{1}{\F(y)}\r]' + \Order(z-y)^2 
+\frac{1}{\F(y)}\r\} \nn\\
&=& \frac{1}{\F(y)} + \frac{z-y}{2}\l[\frac{1}{\F(y)}\r]' + \Order(z-y)^2 
\mlab{1026.1}\\[5mm]
B(y,z) &=& \frac{1}{2(z-y)}\l[\frac{1}{\F(z)}-\frac{1}{\F(y)}\r] \nn\\
&=& \frac{1}{2(z-y)}\l\{\frac{1}{\F(y)}+ (z-y)\l[\frac{1}{\F(y)}\r]'+
\frac{(z-y)^2}{2}\l[\frac{1}{\F(y)}\r]''+ \Order(z-y)^3 
-\frac{1}{\F(y)}\r\} \nn\\
&=& \frac{1}{2}\l[\frac{1}{\F(y)}\r]'+
\frac{(z-y)}{4}\l[\frac{1}{\F(y)}\r]''+ \Order(z-y)^2
\mlab{1026.2}\\[5mm]
C(y,z) &=& -\frac{1}{z-y}\l[\frac{\S(z)}{\F(z)}-\frac{\S(y)}{\F(y)}\r] \nn\\
&=& -\frac{1}{z-y}\l\{\frac{\S(y)}{\F(y)}+(z-y)\l[\frac{\S(y)}{\F(y)}\r]'
+\frac{(z-y)^2}{2}\l[\frac{\S(y)}{\F(y)}\r]''  + \Order(z-y)^3 
-\frac{\S(y)}{\F(y)}\r\} \nn \\
&=& -\l[\frac{\S(y)}{\F(y)}\r]'
- \frac{z-y}{2}\l[\frac{\S(y)}{\F(y)}\r]'' + \Order(z-y)^2 \;. \mlab{1027}
\ea

Substituting Eqs.~(\oref{1024}-\oref{1027}) in the angular functions,
\mref{1023}, and gathering together terms of equal power in $(z-y)$ yields:
\ba
I_A(x,y,\theta) &=& \frac{2y}{\Ds{y}}(1 - 4\cos^2\theta) \mlab{1027.1}\\
&+&  2y(z-y)(1 - 4\cos^2\theta)\l\{
\frac{1}{2}\frac{\F'(y)}{\F(y)}\frac{1}{\Ds{y}}
- \frac{1+2\S(y)\S'(y)}{(\Ds{y})^2}\r\} + \Order(z-y)^2 \nn\\[5mm]
I_B(x,y,\theta) &=& \frac{2y(y-\S^2(y))\F(y)}{\Ds{y}}
\l[\frac{1}{\F(y)}\r]' (1 - 4\cos^2\theta) \mlab{1027.2}\\
&+& 2y(z-y)(1 - 4\cos^2\theta) 
\Bigg\{
\frac{1}{2}\frac{(y-\S^2(y))\F(y)}{\Ds{y}}\l[\frac{1}{\F(y)}\r]'' 
\nn\\
&&+\frac{(y-\S^2(y))\F'(y)}{\Ds{y}} \l[\frac{1}{\F(y)}\r]' \nn\\
&&-\l[\frac{\F(y)(y-\S^2(y))(1+2\S(y)\S'(y))}{(\Ds{y})^2}\r] 
\l[\frac{1}{\F(y)}\r]' 
\nn\\
&& +\frac{1}{2}\frac{\F(y)(1-2\,\S(y)\S'(y))}{\Ds{y}} \l[\frac{1}{\F(y)}\r]'
\Bigg\}  + \Order(z-y)^2 \nn\\[5mm]
I_C(x,y,\theta) &=& \frac{4y\F(y)\S(y)}{\Ds{y}}\l[\frac{\S(y)}{\F(y)}\r]'(1 -
4\cos^2\theta) \mlab{1027.3} \\
&+& 2y(z-y)(1 - 4\cos^2\theta)\l\{\frac{\F(y)\S(y)}{\Ds{y}}
\l[\frac{\S(y)}{\F(y)}\r]'' 
+ \frac{2\F'(y)\S(y)}{\Ds{y}} \l[\frac{\S(y)}{\F(y)}\r]' \r. \nn \\
&& \l. 
-\frac{2\F(y)\S(y)(1+2\S(y)\S'(y))}{\Ds{y}} \l[\frac{\S(y)}{\F(y)}\r]'
+ \frac{\F(y)\S'(y)}{\Ds{y}} \l[\frac{\S(y)}{\F(y)}\r]'\r\} + \Order(z-y)^2
\nn\;.
\ea

Because of the structure of the integrands $J_A$, $J_B$, $J_C$, these have no
potentially quadratic divergent terms. However there can be some remnant linear
divergent bits, which are proportional to $\sqrt{x}$. To separate these, we
only need the first term in the Taylor expansions of the various
components. This gives:
\ba
J_A(x,y,\theta) &=& \frac{6 \sqrt{yx}\cos\theta }{\Ds{y}} + \Order(x) \\[2mm]
J_B(x,y,\theta) &=& \frac{3}{2}\l[\frac{1}{\F(y)}\r]' \frac{\F(y)}{\Ds{y}} 
(4\sqrt{yx}\cos\theta - x)(y-\S^2(y)) + \Order(x) \\[2mm]
J_C(x,y,\theta) &=& \l[\frac{\S(y)}{\F(y)}\r]'\frac{3\F(y)}{\Ds{y}}
\bigg[2\S(y)\sqrt{yx}\cos\theta + (y-z)\S(y)\bigg] + \Order(x) \;.
\ea

\begin{figure}[htbp]
\begin{center}
\mbox{\epsfig{file=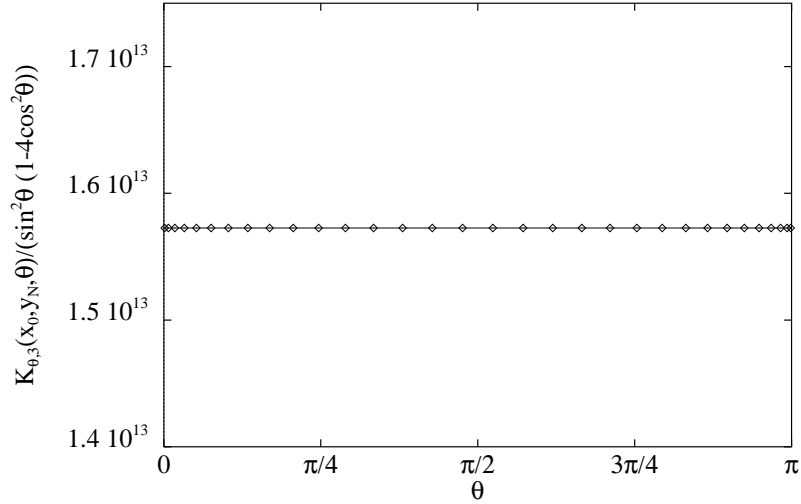,height=8cm,angle=-90}}
\end{center}
\vspace{-0.5cm}
\caption{Angular integrand $K_{\theta,3}(x_0,y_N,\theta)$ for
$\x0$, $\yN$, versus angle $\theta$ from the $\G$-equation with Ball-Chiu
vertex, for $\alpha=1.921$ and $N_f=1$ after removing the trigonometric
factor $\sin^2\theta(1-4\cos^2\theta)$.}
\label{Fig:ti3-nocancel-notrig}
\end{figure}

It is enlightening to study the behaviour of $K_{\theta,3}(x_0,y_N,\theta)$
in more detail. In Fig.~\ref{Fig:tii-nocancel} we see the nice
trigonometric behaviour of the angular kernel as predicted from
\mrefb{1021}{1027.2}. We now show another plot in
Fig.~\ref{Fig:ti3-nocancel-notrig}, where we divide
$K_{R,3}(x_0,y_N,\theta)$ by its trigonometric factor
$\sin^2\theta(1-4\cos^2\theta)$. As expected from the first term of the
Taylor expansion of $I_B$, \mref{1027.2}, the leading order term of the
plotted function is now constant in $\theta$.  Furthermore, because the
integral over $\theta$ of the trigonometric part of the leading order term
of $K_{R,3}(x_0,y_N)$ vanishes, there is no quadratic divergent contribution to
it.

\subsection{Subtracting the leading order term}

To verify that the quadratic divergence is cancelled correctly numerically
and that the remaining result is meaningful we now subtract explicitly the
value of the angular integrand when $x\to 0$, i.e. when $z\to y$. We
define:
\be
\tilde K_{\theta}(x,y,\theta) = \frac{1}{x} 
\l[ x K_{\theta}(x,y,\theta) - \lim_{x\to 0} x K_{\theta}(x,y,\theta) \r].
\mlab{1022}
\ee 

Formally this should not change the value of the angular integral,
\be
K_R(x,y) = \int d\theta \; K_{\theta}(x,y,\theta) = \int d\theta \; \tilde
K_{\theta}(x,y,\theta)
\ee
as we can show analytically that:
\be
\int d\theta \; \l[\lim_{x\to 0} x K_{\theta}(x,y,\theta)\r] = 0 \;.
\ee

Because of the $1/x$ factor in $\rho(x,y)$, \mref{1022} corresponds to the
subtraction of the terms without any $x$-dependency in the angular functions
$I_A, I_B, I_C, J_A, J_B, J_C$. These are exactly the leading order terms
of $I_A$, $I_B$ and $I_C$ in Eqs.~(\oref{1027.1},
\oref{1027.2},
\oref{1027.3}), which we will respectively call $I^0_A$, $I^0_B$, $I^0_C$
and are given by:
\ba
I^0_A(x,y,\theta) &=& \frac{2 y}{\Ds{y}} \; (1 - 4\cos^2\theta)\\
I^0_B(x,y,\theta) &=& \frac{2y(y-\S^2(y))\F(y)}{\Ds{y}} 
\l[\frac{1}{\F(y)}\r]' (1 - 4\cos^2\theta)\\
I^0_C(x,y,\theta) &=& 
\frac{4y \F(y)\S(y)}{\Ds{y}} \l[\frac{\S(y)}{\F(y)}\r]'
(1 - 4\cos^2\theta).
\ea

By subtracting the leading order term, which should vanish anyway after
angular integration, we want to explore what happens to the next
order in the small-$x$ expansions of $I_A$, $I_B$ and $I_C$,
Eqs.~(\oref{1027.1}, \oref{1027.2},
\oref{1027.3}). As there are no potentially quadratic divergences in $J_A$,
$J_B$, $J_C$, we do not subtract any contribution from these kernels.

To compute $I^0_B$ and $I^0_C$ numerically, we need to take the derivatives
of the functions $\F$ and $\S$, which are defined as Chebyshev
expansions. Consider the Chebyshev expansion of a function $f(x)$:
\be
f(x) = \sum_{j=0}^{N-1} c_j T_j(x) - \frac{c_0}{2}, 
\mlab{1030}
\ee
then, its derivative with respect to $x$ is again a Chebyshev expansion:
\be
f'(x) = \sum_{j=0}^{N-1} c'_j T_j(x) - \frac{c'_0}{2}, 
\mlab{1031}
\ee

where the coefficients $c'_j$ are defined by:
\be
c'_{j-1} = c'_{j+1} + 2j c_j , \qquad j=N-1,\ldots,1 ,
\mlab{1032}
\ee
with $c'_N = c'_{N-1} = 0$.

As we expand the functions $\S(x)$, $\F(x)$ and $\G(x)$ in Chebyshev
polynomials of $s(x)$, \mref{828}, instead of $x$, the expansion
\mref{1030} now becomes:
\be
g(x) = f(s(x)) = \sum_{j=0}^{N-1} c_j T_j(s(x)) - \frac{c_0}{2}
\mlab{1033}
\ee

where:
\be
s(x) = \frac{\logten(x/\Lambda\kappa)}{\logten(\Lambda/\kappa)}.
\mlab{1033.1}
\ee

and the derivative is now:
\be
g'(x) = \frac{dg(x)}{dx} = \frac{ds(x)}{dx}\frac{df(s(x))}{ds}
= \frac{f'(s)}{x \ln10 \logten(\Lambda/\kappa)} 
\mlab{1034}
\ee
where $f'(s)$ is defined by \mrefb{1031}{1032}.

Explicitly cancelling these lowest order terms in \mref{1021} will yield
the following angular kernels:
\ba
K_{\theta,1}(x,y,\theta) &=& \rho(x,y) \sin^2\theta \, (I_A - I^0_A)\nn\\
K_{\theta,2}(x,y,\theta) &=& \rho(x,y) \sin^2\theta \, J_A\nn\\
K_{\theta,3}(x,y,\theta) &=& \rho(x,y) \sin^2\theta \, (I_B - I^0_B)\nn\\
K_{\theta,4}(x,y,\theta) &=& \rho(x,y) \sin^2\theta \, J_B\nn\\
K_{\theta,5}(x,y,\theta) &=& \rho(x,y) \sin^2\theta \, (I_C - I^0_C)\nn\\
K_{\theta,6}(x,y,\theta) &=& \rho(x,y) \sin^2\theta \, J_C\nn \;.
\ea

After this cancellation we see that the total radial kernel $K_R(x_0,y_N)$,
after angular integration is still $K_R(x_0,y_N) = -1.82932\ten{+06}$ and
the individual radial kernels $K_{R,i}(x_0,y_N)$, for $i=1,\ldots,6$, are
given in Table~\ref{Tab:K_R-cancel_qd}.

\begin{table}[htbp]
\begin{center}
\begin{tabular}{|l|r|}
\hline
$K_{R,1}$ & $-9.45396\ten{-01}$ \nn\\
$K_{R,2}$ &  $1.44340\ten{+00}$ \nn\\
$K_{R,3}$ & $-1.82932\ten{+06}$ \nn\\
$K_{R,4}$ &  $7.64711\ten{-02}$ \nn\\
$K_{R,5}$ & $-2.35535\ten{-02}$ \nn\\
$K_{R,6}$ & $-3.40508\ten{-09}$ \nn\\
\hline
\end{tabular}
\end{center}
\caption{Radial kernels $K_{R,i}(x_0,y_N)$, for $i=1,\ldots,6$, 
with $\x0$ and $\yN$ with explicit cancellation of the
quadratic divergence.}
\label{Tab:K_R-cancel_qd}
\end{table}

\begin{figure}[htbp]
\begin{center}
\mbox{\epsfig{file=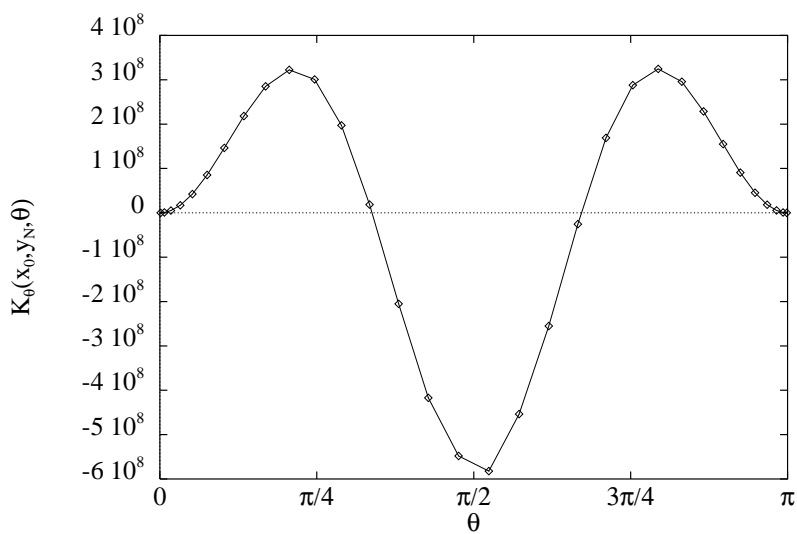,height=8cm,angle=-90}}
\end{center}
\vspace{-0.5cm}
\caption{Angular integrand $K_\theta(x_0,y_N,\theta)$ for
$\x0$, $\yN$, versus angle $\theta$ from the $\G$-equation with Ball-Chiu
vertex, for $\alpha=1.921$ and $N_f=1$ with the explicit cancellation of
the quadratic divergence.}
\label{Fig:ti-cancel_qd}
\end{figure}

\begin{figure}[htbp]
\begin{center}
\mbox{\epsfig{file=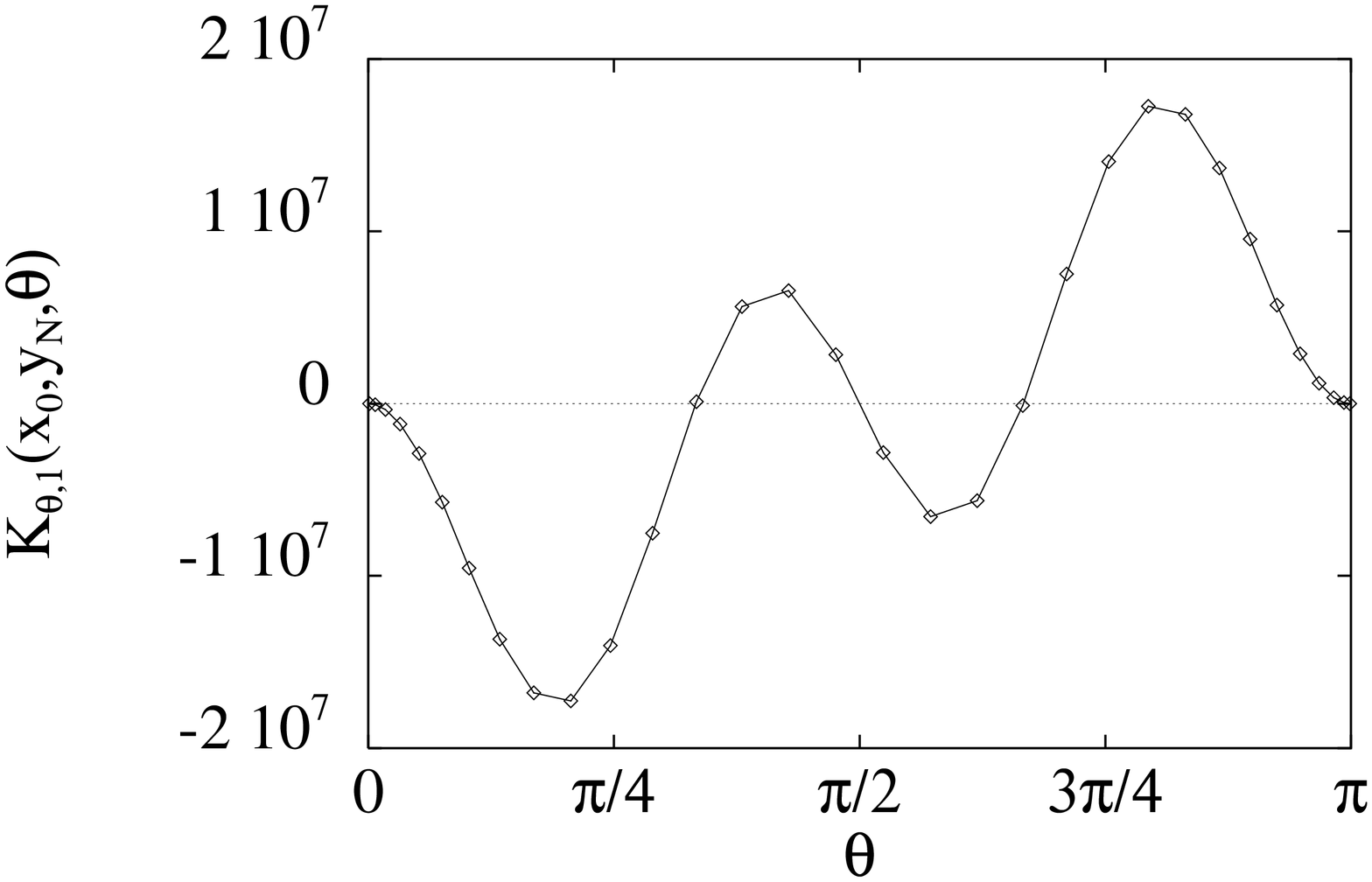,height=5.5cm,angle=-90}
\epsfig{file=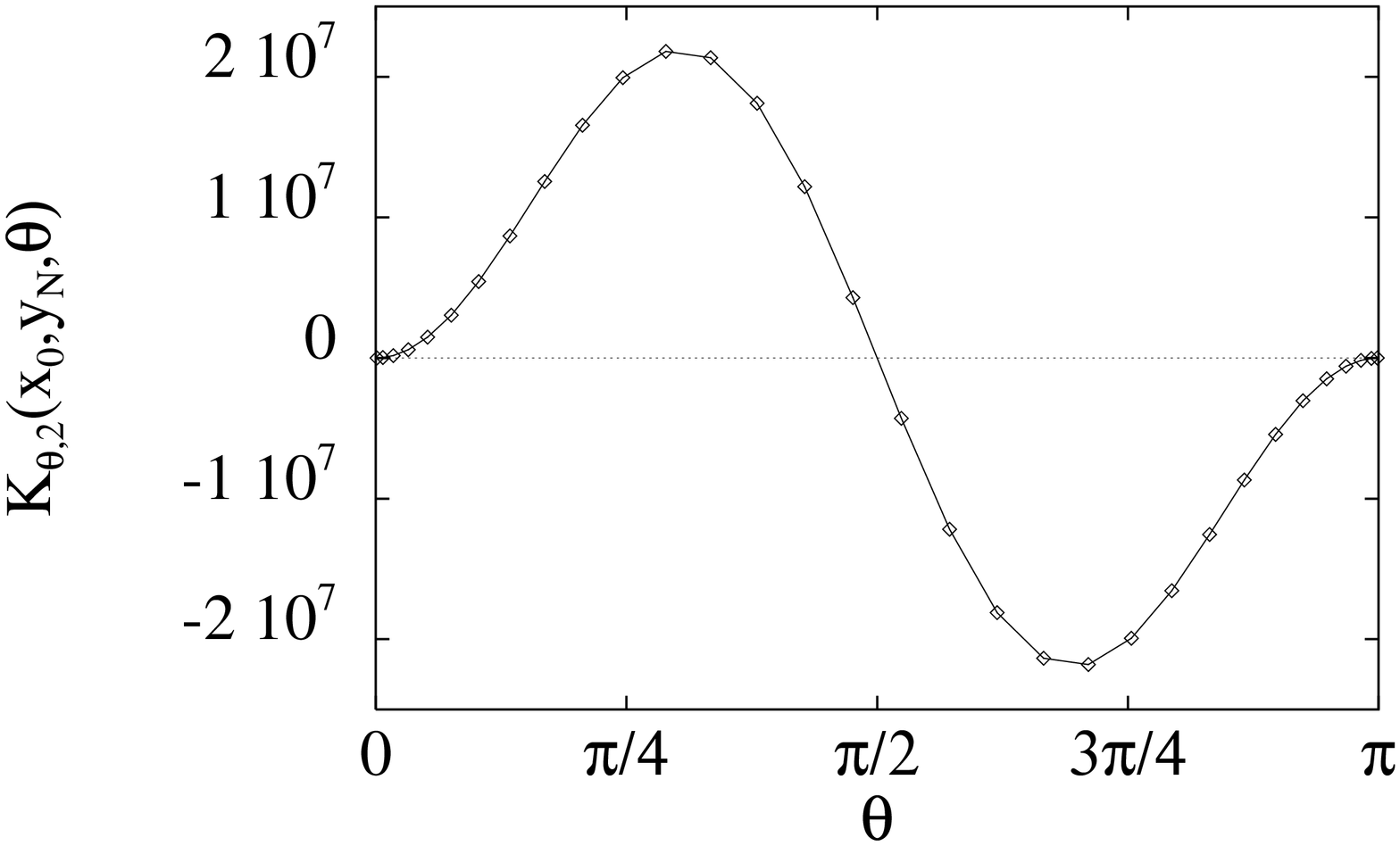,height=5.5cm,angle=-90}}
\mbox{\epsfig{file=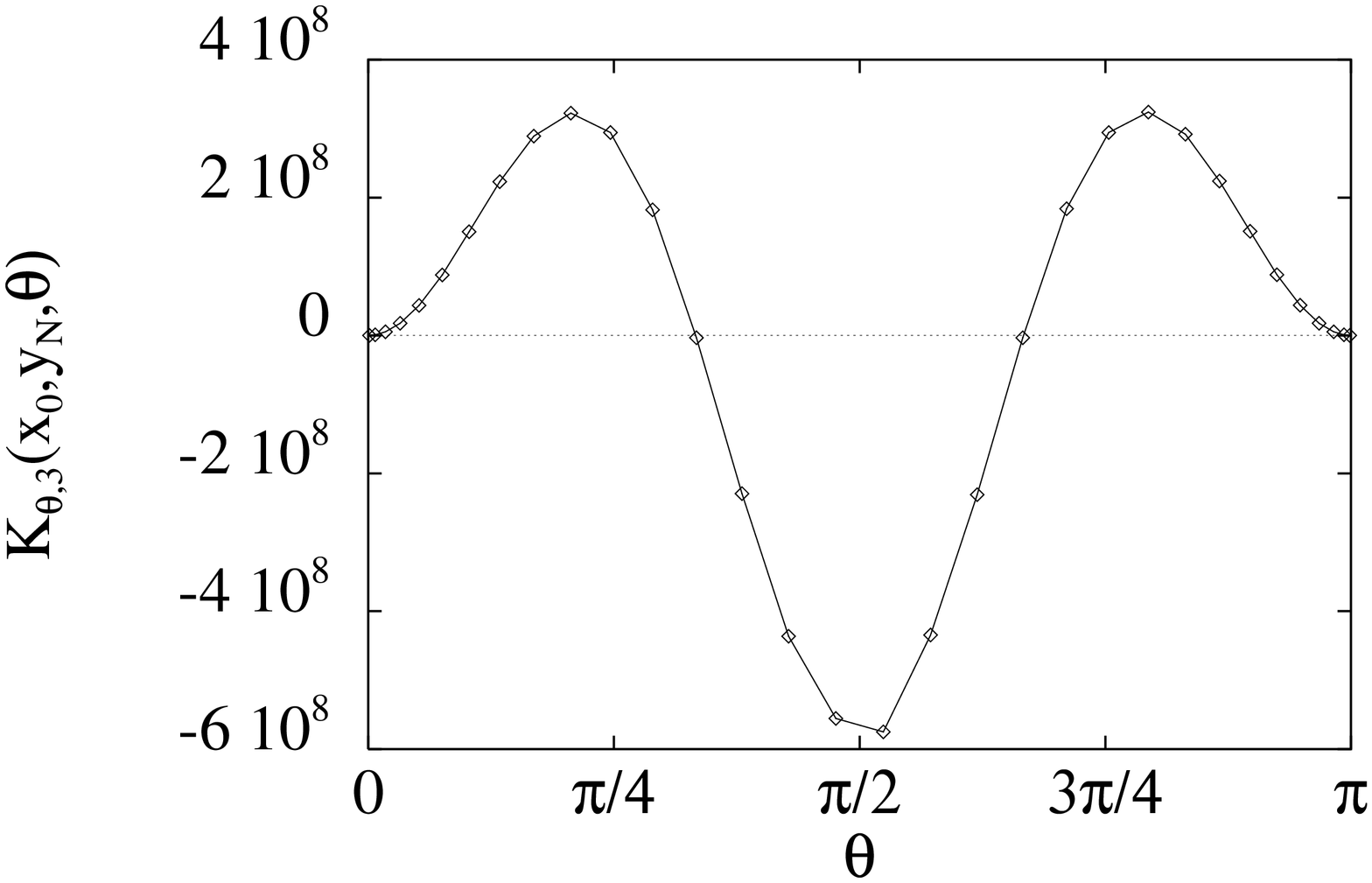,height=5.5cm,angle=-90}
\epsfig{file=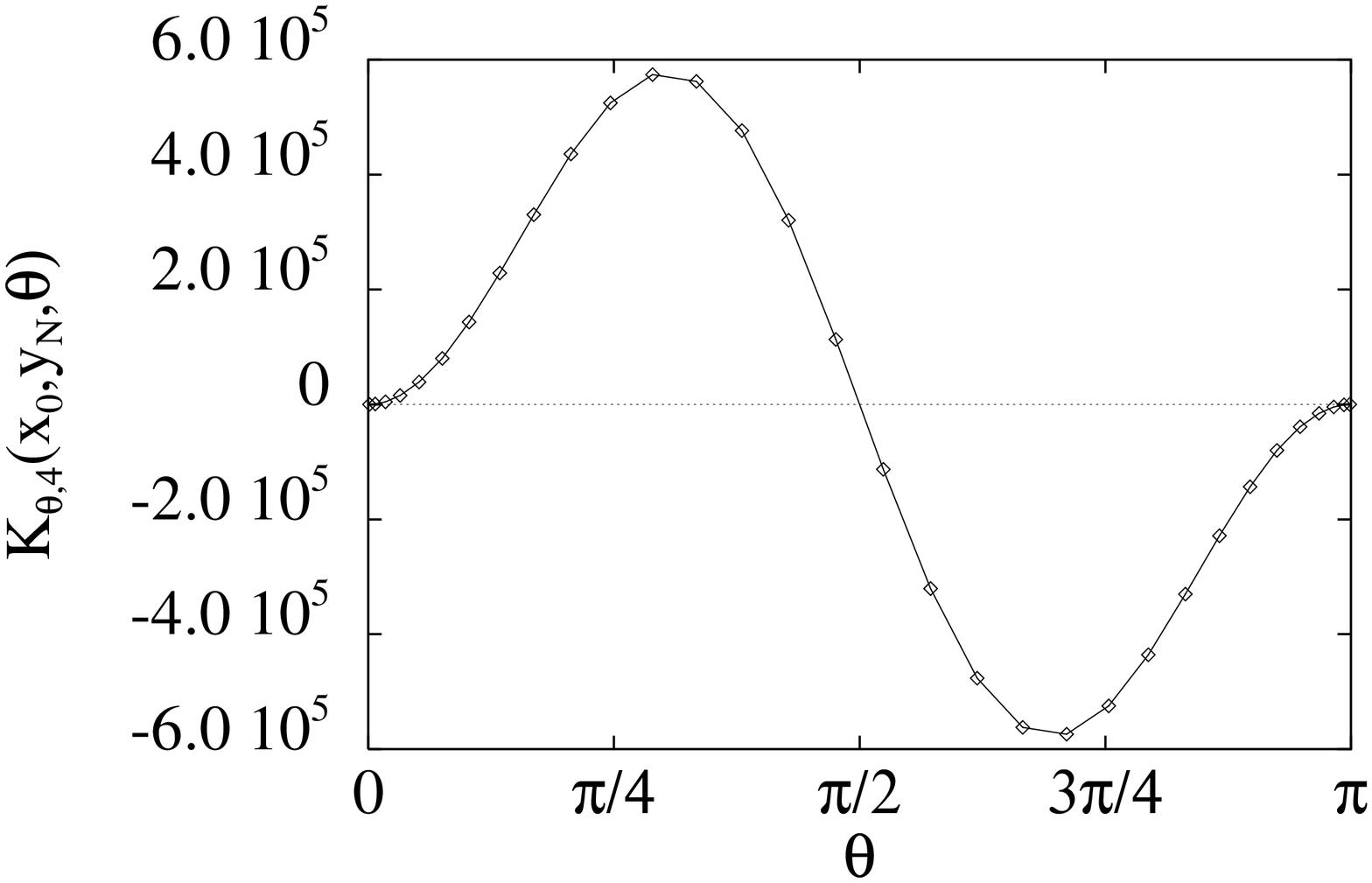,height=5.5cm,angle=-90}}
\mbox{\epsfig{file=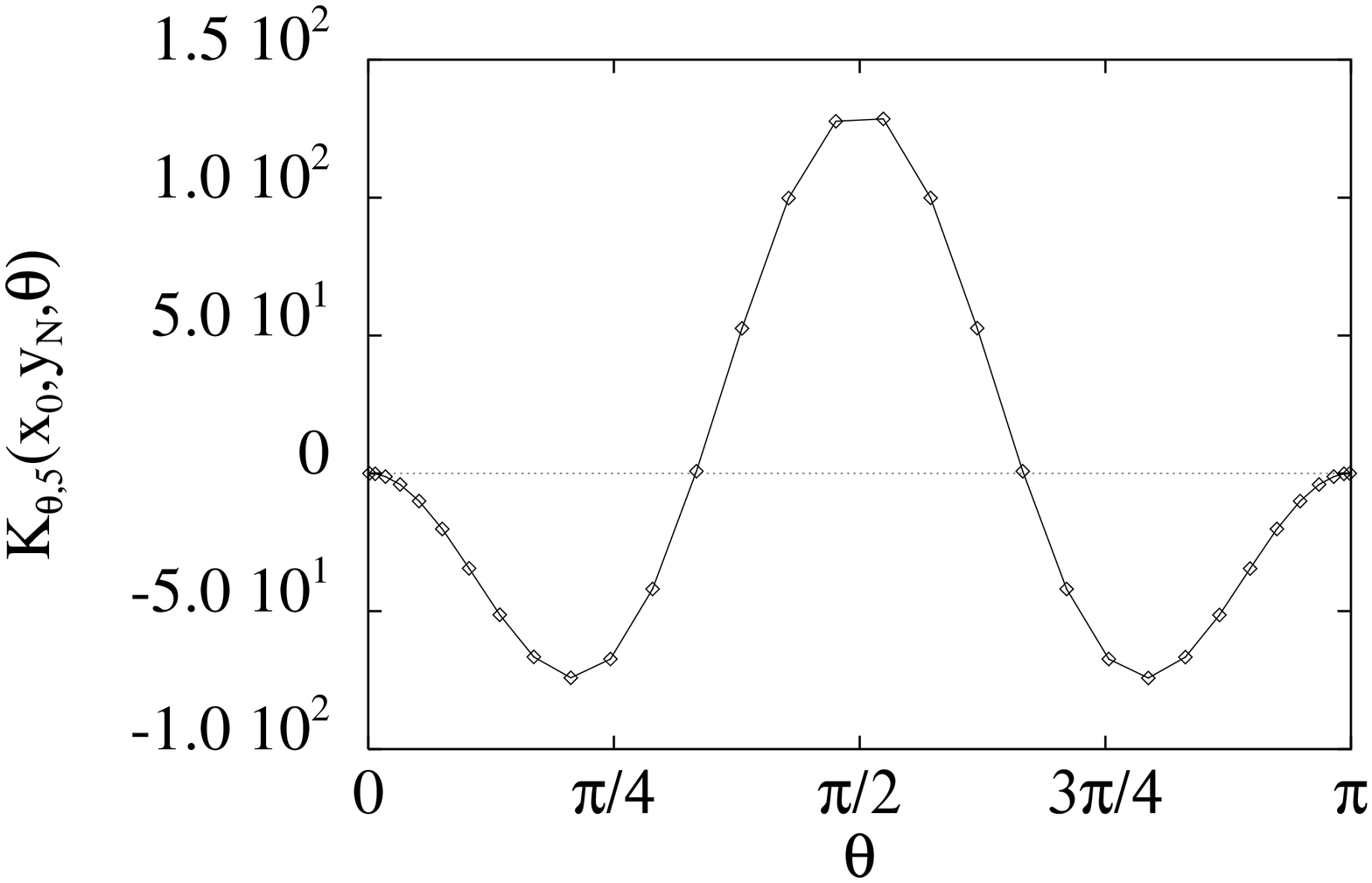,height=5.5cm,angle=-90}
\epsfig{file=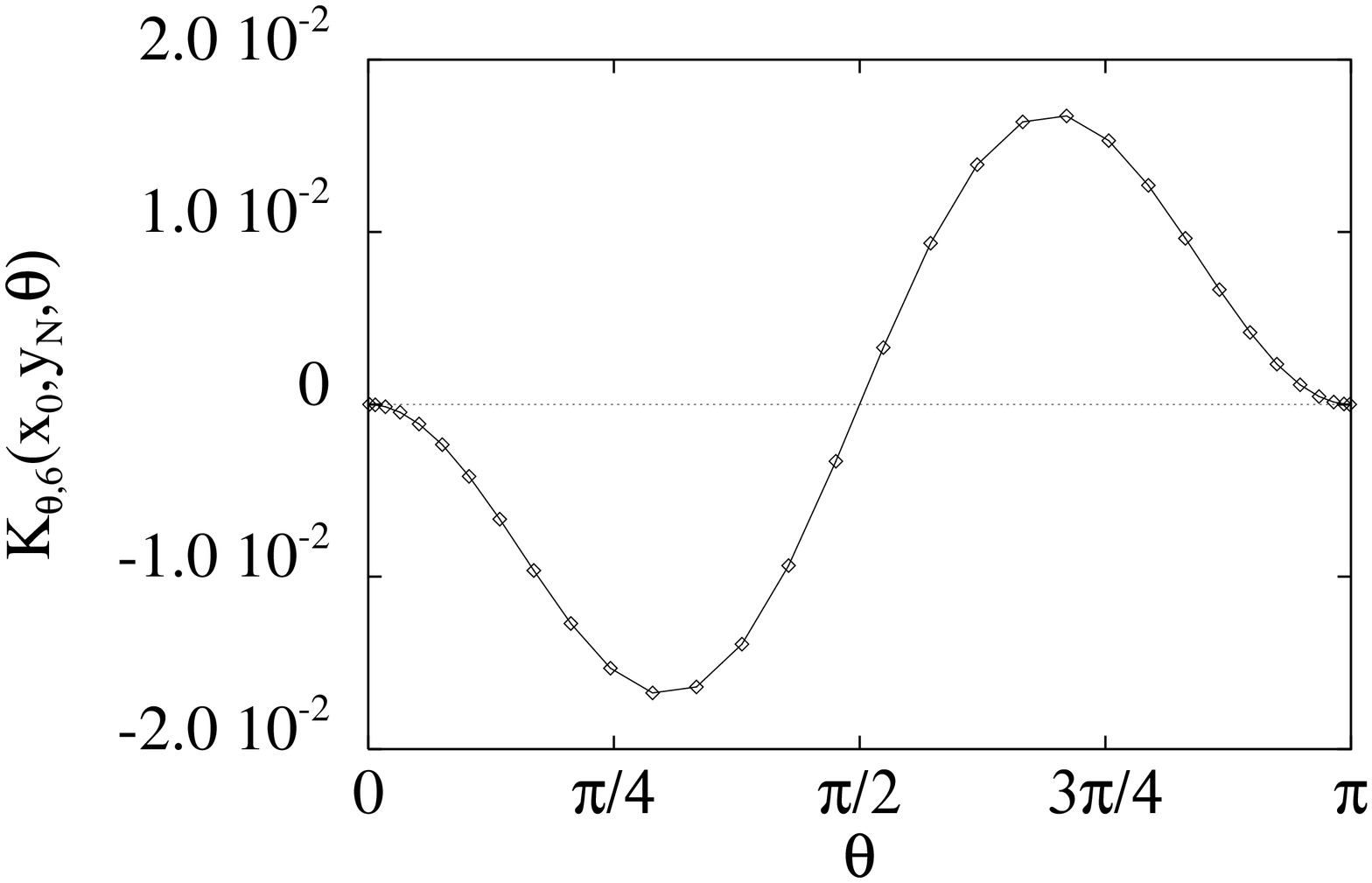,height=5.5cm,angle=-90}}
\end{center}
\vspace{-0.5cm}
\caption{Angular integrands $K_{\theta,i}(x_0,y_N,\theta)$, for $i=1,\ldots,6$,
with $\x0$, $\yN$, versus angle $\theta$ from the $\G$-equation with
Ball-Chiu vertex, for $\alpha=1.921$ and $N_f=1$ with the explicit
cancellation of the quadratic divergence.}
\label{Fig:tii-cancel_qd}
\end{figure}

Although the value of the radial integrands, $K_{R,i}(x_0,y_N)$, shown in
Table~\ref{Tab:K_R-cancel_qd} have not changed much compared to those of
Table~\ref{Tab:K_R-nocancel}, the magnitude of the angular integrands, has
been reduced as can be seen in Fig.~\ref{Fig:ti-cancel_qd} for the total
angular integrand $K_{\theta}(x_0,y_N,\theta)$ and in
Fig.~\ref{Fig:tii-cancel_qd} for the partial ones,
$K_{\theta,i}(x_0,y_N,\theta)$. However the expected reduction factor of
the $\Order(\Lambda/\sqrt{x_0})$ has not been achieved for $K_{\theta,3}$
and $K_{\theta,5}$.

\begin{figure}[htbp]
\begin{center}
\mbox{\epsfig{file=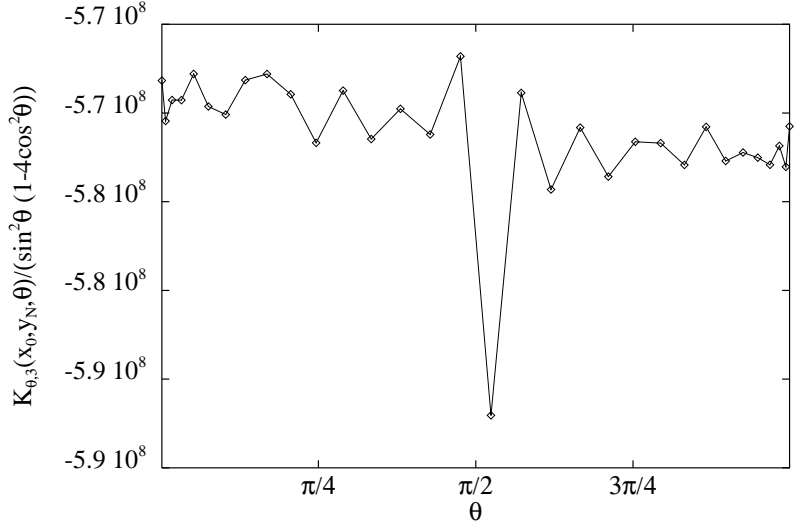,height=8cm,angle=-90}}
\end{center}
\vspace{-0.5cm}
\caption{Angular integrand $K_{\theta,3}(x_0,y_N,\theta)$ for
$\x0$, $\yN$, versus angle $\theta$ from the $\G$-equation with Ball-Chiu
vertex, for $\alpha=1.921$ and $N_f=1$ after removing the trigonometric
factor $\sin^2\theta(1-4\cos^2\theta)$.}
\label{Fig:ti3-cancel_qd-notrig}
\end{figure}

From Fig.~\ref{Fig:tii-nocancel} we saw that, to leading order in $x$, the
angular kernels had the correct trigonometric behaviour. However, from
Fig.~\ref{Fig:tii-cancel_qd}, it seems that, although $K_{\theta,1}$ seems
fine, $K_{\theta,3}$ and $K_{\theta,5}$ do not have the right
next-to-leading order behaviour predicted by the expansions,
Eqs.~(\oref{1027.2}, \oref{1027.3}), which should be proportional to
$\sin^2\theta(1-4\cos^2\theta)\cos\theta$. To emphasize this we show in
Fig.~\ref{Fig:ti3-cancel_qd-notrig} the behaviour of
$K_{\theta,3}(x,y,\theta)$ when we remove the
$\sin^2\theta(1-4\cos^2\theta)$ trigonometric part. Then, we expect
the next-to-leading order term to behave as $\cos\theta$. From
Fig.~\ref{Fig:ti3-cancel_qd-notrig} it is clear that this is not so and
that this part of the angular kernel is varying erratically between
$-5.95\ten{+08}$ and $-5.7\ten{+08}$.

These results indicate that, for some reason as yet unknown, the numerical
accuracy of the $\sqrt{x}$-term in $I_B$ is not very good. However we need
to get this right because these terms will generate a linear divergence in
the vacuum polarization, unless their angular integrals vanish to a high
degree of precision. From the derivation of the Taylor expansion of $I_B$,
\mref{1027.2}, we see that the next-to-leading order term of $I_B$ is
achieved by combining the next-to-leading order terms of its various
components with the leading order terms, except for $B(y,z)$, for which we
need the next-to-next-to-leading order term of the $1/\F(z)$ expansion as
shown in \mref{1026.2}. Of course in the numerical program $B(y,z)$ will be
computed from:
\be
B(y,z) = \frac{1}{2(z-y)}\l[\frac{1}{\F(z)}-\frac{1}{\F(y)}\r] .
\mlab{1035}
\ee

Let us write the Taylor expansion of $1/\F(z)$:
\be
\frac{1}{\F(z)} = \frac{1}{\F(y)}+ (z-y)\l[\frac{1}{\F(y)}\r]'+
\frac{(z-y)^2}{2}\l[\frac{1}{\F(y)}\r]''+ \Order(z-y)^3 .
\mlab{1036}
\ee

Because of the $(z-y)$ denominator of $B(y,z)$, the third term in $1/\F(z)$
will also contribute to the $(z-y)$-term in the expansion of
$B(y,z)$. However, for small values of $x$, subsequent terms in the Taylor
expansion will decrease by a factor of $\Order(\sqrt{x/y})$, which in our
case is of $\Order(3\ten{7})$.  As double precision arithmetic is
accurate to about 16 digits, this means that the $(z-y)^2$ contribution of
\mref{1036} will not be accurate in the numerical evaluation of
$1/\F(z)$, and thus, the next-to-leading order $(z-y)$-term of $B(y,z)$ from
\mref{1026.2} will not behave as it should do. The problem is similar
for $C(y,z)$.

We will now construct a method to get $B(y,z)$ and $C(y,z)$ with sufficient
accuracy by directly evaluating these quantities and cancelling the leading
constant term $1/\F(y)$ and $\S(y)/\F(y)$ of their expansions explicitly .

\subsection{Recurrence formula for difference of Chebyshev expansions}

Assume that $f(y)$ is a Chebyshev expansion and we want to compute the
difference
\ba
\Delta f &\equiv& f(z) - f(y) \nn \\
&=& \l[\sum_{j=0}^{N-1} c_j T_j(z) - \frac{c_0}{2}\r]
- \l[\sum_{j=0}^{N-1} c_j T_j(y) - \frac{c_0}{2}\r] \nn\\
&=& \sum_{j=0}^{N-1} c_j \l[T_j(z)-T_j(y)\r] \;. \nn
\ea

From {\it Clenshaw's} formula, \mref{820.1}, to compute the value of a
Chebyshev expansion, we now derive another, original, recurrence formula
for the difference of two Chebyshev expansions. Subtracting
\mref{820.1} at two arbitrary points, yields:
\ba
f(z) - f(y) &=& \l[ z d_1(z) - d_2(z) + \frac{c_0}{2} \r]
- \l[ y d_1(y) - d_2(y) + \frac{c_0}{2} \r] \nn\\
&=& z d_1(z) - y d_1(y) - (d_2(z) - d_2(y)) 
\mlab{1037} \\
\mbox{and}\hspace{3cm} \nn \\
d_j(z) - d_j(y) &=& \l[2z d_{j+1}(z) - d_{j+2}(z) + c_j\r]
- \l[2y d_{j+1}(y) - d_{j+2}(y) + c_j\r] \hspace{2cm}\nn\\
&=& 2(z d_{j+1}(z) - y d_{j+1}(y)) - (d_{j+2}(z)-d_{j+2}(y)) \;.
\mlab{1038}
\ea

From \mref{1038} it seems logical to look for a recurrence formula for:
\ba
z d_{j}(z) - y d_{j}(y) &=& \l[2z^2 d_{j+1}(z) - z d_{j+2}(z) + z c_j\r]
- \l[2y^2 d_{j+1}(y) - y d_{j+2}(y) + y c_j\r] \nn\\
&=& 2\Big(z^2 d_{j+1}(z)-y^2 d_{j+1}(y)\Big) - 
\Big(z d_{j+2}(y) - y d_{j+2}(y)\Big) + (z-y)c_j \nn\\
&=& 2\bigg[(z+y)\Big(z d_{j+1}(z)-y d_{j+1}(y)\Big) 
- zy\Big(d_{j+1}(z)-d_{j+1}(y)\Big)\bigg] \nn\\
&& - \Big(z d_{j+2}(z) - y d_{j+2}(y)\Big) + (z-y)c_j \mlab{1039}.
\ea

We now define $\alpha_j(z,y)$ and $\beta_j(z,y)$ as:
\ba
\alpha_j(z,y) &\equiv& \frac{d_j(z) - d_j(y)}{z-y}  \nn\\
\beta_j(z,y) &\equiv& \frac{z d_j(z) - y d_j(y)}{z-y} ,
\ea

such that Eqs.~(\oref{1037}, \oref{1038}, \oref{1039}) can be written as
the following recurrence relation:
\ba
\fbox{\parbox{12cm}{\vspace{-3mm}\bann
\Delta f &\equiv& f(z) - f(y) = 
(z-y) \Big(\beta_1(z,y) - \alpha_2(z,y)\Big) \nn\\[2mm]
\alpha_j(z,y) &=&  2 \beta_{j+1}(z,y) - \alpha_{j+2}(z,y) \\[2mm]
\beta_j(z,y) &=&  c_j + 2 (z+y)\beta_{j+1}(z,y) - 2 zy\alpha_{j+1}(z,y) 
- \beta_{j+2}(z,y) \nn
\eann}}
\mlab{1040}
\ea
and $j=N-1,\ldots,1$, $\beta_{N+1}=\beta_N=\alpha_{N+1}=\alpha_N=0$.

This recurrence formula ensures that the leading order term of $\Delta f$
goes as $(z-y)$; the leading order, constant, terms of $f(z)$ and $f(y)$ are
automatically cancelled.

Implementing this in our numerical program, to improve the accuracy of the
computation of $B(y,z)$ and $C(y,z)$, requires these functions to be
written as a difference of Chebyshev expansions. Therefore we will first
construct the Chebyshev expansion of the functions $1/\F(x)$ and
$\S(x)/\F(x)$ which can easily be done using the known expansions for
$\S(x)$ and $\F(x)$. The results achieved with this method are discussed
below.

\begin{figure}[htbp]
\begin{center}
\mbox{\epsfig{file=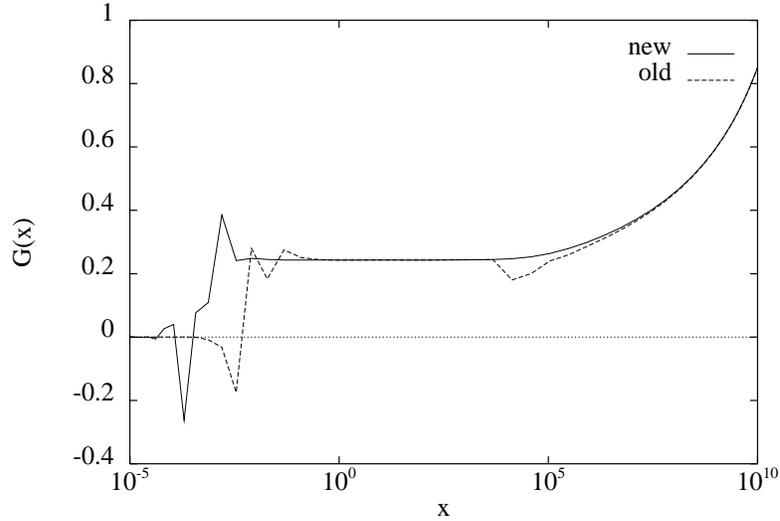,height=8cm,angle=-90}}
\end{center}
\vspace{-0.5cm}
\caption{Photon renormalization function $\G(x)$ versus momentum 
squared $x$ from the $\G$-equation with Ball-Chiu
vertex, for $\alpha=1.921$ and $N_f=1$ using the Chebyshev subtraction
scheme~(new) and without it~(old).}
\label{Fig:G-chebsub}
\end{figure}

In Fig.~\ref{Fig:G-chebsub} we compare the new behaviour of the photon
renormalization function $\G(x)$ with that of Fig.~\ref{Fig:G-nocancel}.
Although we see that for $10^4 < x < 10^7$ and $10^{-2} < x < 1$ the new
calculation for $\G(x)$ is much more stable, the smallest $x$-values still
seem to be problematic. As before we will look at the various radial and
angular integrals to find a clue where the inaccuracy comes from. 

\begin{figure}[htbp]
\begin{center}
\mbox{\epsfig{file=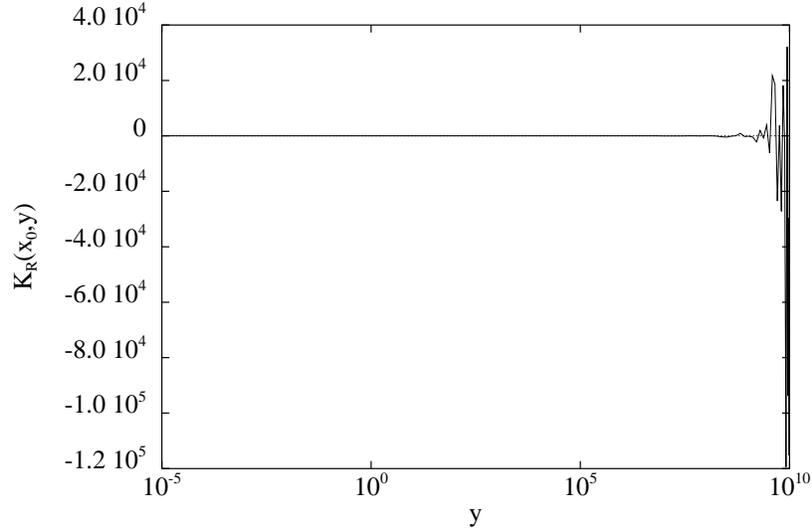,height=8cm,angle=-90}}
\end{center}
\vspace{-0.5cm}
\caption{Radial integrand $K_R(x_0,y)$ for $\x0$ versus 
radial momentum squared $y$ from the $\G$-equation with 
Ball-Chiu vertex, for $\alpha=1.921$ and $N_f=1$ using the Chebyshev
subtraction scheme.}
\label{Fig:radint-chebsub}
\end{figure}

\begin{figure}[htbp]
\begin{center}
\mbox{\epsfig{file=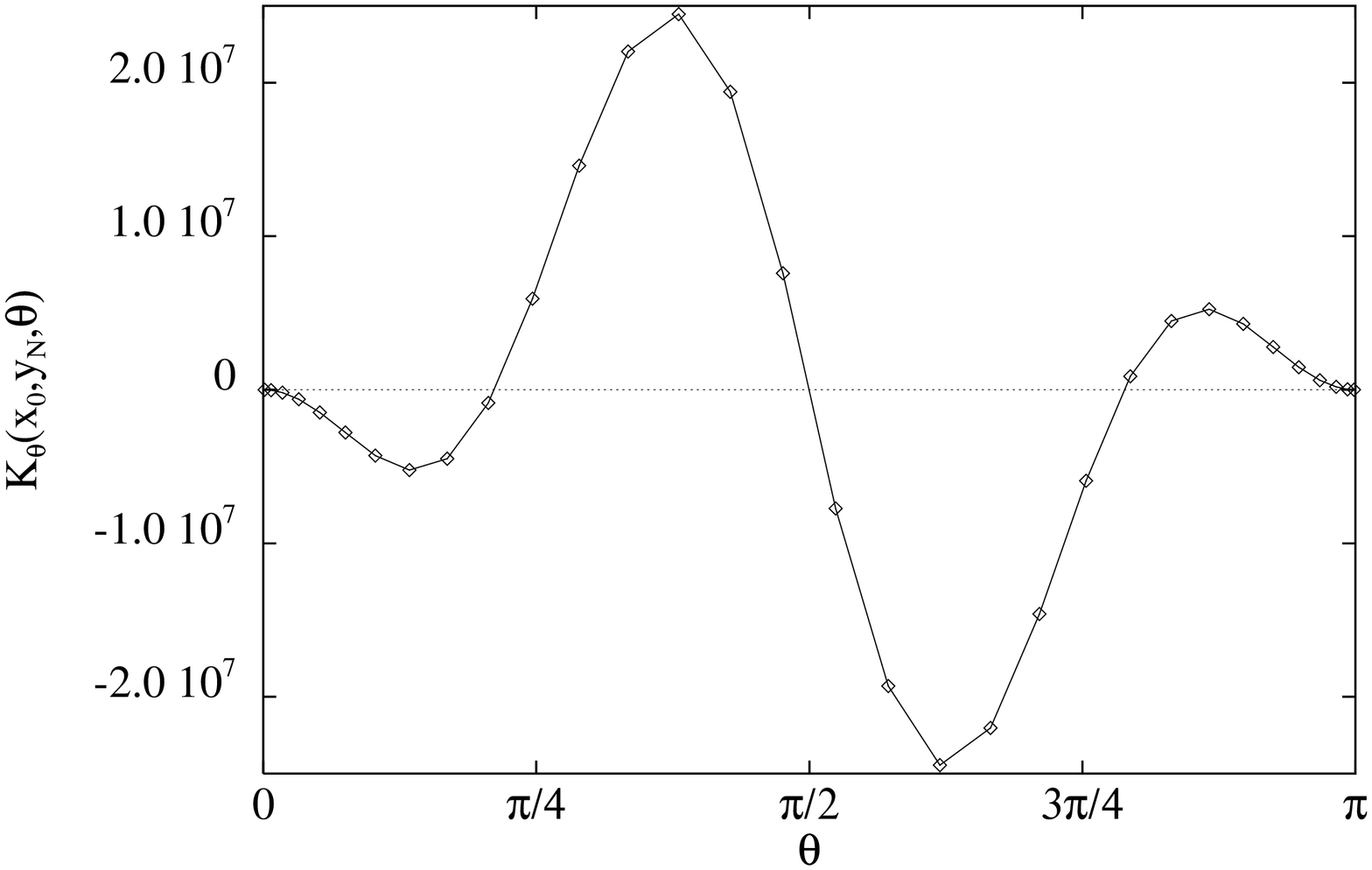,height=8cm,angle=-90}}
\end{center}
\vspace{-0.5cm}
\caption{Angular integrand $K_\theta(x_0,y_N,\theta)$ for
$\x0$, $\yN$, versus angle $\theta$ from the $\G$-equation 
with Ball-Chiu vertex, for $\alpha=1.921$ and $N_f=1$ using the
Chebyshev subtraction scheme.}
\label{Fig:ti-chebsub}
\end{figure}

The radial kernel $K_R(x_0,y)$, for $\x0$, is shown in
Fig.~\ref{Fig:radint-chebsub}. Comparing this with
Fig.~\ref{Fig:radint-nocancel} we see that the new improvement has
decreased the magnitude of the radial kernel. The value of the radial
integrand at $\x0$ and $\yN$ is
$K_R(x_0,y_N)=-834.459$.  This value is the integral of the angular kernel,
$K_{\theta}(x_0,y_N,\theta)$, shown in Fig.~\ref{Fig:ti-chebsub}.

\begin{figure}[htbp]
\begin{center}
\mbox{\epsfig{file=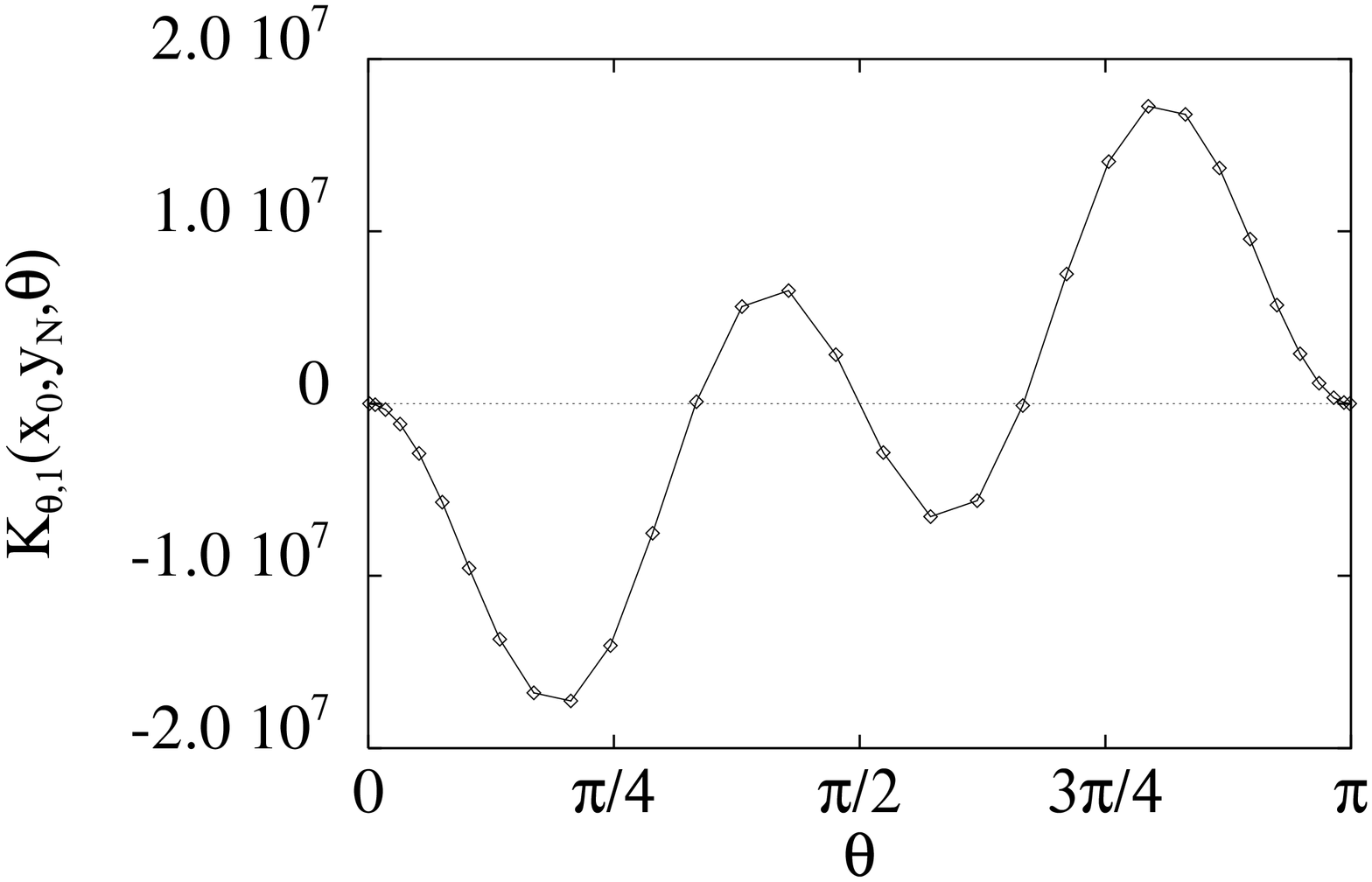,height=5.5cm,angle=-90}
\epsfig{file=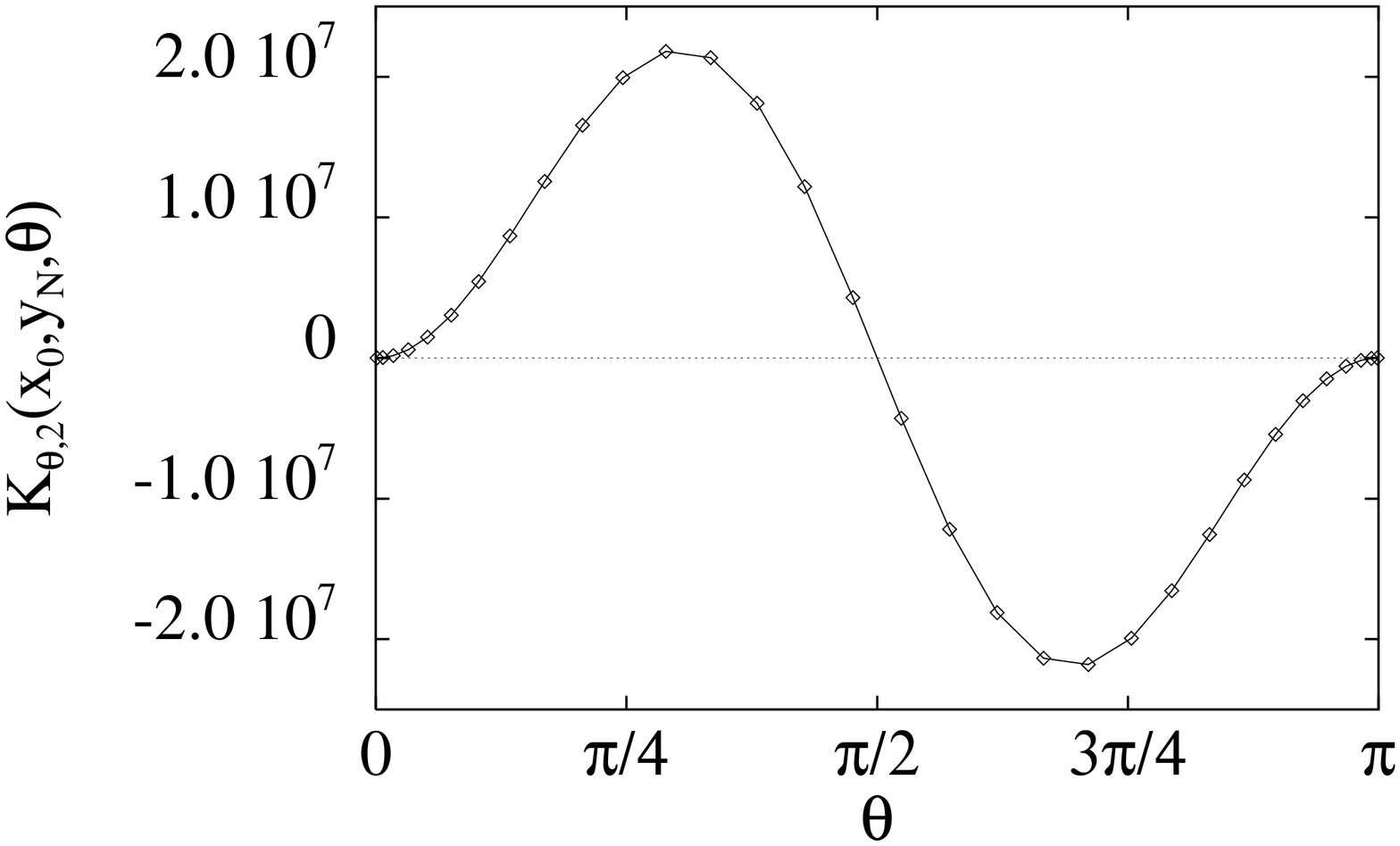,height=5.5cm,angle=-90}}
\mbox{\epsfig{file=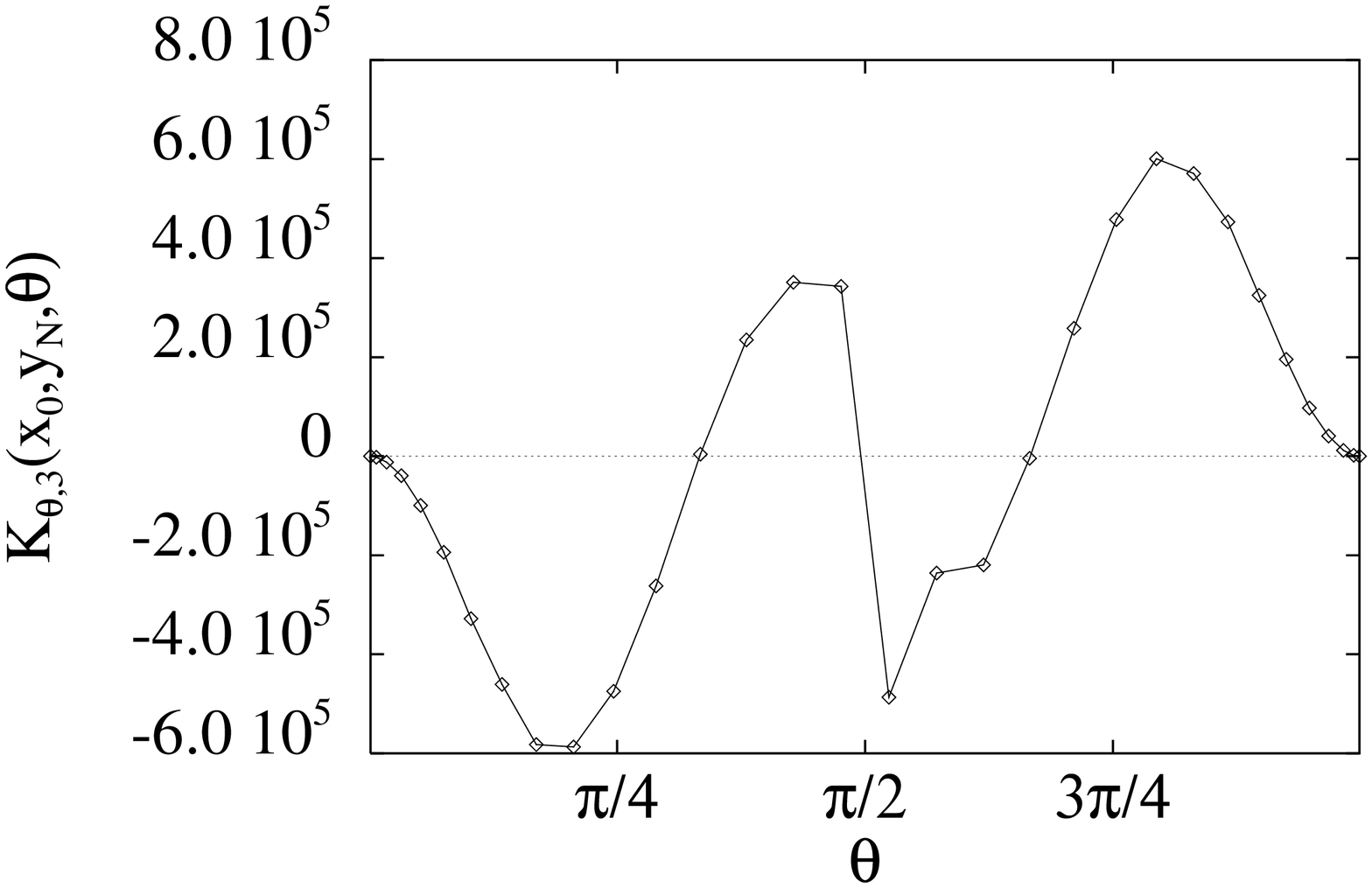,height=5.5cm,angle=-90}
\epsfig{file=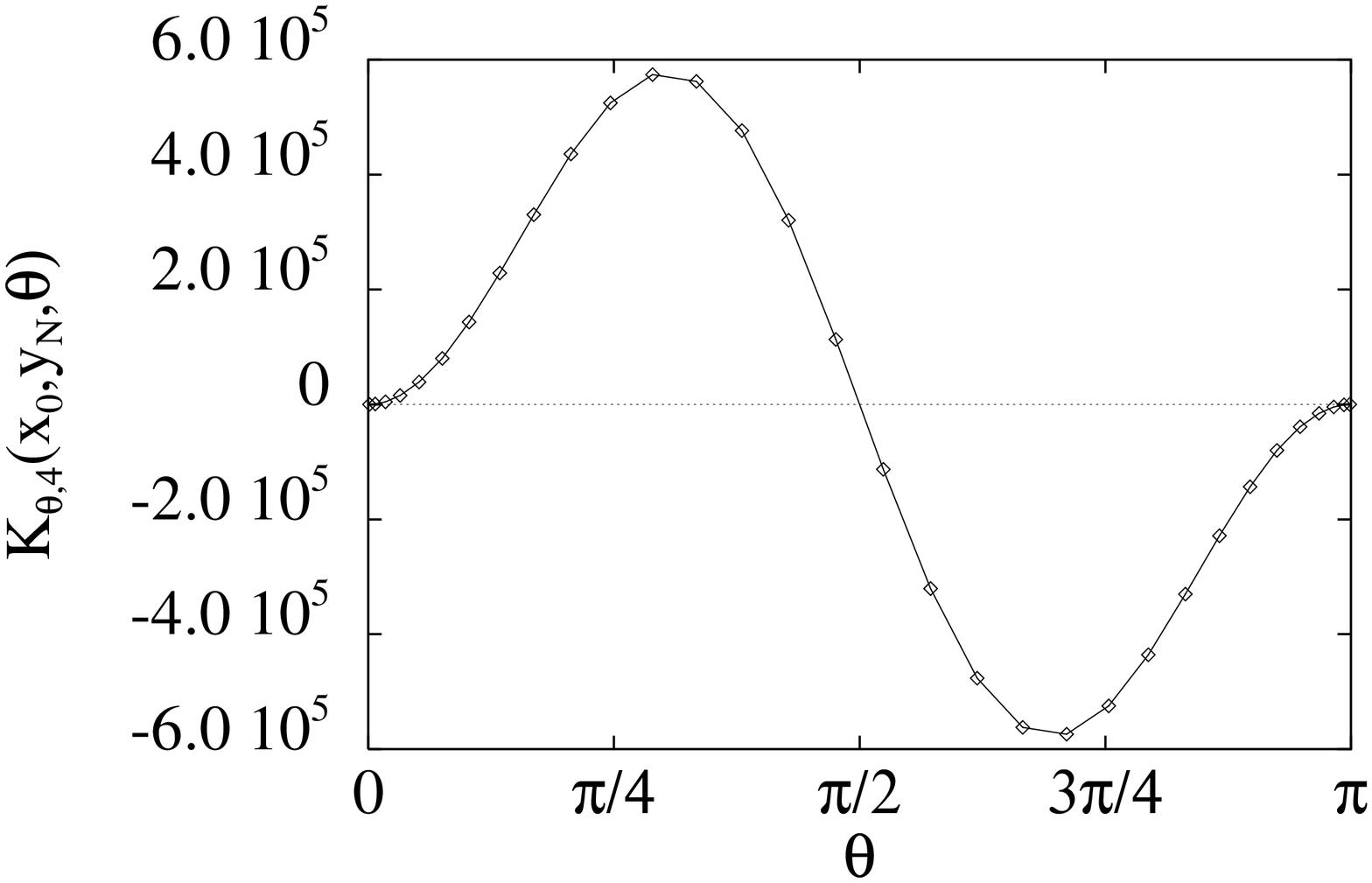,height=5.5cm,angle=-90}}
\mbox{\epsfig{file=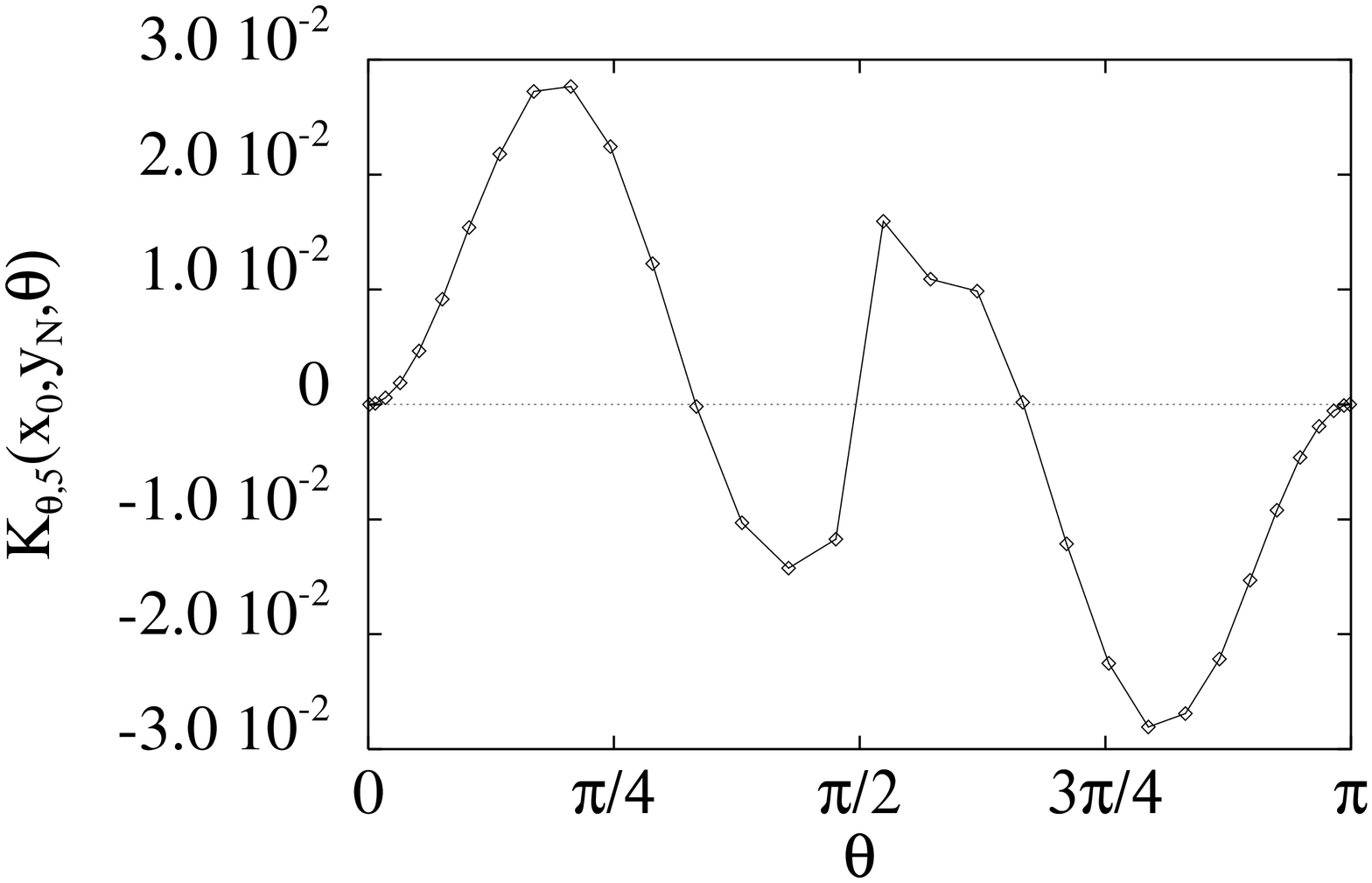,height=5.5cm,angle=-90}
\epsfig{file=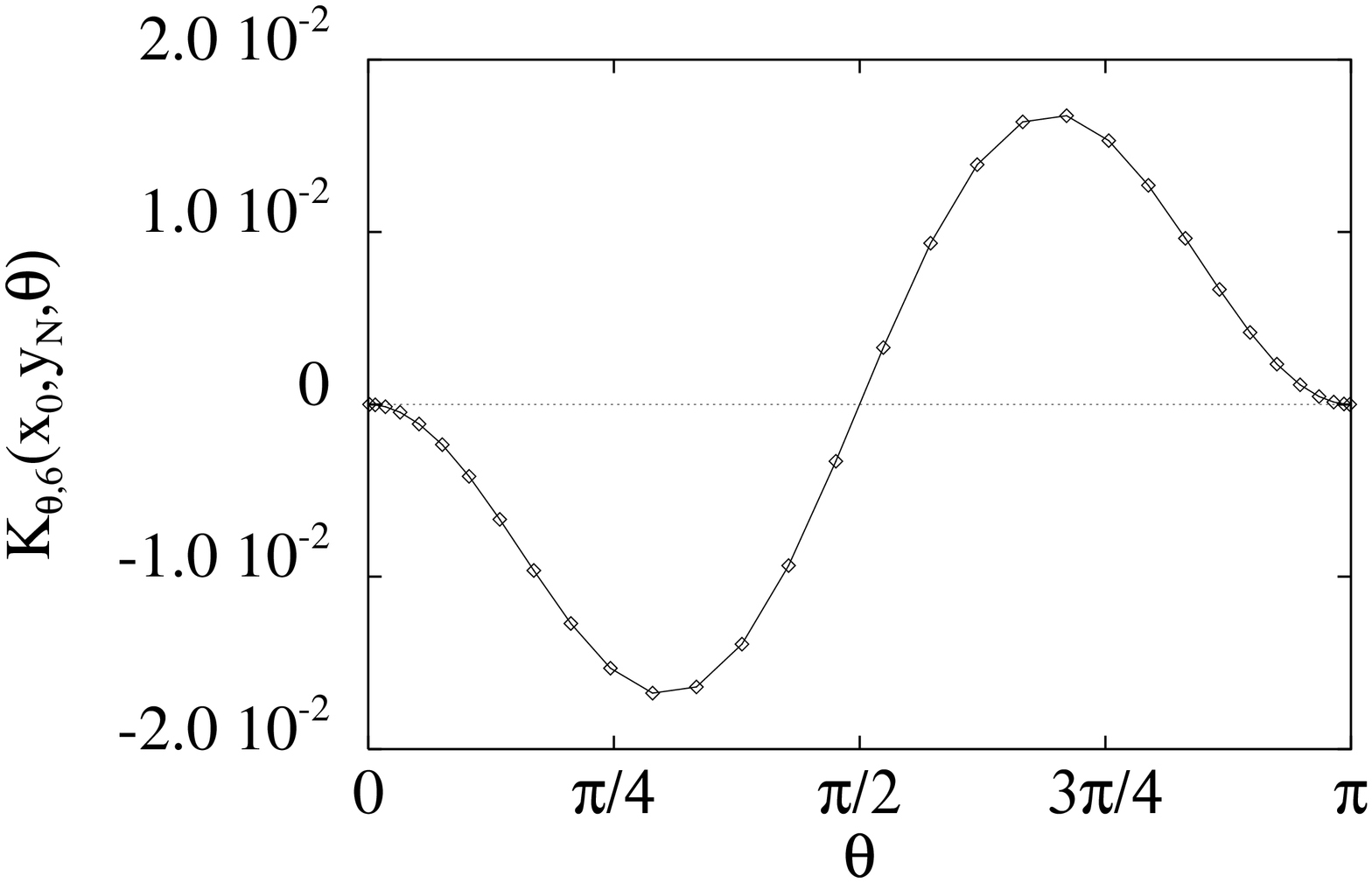,height=5.5cm,angle=-90}}
\end{center}
\vspace{-0.5cm}
\caption{Angular integrands $K_{\theta,i}(x_0,y_N,\theta)$, for $i=1,\ldots,6$,
with $\x0$, $\yN$, versus angle $\theta$ from the $\G$-equation 
with Ball-Chiu vertex, for $\alpha=1.921$ and $N_f=1$ using the Chebyshev
subtraction scheme.}
\label{Fig:tii-chebsub}
\end{figure}

If we again split the total radial integrand into six parts, $K_{R,i}$, for
$i=1,\ldots,6$, their individual values can be found in
Table~\ref{Tab:K_R-chebsub_qd}.
\begin{table}[htbp]
\begin{center}
\begin{tabular}{|l|r|}
\hline
$K_{R,1}$ & $-9.45396\ten{-01}$ \nn\\
$K_{R,2}$ &  $1.44340\ten{+00}$   \nn\\
$K_{R,3}$ & $-8.34999\ten{+02}$  \nn\\
$K_{R,4}$ &  $4.23751\ten{-02}$  \nn\\
$K_{R,5}$ &  $2.43633\ten{-05}$  \nn\\
$K_{R,6}$ & $-1.61445\ten{-09}$  \nn\\
\hline
\end{tabular}
\end{center}
\caption{Radial kernels $K_{R,i}(x_0,y_N)$, for $i=1,\ldots,6$, 
with $\x0$ and $\yN$ using the Chebyshev
subtraction scheme.}
\label{Tab:K_R-chebsub_qd}
\end{table}
The angular kernels from which these integral values are computed are shown
in Fig.~\ref{Fig:tii-chebsub}.

\begin{figure}[htbp]
\begin{center}
\mbox{\epsfig{file=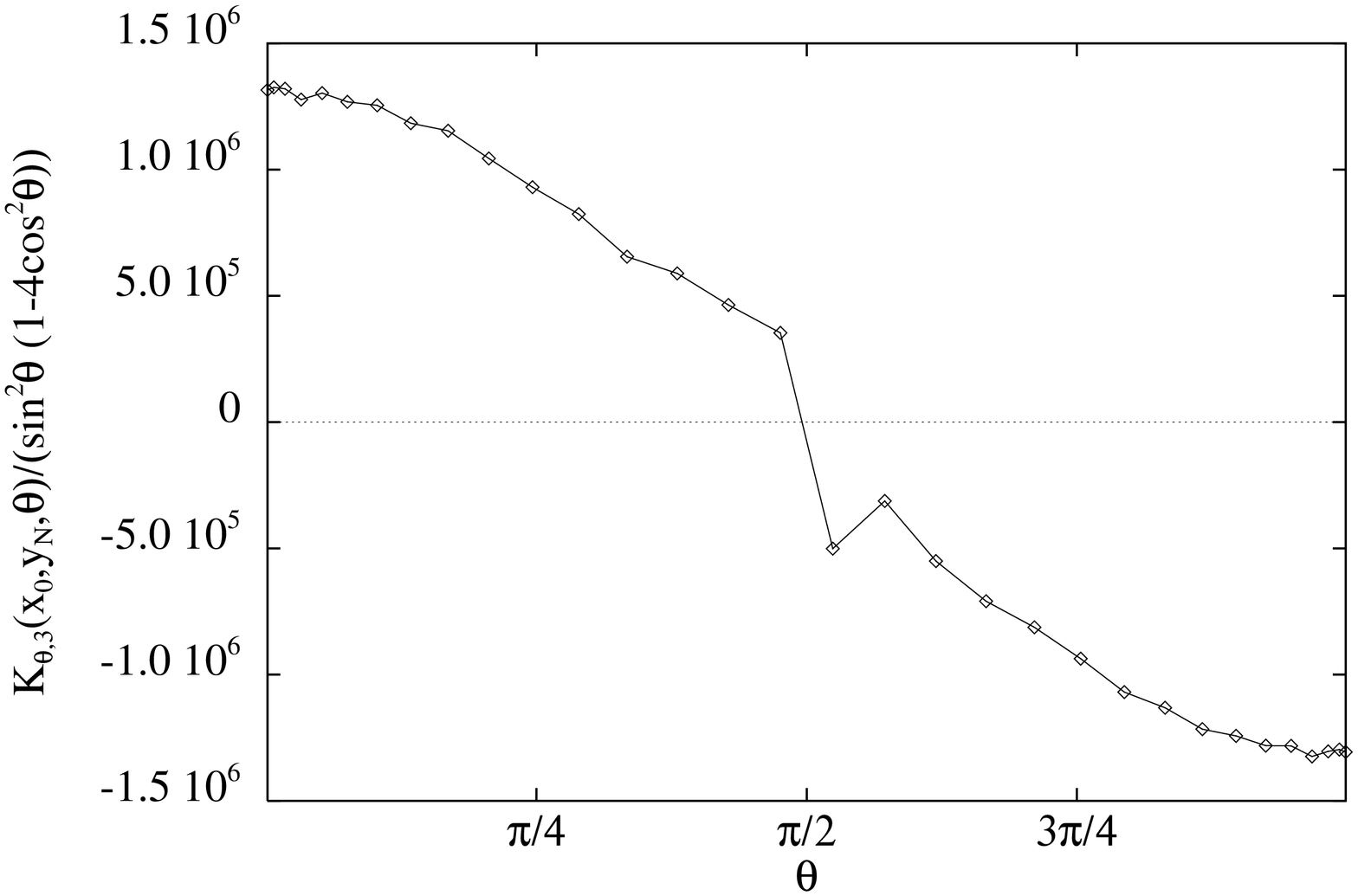,height=8cm,angle=-90}}
\end{center}
\vspace{-0.5cm}
\caption{Angular integrand $K_{\theta,3}(x_0,y_N,\theta)$ for
$\x0$, $\yN$, versus angle $\theta$ from the $\G$-equation 
with Ball-Chiu vertex, for $\alpha=1.921$ and $N_f=1$ after removing
the trigonometric factor $\sin^2\theta(1-4\cos^2\theta)$ using the
Chebyshev subtraction scheme.}
\label{Fig:ti3-chebsub-notrig}
\end{figure}

From Fig.~\ref{Fig:ti3-chebsub-notrig} we see that the angular kernel
$K_{\theta,3}$ after removal of the trigonometric factor
$\sin^2\theta(1-4\cos^2\theta)$ tends to the correct $\cos\theta$-shape.
However, the remaining inaccuracy is still responsible for unacceptable
instabilities in $\G(x)$ for small $x$. Next we will introduce the final
step to improve this.

\subsection{Alternative logarithm calculation}

Using the Chebyshev subtraction scheme \mref{1040} we have ensured that the
difference of two Chebyshev expansions, $f(z)-f(y)$, has a leading term
which will be proportional to $(z-y)$. However, the functions $\S(x)$,
$\F(x)$ and $\G(x)$ are expanded in Chebyshev polynomials of $s(x)$ rather
than $x$. Therefore, the Chebyshev subtraction scheme will ensure that the
leading term of the differences $f(s(z))-f(s(y))$ will be proportional to
$s(z)-s(y)$. Of course, analytically, $s(z)-s(y)$ itself will be
proportional to $(z-y)$ in leading order. However, this is again a possible
source of numerical inaccuracy.  Remember the definition of $s(x)$:
\be
s(x) = \frac{\logten(x/\Lambda\kappa)}{\logten(\Lambda/\kappa)}.
\mlab{1041}
\ee

Then,
\be
s(z)-s(y) = \frac{\logten(z/\Lambda\kappa)}{\logten(\Lambda/\kappa)}
-\frac{\logten(y/\Lambda\kappa)}{\logten(\Lambda/\kappa)}
= \frac{\logten(z/y)}{\logten(\Lambda/\kappa)}.
\ee

From Figs.~\ref{Fig:tii-chebsub}, \ref{Fig:ti3-chebsub-notrig} we see that
the problem resides around $\theta=\pi/2$. 
We can write, 
\[\frac{z}{y}=1+\frac{x-2\sqrt{xy}\cos\theta}{y}\]
and if $x$ is very small and $\theta\approx\pi/2$, the second term in this
last expression will be much smaller than one and its accuracy can be
completely lost. However it is exactly this bit of the expression which
determines completely the answer of $\logten(z/y)$ and therefore its
inaccuracy will be responsible for the incorrect cancellation in the
angular integrals. To improve on this we cannot anymore use the logarithm
function of the standard mathematical library of the computer; instead we
will implement our own routine for values of $z$ very close to $y$. We will
use the following Taylor series~\cite{Spiegel}:
\be
\hspace{3cm}
\frac{1}{2}\ln\frac{1+u}{1-u} = u + \frac{u^3}{3} + \frac{u^5}{5} +
\frac{u^7}{7} + \cdots , \qquad -1<u<1.
\mlab{1042}
\ee

If we define $u \equiv \D\frac{z-y}{z+y}$ then,
\be
\frac{z}{y} = \frac{1+u}{1-u} .
\ee

\mref{1042} ensures that the leading term of $\logten(z/y)$ will be
proportional to $(z-y)$.  We implement the Taylor series up to $u^7$-terms
and use it when $|u| < 10^{-3}$, otherwise we use the standard logarithm
routine. 

\begin{figure}[htbp]
\begin{center}
\mbox{\epsfig{file=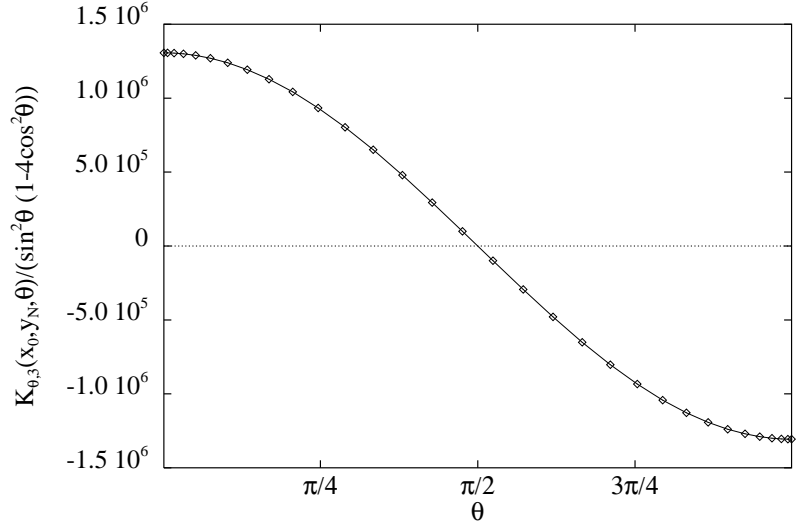,height=8cm,angle=-90}}
\end{center}
\vspace{-0.5cm}
\caption{Angular integrand $K_{\theta,3}(x_0,y_N,\theta)$ for
$\x0$, $\yN$, versus angle $\theta$ from the $\G$-equation with Ball-Chiu
vertex, for $\alpha=1.921$ and $N_f=1$ after removing the trigonometric
factor $\sin^2\theta(1-4\cos^2\theta)$ using the alternative logarithm
calculation.}
\label{Fig:ti3-ldif-notrig}
\end{figure}

\begin{figure}[htbp]
\begin{center}
\mbox{\epsfig{file=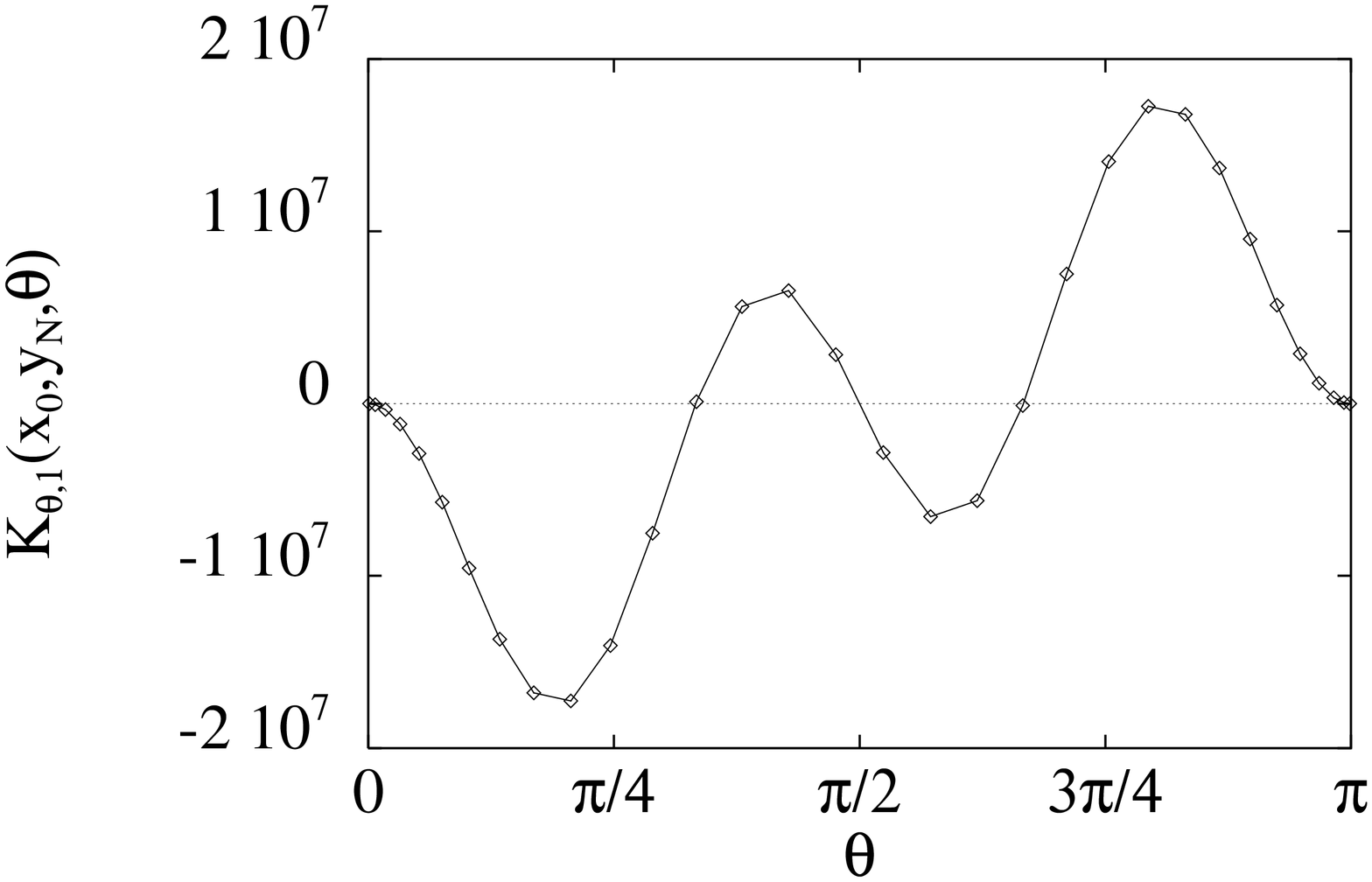,height=5.5cm,angle=-90}
\epsfig{file=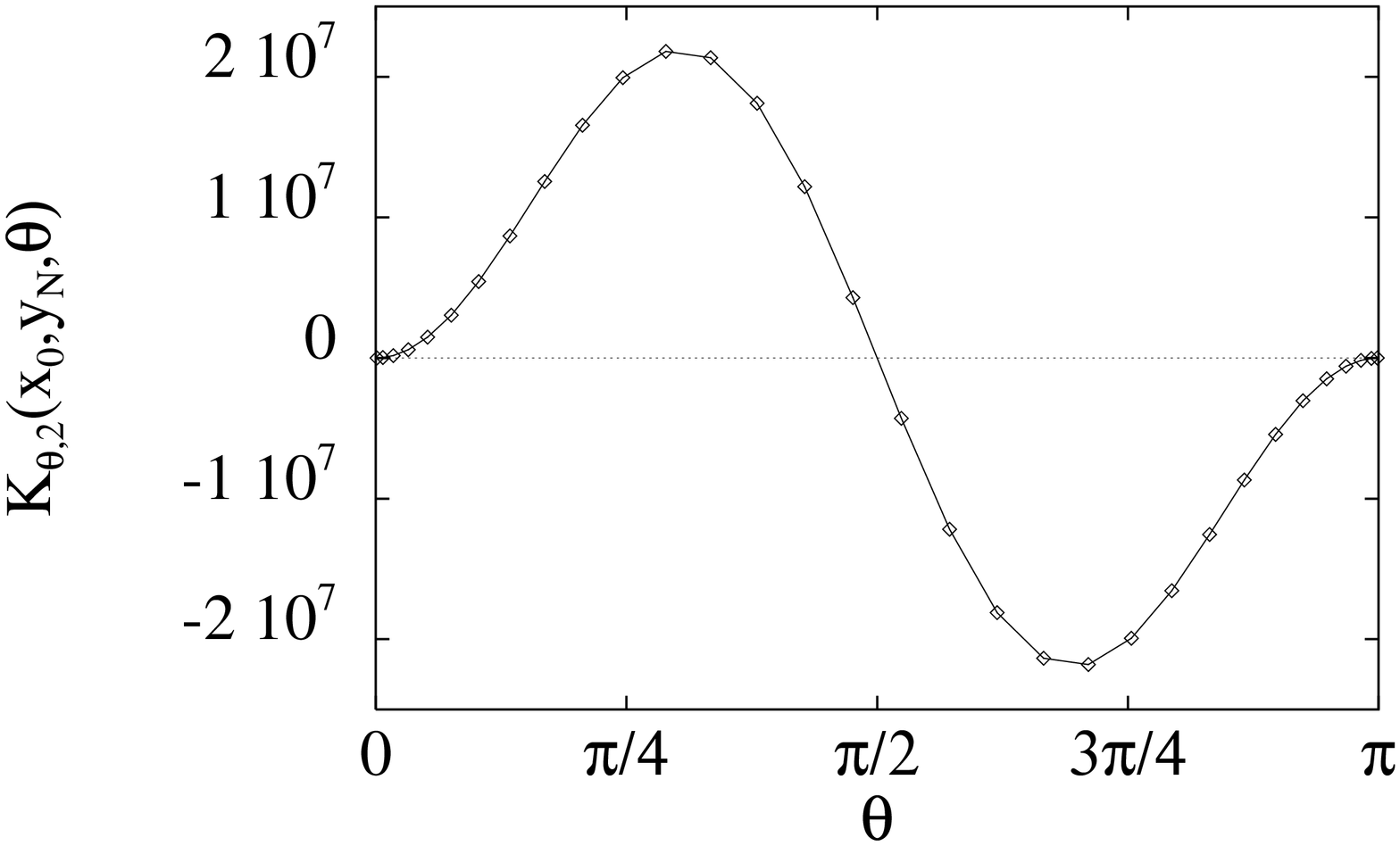,height=5.5cm,angle=-90}}
\mbox{\epsfig{file=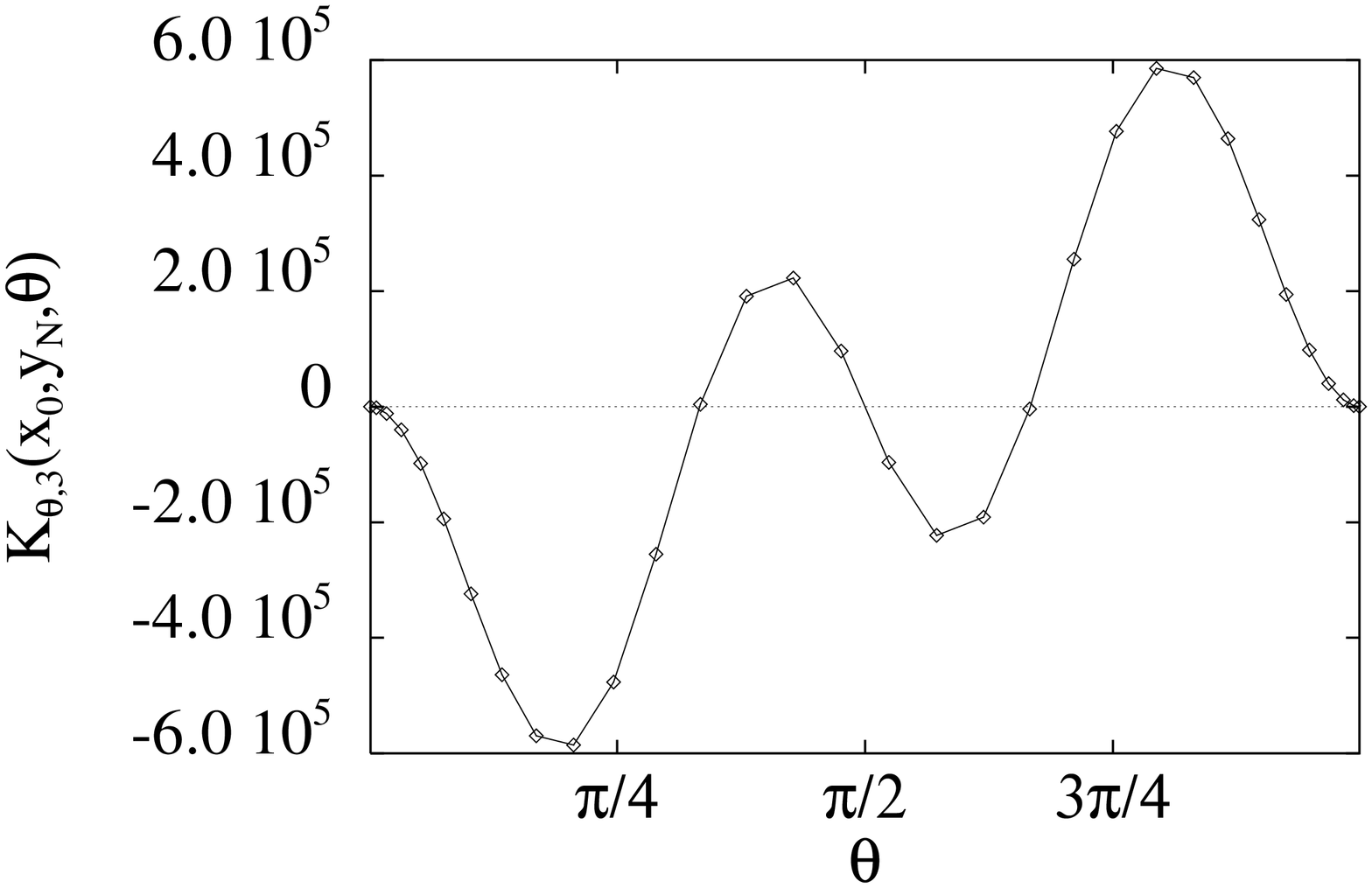,height=5.5cm,angle=-90}
\epsfig{file=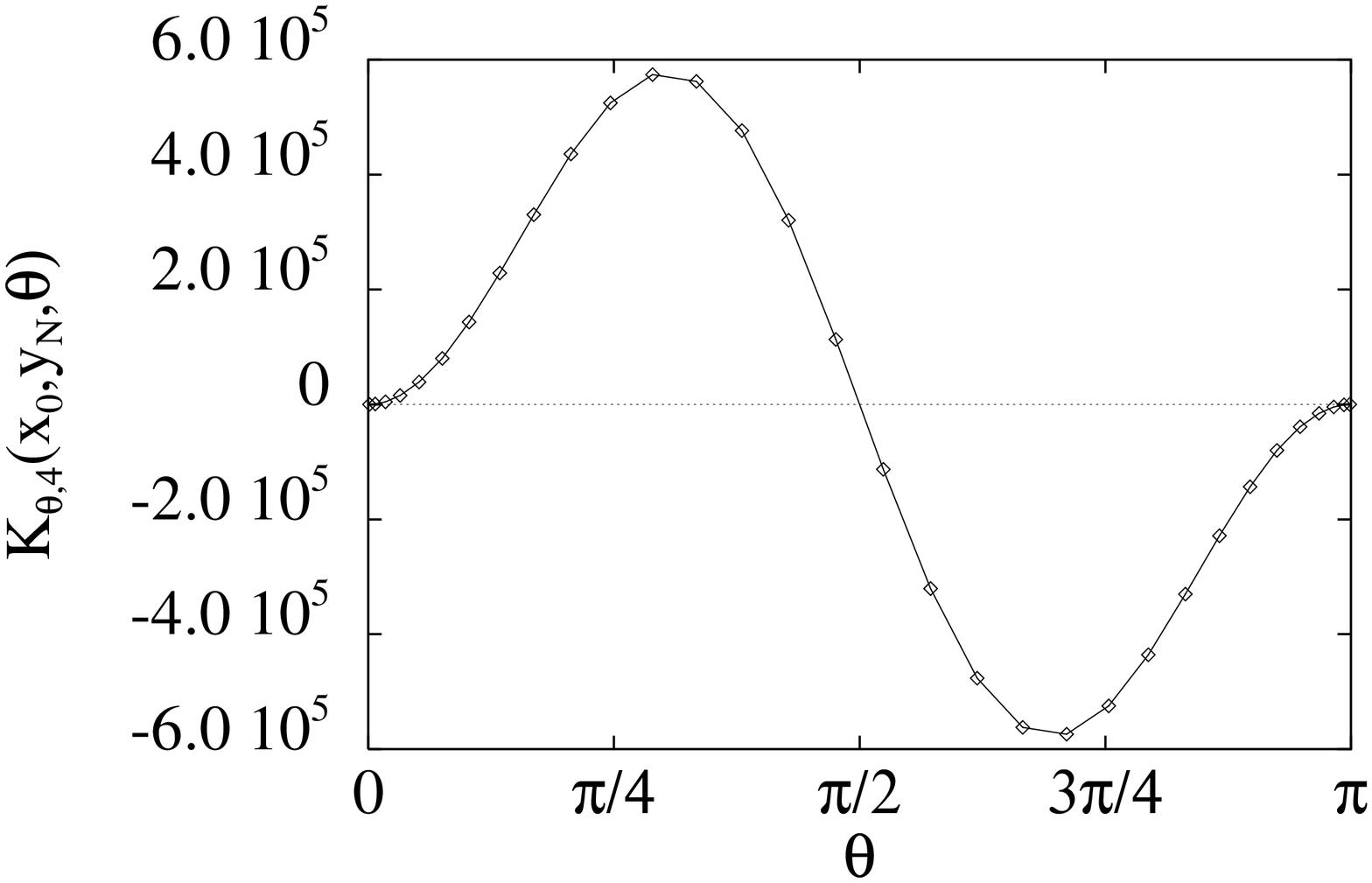,height=5.5cm,angle=-90}}
\mbox{\epsfig{file=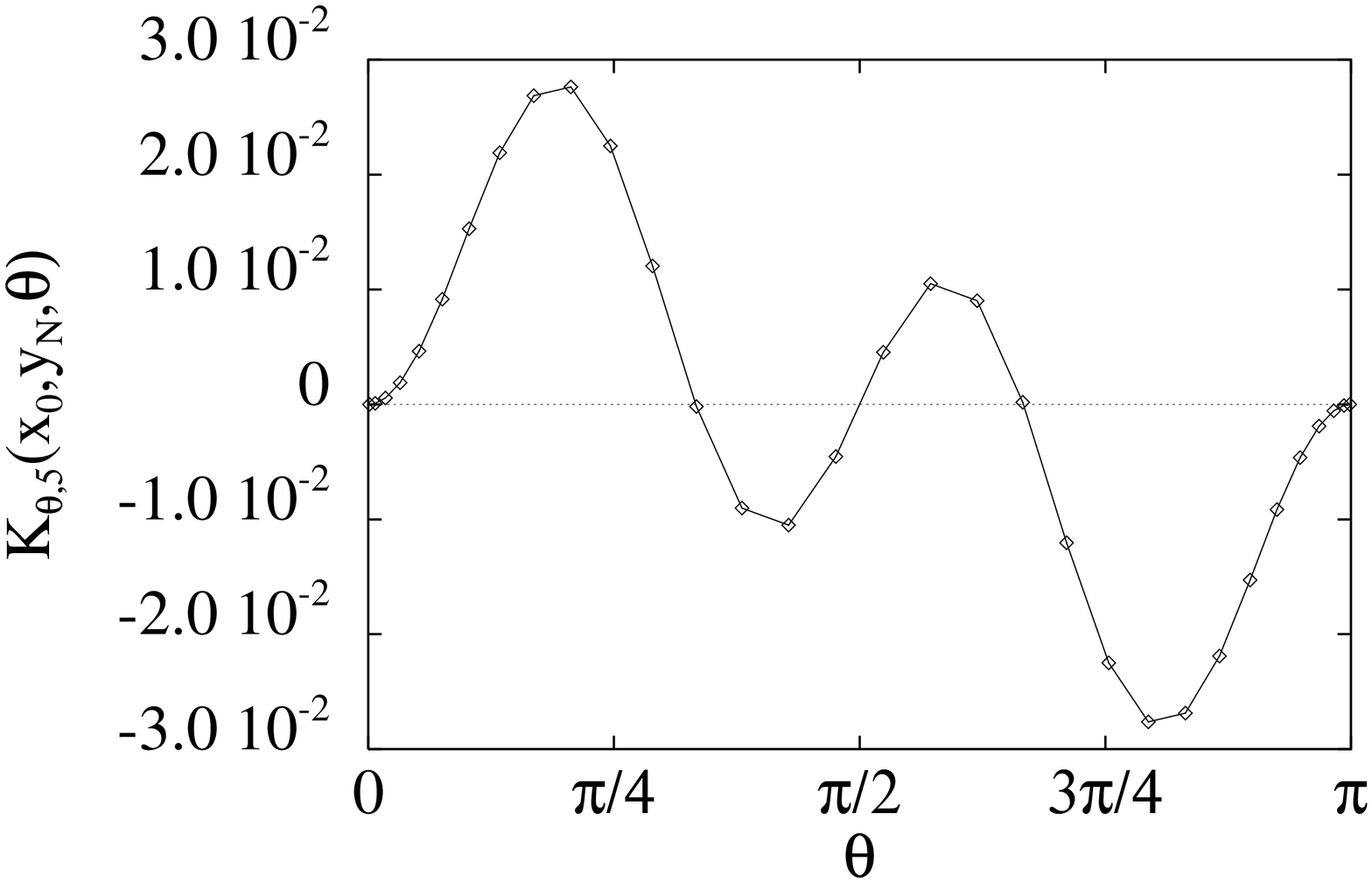,height=5.5cm,angle=-90}
\epsfig{file=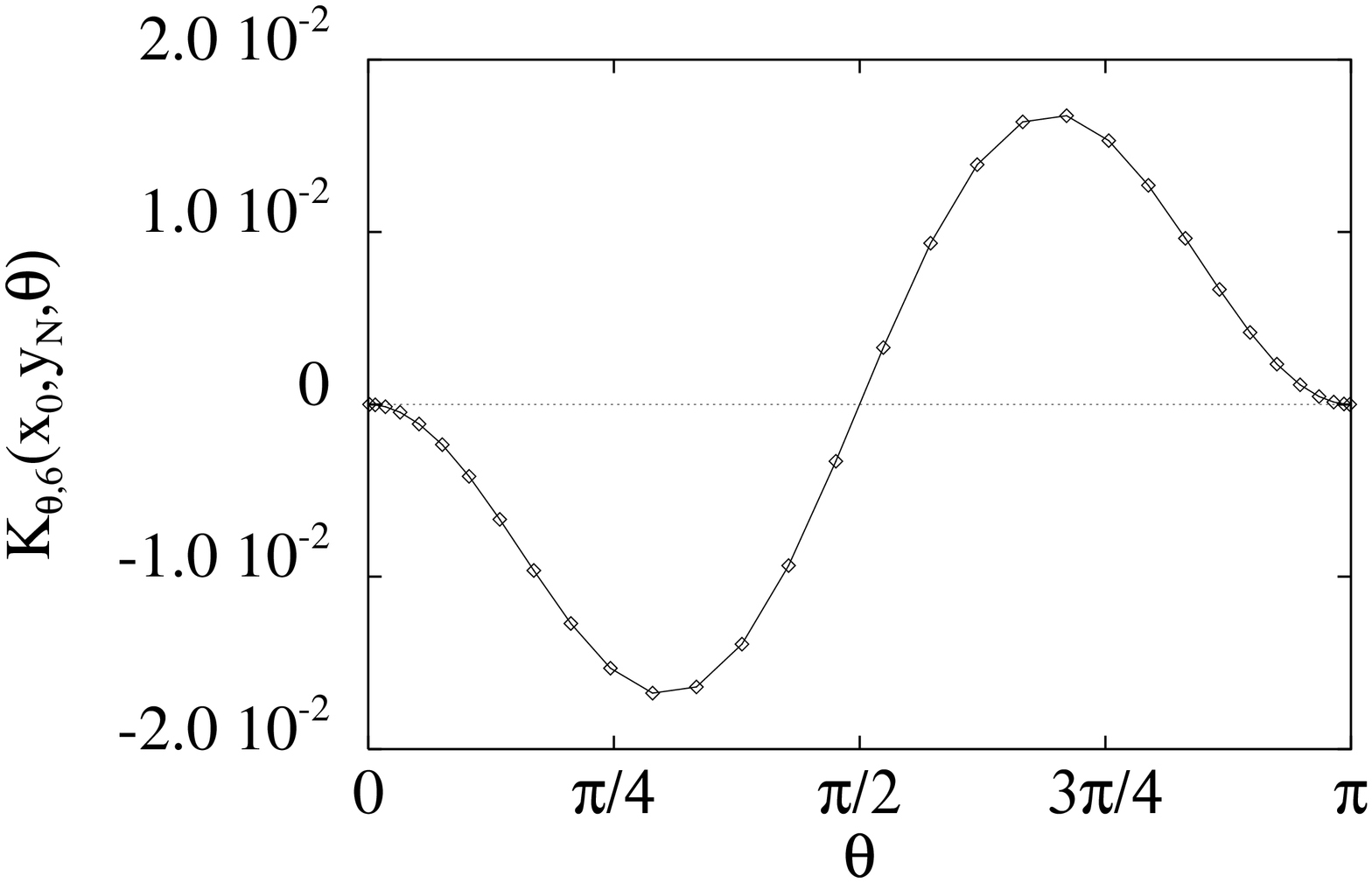,height=5.5cm,angle=-90}}
\end{center}
\vspace{-0.5cm}
\caption{Angular integrands $K_{\theta,i}(x_0,y_N,\theta)$, for $i=1,\ldots,6$,
with $\x0$, $\yN$, versus angle $\theta$ from the $\G$-equation 
with Ball-Chiu vertex, for $\alpha=1.921$ and $N_f=1$ using the alternative 
logarithm calculation.}
\label{Fig:tii-ldif}
\end{figure}

\begin{figure}[htbp]
\begin{center}
\mbox{\epsfig{file=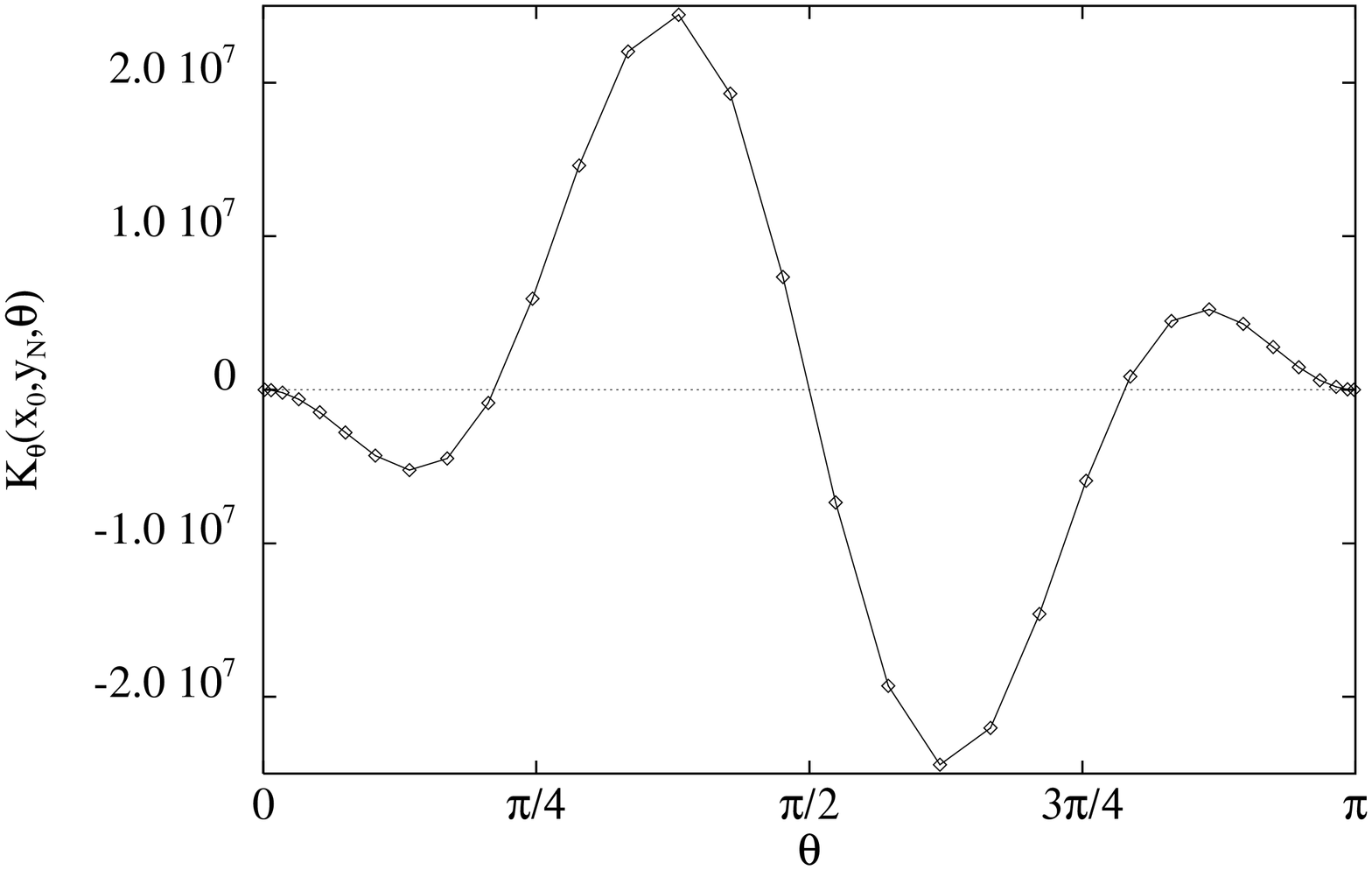,height=8cm,angle=-90}}
\end{center}
\vspace{-0.5cm}
\caption{Angular integrand $K_\theta(x_0,y_N,\theta)$ for
$\x0$, $\yN$, versus angle $\theta$ from the $\G$-equation 
with Ball-Chiu vertex, for $\alpha=1.921$ and $N_f=1$ using the
alternative logarithm calculation.}
\label{Fig:ti-ldif}
\end{figure}

In the following figures we will show the various results after
this improvement has been implemented. The angular integral
$K_{\theta,3}/\sin^2\theta(1-4\cos^2\theta)$, after removal of the
trigonometric part is shown in Fig.~\ref{Fig:ti3-ldif-notrig} and has the
expected $\cos\theta$ behaviour.
The influence on the angular integrands with their complete trigonometric
behaviour can be seen in Fig.~\ref{Fig:tii-ldif}. The values of these six
angular integrals are given in Table~\ref{Tab:K_R-ldif}.  The total angular
integrand is shown in Fig.~\ref{Fig:ti-ldif} and the integral value is
$K_R(x_0,y_N) = 0.524575$ for $\x0$ and $\yN$.

\begin{table}[htbp]
\begin{center}
\begin{tabular}{|l|r|}
\hline
$K_{R,1}$ & $-9.45396\ten{-01}$    \nn\\
$K_{R,2}$ & $1.44340\ten{+00}$       \nn\\
$K_{R,3}$ & $-1.30249\ten{-02}$   \nn\\
$K_{R,4}$ & $3.95985\ten{-02}$    \nn\\
$K_{R,5}$ & $2.63095\ten{-09}$  \nn\\
$K_{R,6}$ & $-1.53344\ten{-09}$ \nn\\
\hline
\end{tabular}
\end{center}
\caption{Radial kernels $K_{R,i}(x_0,y_N)$, for $i=1,\ldots,6$, 
with $\x0$ and $\yN$ using the alternative logarithm calculation.}
\label{Tab:K_R-ldif}
\end{table}

\begin{figure}[htbp]
\begin{center}
\mbox{\epsfig{file=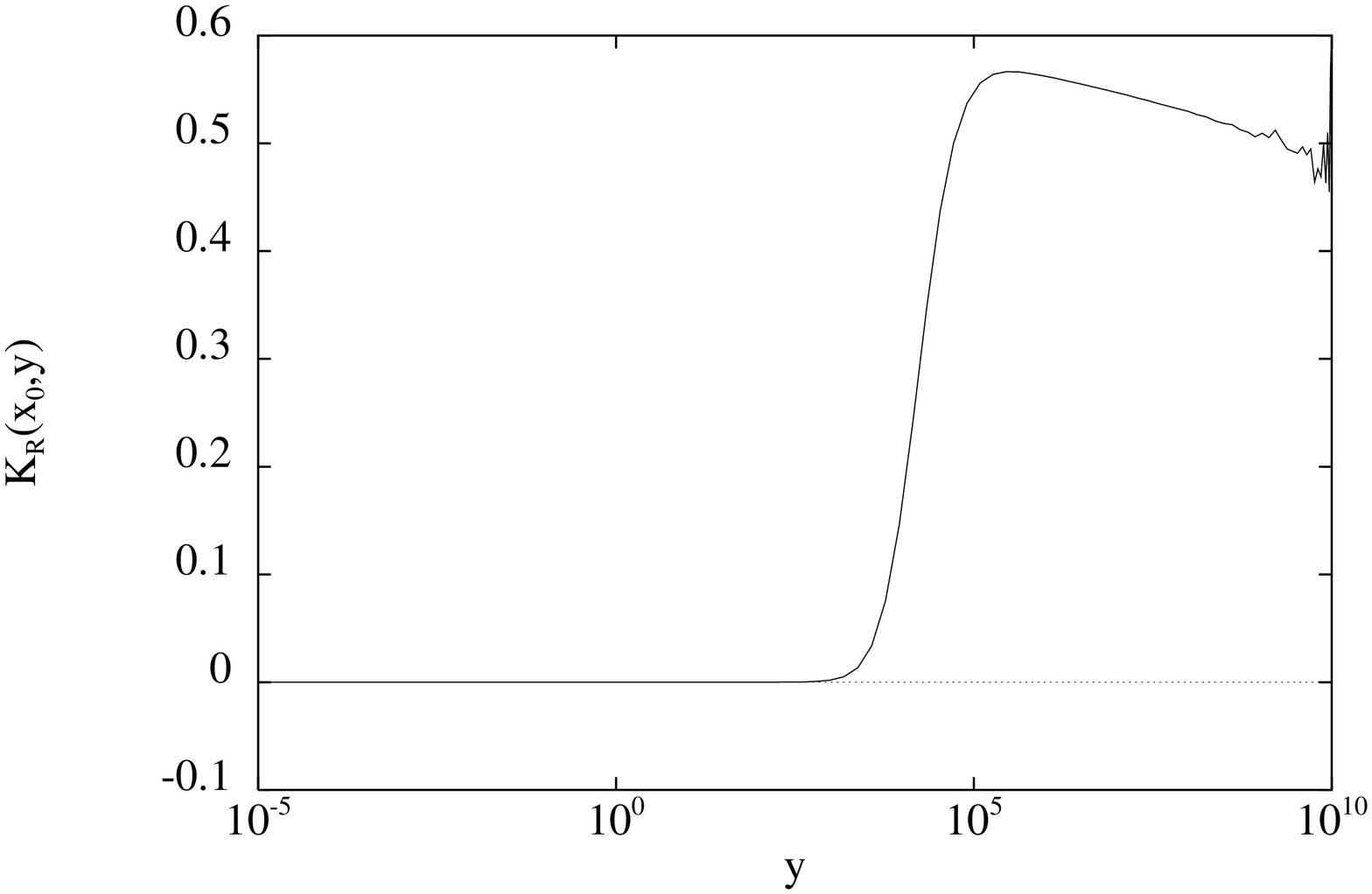,height=8cm,angle=-90}}
\end{center}
\vspace{-0.5cm}
\caption{Radial integrand $K_R(x_0,y)$ for $\x0$ versus radial momentum 
squared $y$ from the $\G$-equation from the $\G$-equation with Ball-Chiu
vertex, for $\alpha=1.921$ and $N_f=1$ using the alternative logarithm
calculation.}
\label{Fig:radint-ldif}
\end{figure}

\begin{figure}[htbp]
\begin{center}
\mbox{\epsfig{file=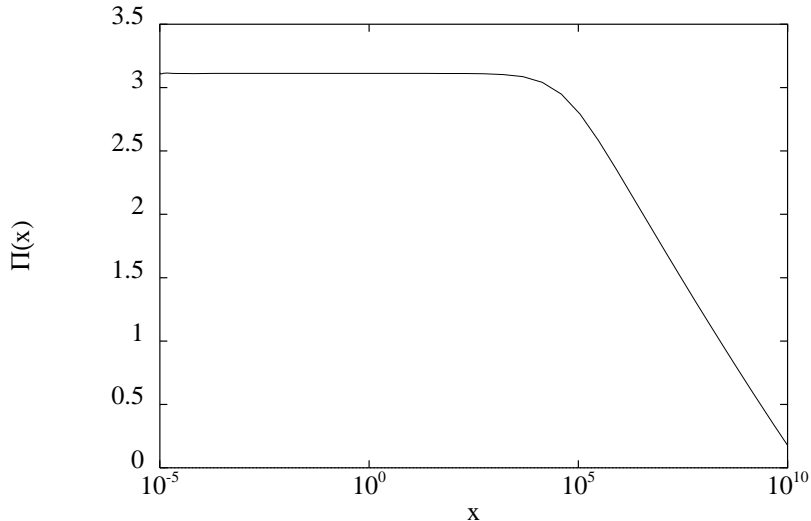,height=8cm,angle=-90}}
\end{center}
\vspace{-0.5cm}
\caption{Vacuum polarization $\Pi(x)$ versus momentum 
squared $x$ from the $\G$-equation with Ball-Chiu
vertex, for $\alpha=1.921$ and $N_f=1$ using the alternative logarithm
calculation.}
\label{Fig:vp-ldif}
\end{figure}

The radial integrand for $\x0$ is shown in Fig.~\ref{Fig:radint-ldif}. We
see that there is only a very light wriggle left for very large y-values
and this does not alter the fundamental behaviour of the vacuum polarization
function. The vacuum polarization at $\x0$ is $\Pi(x_0) = 3.11466$. 
After performing the radial integrals, the evolution of the vacuum
polarization integral $\Pi(x)$ with momentum is shown in
Fig.~\ref{Fig:vp-ldif}.
 
\begin{figure}[htbp]
\begin{center}
\mbox{\epsfig{file=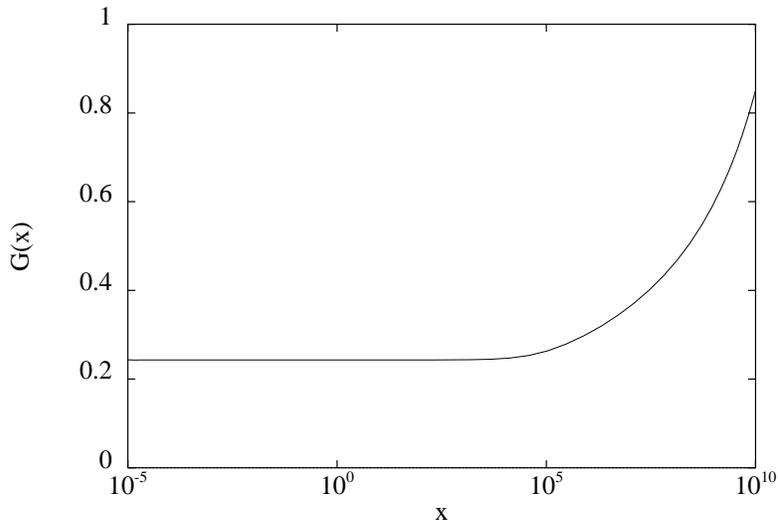,height=8cm,angle=-90}}
\end{center}
\vspace{-0.5cm}
\caption{Photon renormalization function $\G(x)$ versus momentum 
squared $x$ from the $\G$-equation with Ball-Chiu
vertex, for $\alpha=1.921$ and $N_f=1$ using the alternative logarithm
calculation.}
\label{Fig:G-ldif}
\end{figure}

From the vacuum polarization $\Pi(x)$, we now compute the photon
renormalization function $\G(x)$ and plot the result in
Fig.~\ref{Fig:G-ldif}. 
We see that $\G(x)$ behaves perfectly well now; the
unphysicalities due to numerical inaccuracies have been worked away, down
to photon momenta of the order $x/\Lambda^2 \approx \Order(10^{-15})$.

\subsection{Numerical results}

\begin{figure}[htbp]
\begin{center}
\mbox{\epsfig{file=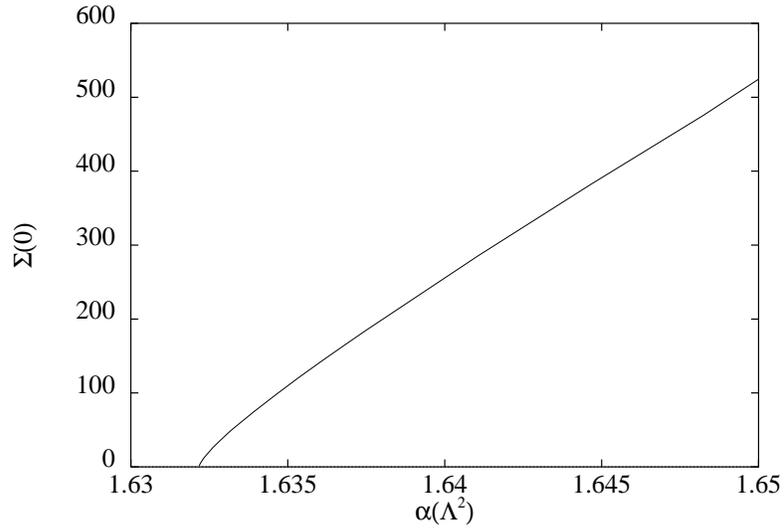,height=8cm,angle=-90}}
\end{center}
\vspace{-0.5cm}
\caption{Generated fermion mass $\S(0)$ versus running coupling 
$\alpha(\Lambda^2)$ for the coupled $(\S, \F, \G)$-system with the
Ball-Chiu vertex, for $N_f=1$.}
\label{Fig:fmg-BFG-BC-rc}
\end{figure}

Having done this, we can now apply the iterative procedure to solve the
($\S$, $\F$, $\G$)-system and determine the critical coupling in unquenched
QED with Ball-Chiu vertex.  The evolution of the generated fermion mass,
$\S(0)$, versus the running coupling at the UV-cutoff, $\alpha(\Lambda^2)$
is shown in Fig.~\ref{Fig:fmg-BFG-BC-rc}. The critical coupling is
\fbox{$\alpha_c(\Lambda^2, N_f=1) = 1.63218$}. We remark that the program only
converges if the starting guesses for the unknown functions are close to
the solutions.  The logical choice for this is always the solutions of the
system of equations for another value of the coupling, which is a little
bit larger than the current one. The more difficult it is to achieve
convergence, the closer we have to choose the subsequent couplings for
which we compute the generated fermion mass. The reason for this
convergence problem lies in the global iterative procedure connecting the
coupled ($\S$, $\F$)-system to the $\G$-equation. It is very likely that
convergence would be achieved more consistently and we could reach the
critical point with fewer computations if the complete ($\S$, $\F$,
$\G$)-system were treated with Newton's method in a unified way.

\begin{figure}[htbp]
\begin{center}
\mbox{\epsfig{file=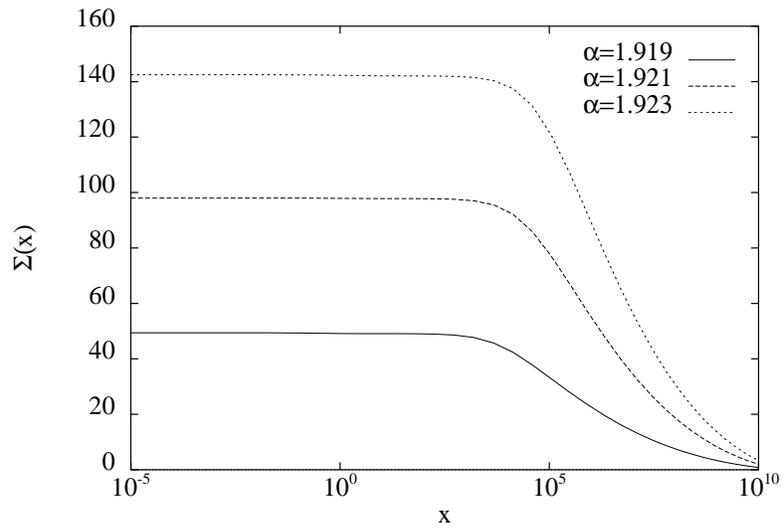,height=8cm,angle=-90}}
\end{center}
\vspace{-0.5cm}
\caption{Dynamical fermion mass $\S(x)$ versus momentum squared $x$ for
the coupled $(\S, \F, \G)$-system with the Ball-Chiu vertex, for $N_f=1$
and $\alpha=1.919, 1.921, 1.923$.}
\label{Fig:B-BFG-BC-vertex}
\end{figure}

\begin{figure}[htbp]
\begin{center}
\mbox{\epsfig{file=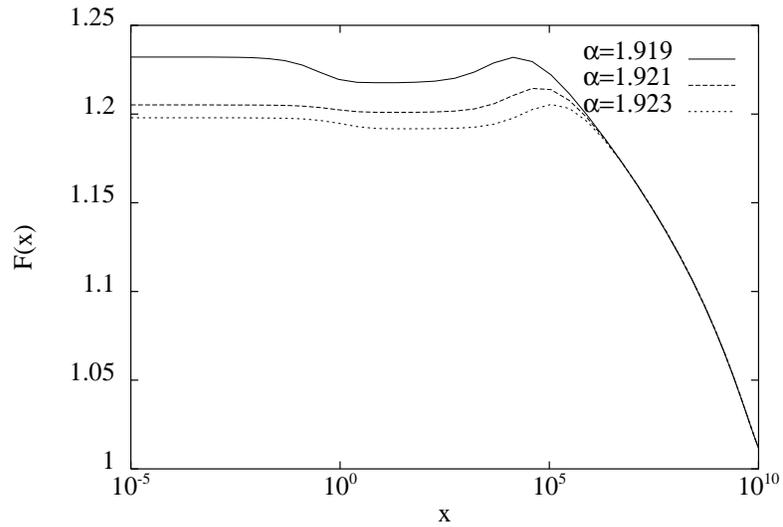,height=8cm,angle=-90}}
\end{center}
\vspace{-0.5cm}
\caption{Fermion wavefunction renormalization $\F(x)$ versus momentum 
squared $x$ for the coupled $(\S, \F, \G)$-system with the Ball-Chiu
vertex, for $N_f=1$ and $\alpha=1.919, 1.921, 1.923$.}
\label{Fig:F-BFG-BC-vertex}
\end{figure}

\begin{figure}[htbp]
\begin{center}
\mbox{\epsfig{file=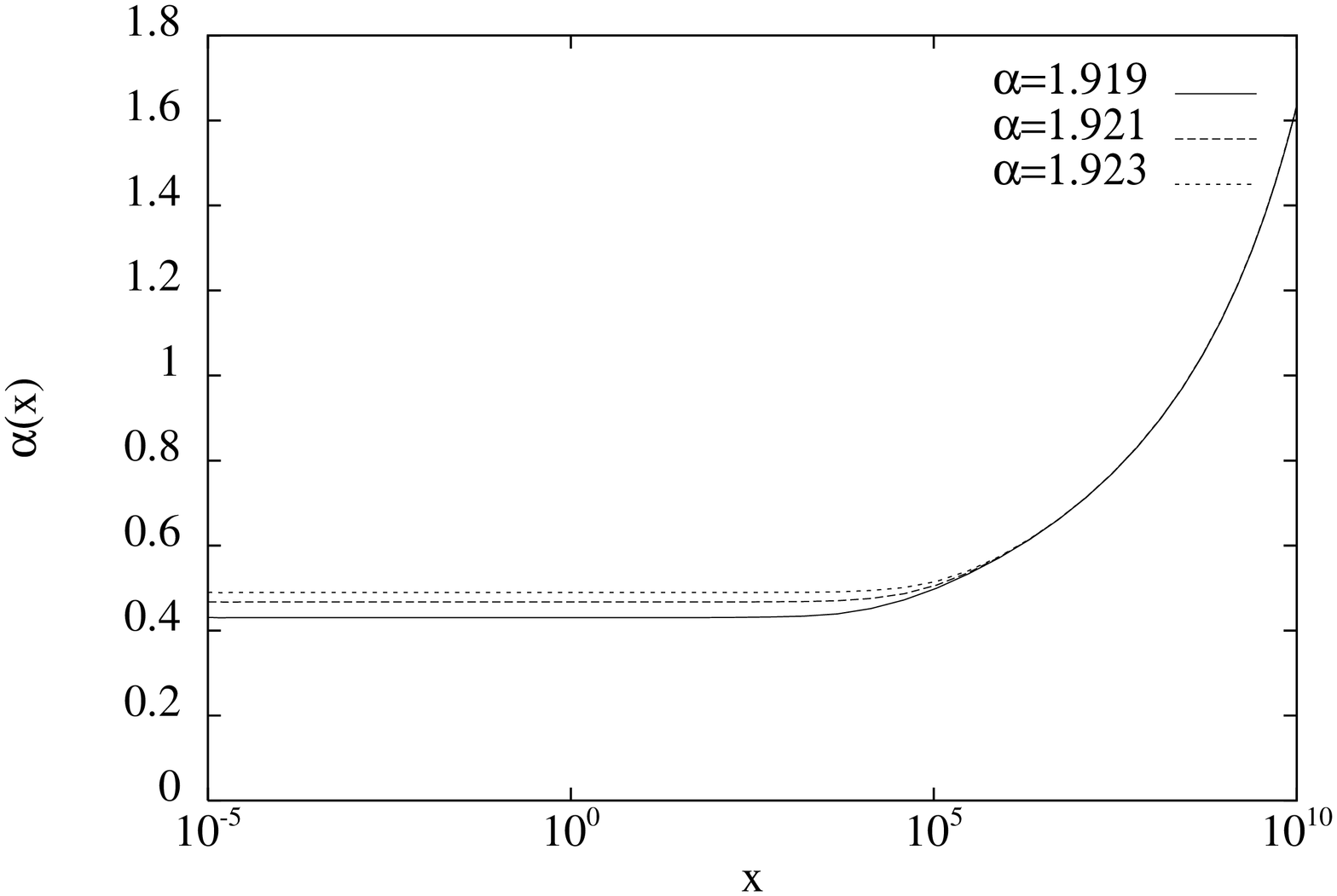,height=8cm,angle=-90}}
\end{center}
\vspace{-0.5cm}
\caption{Running coupling $\alpha(x)$ versus momentum squared $x$ for
the coupled $(\S, \F, \G)$-system with the Ball-Chiu vertex, for $N_f=1$
and $\alpha=1.919, 1.921, 1.923$.}
\label{Fig:rc-BFG-BC-vertex}
\end{figure}

Typical plots of $\S(x)$, $\F(x)$ and $\alpha(x)=\alpha \, \G(x)$ are shown
in Fig.~\ref{Fig:B-BFG-BC-vertex}, Fig.~\ref{Fig:F-BFG-BC-vertex} and
Fig.~\ref{Fig:rc-BFG-BC-vertex}. From Fig.~\ref{Fig:F-BFG-BC-vertex} we see
that the wavefunction renormalization tends to have a peculiar behaviour
with the Ball-Chiu vertex when the coupling is close to its critical value.

For $N_f=2$ the situation is even more delicate as the procedure initially
does not converge for any value of the coupling. To make it converge it
is important to start from a realistic set of functions. This can be
achieved by using the functions $\S(x)$, $\F(x)$ and $\G(x)$ obtained with
the bare vertex approximation for $\alpha=5$ and using these as starting
values to find the results for the Ball-Chiu vertex for the same
coupling. We then slowly work our way down to smaller values of the
coupling. Unfortunately the program does not converge anymore for $\S(0) <
\Order(500)$ (for $\Lambda=1\ten{5}$) and further investigation would be 
needed to understand if this is because of numerical or physical reasons.

\clearpage
\section{Curtis-Pennington vertex}

We recall the Curtis-Pennington vertex Ansatz as introduced in
Section~\ref{Sec:CP-vertex}:
\ba
\Gamma^\mu_{CP}(k,p) &=&
\frac{1}{2}\l[\frac{1}{\F(k^2)}+\frac{1}{\F(p^2)}\r]\gamma^\mu
+ \frac{1}{2}\l[\frac{1}{\F(k^2)}-\frac{1}{\F(p^2)}\r]
\frac{(k+p)^\mu(\slash{k}+\slash{p})}{k^2-p^2} \mlab{1005} \\[1mm]
&& - \l[\frac{\S(k^2)}{\F(k^2)} - \frac{\S(p^2)}{\F(p^2)}\r]
\frac{(k+p)^\mu}{k^2-p^2} \nn\\
&& + \frac{1}{2}\l[\frac{1}{\F(k^2)}-\frac{1}{\F(p^2)}\r]
\frac{(k^2+p^2)\l[\gamma^{\mu}(k^2-p^2)-(k+p)^{\mu}(\slash{k}-\slash{p})\r]}
{(k^2-p^2)^2+\l(\S^2(k^2)+\S^2(p^2)\r)^2} \nn .
\ea

This vertex Ansatz has been used in Section~\ref{Sec:abgpr} and in
Ref.~\cite{CP91,CP92,CP93,Atk93,ABGPR} to study the behaviour of the
fermion propagator and the dynamical generation of fermion mass in 
quenched QED.

We now want to investigate the possibility of dynamical fermion mass
generation in unquenched QED with the Curtis-Pennington vertex. We recall
the set of coupled integral equations (in Euclidean space),
Eqs.~(\oref{1.1006}, \oref{1.1007}, \oref{1.1008}) with $m_0=0$ and in the
Landau gauge:
\ba
\frac{\S(x)}{\F(x)} &=& \frac{\alpha}{2\pi^2} \int dy \, 
\frac{y\F(y)}{\Ds{y}} \int d\theta \sin^2\theta \, \G(z) \mlab{1006} \\
&& \hspace{10mm} \times \l\{ \Big[A(y,x) + \tau_6(y,x)(y-x)\Big] 
\frac{3\S(y)}{z} 
- \frac{\S(y)-\S(x)}{\F(x)(y-x)}\frac{2yx\sin^2\theta}{z^2} \r\} \nn\\[3mm]
\frac{1}{\F(x)} &=& 1 - \frac{\alpha}{2\pi^2 x} \int dy \,
\frac{y\F(y)}{y+\S^2(y)} \int d\theta \, \sin^2 \theta \, \G(z) \mlab{1007} \\
&& \times \Bigg\{
A(y,x)\l[\frac{2yx\sin^2\theta}{z^2} - \frac{3\sqrt{yx}\cos\theta}{z}\r] \nn \\
&& \hspace{4mm}+\Big[B(y,x)(y+x)-C(y,x)\S(y)\Big]\frac{2yx\sin^2\theta}{z^2} 
- \tau_6(y,x)(y-x)\frac{3\sqrt{yx}\cos\theta}{z} 
\Bigg\} \nn\\[3mm]
\frac{1}{\G(x)} &=& 1 + \frac{2N_f\alpha}{3\pi^2 x} \int dy \, 
\frac{y\F(y)}{\Ds{y}} \int d\theta \, \sin^2\theta \,
\frac{\F(z)}{\Ds{z}} \mlab{1008}\\
&& \times \Bigg\{ 2A(y,z)\bigg[y(1-4\cos^2\theta) 
+ 3\sqrt{yx}\cos\theta\bigg] \nn\\
&& \hspace{5mm} + 
B(y,z) \bigg[\Big(y+z-2\S(y)\S(z)\Big)\,
\Big(2y(1-4\cos^2\theta)+3\sqrt{yx}\cos\theta\Big) \nn\\
&& \hspace{22mm} + 3(y-z)\Big(y-\S(y)\S(z)\Big)\bigg] \nn\\
&& \hspace{5mm} - C(y,z) 
\bigg[\Big(\S(y)+\S(z)\Big)\Big(2y(1-4\cos^2\theta)+3\sqrt{yx}\cos\theta\Big) 
+ 3(y-z)\S(y)\bigg]\nn\\
&& \hspace{5mm} - 3\tau_6(y,z) (y-z)\Big(y-\sqrt{yx}\cos\theta+\S(y)\S(z)\Big)
\Bigg\} \nn
\ea
where 
\ba
A(y,x) &=& \frac{1}{2}\l[\frac{1}{\F(y)}+\frac{1}{\F(x)}\r] \nn\\
B(y,x) &=& \frac{1}{2(y-x)}\l[\frac{1}{\F(y)}-\frac{1}{\F(x)}\r] \nn\\
C(y,x) &=& -\frac{1}{y-x}\l[\frac{\S(y)}{\F(y)}-\frac{\S(x)}{\F(x)}\r] \nn\\
\tau_6(y,x) &=& \frac{y+x}{2\l[(y-x)^2+(\S^2(y)+\S^2(x))^2\r]}
\l[\frac{1}{\F(y)}-\frac{1}{\F(x)}\r] \;. \nn
\ea

\begin{figure}[htbp]
\begin{center}
\mbox{\epsfig{file=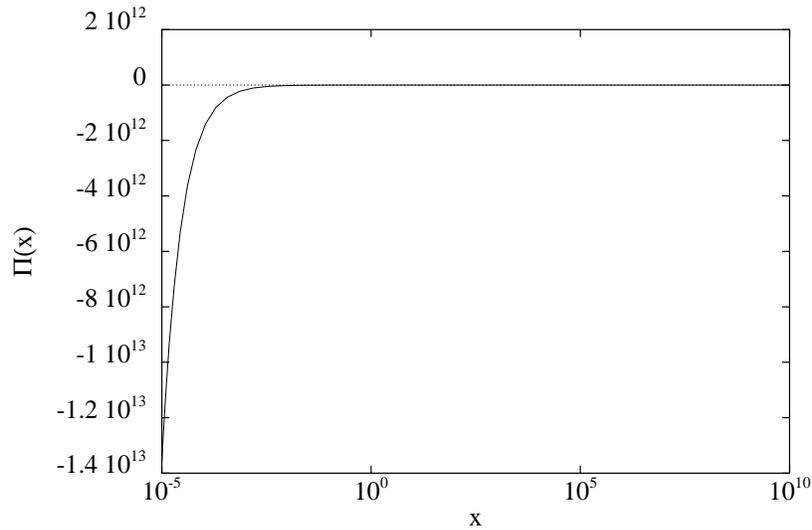,height=8cm,angle=-90}}
\end{center}
\vspace{-0.5cm}
\caption{Vacuum polarization $\Pi(x)$ versus momentum 
squared $x$ from the $\G$-equation with Curtis-Pennington vertex, for
$\alpha=1.921$ and $N_f=1$.}
\label{Fig:vp-CP}
\end{figure}

If we now implement the system of coupled integral equations with the
Curtis-Pennington vertex, Eqs.~(\oref{1006}, \oref{1007}, \oref{1008}), the
numerical program does not seem to converge. A check of the vacuum
polarization integral calculated from realistic input functions is shown in
Fig.~\ref{Fig:vp-CP}.

The huge negative value of the vacuum polarization for decreasing value of
momenta is unphysical. It is clear that the CP-vertex gives rise to a
quadratic divergence in the vacuum polarization in the massless case. In
the case where mass is generated dynamically the situation is more
complicated, but Fig.~\ref{Fig:vp-CP} shows that the CP-vertex does not give
physical results for unquenched QED.

\section{Hybrid method}

\vspace{-2mm}
In the previous section we have seen that the Curtis-Pennington vertex did
not yield physical results when applied to the photon equation.
Nevertheless it is useful to use the CP-vertex for the fermion equations as
it ensures that the fermion propagator is multiplicatively renormalizable
in the quenched case. Therefore we will now introduce a hybrid method where
we use the CP-vertex in the fermion equations, yielding \mrefb{1006}{1007},
and the Ball-Chiu vertex in the photon equation, giving \mref{1015}, in
order to avoid quadratic divergences in the vacuum polarization.

\begin{figure}[htbp]
\begin{center}
\mbox{\epsfig{file=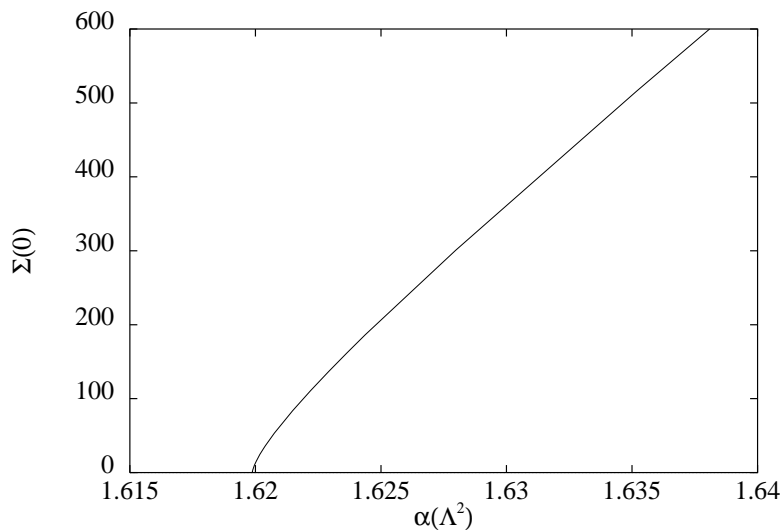,height=8cm,angle=-90}}
\end{center}
\vspace{-9mm}
\caption{Generated fermion mass $\S(0)$ versus running coupling 
$\alpha(\Lambda^2)$ for the coupled $(\S, \F, \G)$-system with the
hybrid method, for $N_f=1$.}
\label{Fig:fmg-BFG-hybrid-rc}
\end{figure}

\begin{figure}[htbp]
\begin{center}
\mbox{\epsfig{file=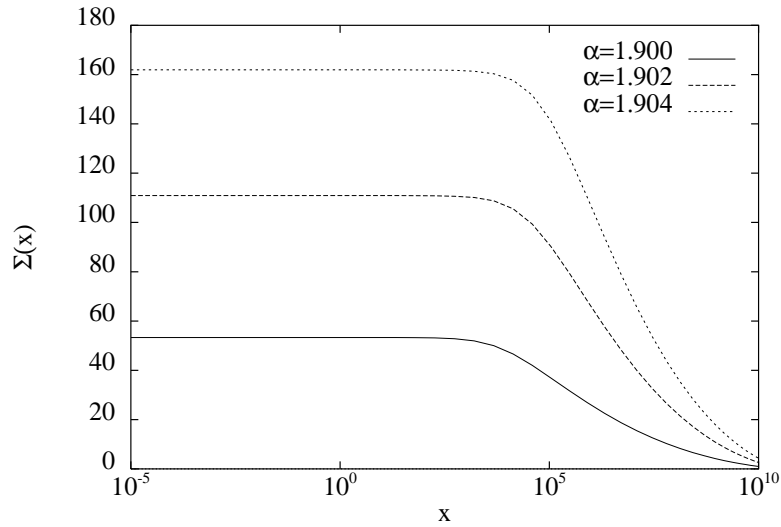,height=8cm,angle=-90}}
\end{center}
\vspace{-0.5cm}
\caption{Dynamical fermion mass $\S(x)$ versus momentum squared $x$ for
the coupled $(\S, \F, \G)$-system with the hybrid method, for
$N_f=1$ and $\alpha=1.900, 1.902, 1.904$.}
\label{Fig:B-BFG-hybrid}
\end{figure}

\begin{figure}[htbp]
\begin{center}
\mbox{\epsfig{file=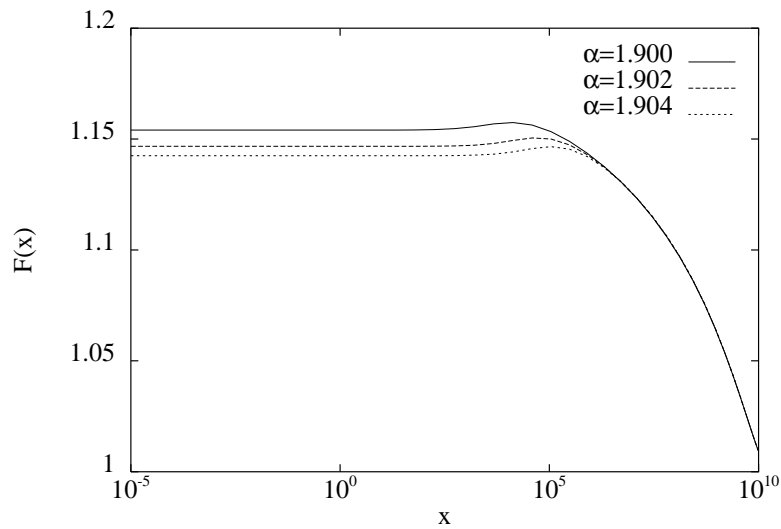,height=8cm,angle=-90}}
\end{center}
\vspace{-0.5cm}
\caption{Fermion wavefunction renormalization $\F(x)$ versus momentum 
squared $x$ for the coupled $(\S, \F, \G)$-system with the hybrid
method, for $N_f=1$ and $\alpha=1.900, 1.902, 1.904$.}
\label{Fig:F-BFG-hybrid}
\end{figure}

\begin{figure}[htbp]
\begin{center}
\mbox{\epsfig{file=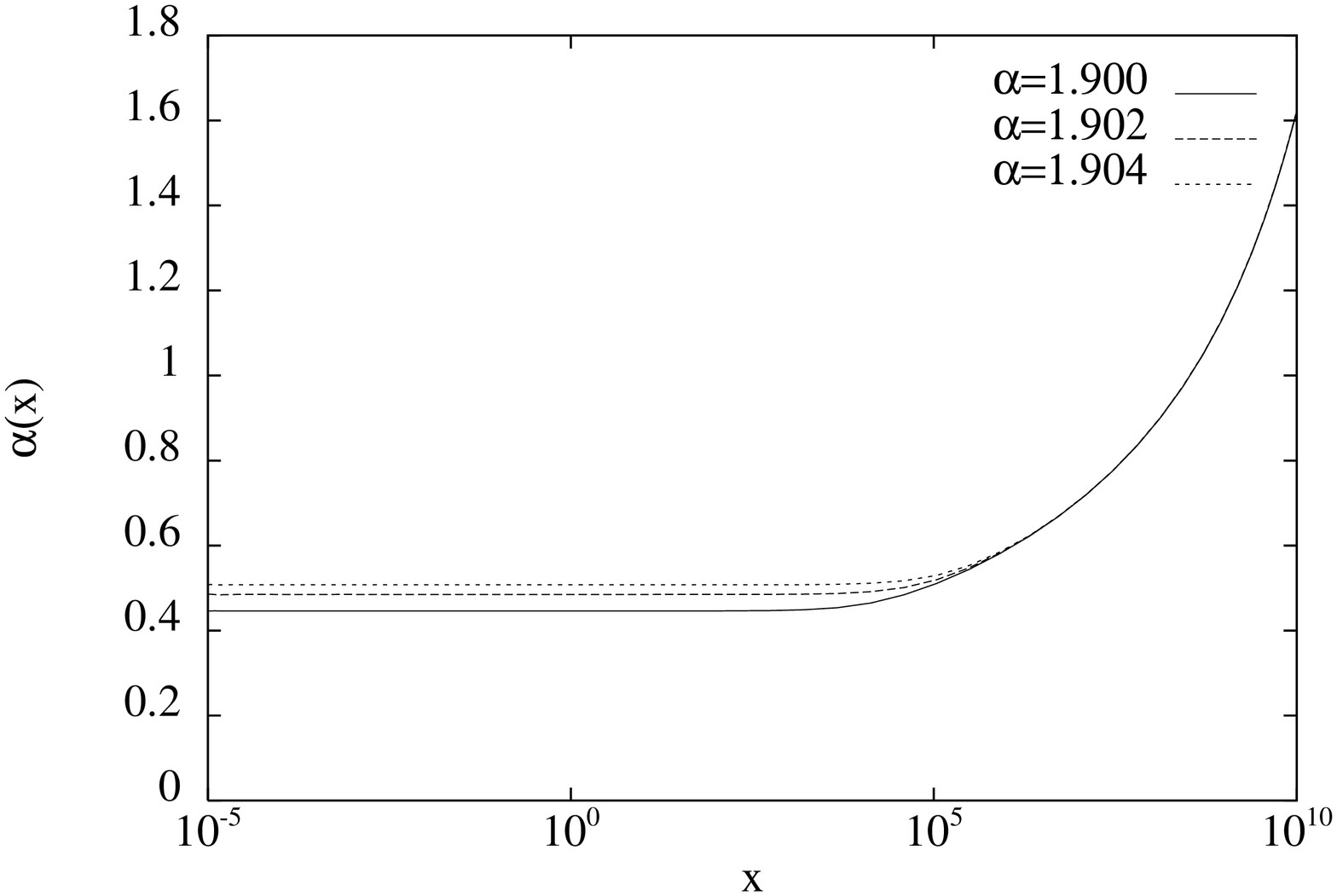,height=8cm,angle=-90}}
\end{center}
\vspace{-0.5cm}
\caption{Running coupling $\alpha(x)$ versus momentum squared $x$ for
the coupled $(\S, \F, \G)$-system with the hybrid method, for
$N_f=1$ and $\alpha=1.900, 1.902, 1.904$.}
\label{Fig:rc-BFG-hybrid}
\end{figure}

The evolution of the generated fermion mass, $\S(0)$, versus the running
coupling, $\alpha_c(\Lambda^2)$, is shown in
Fig.~\ref{Fig:fmg-BFG-hybrid-rc}. The value of the critical coupling at the
UV-cutoff is \fbox{$\alpha_c(\Lambda^2, N_f=1) = 1.61988$}.
Plots of $\S(x)$, $\F(x)$ and $\alpha(x)=\alpha \, \G(x)$ are shown
in Fig.~\ref{Fig:B-BFG-hybrid}, Fig.~\ref{Fig:F-BFG-hybrid} and
Fig.~\ref{Fig:rc-BFG-hybrid}. 

\begin{figure}[htbp]
\begin{center}
\mbox{\epsfig{file=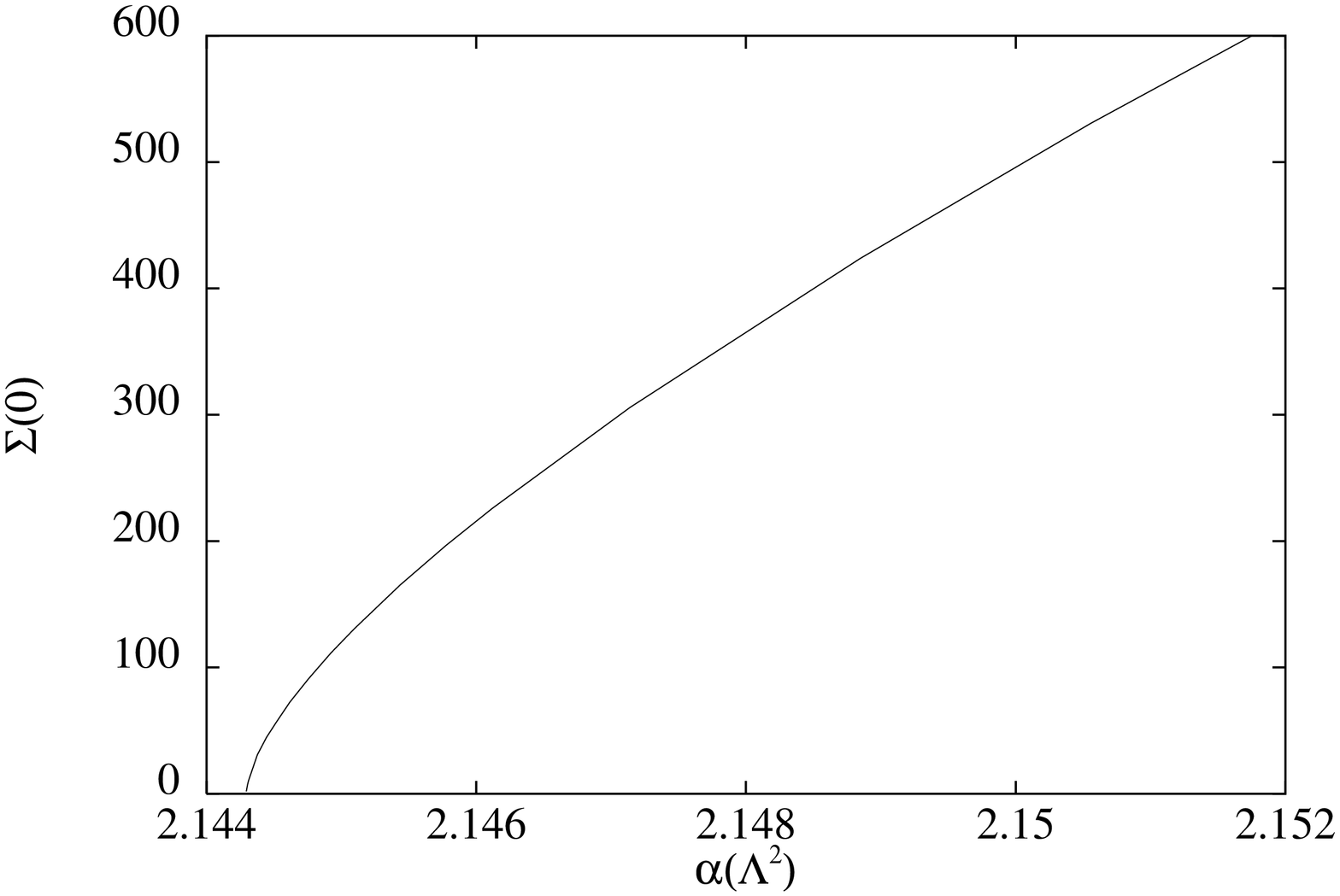,height=8cm,angle=-90}}
\end{center}
\vspace{-0.5cm}
\caption{Generated fermion mass $\S(0)$ versus running coupling 
$\alpha(\Lambda^2)$ for the coupled $(\S, \F, \G)$-system with the
hybrid method, for $N_f=2$.}
\label{Fig:fmg-BFG-hybrid-rc-N2}
\end{figure}

For $N_f=2$ the generated fermion mass is plotted against the coupling in
Fig.\ref{Fig:fmg-BFG-hybrid-rc-N2}. The critical coupling is
\fbox{$\alpha_c(\Lambda^2, N_f=2) = 2.14429$}.

Although for $N_f=1$, the critical coupling only changes with less than 1\%
between the Ball-Chiu vertex and the hybrid method, the latter definitely
improves the convergence of the numerical method. This is especially true
for $N_f=2$ where we could not locate the critical point with the Ball-Chiu
vertex, while it is easily done with the hybrid method.

\section{About the renormalization of the SD equations}

\begin{figure}[htbp]
\begin{center}
\mbox{\epsfig{file=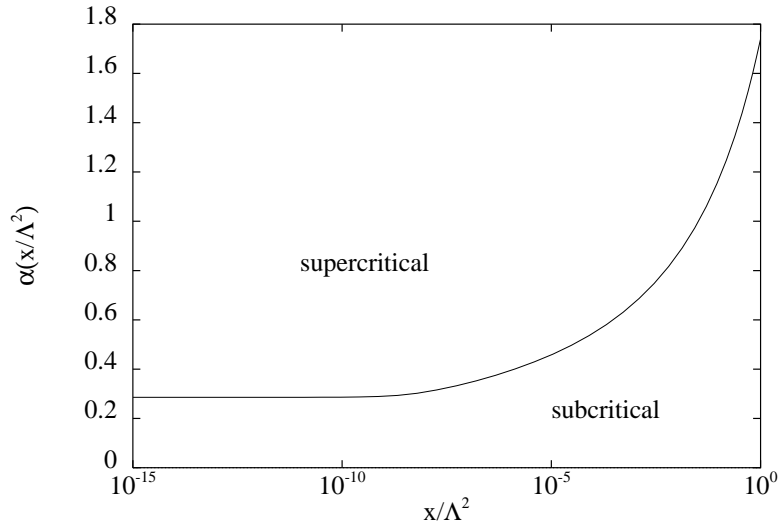,height=8cm,angle=-90}}
\end{center}
\vspace{-0.5cm}
\caption{Running coupling $\alpha(x/\Lambda^2)$ versus momentum squared
 $x/\Lambda^2$ for the coupled $(\S, \F, \G)$-system, for $N_f=1$ for
$\alpha=2.0825$.}
\label{Fig:rcc-BFG}
\end{figure}

In this study we have always used the bare Schwinger-Dyson equations,
regularized by an UV-cutoff $\Lambda$. From the plots of the running
coupling $\alpha(x)$ versus $x$ in the previous result sections, we
understand that in the critical point we will be able to define a running
critical coupling $\alpha_c(x)$ which depends only on the relative position
of $x$ with respect to $\Lambda^2$, i.e. we have a fixed line
$\alpha_c(x/\Lambda^2)$. This is shown in Fig.~\ref{Fig:rcc-BFG} where we
plot the running coupling $\alpha_c(x/\Lambda^2)$ for $\alpha$ close to the
critical point ($\S(0)/\Lambda \approx 1\ten{-5}$). Note that if we could
go even closer to the critical point we would see that
$\alpha_c(x/\Lambda^2)\to 0$ when $x\to 0$, as the flat low momentum
behaviour of Fig.~\ref{Fig:rcc-BFG} is entirely due to the small generated
fermion mass.

We now try to relate this intuitively with the physical world, which is
described by the renormalized theory. Suppose we fix the value of the
coupling at a certain renormalization scale, $\alpha_R(\mu^2)$.
\Comment{Funny remark: renormalizing the coupling should mean 
$\alpha(x/\Lambda^2)\to \alpha(x/\mu^2)$. Still the critical line depends
on $x/\Lambda^2$!!}  If we now look at the generation of fermion mass, the
following scenario happens depending on the values of $\Lambda$. If we take
$\Lambda$ such that $\alpha_R(\mu^2)$ lies below the critical line in
Fig.~\ref{Fig:rcc-BFG}, no mass will be generated. When we increase
$\Lambda$, $\mu^2/\Lambda^2$ decreases, and as we are traveling from right
to left on a horizontal line in Fig.~\ref{Fig:rcc-BFG},
$\alpha_c(\mu^2/\Lambda^2)$ decreases such that the renormalized coupling
$\alpha_R(\mu^2)$ which started subcritical becomes critical for some
critical UV-cutoff $\Lambda_c$. If we increase $\Lambda$ further we enter
the supercritical phase where fermion mass is generated. However there is a
serious problem related to the Landau pole.  When we shift the renormalized
coupling from right to left with respect to the critical line in
Fig.~\ref{Fig:rcc-BFG}, corresponding to increasing $\Lambda$, it is the
whole running renormalized coupling which is displaced. If the Landau pole
of the running renormalized coupling moves into the integration region
[0,$\Lambda$], the SD equations are not solvable anymore. Therefore,
$\Lambda$ cannot be taken to infinity and the existence of the Landau pole
does not allow us to make a consistent discussion of fermion mass
generation in renormalized QED. However, if another theory would take over
at some scale $\Omega$, we could compare this scale with $\Lambda_c$ and
fermion mass generation could then be possible in a consistent way. A more
detailed numerical analysis of the above described scenario could reveal
more about this.

\clearpage

\section{Summary}

From the results of this chapter we can conclude that dynamical fermion
mass generation does occur in unquenched QED, for $N_f=1$ and $N_f=2$. The
exact value of the critical coupling is dependent on the vertex Ansatz.

In Table~\ref{Tab:alphac-globalcomp} we compare the various
values obtained for the critical coupling for $N_f=1$ and $N_f=2$ in the
($\S$, $\F$, $\G$)-system with different vertex approximations in the
previous sections and in Section~\ref{BFG}. The critical coupling varies at
most 18\% for $N_f=1$ and 21\% for $N_f=2$.

\begin{table}[htbp]
\begin{center}
\begin{tabular}{|c||c|c|}
\hline
vertex & $\alpha_c(N_f=1)$ & $\alpha_c(N_f=2)$ \\
\hline
bare          & 1.74102 & 2.22948 \\
$1/\F$-vertex & 1.90911 & 2.59578 \\
Ball-Chiu     & 1.63218 &  ---\\
hybrid        & 1.61988 & 2.14429 \\
\hline
\end{tabular}
\caption[Critical coupling $\alpha_c$ for $N_f=1$ and $N_f=2$, for the
($\S,\F,\G$)-system with various vertex approximations] {Critical coupling
$\alpha_c$ for $N_f=1$ and $N_f=2$ for the
($\S,\F,\G$)-system with various vertex approximations.}
\label{Tab:alphac-globalcomp}
\end{center}
\end{table}

The study in this chapter has been the very first attempt to introduce an
improved vertex in the study of dynamical fermion mass generation in
unquenched QED. More work has to be done to construct a vertex Ansatz which
ensures the multiplicative renormalizability of the photon propagator in
addition to that of the fermion propagator~\cite{Ayse}. However, although
the vertex approximations used in this work were not designed specifically
for unquenched QED, the numerical method developed here will prove very
helpful when such vertices will be available. Having achieved
the proper numerical cancellation of the quadratic divergence in the
vacuum polarization integral with the Ball-Chiu vertex is important as this
vertex is the uniquely determined longitudinal part of the QED-vertex, and
as such will have to be present in any realistic improvement of the vertex
in unquenched QED.

\chapter{Conclusions}
\label{Concl}


We have investigated the dynamical generation of fermion mass
in QED, using the Schwinger-Dyson equations to approach this
non-perturbative phenomenon of quantum field theory. This infinite set of
equations was then truncated by introducing various approximations.

Bifurcation analysis was applied to determine the critical point in the
Curtis-Pennington approximation to {\it quenched} QED.  We computed the
critical coupling, above which fermion mass is generated dynamically, for a
large range of the covariant gauge parameter and concluded that the
critical coupling has a much smaller dependence on the gauge parameter than
in the bare vertex approximation. The critical coupling in the Landau
gauge, which is $\alpha_c=\pi/3$ in the bare vertex approximation, becomes
$\alpha_c=0.933667$ with the CP-vertex. Furthermore we have shown that the
generated fermion mass follows the Miransky-scaling law in the
neighbourhood of the critical point, with the bare vertex as well as with
the Curtis-Pennington vertex.

We went on to discuss dynamical fermion mass generation in {\it unquenched}
QED and developed a sophisticated computer program to investigate this
numerically. We derived the formalism to solve the coupled system of
non-linear integral equations by approximating the integrals by suitable
quadrature formulae, paying special attention to the kink in the radial
integrand, and solving the resulting system of coupled non-linear algebraic
equations with an iteration method. Using Newton's iterative procedure
ensured the quadratic convergence of the method.

When applying this method to the coupled equations for the dynamical mass
$\S$ and the photon renormalization function $\G$, we observed that $\G(x)$
had an unphysical behaviour for intermediate small values of $x$. A
detailed analysis exposed a problem in the numerical cancellation of the
quadratic divergence in the vacuum polarization integral. It was suggested
that smooth approximations to $\S$, $\F$ and $\G$ would be preferable in
contrast to the discretized approach used so far.

Motivated by this observation, we introduced Chebyshev expansions for $\S$,
$\F$ and $\G$ and adapted the previously developed method to solve the
integral equations in the most advantageous way. The introduction of the
Chebyshev expansion method proved to be very powerful as the smoothness of
the functions is automatically guaranteed, which avoids all interpolation
problems. Furthermore higher order numerical quadrature formulae can now be
used, enhancing the accuracy and speeding up the calculations.

We went on applying the Chebyshev solution method to various approximations
to unquenched QED in the bare vertex approximation. We computed the value
of the critical coupling for one and two flavours and compared these with
the results found in the literature. First in the 1-loop approximation to
$\G$ using three variants: the LAK-approximation, the $\F\equiv 1$
approximation and the coupled ($\S$, $\F$)-system. Then we solved the
coupled ($\S$, $\G$)-system for $\F\equiv 1$ and finally we computed the
solution to the full ($\S$, $\F$, $\G$)-system. Where the calculations had
been performed previously we found very good agreement with these results,
supporting our claim to have implemented a highly accurate method of
solution.  Furthermore, it is the first time that values for the critical
coupling are produced in a system where the self-energy corrections to the
wavefunction renormalization $\F$ are taken into account in a consistent
way.

Since the previous calculations have all been performed in the bare vertex
approximation, we decided it was important to explore the influence of
improved vertices on the dynamical generation of fermion mass in unquenched
QED. A first, simple extension to the bare vertex consisted to include a
$1/\F$-fermion wavefunction dependence in the vertex. We observed chiral
symmetry breaking and computed the critical coupling for one and two
flavours. Then the Ball-Chiu vertex was implemented as this is the correct,
non-perturbative, longitudinal part of the QED vertex, uniquely determined
by requiring the satisfaction of the Ward-Takahashi identity. Although
formally this vertex ensures that the quadratic divergences in the vacuum
polarization integral vanish, the precise form of the Ball-Chiu vertex is a
source for many numerical accuracy problems affecting the correct numerical
cancellation of the quadratic divergence. In a detailed numerical
investigation of the kernels of the angular integrals we were able to
locate the various sources of inaccuracies in successive steps and to
recover the accuracy necessary to cancel the quadratic divergence
correctly. Then fermion mass generation was found and the critical coupling
was determined. For the Curtis-Pennington vertex we could not find
solutions to the integral equations because the transverse part of the
vertex intrinsically leads to a quadratic divergence in the vacuum
polarization.  Therefore we considered a hybrid method where the
Curtis-Pennington vertex is used in the fermion equations and the Ball-Chiu
vertex is used on the photon equation to avoid unphysical quadratic
divergences. We found dynamical fermion mass generation and determined the
value of the critical coupling.


We now give some considerations about future studies. An extension to our
investigation would be to implement more sophisticated vertices, which not
only satisfy the Ward-Takahashi identities but, also ensure the
multiplicative renormalizability of the fermion and photon propagators.
This could be done by merging the work performed by
A.~K{\i}z{\i}lers\"u~\cite{Ayse} on the QED-vertex with the numerical
program developed here.  This is especially wanted if we are to study the
fermion mass generation with the renormalized Schwinger-Dyson equations, as
these are derived by making explicit use of the multiplicative
renormalizability of the fermion and photon propagators.

An important improvement to the numerical program would be to implement
Newton's iteration method to solve the complete system of non-linear
equations for $\S$, $\F$ and $\G$. As we explained before, limitations on
memory and computing time forced us to implement a hybrid method, using
Newton's iteration method on the coupled ($\S$, $\F$)-system, while the much
slower, natural iterative procedure is used to couple this system to the
$\G$-equation. The use of a single Newton's method on the whole system would
surely improve the convergence rate and the accuracy of the solutions,
especially if we use more complicated vertices.

All the present results tend to show that the generation of fermion mass
sets in, if the QED coupling is sufficiently strong. Although QED with a
strong coupling seems fictitious, the experimental situation in heavy ion
collisions, where the normal weak coupling QED is submerged in a very
strong, rapidly varying background field, could approximate it quite
well. An extensive numerical investigation of this situation could teach us
more about the possibility of a new phase transition which might already
have been discovered experimentally~\cite{Sch89,Cal89}.  For this purpose
we could start with the same computer program developed here.  The major
task would be to reformulate the Schwinger-Dyson formalism in the presence
of realistic background fields. Interestingly, Gusynin et al.\ have recently
shown in an analytic study~\cite{Gus95} that chiral symmetry is
spontaneously broken by a constant magnetic field in QED.

Furthermore, there is no reason to restrict the use of the numerical
method to the study of QED. For instance, we could add a four-fermion
interaction in the original QED Lagrangian. This is motivated by quenched
QED, where the four-fermion operator becomes renormalizable because of its
large anomalous dimension and should therefore be included in the
theory~\cite{Bar86}. It could also be used to study the propagators in
non-Abelian theories as QCD. There we could, for instance, study the
interplay between the infrared behaviour of the QCD-coupling and the
dynamical generated fermion mass to enhance our understanding of
confinement.


Finally, we conclude from our study of the coupled Schwinger-Dyson
equations for the fermion and photon propagator that fermion mass is
generated dynamically in quenched QED and in unquenched QED with one or two
flavours, provided the coupling is larger than a critical value which
depends on the approximations introduced. We developed a powerful, very
accurate numerical method which avoids the many pitfalls encountered in
solving the complicated system of non-linear coupled integral equations
describing the mass generation. We are convinced that our numerical
method will be very valuable for future investigations.

\begin{appendix}
\chapter{Angular integrals}
\label{App:angint}

In this Appendix we derive the following angular integrals:
\ba
\fbox{\parbox{10cm}{\[
\int_0^\pi d\theta \; \frac{\sin^2\theta}{z} = 
\frac{\pi}{2}\l[\frac{\theta(x-y)}{x}+\frac{\theta(y-x)}{y}\r] \]}}
\mlab{A1} \\
\fbox{\parbox{10cm}{\[
\int_0^\pi d\theta\, \frac{\sin^2\theta}{z^2}
= \frac{\pi}{2}
\l[\frac{\theta(x-y)}{x(x-y)} + \frac{\theta(y-x)}{y(y-x)} \r] \]}}
\mlab{A10} \\
\fbox{\parbox{10cm}{\[
\int_0^\pi d\theta \; \frac{\sin^4\theta}{z^2} = 
\frac{3\pi}{8}\l[\frac{\theta(x-y)}{x^2}+\frac{\theta(y-x)}{y^2}\r] \]}}
\mlab{A2} \\
\fbox{\parbox{10cm}{\[
\int_0^\pi d\theta \; \frac{\sin^2\theta \cos\theta}{z} = \frac{\pi}{4\sqrt{yx}}
\l[\frac{y}{x}\theta(x-y)+\frac{x}{y}\theta(y-x)\r] \]}}
\mlab{A3} \\
\fbox{\parbox{10cm}{\[
\int_0^\pi d\theta \; \frac{\sin^2\theta \cos\theta}{z^2} 
= \frac{\pi}{2\sqrt{xy}} \l[\frac{y}{x} 
\frac{\theta(x-y)}{x-y} + \frac{x}{y}\frac{\theta(y-x)}{y-x} \r]\;. \]  }}
\mlab{A11}
\ea

To evaluate these angular integrals we will make use of the integral,
Eq.~(3.665.2) of Ref.~\cite{Grads}:
\be
\int_0^\pi d\theta\, \frac{\sin^{2r}\theta}{(1+2a\cos\theta+a^2)^n} =
\Beta\left(r+\frac{1}{2},\frac{1}{2}\right) 
F\left(n,n-r;r+1;a^2\right) \, ,
\mlab{3.27}
\ee
where $\mbox{Re}(r)>-1/2$, $|a|<1$.

The beta function can be written as:
\be
\Beta(m,n) = \frac{\Gamma(m)\Gamma(n)}{\Gamma(m+n)} \, ,
\mlab{3.28}
\ee

while the hypergeometric function $F(\alpha,\beta;\gamma;z)$ is given
by Eq.~(9.100) of Ref.~\cite{Grads}:
\ba
F(\alpha,\beta;\gamma;z) &=& 1 + \frac{\alpha\beta}{\gamma.1} z
+ \frac{\alpha(\alpha+1)\beta(\beta+1)}{\gamma(\gamma+1).1.2} z^2
+ \frac{\alpha(\alpha+1)(\alpha+2)\beta(\beta+1)(\beta+2)}
{\gamma(\gamma+1)(\gamma+2).1.2.3} z^3 + \cdots \nn \\
&=& \frac{\Gamma(\gamma)}{\Gamma(\alpha)\Gamma(\beta)}
\sum_{n=0}^\infty \frac{\Gamma(\alpha+n)\Gamma(\beta+n)}{\Gamma(\gamma+n)\,n!}
z^n \, .
\mlab{3.29}
\ea

If either $\alpha$ or $\beta$ is negative the series terminates after
a finite number of terms.

We compute the following integral:
\ba
\int_0^\pi d\theta \; \frac{\sin^2\theta}{z} &=& 
\int_0^\pi d\theta \; \frac{\sin^2\theta}{x+y-2\sqrt{xy}\cos\theta} \nn\\
&=& \frac{\theta(x-y)}{x} \int_0^\pi d\theta \; 
\frac{\sin^2\theta}{1-2\sqrt{y/x}\cos\theta+y/x}
+ ( y \leftrightarrow x ) \;.
\mlab{3.29.0}
\ea

Applying \mref{3.27} with $r=1$ and $n=1$ to \mref{3.29.0} yields:
\be
\int_0^\pi d\theta\, \frac{\sin^2\theta}{z}
= \Beta\l(\frac{3}{2},\frac{1}{2}\r)
\l[\frac{\theta(x-y)}{x} F\l(1,0;2;\frac{y}{x}\r)
+ ( y \leftrightarrow x ) \r] \, . 
\mlab{3.31}
\ee

Note from \mref{3.28}:
\be
\Beta\l(\frac{3}{2},\frac{1}{2}\r) = 
\frac{\Gamma\l(\frac{3}{2}\r)\Gamma\l(\frac{1}{2}\r)}{\Gamma(2)}
= \frac{\pi}{2} \,
\mlab{3.32} 
\ee
because $\Gamma\l(\frac{1}{2}\r) = \sqrt\pi$. From \mref{3.29} we find:
\be
F\l(1,0;2;z\r) = 1 \, .
\mlab{3.33}
\ee

Substituting \mrefb{3.32}{3.33} in \mref{3.31} finally yields \mref{A1}:
\[
\int_0^\pi d\theta \; \frac{\sin^2\theta}{z} = 
\frac{\pi}{2}\l[\frac{\theta(x-y)}{x}+\frac{\theta(y-x)}{y}\r] \;.
\]

Next we compute:
\ba
\int_0^\pi d\theta \; \frac{\sin^2\theta}{z^2} &=& 
\int_0^\pi d\theta \; \frac{\sin^2\theta}{(x+y-2\sqrt{xy}\cos\theta)^2} \nn\\
&=& \frac{\theta(x-y)}{x^2} \int_0^\pi d\theta \; 
\frac{\sin^2\theta}{(1-2\sqrt{y/x}\cos\theta+y/x)^2}
+ ( y \leftrightarrow x ) \;. \mlab{3.29.1}
\ea

We now apply \mref{3.27} with $r=1$ and $n=2$ to \mref{3.29.1}. This gives:
\be
\int_0^\pi d\theta\, \frac{\sin^2\theta}{z^2}
= \Beta\l(\frac{3}{2},\frac{1}{2}\r)
\l[\frac{\theta(x-y)}{x^2} F\l(2,1;2;\frac{y}{x}\r)
+ ( y \leftrightarrow x ) \r] \, . 
\mlab{A7}
\ee

From \mref{3.29} we compute:
\be
F\l(2,1;2;z\r) = 1 + z + z^2 + z^3 + \cdots = \frac{1}{1-z}.
\mlab{A8} 
\ee

Substituting \mrefb{3.32}{A8} in \mref{A7} yields \mref{A10}:
\[
\int_0^\pi d\theta\, \frac{\sin^2\theta}{z^2}
= \frac{\pi}{2}
\l[\frac{\theta(x-y)}{x(x-y)} + \frac{\theta(y-x)}{y(y-x)} \r] \;.
\]

Next we compute:
\ba
\int_0^\pi d\theta \; \frac{\sin^4\theta}{z^2} 
&=& \int_0^\pi d\theta \; \frac{\sin^4\theta}{(x+y-2\sqrt{xy}\cos\theta)^2} \nn\\
&=& \frac{\theta(x-y)}{x^2}
\int_0^\pi d\theta \; \frac{\sin^4\theta}{(1-2\sqrt{y/x}\cos\theta+y/x)^2}
+ ( y \leftrightarrow x ) \;. \mlab{3.29.2}
\ea

Applying \mref{3.27} with $r=2$ and $n=2$ yields:
\be
\int_0^\pi d\theta \; \frac{\sin^4\theta}{z^2} 
= B\l(\frac{5}{2},\frac{1}{2}\r)
\l[F\l(2,0;3;\frac{y}{x}\r)\frac{\theta(x-y)}{x^2}
+ ( y \leftrightarrow x ) \r] \;.
\mlab{1.1011}
\ee

We know that:
\be
B\l(\frac{5}{2},\frac{1}{2}\r) =
\frac{\Gamma(\frac{5}{2})\Gamma(\frac{1}{2})}{\Gamma(3)} = \frac{3\pi}{8}
\mlab{1.1012}
\ee

and
\be
F\l(2,0;3;z\r) = 1 ,
\ee

such that \mref{1.1011} becomes \mref{A2}:
\[
\int_0^\pi d\theta \; \frac{\sin^4\theta}{z^2} = 
\frac{3\pi}{8}\l[\frac{\theta(x-y)}{x^2}+\frac{\theta(y-x)}{y^2}\r] \;.
\]

We now compute:
\ba
\int_0^\pi d\theta \; \frac{\sin^2\theta \cos\theta}{z} 
&=& \frac{1}{2\sqrt{xy}} \int_0^\pi d\theta \; 
\frac{\sin^2\theta (x+y-z)}{z} \nn\\
&=& \frac{1}{2\sqrt{xy}} \l[(x+y) \int_0^\pi d\theta \; 
\frac{\sin^2\theta}{z} - \int_0^\pi d\theta \; \sin^2\theta \r]
\mlab{A4}
\ea
where we used $z = x+y-2\sqrt{xy}\cos\theta$.

From Eq.~(14.347) of Ref.~\cite{Spiegel} we know that:
\be
\int_0^\pi d\theta \; \sin^2\theta = \frac{\pi}{2} .
\mlab{A5}
\ee

Substituting \mrefb{A1}{A5} in \mref{A4} yields:
\be
\int_0^\pi d\theta \; \frac{\sin^2\theta \cos\theta}{z} 
= \frac{\pi}{4\sqrt{xy}} \l\{
(x+y) \l[\frac{\theta(x-y)}{x}+\frac{\theta(y-x)}{y}\r]
- 1 \r\} ,
\ee
which proves \mref{A3}:
\[
\int_0^\pi d\theta \; \frac{\sin^2\theta \cos\theta}{z} = \frac{\pi}{4\sqrt{yx}}
\l[\frac{y}{x}\theta(x-y)+\frac{x}{y}\theta(y-x)\r] \;.
\]

In an analogous way we compute:
\ba
\int_0^\pi d\theta \; \frac{\sin^2\theta \cos\theta}{z^2} 
&=& \frac{1}{2\sqrt{xy}} \int_0^\pi d\theta \; 
\frac{\sin^2\theta (x+y-z)}{z^2} \nn\\
&=& \frac{1}{2\sqrt{xy}} \l[(x+y) \int_0^\pi d\theta \; 
\frac{\sin^2\theta}{z^2} - \int_0^\pi d\theta \; \frac{\sin^2\theta}{z} \r] \;.
\mlab{A6}
\ea

We substitute \mrefb{A10}{A1} in \mref{A6}, proving \mref{A11}:
\[
\int_0^\pi d\theta \; \frac{\sin^2\theta \cos\theta}{z^2} 
= \frac{\pi}{2\sqrt{xy}} \l[\frac{y}{x} 
\frac{\theta(x-y)}{x-y} + \frac{x}{y}\frac{\theta(y-x)}{y-x} \r] \;.
\]

\end{appendix}

\addcontentsline{toc}{chapter}{Bibliography}

\raggedright
\bibliographystyle{unsrt}
\bibliography{Biblio}

\end{document}